\documentclass[10pt,journal,letterpaper]{IEEEtran}
\usepackage[letterpaper,margin=1in]{geometry}

\usepackage[T1]{fontenc}
\pdfglyphtounicode{A}{0041}
\pdfglyphtounicode{AE}{00C6}
\pdfglyphtounicode{AEacute}{01FC}
\pdfglyphtounicode{AEmacron}{01E2}
\pdfglyphtounicode{AEsmall}{00E6}
\pdfglyphtounicode{Aacute}{00C1}
\pdfglyphtounicode{Aacutesmall}{00E1}
\pdfglyphtounicode{Abreve}{0102}
\pdfglyphtounicode{Abreveacute}{1EAE}
\pdfglyphtounicode{Abrevecyrillic}{04D0}
\pdfglyphtounicode{Abrevedotbelow}{1EB6}
\pdfglyphtounicode{Abrevegrave}{1EB0}
\pdfglyphtounicode{Abrevehookabove}{1EB2}
\pdfglyphtounicode{Abrevetilde}{1EB4}
\pdfglyphtounicode{Acaron}{01CD}
\pdfglyphtounicode{Acircle}{24B6}
\pdfglyphtounicode{Acircumflex}{00C2}
\pdfglyphtounicode{Acircumflexacute}{1EA4}
\pdfglyphtounicode{Acircumflexdotbelow}{1EAC}
\pdfglyphtounicode{Acircumflexgrave}{1EA6}
\pdfglyphtounicode{Acircumflexhookabove}{1EA8}
\pdfglyphtounicode{Acircumflexsmall}{00E2}
\pdfglyphtounicode{Acircumflextilde}{1EAA}
\pdfglyphtounicode{Acute}{00B4}
\pdfglyphtounicode{Acutesmall}{00B4}
\pdfglyphtounicode{Acyrillic}{0410}
\pdfglyphtounicode{Adblgrave}{0200}
\pdfglyphtounicode{Adieresis}{00C4}
\pdfglyphtounicode{Adieresiscyrillic}{04D2}
\pdfglyphtounicode{Adieresismacron}{01DE}
\pdfglyphtounicode{Adieresissmall}{00E4}
\pdfglyphtounicode{Adotbelow}{1EA0}
\pdfglyphtounicode{Adotmacron}{01E0}
\pdfglyphtounicode{Agrave}{00C0}
\pdfglyphtounicode{Agravesmall}{00E0}
\pdfglyphtounicode{Ahookabove}{1EA2}
\pdfglyphtounicode{Aiecyrillic}{04D4}
\pdfglyphtounicode{Ainvertedbreve}{0202}
\pdfglyphtounicode{Alpha}{0391}
\pdfglyphtounicode{Alphatonos}{0386}
\pdfglyphtounicode{Amacron}{0100}
\pdfglyphtounicode{Amonospace}{FF21}
\pdfglyphtounicode{Aogonek}{0104}
\pdfglyphtounicode{Aring}{00C5}
\pdfglyphtounicode{Aringacute}{01FA}
\pdfglyphtounicode{Aringbelow}{1E00}
\pdfglyphtounicode{Aringsmall}{00E5}
\pdfglyphtounicode{Asmall}{0061}
\pdfglyphtounicode{Atilde}{00C3}
\pdfglyphtounicode{Atildesmall}{00E3}
\pdfglyphtounicode{Aybarmenian}{0531}
\pdfglyphtounicode{B}{0042}
\pdfglyphtounicode{Bcircle}{24B7}
\pdfglyphtounicode{Bdotaccent}{1E02}
\pdfglyphtounicode{Bdotbelow}{1E04}
\pdfglyphtounicode{Becyrillic}{0411}
\pdfglyphtounicode{Benarmenian}{0532}
\pdfglyphtounicode{Beta}{0392}
\pdfglyphtounicode{Bhook}{0181}
\pdfglyphtounicode{Blinebelow}{1E06}
\pdfglyphtounicode{Bmonospace}{FF22}
\pdfglyphtounicode{Brevesmall}{02D8}
\pdfglyphtounicode{Bsmall}{0062}
\pdfglyphtounicode{Btopbar}{0182}
\pdfglyphtounicode{C}{0043}
\pdfglyphtounicode{Caarmenian}{053E}
\pdfglyphtounicode{Cacute}{0106}
\pdfglyphtounicode{Caron}{02C7}
\pdfglyphtounicode{Caronsmall}{02C7}
\pdfglyphtounicode{Ccaron}{010C}
\pdfglyphtounicode{Ccedilla}{00C7}
\pdfglyphtounicode{Ccedillaacute}{1E08}
\pdfglyphtounicode{Ccedillasmall}{00E7}
\pdfglyphtounicode{Ccircle}{24B8}
\pdfglyphtounicode{Ccircumflex}{0108}
\pdfglyphtounicode{Cdot}{010A}
\pdfglyphtounicode{Cdotaccent}{010A}
\pdfglyphtounicode{Cedillasmall}{00B8}
\pdfglyphtounicode{Chaarmenian}{0549}
\pdfglyphtounicode{Cheabkhasiancyrillic}{04BC}
\pdfglyphtounicode{Checyrillic}{0427}
\pdfglyphtounicode{Chedescenderabkhasiancyrillic}{04BE}
\pdfglyphtounicode{Chedescendercyrillic}{04B6}
\pdfglyphtounicode{Chedieresiscyrillic}{04F4}
\pdfglyphtounicode{Cheharmenian}{0543}
\pdfglyphtounicode{Chekhakassiancyrillic}{04CB}
\pdfglyphtounicode{Cheverticalstrokecyrillic}{04B8}
\pdfglyphtounicode{Chi}{03A7}
\pdfglyphtounicode{Chook}{0187}
\pdfglyphtounicode{Circumflexsmall}{02C6}
\pdfglyphtounicode{Cmonospace}{FF23}
\pdfglyphtounicode{Coarmenian}{0551}
\pdfglyphtounicode{Csmall}{0063}
\pdfglyphtounicode{D}{0044}
\pdfglyphtounicode{DZ}{01F1}
\pdfglyphtounicode{DZcaron}{01C4}
\pdfglyphtounicode{Daarmenian}{0534}
\pdfglyphtounicode{Dafrican}{0189}
\pdfglyphtounicode{Dbar}{0110}
\pdfglyphtounicode{Dcaron}{010E}
\pdfglyphtounicode{Dcedilla}{1E10}
\pdfglyphtounicode{Dcircle}{24B9}
\pdfglyphtounicode{Dcircumflexbelow}{1E12}
\pdfglyphtounicode{Dcroat}{0110}
\pdfglyphtounicode{Ddotaccent}{1E0A}
\pdfglyphtounicode{Ddotbelow}{1E0C}
\pdfglyphtounicode{Decyrillic}{0414}
\pdfglyphtounicode{Deicoptic}{03EE}
\pdfglyphtounicode{Delta}{2206}
\pdfglyphtounicode{Deltagreek}{0394}
\pdfglyphtounicode{Dhook}{018A}
\pdfglyphtounicode{Dieresis}{00A8}
\pdfglyphtounicode{DieresisAcute}{F6CC}
\pdfglyphtounicode{DieresisGrave}{F6CD}
\pdfglyphtounicode{Dieresissmall}{00A8}
\pdfglyphtounicode{Digamma}{D875 DFCB}
\pdfglyphtounicode{Digammagreek}{03DC}
\pdfglyphtounicode{Djecyrillic}{0402}
\pdfglyphtounicode{Dlinebelow}{1E0E}
\pdfglyphtounicode{Dmonospace}{FF24}
\pdfglyphtounicode{Dotaccentsmall}{02D9}
\pdfglyphtounicode{Dslash}{0110}
\pdfglyphtounicode{Dsmall}{0064}
\pdfglyphtounicode{Dtopbar}{018B}
\pdfglyphtounicode{Dz}{01F2}
\pdfglyphtounicode{Dzcaron}{01C5}
\pdfglyphtounicode{Dzeabkhasiancyrillic}{04E0}
\pdfglyphtounicode{Dzecyrillic}{0405}
\pdfglyphtounicode{Dzhecyrillic}{040F}
\pdfglyphtounicode{E}{0045}
\pdfglyphtounicode{Eacute}{00C9}
\pdfglyphtounicode{Eacutesmall}{00E9}
\pdfglyphtounicode{Ebreve}{0114}
\pdfglyphtounicode{Ecaron}{011A}
\pdfglyphtounicode{Ecedillabreve}{1E1C}
\pdfglyphtounicode{Echarmenian}{0535}
\pdfglyphtounicode{Ecircle}{24BA}
\pdfglyphtounicode{Ecircumflex}{00CA}
\pdfglyphtounicode{Ecircumflexacute}{1EBE}
\pdfglyphtounicode{Ecircumflexbelow}{1E18}
\pdfglyphtounicode{Ecircumflexdotbelow}{1EC6}
\pdfglyphtounicode{Ecircumflexgrave}{1EC0}
\pdfglyphtounicode{Ecircumflexhookabove}{1EC2}
\pdfglyphtounicode{Ecircumflexsmall}{00EA}
\pdfglyphtounicode{Ecircumflextilde}{1EC4}
\pdfglyphtounicode{Ecyrillic}{0404}
\pdfglyphtounicode{Edblgrave}{0204}
\pdfglyphtounicode{Edieresis}{00CB}
\pdfglyphtounicode{Edieresissmall}{00EB}
\pdfglyphtounicode{Edot}{0116}
\pdfglyphtounicode{Edotaccent}{0116}
\pdfglyphtounicode{Edotbelow}{1EB8}
\pdfglyphtounicode{Efcyrillic}{0424}
\pdfglyphtounicode{Egrave}{00C8}
\pdfglyphtounicode{Egravesmall}{00E8}
\pdfglyphtounicode{Eharmenian}{0537}
\pdfglyphtounicode{Ehookabove}{1EBA}
\pdfglyphtounicode{Eightroman}{2167}
\pdfglyphtounicode{Einvertedbreve}{0206}
\pdfglyphtounicode{Eiotifiedcyrillic}{0464}
\pdfglyphtounicode{Elcyrillic}{041B}
\pdfglyphtounicode{Elevenroman}{216A}
\pdfglyphtounicode{Emacron}{0112}
\pdfglyphtounicode{Emacronacute}{1E16}
\pdfglyphtounicode{Emacrongrave}{1E14}
\pdfglyphtounicode{Emcyrillic}{041C}
\pdfglyphtounicode{Emonospace}{FF25}
\pdfglyphtounicode{Encyrillic}{041D}
\pdfglyphtounicode{Endescendercyrillic}{04A2}
\pdfglyphtounicode{Eng}{014A}
\pdfglyphtounicode{Enghecyrillic}{04A4}
\pdfglyphtounicode{Enhookcyrillic}{04C7}
\pdfglyphtounicode{Eogonek}{0118}
\pdfglyphtounicode{Eopen}{0190}
\pdfglyphtounicode{Epsilon}{0395}
\pdfglyphtounicode{Epsilontonos}{0388}
\pdfglyphtounicode{Ercyrillic}{0420}
\pdfglyphtounicode{Ereversed}{018E}
\pdfglyphtounicode{Ereversedcyrillic}{042D}
\pdfglyphtounicode{Escyrillic}{0421}
\pdfglyphtounicode{Esdescendercyrillic}{04AA}
\pdfglyphtounicode{Esh}{01A9}
\pdfglyphtounicode{Esmall}{0065}
\pdfglyphtounicode{Eta}{0397}
\pdfglyphtounicode{Etarmenian}{0538}
\pdfglyphtounicode{Etatonos}{0389}
\pdfglyphtounicode{Eth}{00D0}
\pdfglyphtounicode{Ethsmall}{00F0}
\pdfglyphtounicode{Etilde}{1EBC}
\pdfglyphtounicode{Etildebelow}{1E1A}
\pdfglyphtounicode{Euro}{20AC}
\pdfglyphtounicode{Ezh}{01B7}
\pdfglyphtounicode{Ezhcaron}{01EE}
\pdfglyphtounicode{Ezhreversed}{01B8}
\pdfglyphtounicode{F}{0046}
\pdfglyphtounicode{FFIsmall}{0066 0066 0069}
\pdfglyphtounicode{FFLsmall}{0066 0066 006C}
\pdfglyphtounicode{FFsmall}{0066 0066}
\pdfglyphtounicode{FIsmall}{0066 0069}
\pdfglyphtounicode{FLsmall}{0066 006C}
\pdfglyphtounicode{Fcircle}{24BB}
\pdfglyphtounicode{Fdotaccent}{1E1E}
\pdfglyphtounicode{Feharmenian}{0556}
\pdfglyphtounicode{Feicoptic}{03E4}
\pdfglyphtounicode{Fhook}{0191}
\pdfglyphtounicode{Finv}{2132}
\pdfglyphtounicode{Fitacyrillic}{0472}
\pdfglyphtounicode{Fiveroman}{2164}
\pdfglyphtounicode{Fmonospace}{FF26}
\pdfglyphtounicode{Fourroman}{2163}
\pdfglyphtounicode{Fsmall}{0066}
\pdfglyphtounicode{G}{0047}
\pdfglyphtounicode{GBsquare}{3387}
\pdfglyphtounicode{Gacute}{01F4}
\pdfglyphtounicode{Gamma}{0393}
\pdfglyphtounicode{Gammaafrican}{0194}
\pdfglyphtounicode{Gangiacoptic}{03EA}
\pdfglyphtounicode{Gbreve}{011E}
\pdfglyphtounicode{Gcaron}{01E6}
\pdfglyphtounicode{Gcedilla}{0122}
\pdfglyphtounicode{Gcircle}{24BC}
\pdfglyphtounicode{Gcircumflex}{011C}
\pdfglyphtounicode{Gcommaaccent}{0122}
\pdfglyphtounicode{Gdot}{0120}
\pdfglyphtounicode{Gdotaccent}{0120}
\pdfglyphtounicode{Gecyrillic}{0413}
\pdfglyphtounicode{Germandbls}{0053 0053}
\pdfglyphtounicode{Germandblssmall}{0073 0073}
\pdfglyphtounicode{Ghadarmenian}{0542}
\pdfglyphtounicode{Ghemiddlehookcyrillic}{0494}
\pdfglyphtounicode{Ghestrokecyrillic}{0492}
\pdfglyphtounicode{Gheupturncyrillic}{0490}
\pdfglyphtounicode{Ghook}{0193}
\pdfglyphtounicode{Gimarmenian}{0533}
\pdfglyphtounicode{Gjecyrillic}{0403}
\pdfglyphtounicode{Gmacron}{1E20}
\pdfglyphtounicode{Gmir}{2141}
\pdfglyphtounicode{Gmonospace}{FF27}
\pdfglyphtounicode{Grave}{0060}
\pdfglyphtounicode{Gravesmall}{0060}
\pdfglyphtounicode{Gsmall}{0067}
\pdfglyphtounicode{Gsmallhook}{029B}
\pdfglyphtounicode{Gstroke}{01E4}
\pdfglyphtounicode{H}{0048}
\pdfglyphtounicode{H18533}{25CF}
\pdfglyphtounicode{H18543}{25AA}
\pdfglyphtounicode{H18551}{25AB}
\pdfglyphtounicode{H22073}{25A1}
\pdfglyphtounicode{HPsquare}{33CB}
\pdfglyphtounicode{Haabkhasiancyrillic}{04A8}
\pdfglyphtounicode{Hadescendercyrillic}{04B2}
\pdfglyphtounicode{Hardsigncyrillic}{042A}
\pdfglyphtounicode{Hbar}{0126}
\pdfglyphtounicode{Hbrevebelow}{1E2A}
\pdfglyphtounicode{Hcedilla}{1E28}
\pdfglyphtounicode{Hcircle}{24BD}
\pdfglyphtounicode{Hcircumflex}{0124}
\pdfglyphtounicode{Hdieresis}{1E26}
\pdfglyphtounicode{Hdotaccent}{1E22}
\pdfglyphtounicode{Hdotbelow}{1E24}
\pdfglyphtounicode{Hmonospace}{FF28}
\pdfglyphtounicode{Hoarmenian}{0540}
\pdfglyphtounicode{Horicoptic}{03E8}
\pdfglyphtounicode{Hsmall}{0068}
\pdfglyphtounicode{Hungarumlaut}{02DD}
\pdfglyphtounicode{Hungarumlautsmall}{02DD}
\pdfglyphtounicode{Hzsquare}{3390}
\pdfglyphtounicode{I}{0049}
\pdfglyphtounicode{IAcyrillic}{042F}
\pdfglyphtounicode{IJ}{0132}
\pdfglyphtounicode{IUcyrillic}{042E}
\pdfglyphtounicode{Iacute}{00CD}
\pdfglyphtounicode{Iacutesmall}{00ED}
\pdfglyphtounicode{Ibreve}{012C}
\pdfglyphtounicode{Icaron}{01CF}
\pdfglyphtounicode{Icircle}{24BE}
\pdfglyphtounicode{Icircumflex}{00CE}
\pdfglyphtounicode{Icircumflexsmall}{00EE}
\pdfglyphtounicode{Icyrillic}{0406}
\pdfglyphtounicode{Idblgrave}{0208}
\pdfglyphtounicode{Idieresis}{00CF}
\pdfglyphtounicode{Idieresisacute}{1E2E}
\pdfglyphtounicode{Idieresiscyrillic}{04E4}
\pdfglyphtounicode{Idieresissmall}{00EF}
\pdfglyphtounicode{Idot}{0130}
\pdfglyphtounicode{Idotaccent}{0130}
\pdfglyphtounicode{Idotbelow}{1ECA}
\pdfglyphtounicode{Iebrevecyrillic}{04D6}
\pdfglyphtounicode{Iecyrillic}{0415}
\pdfglyphtounicode{Ifractur}{2111}
\pdfglyphtounicode{Ifraktur}{2111}
\pdfglyphtounicode{Igrave}{00CC}
\pdfglyphtounicode{Igravesmall}{00EC}
\pdfglyphtounicode{Ihookabove}{1EC8}
\pdfglyphtounicode{Iicyrillic}{0418}
\pdfglyphtounicode{Iinvertedbreve}{020A}
\pdfglyphtounicode{Iishortcyrillic}{0419}
\pdfglyphtounicode{Imacron}{012A}
\pdfglyphtounicode{Imacroncyrillic}{04E2}
\pdfglyphtounicode{Imonospace}{FF29}
\pdfglyphtounicode{Iniarmenian}{053B}
\pdfglyphtounicode{Iocyrillic}{0401}
\pdfglyphtounicode{Iogonek}{012E}
\pdfglyphtounicode{Iota}{0399}
\pdfglyphtounicode{Iotaafrican}{0196}
\pdfglyphtounicode{Iotadieresis}{03AA}
\pdfglyphtounicode{Iotatonos}{038A}
\pdfglyphtounicode{Ismall}{0069}
\pdfglyphtounicode{Istroke}{0197}
\pdfglyphtounicode{Itilde}{0128}
\pdfglyphtounicode{Itildebelow}{1E2C}
\pdfglyphtounicode{Izhitsacyrillic}{0474}
\pdfglyphtounicode{Izhitsadblgravecyrillic}{0476}
\pdfglyphtounicode{J}{004A}
\pdfglyphtounicode{Jaarmenian}{0541}
\pdfglyphtounicode{Jcircle}{24BF}
\pdfglyphtounicode{Jcircumflex}{0134}
\pdfglyphtounicode{Jecyrillic}{0408}
\pdfglyphtounicode{Jheharmenian}{054B}
\pdfglyphtounicode{Jmonospace}{FF2A}
\pdfglyphtounicode{Jsmall}{006A}
\pdfglyphtounicode{K}{004B}
\pdfglyphtounicode{KBsquare}{3385}
\pdfglyphtounicode{KKsquare}{33CD}
\pdfglyphtounicode{Kabashkircyrillic}{04A0}
\pdfglyphtounicode{Kacute}{1E30}
\pdfglyphtounicode{Kacyrillic}{041A}
\pdfglyphtounicode{Kadescendercyrillic}{049A}
\pdfglyphtounicode{Kahookcyrillic}{04C3}
\pdfglyphtounicode{Kappa}{039A}
\pdfglyphtounicode{Kastrokecyrillic}{049E}
\pdfglyphtounicode{Kaverticalstrokecyrillic}{049C}
\pdfglyphtounicode{Kcaron}{01E8}
\pdfglyphtounicode{Kcedilla}{0136}
\pdfglyphtounicode{Kcircle}{24C0}
\pdfglyphtounicode{Kcommaaccent}{0136}
\pdfglyphtounicode{Kdotbelow}{1E32}
\pdfglyphtounicode{Keharmenian}{0554}
\pdfglyphtounicode{Kenarmenian}{053F}
\pdfglyphtounicode{Khacyrillic}{0425}
\pdfglyphtounicode{Kheicoptic}{03E6}
\pdfglyphtounicode{Khook}{0198}
\pdfglyphtounicode{Kjecyrillic}{040C}
\pdfglyphtounicode{Klinebelow}{1E34}
\pdfglyphtounicode{Kmonospace}{FF2B}
\pdfglyphtounicode{Koppacyrillic}{0480}
\pdfglyphtounicode{Koppagreek}{03DE}
\pdfglyphtounicode{Ksicyrillic}{046E}
\pdfglyphtounicode{Ksmall}{006B}
\pdfglyphtounicode{L}{004C}
\pdfglyphtounicode{LJ}{01C7}
\pdfglyphtounicode{LL}{004C 004C}
\pdfglyphtounicode{Lacute}{0139}
\pdfglyphtounicode{Lambda}{039B}
\pdfglyphtounicode{Lcaron}{013D}
\pdfglyphtounicode{Lcedilla}{013B}
\pdfglyphtounicode{Lcircle}{24C1}
\pdfglyphtounicode{Lcircumflexbelow}{1E3C}
\pdfglyphtounicode{Lcommaaccent}{013B}
\pdfglyphtounicode{Ldot}{013F}
\pdfglyphtounicode{Ldotaccent}{013F}
\pdfglyphtounicode{Ldotbelow}{1E36}
\pdfglyphtounicode{Ldotbelowmacron}{1E38}
\pdfglyphtounicode{Liwnarmenian}{053C}
\pdfglyphtounicode{Lj}{01C8}
\pdfglyphtounicode{Ljecyrillic}{0409}
\pdfglyphtounicode{Llinebelow}{1E3A}
\pdfglyphtounicode{Lmonospace}{FF2C}
\pdfglyphtounicode{Lslash}{0141}
\pdfglyphtounicode{Lslashsmall}{0142}
\pdfglyphtounicode{Lsmall}{006C}
\pdfglyphtounicode{M}{004D}
\pdfglyphtounicode{MBsquare}{3386}
\pdfglyphtounicode{Macron}{00AF}
\pdfglyphtounicode{Macronsmall}{00AF}
\pdfglyphtounicode{Macute}{1E3E}
\pdfglyphtounicode{Mcircle}{24C2}
\pdfglyphtounicode{Mdotaccent}{1E40}
\pdfglyphtounicode{Mdotbelow}{1E42}
\pdfglyphtounicode{Menarmenian}{0544}
\pdfglyphtounicode{Mmonospace}{FF2D}
\pdfglyphtounicode{Msmall}{006D}
\pdfglyphtounicode{Mturned}{019C}
\pdfglyphtounicode{Mu}{039C}
\pdfglyphtounicode{N}{004E}
\pdfglyphtounicode{NJ}{01CA}
\pdfglyphtounicode{Nacute}{0143}
\pdfglyphtounicode{Ncaron}{0147}
\pdfglyphtounicode{Ncedilla}{0145}
\pdfglyphtounicode{Ncircle}{24C3}
\pdfglyphtounicode{Ncircumflexbelow}{1E4A}
\pdfglyphtounicode{Ncommaaccent}{0145}
\pdfglyphtounicode{Ndotaccent}{1E44}
\pdfglyphtounicode{Ndotbelow}{1E46}
\pdfglyphtounicode{Ng}{014A}
\pdfglyphtounicode{Nhookleft}{019D}
\pdfglyphtounicode{Nineroman}{2168}
\pdfglyphtounicode{Nj}{01CB}
\pdfglyphtounicode{Njecyrillic}{040A}
\pdfglyphtounicode{Nlinebelow}{1E48}
\pdfglyphtounicode{Nmonospace}{FF2E}
\pdfglyphtounicode{Nowarmenian}{0546}
\pdfglyphtounicode{Nsmall}{006E}
\pdfglyphtounicode{Ntilde}{00D1}
\pdfglyphtounicode{Ntildesmall}{00F1}
\pdfglyphtounicode{Nu}{039D}
\pdfglyphtounicode{O}{004F}
\pdfglyphtounicode{OE}{0152}
\pdfglyphtounicode{OEsmall}{0153}
\pdfglyphtounicode{Oacute}{00D3}
\pdfglyphtounicode{Oacutesmall}{00F3}
\pdfglyphtounicode{Obarredcyrillic}{04E8}
\pdfglyphtounicode{Obarreddieresiscyrillic}{04EA}
\pdfglyphtounicode{Obreve}{014E}
\pdfglyphtounicode{Ocaron}{01D1}
\pdfglyphtounicode{Ocenteredtilde}{019F}
\pdfglyphtounicode{Ocircle}{24C4}
\pdfglyphtounicode{Ocircumflex}{00D4}
\pdfglyphtounicode{Ocircumflexacute}{1ED0}
\pdfglyphtounicode{Ocircumflexdotbelow}{1ED8}
\pdfglyphtounicode{Ocircumflexgrave}{1ED2}
\pdfglyphtounicode{Ocircumflexhookabove}{1ED4}
\pdfglyphtounicode{Ocircumflexsmall}{00F4}
\pdfglyphtounicode{Ocircumflextilde}{1ED6}
\pdfglyphtounicode{Ocyrillic}{041E}
\pdfglyphtounicode{Odblacute}{0150}
\pdfglyphtounicode{Odblgrave}{020C}
\pdfglyphtounicode{Odieresis}{00D6}
\pdfglyphtounicode{Odieresiscyrillic}{04E6}
\pdfglyphtounicode{Odieresissmall}{00F6}
\pdfglyphtounicode{Odotbelow}{1ECC}
\pdfglyphtounicode{Ogoneksmall}{02DB}
\pdfglyphtounicode{Ograve}{00D2}
\pdfglyphtounicode{Ogravesmall}{00F2}
\pdfglyphtounicode{Oharmenian}{0555}
\pdfglyphtounicode{Ohm}{2126}
\pdfglyphtounicode{Ohookabove}{1ECE}
\pdfglyphtounicode{Ohorn}{01A0}
\pdfglyphtounicode{Ohornacute}{1EDA}
\pdfglyphtounicode{Ohorndotbelow}{1EE2}
\pdfglyphtounicode{Ohorngrave}{1EDC}
\pdfglyphtounicode{Ohornhookabove}{1EDE}
\pdfglyphtounicode{Ohorntilde}{1EE0}
\pdfglyphtounicode{Ohungarumlaut}{0150}
\pdfglyphtounicode{Oi}{01A2}
\pdfglyphtounicode{Oinvertedbreve}{020E}
\pdfglyphtounicode{Omacron}{014C}
\pdfglyphtounicode{Omacronacute}{1E52}
\pdfglyphtounicode{Omacrongrave}{1E50}
\pdfglyphtounicode{Omega}{2126}
\pdfglyphtounicode{Omegacyrillic}{0460}
\pdfglyphtounicode{Omegagreek}{03A9}
\pdfglyphtounicode{Omegainv}{2127}
\pdfglyphtounicode{Omegaroundcyrillic}{047A}
\pdfglyphtounicode{Omegatitlocyrillic}{047C}
\pdfglyphtounicode{Omegatonos}{038F}
\pdfglyphtounicode{Omicron}{039F}
\pdfglyphtounicode{Omicrontonos}{038C}
\pdfglyphtounicode{Omonospace}{FF2F}
\pdfglyphtounicode{Oneroman}{2160}
\pdfglyphtounicode{Oogonek}{01EA}
\pdfglyphtounicode{Oogonekmacron}{01EC}
\pdfglyphtounicode{Oopen}{0186}
\pdfglyphtounicode{Oslash}{00D8}
\pdfglyphtounicode{Oslashacute}{01FE}
\pdfglyphtounicode{Oslashsmall}{00F8}
\pdfglyphtounicode{Osmall}{006F}
\pdfglyphtounicode{Ostrokeacute}{01FE}
\pdfglyphtounicode{Otcyrillic}{047E}
\pdfglyphtounicode{Otilde}{00D5}
\pdfglyphtounicode{Otildeacute}{1E4C}
\pdfglyphtounicode{Otildedieresis}{1E4E}
\pdfglyphtounicode{Otildesmall}{00F5}
\pdfglyphtounicode{P}{0050}
\pdfglyphtounicode{Pacute}{1E54}
\pdfglyphtounicode{Pcircle}{24C5}
\pdfglyphtounicode{Pdotaccent}{1E56}
\pdfglyphtounicode{Pecyrillic}{041F}
\pdfglyphtounicode{Peharmenian}{054A}
\pdfglyphtounicode{Pemiddlehookcyrillic}{04A6}
\pdfglyphtounicode{Phi}{03A6}
\pdfglyphtounicode{Phook}{01A4}
\pdfglyphtounicode{Pi}{03A0}
\pdfglyphtounicode{Piwrarmenian}{0553}
\pdfglyphtounicode{Pmonospace}{FF30}
\pdfglyphtounicode{Psi}{03A8}
\pdfglyphtounicode{Psicyrillic}{0470}
\pdfglyphtounicode{Psmall}{0070}
\pdfglyphtounicode{Q}{0051}
\pdfglyphtounicode{Qcircle}{24C6}
\pdfglyphtounicode{Qmonospace}{FF31}
\pdfglyphtounicode{Qsmall}{0071}
\pdfglyphtounicode{R}{0052}
\pdfglyphtounicode{Raarmenian}{054C}
\pdfglyphtounicode{Racute}{0154}
\pdfglyphtounicode{Rcaron}{0158}
\pdfglyphtounicode{Rcedilla}{0156}
\pdfglyphtounicode{Rcircle}{24C7}
\pdfglyphtounicode{Rcommaaccent}{0156}
\pdfglyphtounicode{Rdblgrave}{0210}
\pdfglyphtounicode{Rdotaccent}{1E58}
\pdfglyphtounicode{Rdotbelow}{1E5A}
\pdfglyphtounicode{Rdotbelowmacron}{1E5C}
\pdfglyphtounicode{Reharmenian}{0550}
\pdfglyphtounicode{Rfractur}{211C}
\pdfglyphtounicode{Rfraktur}{211C}
\pdfglyphtounicode{Rho}{03A1}
\pdfglyphtounicode{Ringsmall}{02DA}
\pdfglyphtounicode{Rinvertedbreve}{0212}
\pdfglyphtounicode{Rlinebelow}{1E5E}
\pdfglyphtounicode{Rmonospace}{FF32}
\pdfglyphtounicode{Rsmall}{0072}
\pdfglyphtounicode{Rsmallinverted}{0281}
\pdfglyphtounicode{Rsmallinvertedsuperior}{02B6}
\pdfglyphtounicode{S}{0053}
\pdfglyphtounicode{SF010000}{250C}
\pdfglyphtounicode{SF020000}{2514}
\pdfglyphtounicode{SF030000}{2510}
\pdfglyphtounicode{SF040000}{2518}
\pdfglyphtounicode{SF050000}{253C}
\pdfglyphtounicode{SF060000}{252C}
\pdfglyphtounicode{SF070000}{2534}
\pdfglyphtounicode{SF080000}{251C}
\pdfglyphtounicode{SF090000}{2524}
\pdfglyphtounicode{SF100000}{2500}
\pdfglyphtounicode{SF110000}{2502}
\pdfglyphtounicode{SF190000}{2561}
\pdfglyphtounicode{SF200000}{2562}
\pdfglyphtounicode{SF210000}{2556}
\pdfglyphtounicode{SF220000}{2555}
\pdfglyphtounicode{SF230000}{2563}
\pdfglyphtounicode{SF240000}{2551}
\pdfglyphtounicode{SF250000}{2557}
\pdfglyphtounicode{SF260000}{255D}
\pdfglyphtounicode{SF270000}{255C}
\pdfglyphtounicode{SF280000}{255B}
\pdfglyphtounicode{SF360000}{255E}
\pdfglyphtounicode{SF370000}{255F}
\pdfglyphtounicode{SF380000}{255A}
\pdfglyphtounicode{SF390000}{2554}
\pdfglyphtounicode{SF400000}{2569}
\pdfglyphtounicode{SF410000}{2566}
\pdfglyphtounicode{SF420000}{2560}
\pdfglyphtounicode{SF430000}{2550}
\pdfglyphtounicode{SF440000}{256C}
\pdfglyphtounicode{SF450000}{2567}
\pdfglyphtounicode{SF460000}{2568}
\pdfglyphtounicode{SF470000}{2564}
\pdfglyphtounicode{SF480000}{2565}
\pdfglyphtounicode{SF490000}{2559}
\pdfglyphtounicode{SF500000}{2558}
\pdfglyphtounicode{SF510000}{2552}
\pdfglyphtounicode{SF520000}{2553}
\pdfglyphtounicode{SF530000}{256B}
\pdfglyphtounicode{SF540000}{256A}
\pdfglyphtounicode{SS}{0053 0053}
\pdfglyphtounicode{SSsmall}{0073 0073}
\pdfglyphtounicode{Sacute}{015A}
\pdfglyphtounicode{Sacutedotaccent}{1E64}
\pdfglyphtounicode{Sampigreek}{03E0}
\pdfglyphtounicode{Scaron}{0160}
\pdfglyphtounicode{Scarondotaccent}{1E66}
\pdfglyphtounicode{Scaronsmall}{0161}
\pdfglyphtounicode{Scedilla}{015E}
\pdfglyphtounicode{Schwa}{018F}
\pdfglyphtounicode{Schwacyrillic}{04D8}
\pdfglyphtounicode{Schwadieresiscyrillic}{04DA}
\pdfglyphtounicode{Scircle}{24C8}
\pdfglyphtounicode{Scircumflex}{015C}
\pdfglyphtounicode{Scommaaccent}{0218}
\pdfglyphtounicode{Sdotaccent}{1E60}
\pdfglyphtounicode{Sdotbelow}{1E62}
\pdfglyphtounicode{Sdotbelowdotaccent}{1E68}
\pdfglyphtounicode{Seharmenian}{054D}
\pdfglyphtounicode{Sevenroman}{2166}
\pdfglyphtounicode{Shaarmenian}{0547}
\pdfglyphtounicode{Shacyrillic}{0428}
\pdfglyphtounicode{Shchacyrillic}{0429}
\pdfglyphtounicode{Sheicoptic}{03E2}
\pdfglyphtounicode{Shhacyrillic}{04BA}
\pdfglyphtounicode{Shimacoptic}{03EC}
\pdfglyphtounicode{Sigma}{03A3}
\pdfglyphtounicode{Sixroman}{2165}
\pdfglyphtounicode{Smonospace}{FF33}
\pdfglyphtounicode{Softsigncyrillic}{042C}
\pdfglyphtounicode{Ssmall}{0073}
\pdfglyphtounicode{Stigmagreek}{03DA}
\pdfglyphtounicode{T}{0054}
\pdfglyphtounicode{Tau}{03A4}
\pdfglyphtounicode{Tbar}{0166}
\pdfglyphtounicode{Tcaron}{0164}
\pdfglyphtounicode{Tcedilla}{0162}
\pdfglyphtounicode{Tcircle}{24C9}
\pdfglyphtounicode{Tcircumflexbelow}{1E70}
\pdfglyphtounicode{Tcommaaccent}{0162}
\pdfglyphtounicode{Tdotaccent}{1E6A}
\pdfglyphtounicode{Tdotbelow}{1E6C}
\pdfglyphtounicode{Tecyrillic}{0422}
\pdfglyphtounicode{Tedescendercyrillic}{04AC}
\pdfglyphtounicode{Tenroman}{2169}
\pdfglyphtounicode{Tetsecyrillic}{04B4}
\pdfglyphtounicode{Theta}{0398}
\pdfglyphtounicode{Thook}{01AC}
\pdfglyphtounicode{Thorn}{00DE}
\pdfglyphtounicode{Thornsmall}{00FE}
\pdfglyphtounicode{Threeroman}{2162}
\pdfglyphtounicode{Tildesmall}{02DC}
\pdfglyphtounicode{Tiwnarmenian}{054F}
\pdfglyphtounicode{Tlinebelow}{1E6E}
\pdfglyphtounicode{Tmonospace}{FF34}
\pdfglyphtounicode{Toarmenian}{0539}
\pdfglyphtounicode{Tonefive}{01BC}
\pdfglyphtounicode{Tonesix}{0184}
\pdfglyphtounicode{Tonetwo}{01A7}
\pdfglyphtounicode{Tretroflexhook}{01AE}
\pdfglyphtounicode{Tsecyrillic}{0426}
\pdfglyphtounicode{Tshecyrillic}{040B}
\pdfglyphtounicode{Tsmall}{0074}
\pdfglyphtounicode{Twelveroman}{216B}
\pdfglyphtounicode{Tworoman}{2161}
\pdfglyphtounicode{U}{0055}
\pdfglyphtounicode{Uacute}{00DA}
\pdfglyphtounicode{Uacutesmall}{00FA}
\pdfglyphtounicode{Ubreve}{016C}
\pdfglyphtounicode{Ucaron}{01D3}
\pdfglyphtounicode{Ucircle}{24CA}
\pdfglyphtounicode{Ucircumflex}{00DB}
\pdfglyphtounicode{Ucircumflexbelow}{1E76}
\pdfglyphtounicode{Ucircumflexsmall}{00FB}
\pdfglyphtounicode{Ucyrillic}{0423}
\pdfglyphtounicode{Udblacute}{0170}
\pdfglyphtounicode{Udblgrave}{0214}
\pdfglyphtounicode{Udieresis}{00DC}
\pdfglyphtounicode{Udieresisacute}{01D7}
\pdfglyphtounicode{Udieresisbelow}{1E72}
\pdfglyphtounicode{Udieresiscaron}{01D9}
\pdfglyphtounicode{Udieresiscyrillic}{04F0}
\pdfglyphtounicode{Udieresisgrave}{01DB}
\pdfglyphtounicode{Udieresismacron}{01D5}
\pdfglyphtounicode{Udieresissmall}{00FC}
\pdfglyphtounicode{Udotbelow}{1EE4}
\pdfglyphtounicode{Ugrave}{00D9}
\pdfglyphtounicode{Ugravesmall}{00F9}
\pdfglyphtounicode{Uhookabove}{1EE6}
\pdfglyphtounicode{Uhorn}{01AF}
\pdfglyphtounicode{Uhornacute}{1EE8}
\pdfglyphtounicode{Uhorndotbelow}{1EF0}
\pdfglyphtounicode{Uhorngrave}{1EEA}
\pdfglyphtounicode{Uhornhookabove}{1EEC}
\pdfglyphtounicode{Uhorntilde}{1EEE}
\pdfglyphtounicode{Uhungarumlaut}{0170}
\pdfglyphtounicode{Uhungarumlautcyrillic}{04F2}
\pdfglyphtounicode{Uinvertedbreve}{0216}
\pdfglyphtounicode{Ukcyrillic}{0478}
\pdfglyphtounicode{Umacron}{016A}
\pdfglyphtounicode{Umacroncyrillic}{04EE}
\pdfglyphtounicode{Umacrondieresis}{1E7A}
\pdfglyphtounicode{Umonospace}{FF35}
\pdfglyphtounicode{Uogonek}{0172}
\pdfglyphtounicode{Upsilon}{03A5}
\pdfglyphtounicode{Upsilon1}{03D2}
\pdfglyphtounicode{Upsilonacutehooksymbolgreek}{03D3}
\pdfglyphtounicode{Upsilonafrican}{01B1}
\pdfglyphtounicode{Upsilondieresis}{03AB}
\pdfglyphtounicode{Upsilondieresishooksymbolgreek}{03D4}
\pdfglyphtounicode{Upsilonhooksymbol}{03D2}
\pdfglyphtounicode{Upsilontonos}{038E}
\pdfglyphtounicode{Uring}{016E}
\pdfglyphtounicode{Ushortcyrillic}{040E}
\pdfglyphtounicode{Usmall}{0075}
\pdfglyphtounicode{Ustraightcyrillic}{04AE}
\pdfglyphtounicode{Ustraightstrokecyrillic}{04B0}
\pdfglyphtounicode{Utilde}{0168}
\pdfglyphtounicode{Utildeacute}{1E78}
\pdfglyphtounicode{Utildebelow}{1E74}
\pdfglyphtounicode{V}{0056}
\pdfglyphtounicode{Vcircle}{24CB}
\pdfglyphtounicode{Vdotbelow}{1E7E}
\pdfglyphtounicode{Vecyrillic}{0412}
\pdfglyphtounicode{Vewarmenian}{054E}
\pdfglyphtounicode{Vhook}{01B2}
\pdfglyphtounicode{Vmonospace}{FF36}
\pdfglyphtounicode{Voarmenian}{0548}
\pdfglyphtounicode{Vsmall}{0076}
\pdfglyphtounicode{Vtilde}{1E7C}
\pdfglyphtounicode{W}{0057}
\pdfglyphtounicode{Wacute}{1E82}
\pdfglyphtounicode{Wcircle}{24CC}
\pdfglyphtounicode{Wcircumflex}{0174}
\pdfglyphtounicode{Wdieresis}{1E84}
\pdfglyphtounicode{Wdotaccent}{1E86}
\pdfglyphtounicode{Wdotbelow}{1E88}
\pdfglyphtounicode{Wgrave}{1E80}
\pdfglyphtounicode{Wmonospace}{FF37}
\pdfglyphtounicode{Wsmall}{0077}
\pdfglyphtounicode{X}{0058}
\pdfglyphtounicode{Xcircle}{24CD}
\pdfglyphtounicode{Xdieresis}{1E8C}
\pdfglyphtounicode{Xdotaccent}{1E8A}
\pdfglyphtounicode{Xeharmenian}{053D}
\pdfglyphtounicode{Xi}{039E}
\pdfglyphtounicode{Xmonospace}{FF38}
\pdfglyphtounicode{Xsmall}{0078}
\pdfglyphtounicode{Y}{0059}
\pdfglyphtounicode{Yacute}{00DD}
\pdfglyphtounicode{Yacutesmall}{00FD}
\pdfglyphtounicode{Yatcyrillic}{0462}
\pdfglyphtounicode{Ycircle}{24CE}
\pdfglyphtounicode{Ycircumflex}{0176}
\pdfglyphtounicode{Ydieresis}{0178}
\pdfglyphtounicode{Ydieresissmall}{00FF}
\pdfglyphtounicode{Ydotaccent}{1E8E}
\pdfglyphtounicode{Ydotbelow}{1EF4}
\pdfglyphtounicode{Yen}{00A5}
\pdfglyphtounicode{Yericyrillic}{042B}
\pdfglyphtounicode{Yerudieresiscyrillic}{04F8}
\pdfglyphtounicode{Ygrave}{1EF2}
\pdfglyphtounicode{Yhook}{01B3}
\pdfglyphtounicode{Yhookabove}{1EF6}
\pdfglyphtounicode{Yiarmenian}{0545}
\pdfglyphtounicode{Yicyrillic}{0407}
\pdfglyphtounicode{Yiwnarmenian}{0552}
\pdfglyphtounicode{Ymonospace}{FF39}
\pdfglyphtounicode{Ysmall}{0079}
\pdfglyphtounicode{Ytilde}{1EF8}
\pdfglyphtounicode{Yusbigcyrillic}{046A}
\pdfglyphtounicode{Yusbigiotifiedcyrillic}{046C}
\pdfglyphtounicode{Yuslittlecyrillic}{0466}
\pdfglyphtounicode{Yuslittleiotifiedcyrillic}{0468}
\pdfglyphtounicode{Z}{005A}
\pdfglyphtounicode{Zaarmenian}{0536}
\pdfglyphtounicode{Zacute}{0179}
\pdfglyphtounicode{Zcaron}{017D}
\pdfglyphtounicode{Zcaronsmall}{017E}
\pdfglyphtounicode{Zcircle}{24CF}
\pdfglyphtounicode{Zcircumflex}{1E90}
\pdfglyphtounicode{Zdot}{017B}
\pdfglyphtounicode{Zdotaccent}{017B}
\pdfglyphtounicode{Zdotbelow}{1E92}
\pdfglyphtounicode{Zecyrillic}{0417}
\pdfglyphtounicode{Zedescendercyrillic}{0498}
\pdfglyphtounicode{Zedieresiscyrillic}{04DE}
\pdfglyphtounicode{Zeta}{0396}
\pdfglyphtounicode{Zhearmenian}{053A}
\pdfglyphtounicode{Zhebrevecyrillic}{04C1}
\pdfglyphtounicode{Zhecyrillic}{0416}
\pdfglyphtounicode{Zhedescendercyrillic}{0496}
\pdfglyphtounicode{Zhedieresiscyrillic}{04DC}
\pdfglyphtounicode{Zlinebelow}{1E94}
\pdfglyphtounicode{Zmonospace}{FF3A}
\pdfglyphtounicode{Zsmall}{007A}
\pdfglyphtounicode{Zstroke}{01B5}
\pdfglyphtounicode{a}{0061}
\pdfglyphtounicode{aabengali}{0986}
\pdfglyphtounicode{aacute}{00E1}
\pdfglyphtounicode{aadeva}{0906}
\pdfglyphtounicode{aagujarati}{0A86}
\pdfglyphtounicode{aagurmukhi}{0A06}
\pdfglyphtounicode{aamatragurmukhi}{0A3E}
\pdfglyphtounicode{aarusquare}{3303}
\pdfglyphtounicode{aavowelsignbengali}{09BE}
\pdfglyphtounicode{aavowelsigndeva}{093E}
\pdfglyphtounicode{aavowelsigngujarati}{0ABE}
\pdfglyphtounicode{abbreviationmarkarmenian}{055F}
\pdfglyphtounicode{abbreviationsigndeva}{0970}
\pdfglyphtounicode{abengali}{0985}
\pdfglyphtounicode{abopomofo}{311A}
\pdfglyphtounicode{abreve}{0103}
\pdfglyphtounicode{abreveacute}{1EAF}
\pdfglyphtounicode{abrevecyrillic}{04D1}
\pdfglyphtounicode{abrevedotbelow}{1EB7}
\pdfglyphtounicode{abrevegrave}{1EB1}
\pdfglyphtounicode{abrevehookabove}{1EB3}
\pdfglyphtounicode{abrevetilde}{1EB5}
\pdfglyphtounicode{acaron}{01CE}
\pdfglyphtounicode{acircle}{24D0}
\pdfglyphtounicode{acircumflex}{00E2}
\pdfglyphtounicode{acircumflexacute}{1EA5}
\pdfglyphtounicode{acircumflexdotbelow}{1EAD}
\pdfglyphtounicode{acircumflexgrave}{1EA7}
\pdfglyphtounicode{acircumflexhookabove}{1EA9}
\pdfglyphtounicode{acircumflextilde}{1EAB}
\pdfglyphtounicode{acute}{00B4}
\pdfglyphtounicode{acutebelowcmb}{0317}
\pdfglyphtounicode{acutecmb}{0301}
\pdfglyphtounicode{acutecomb}{0301}
\pdfglyphtounicode{acutedeva}{0954}
\pdfglyphtounicode{acutelowmod}{02CF}
\pdfglyphtounicode{acutetonecmb}{0341}
\pdfglyphtounicode{acyrillic}{0430}
\pdfglyphtounicode{adblgrave}{0201}
\pdfglyphtounicode{addakgurmukhi}{0A71}
\pdfglyphtounicode{adeva}{0905}
\pdfglyphtounicode{adieresis}{00E4}
\pdfglyphtounicode{adieresiscyrillic}{04D3}
\pdfglyphtounicode{adieresismacron}{01DF}
\pdfglyphtounicode{adotbelow}{1EA1}
\pdfglyphtounicode{adotmacron}{01E1}
\pdfglyphtounicode{ae}{00E6}
\pdfglyphtounicode{aeacute}{01FD}
\pdfglyphtounicode{aekorean}{3150}
\pdfglyphtounicode{aemacron}{01E3}
\pdfglyphtounicode{afii00208}{2015}
\pdfglyphtounicode{afii08941}{20A4}
\pdfglyphtounicode{afii10017}{0410}
\pdfglyphtounicode{afii10018}{0411}
\pdfglyphtounicode{afii10019}{0412}
\pdfglyphtounicode{afii10020}{0413}
\pdfglyphtounicode{afii10021}{0414}
\pdfglyphtounicode{afii10022}{0415}
\pdfglyphtounicode{afii10023}{0401}
\pdfglyphtounicode{afii10024}{0416}
\pdfglyphtounicode{afii10025}{0417}
\pdfglyphtounicode{afii10026}{0418}
\pdfglyphtounicode{afii10027}{0419}
\pdfglyphtounicode{afii10028}{041A}
\pdfglyphtounicode{afii10029}{041B}
\pdfglyphtounicode{afii10030}{041C}
\pdfglyphtounicode{afii10031}{041D}
\pdfglyphtounicode{afii10032}{041E}
\pdfglyphtounicode{afii10033}{041F}
\pdfglyphtounicode{afii10034}{0420}
\pdfglyphtounicode{afii10035}{0421}
\pdfglyphtounicode{afii10036}{0422}
\pdfglyphtounicode{afii10037}{0423}
\pdfglyphtounicode{afii10038}{0424}
\pdfglyphtounicode{afii10039}{0425}
\pdfglyphtounicode{afii10040}{0426}
\pdfglyphtounicode{afii10041}{0427}
\pdfglyphtounicode{afii10042}{0428}
\pdfglyphtounicode{afii10043}{0429}
\pdfglyphtounicode{afii10044}{042A}
\pdfglyphtounicode{afii10045}{042B}
\pdfglyphtounicode{afii10046}{042C}
\pdfglyphtounicode{afii10047}{042D}
\pdfglyphtounicode{afii10048}{042E}
\pdfglyphtounicode{afii10049}{042F}
\pdfglyphtounicode{afii10050}{0490}
\pdfglyphtounicode{afii10051}{0402}
\pdfglyphtounicode{afii10052}{0403}
\pdfglyphtounicode{afii10053}{0404}
\pdfglyphtounicode{afii10054}{0405}
\pdfglyphtounicode{afii10055}{0406}
\pdfglyphtounicode{afii10056}{0407}
\pdfglyphtounicode{afii10057}{0408}
\pdfglyphtounicode{afii10058}{0409}
\pdfglyphtounicode{afii10059}{040A}
\pdfglyphtounicode{afii10060}{040B}
\pdfglyphtounicode{afii10061}{040C}
\pdfglyphtounicode{afii10062}{040E}
\pdfglyphtounicode{afii10063}{F6C4}
\pdfglyphtounicode{afii10064}{F6C5}
\pdfglyphtounicode{afii10065}{0430}
\pdfglyphtounicode{afii10066}{0431}
\pdfglyphtounicode{afii10067}{0432}
\pdfglyphtounicode{afii10068}{0433}
\pdfglyphtounicode{afii10069}{0434}
\pdfglyphtounicode{afii10070}{0435}
\pdfglyphtounicode{afii10071}{0451}
\pdfglyphtounicode{afii10072}{0436}
\pdfglyphtounicode{afii10073}{0437}
\pdfglyphtounicode{afii10074}{0438}
\pdfglyphtounicode{afii10075}{0439}
\pdfglyphtounicode{afii10076}{043A}
\pdfglyphtounicode{afii10077}{043B}
\pdfglyphtounicode{afii10078}{043C}
\pdfglyphtounicode{afii10079}{043D}
\pdfglyphtounicode{afii10080}{043E}
\pdfglyphtounicode{afii10081}{043F}
\pdfglyphtounicode{afii10082}{0440}
\pdfglyphtounicode{afii10083}{0441}
\pdfglyphtounicode{afii10084}{0442}
\pdfglyphtounicode{afii10085}{0443}
\pdfglyphtounicode{afii10086}{0444}
\pdfglyphtounicode{afii10087}{0445}
\pdfglyphtounicode{afii10088}{0446}
\pdfglyphtounicode{afii10089}{0447}
\pdfglyphtounicode{afii10090}{0448}
\pdfglyphtounicode{afii10091}{0449}
\pdfglyphtounicode{afii10092}{044A}
\pdfglyphtounicode{afii10093}{044B}
\pdfglyphtounicode{afii10094}{044C}
\pdfglyphtounicode{afii10095}{044D}
\pdfglyphtounicode{afii10096}{044E}
\pdfglyphtounicode{afii10097}{044F}
\pdfglyphtounicode{afii10098}{0491}
\pdfglyphtounicode{afii10099}{0452}
\pdfglyphtounicode{afii10100}{0453}
\pdfglyphtounicode{afii10101}{0454}
\pdfglyphtounicode{afii10102}{0455}
\pdfglyphtounicode{afii10103}{0456}
\pdfglyphtounicode{afii10104}{0457}
\pdfglyphtounicode{afii10105}{0458}
\pdfglyphtounicode{afii10106}{0459}
\pdfglyphtounicode{afii10107}{045A}
\pdfglyphtounicode{afii10108}{045B}
\pdfglyphtounicode{afii10109}{045C}
\pdfglyphtounicode{afii10110}{045E}
\pdfglyphtounicode{afii10145}{040F}
\pdfglyphtounicode{afii10146}{0462}
\pdfglyphtounicode{afii10147}{0472}
\pdfglyphtounicode{afii10148}{0474}
\pdfglyphtounicode{afii10192}{F6C6}
\pdfglyphtounicode{afii10193}{045F}
\pdfglyphtounicode{afii10194}{0463}
\pdfglyphtounicode{afii10195}{0473}
\pdfglyphtounicode{afii10196}{0475}
\pdfglyphtounicode{afii10831}{F6C7}
\pdfglyphtounicode{afii10832}{F6C8}
\pdfglyphtounicode{afii10846}{04D9}
\pdfglyphtounicode{afii299}{200E}
\pdfglyphtounicode{afii300}{200F}
\pdfglyphtounicode{afii301}{200D}
\pdfglyphtounicode{afii57381}{066A}
\pdfglyphtounicode{afii57388}{060C}
\pdfglyphtounicode{afii57392}{0660}
\pdfglyphtounicode{afii57393}{0661}
\pdfglyphtounicode{afii57394}{0662}
\pdfglyphtounicode{afii57395}{0663}
\pdfglyphtounicode{afii57396}{0664}
\pdfglyphtounicode{afii57397}{0665}
\pdfglyphtounicode{afii57398}{0666}
\pdfglyphtounicode{afii57399}{0667}
\pdfglyphtounicode{afii57400}{0668}
\pdfglyphtounicode{afii57401}{0669}
\pdfglyphtounicode{afii57403}{061B}
\pdfglyphtounicode{afii57407}{061F}
\pdfglyphtounicode{afii57409}{0621}
\pdfglyphtounicode{afii57410}{0622}
\pdfglyphtounicode{afii57411}{0623}
\pdfglyphtounicode{afii57412}{0624}
\pdfglyphtounicode{afii57413}{0625}
\pdfglyphtounicode{afii57414}{0626}
\pdfglyphtounicode{afii57415}{0627}
\pdfglyphtounicode{afii57416}{0628}
\pdfglyphtounicode{afii57417}{0629}
\pdfglyphtounicode{afii57418}{062A}
\pdfglyphtounicode{afii57419}{062B}
\pdfglyphtounicode{afii57420}{062C}
\pdfglyphtounicode{afii57421}{062D}
\pdfglyphtounicode{afii57422}{062E}
\pdfglyphtounicode{afii57423}{062F}
\pdfglyphtounicode{afii57424}{0630}
\pdfglyphtounicode{afii57425}{0631}
\pdfglyphtounicode{afii57426}{0632}
\pdfglyphtounicode{afii57427}{0633}
\pdfglyphtounicode{afii57428}{0634}
\pdfglyphtounicode{afii57429}{0635}
\pdfglyphtounicode{afii57430}{0636}
\pdfglyphtounicode{afii57431}{0637}
\pdfglyphtounicode{afii57432}{0638}
\pdfglyphtounicode{afii57433}{0639}
\pdfglyphtounicode{afii57434}{063A}
\pdfglyphtounicode{afii57440}{0640}
\pdfglyphtounicode{afii57441}{0641}
\pdfglyphtounicode{afii57442}{0642}
\pdfglyphtounicode{afii57443}{0643}
\pdfglyphtounicode{afii57444}{0644}
\pdfglyphtounicode{afii57445}{0645}
\pdfglyphtounicode{afii57446}{0646}
\pdfglyphtounicode{afii57448}{0648}
\pdfglyphtounicode{afii57449}{0649}
\pdfglyphtounicode{afii57450}{064A}
\pdfglyphtounicode{afii57451}{064B}
\pdfglyphtounicode{afii57452}{064C}
\pdfglyphtounicode{afii57453}{064D}
\pdfglyphtounicode{afii57454}{064E}
\pdfglyphtounicode{afii57455}{064F}
\pdfglyphtounicode{afii57456}{0650}
\pdfglyphtounicode{afii57457}{0651}
\pdfglyphtounicode{afii57458}{0652}
\pdfglyphtounicode{afii57470}{0647}
\pdfglyphtounicode{afii57505}{06A4}
\pdfglyphtounicode{afii57506}{067E}
\pdfglyphtounicode{afii57507}{0686}
\pdfglyphtounicode{afii57508}{0698}
\pdfglyphtounicode{afii57509}{06AF}
\pdfglyphtounicode{afii57511}{0679}
\pdfglyphtounicode{afii57512}{0688}
\pdfglyphtounicode{afii57513}{0691}
\pdfglyphtounicode{afii57514}{06BA}
\pdfglyphtounicode{afii57519}{06D2}
\pdfglyphtounicode{afii57534}{06D5}
\pdfglyphtounicode{afii57636}{20AA}
\pdfglyphtounicode{afii57645}{05BE}
\pdfglyphtounicode{afii57658}{05C3}
\pdfglyphtounicode{afii57664}{05D0}
\pdfglyphtounicode{afii57665}{05D1}
\pdfglyphtounicode{afii57666}{05D2}
\pdfglyphtounicode{afii57667}{05D3}
\pdfglyphtounicode{afii57668}{05D4}
\pdfglyphtounicode{afii57669}{05D5}
\pdfglyphtounicode{afii57670}{05D6}
\pdfglyphtounicode{afii57671}{05D7}
\pdfglyphtounicode{afii57672}{05D8}
\pdfglyphtounicode{afii57673}{05D9}
\pdfglyphtounicode{afii57674}{05DA}
\pdfglyphtounicode{afii57675}{05DB}
\pdfglyphtounicode{afii57676}{05DC}
\pdfglyphtounicode{afii57677}{05DD}
\pdfglyphtounicode{afii57678}{05DE}
\pdfglyphtounicode{afii57679}{05DF}
\pdfglyphtounicode{afii57680}{05E0}
\pdfglyphtounicode{afii57681}{05E1}
\pdfglyphtounicode{afii57682}{05E2}
\pdfglyphtounicode{afii57683}{05E3}
\pdfglyphtounicode{afii57684}{05E4}
\pdfglyphtounicode{afii57685}{05E5}
\pdfglyphtounicode{afii57686}{05E6}
\pdfglyphtounicode{afii57687}{05E7}
\pdfglyphtounicode{afii57688}{05E8}
\pdfglyphtounicode{afii57689}{05E9}
\pdfglyphtounicode{afii57690}{05EA}
\pdfglyphtounicode{afii57694}{FB2A}
\pdfglyphtounicode{afii57695}{FB2B}
\pdfglyphtounicode{afii57700}{FB4B}
\pdfglyphtounicode{afii57705}{FB1F}
\pdfglyphtounicode{afii57716}{05F0}
\pdfglyphtounicode{afii57717}{05F1}
\pdfglyphtounicode{afii57718}{05F2}
\pdfglyphtounicode{afii57723}{FB35}
\pdfglyphtounicode{afii57793}{05B4}
\pdfglyphtounicode{afii57794}{05B5}
\pdfglyphtounicode{afii57795}{05B6}
\pdfglyphtounicode{afii57796}{05BB}
\pdfglyphtounicode{afii57797}{05B8}
\pdfglyphtounicode{afii57798}{05B7}
\pdfglyphtounicode{afii57799}{05B0}
\pdfglyphtounicode{afii57800}{05B2}
\pdfglyphtounicode{afii57801}{05B1}
\pdfglyphtounicode{afii57802}{05B3}
\pdfglyphtounicode{afii57803}{05C2}
\pdfglyphtounicode{afii57804}{05C1}
\pdfglyphtounicode{afii57806}{05B9}
\pdfglyphtounicode{afii57807}{05BC}
\pdfglyphtounicode{afii57839}{05BD}
\pdfglyphtounicode{afii57841}{05BF}
\pdfglyphtounicode{afii57842}{05C0}
\pdfglyphtounicode{afii57929}{02BC}
\pdfglyphtounicode{afii61248}{2105}
\pdfglyphtounicode{afii61289}{2113}
\pdfglyphtounicode{afii61352}{2116}
\pdfglyphtounicode{afii61573}{202C}
\pdfglyphtounicode{afii61574}{202D}
\pdfglyphtounicode{afii61575}{202E}
\pdfglyphtounicode{afii61664}{200C}
\pdfglyphtounicode{afii63167}{066D}
\pdfglyphtounicode{afii64937}{02BD}
\pdfglyphtounicode{agrave}{00E0}
\pdfglyphtounicode{agujarati}{0A85}
\pdfglyphtounicode{agurmukhi}{0A05}
\pdfglyphtounicode{ahiragana}{3042}
\pdfglyphtounicode{ahookabove}{1EA3}
\pdfglyphtounicode{aibengali}{0990}
\pdfglyphtounicode{aibopomofo}{311E}
\pdfglyphtounicode{aideva}{0910}
\pdfglyphtounicode{aiecyrillic}{04D5}
\pdfglyphtounicode{aigujarati}{0A90}
\pdfglyphtounicode{aigurmukhi}{0A10}
\pdfglyphtounicode{aimatragurmukhi}{0A48}
\pdfglyphtounicode{ainarabic}{0639}
\pdfglyphtounicode{ainfinalarabic}{FECA}
\pdfglyphtounicode{aininitialarabic}{FECB}
\pdfglyphtounicode{ainmedialarabic}{FECC}
\pdfglyphtounicode{ainvertedbreve}{0203}
\pdfglyphtounicode{aivowelsignbengali}{09C8}
\pdfglyphtounicode{aivowelsigndeva}{0948}
\pdfglyphtounicode{aivowelsigngujarati}{0AC8}
\pdfglyphtounicode{akatakana}{30A2}
\pdfglyphtounicode{akatakanahalfwidth}{FF71}
\pdfglyphtounicode{akorean}{314F}
\pdfglyphtounicode{alef}{05D0}
\pdfglyphtounicode{alefarabic}{0627}
\pdfglyphtounicode{alefdageshhebrew}{FB30}
\pdfglyphtounicode{aleffinalarabic}{FE8E}
\pdfglyphtounicode{alefhamzaabovearabic}{0623}
\pdfglyphtounicode{alefhamzaabovefinalarabic}{FE84}
\pdfglyphtounicode{alefhamzabelowarabic}{0625}
\pdfglyphtounicode{alefhamzabelowfinalarabic}{FE88}
\pdfglyphtounicode{alefhebrew}{05D0}
\pdfglyphtounicode{aleflamedhebrew}{FB4F}
\pdfglyphtounicode{alefmaddaabovearabic}{0622}
\pdfglyphtounicode{alefmaddaabovefinalarabic}{FE82}
\pdfglyphtounicode{alefmaksuraarabic}{0649}
\pdfglyphtounicode{alefmaksurafinalarabic}{FEF0}
\pdfglyphtounicode{alefmaksurainitialarabic}{FEF3}
\pdfglyphtounicode{alefmaksuramedialarabic}{FEF4}
\pdfglyphtounicode{alefpatahhebrew}{FB2E}
\pdfglyphtounicode{alefqamatshebrew}{FB2F}
\pdfglyphtounicode{aleph}{2135}
\pdfglyphtounicode{allequal}{224C}
\pdfglyphtounicode{alpha}{03B1}
\pdfglyphtounicode{alphatonos}{03AC}
\pdfglyphtounicode{amacron}{0101}
\pdfglyphtounicode{amonospace}{FF41}
\pdfglyphtounicode{ampersand}{0026}
\pdfglyphtounicode{ampersandmonospace}{FF06}
\pdfglyphtounicode{ampersandsmall}{0026}
\pdfglyphtounicode{amsquare}{33C2}
\pdfglyphtounicode{anbopomofo}{3122}
\pdfglyphtounicode{angbopomofo}{3124}
\pdfglyphtounicode{angbracketleft}{27E8}
\pdfglyphtounicode{angbracketright}{27E9}
\pdfglyphtounicode{angkhankhuthai}{0E5A}
\pdfglyphtounicode{angle}{2220}
\pdfglyphtounicode{anglebracketleft}{3008}
\pdfglyphtounicode{anglebracketleftvertical}{FE3F}
\pdfglyphtounicode{anglebracketright}{3009}
\pdfglyphtounicode{anglebracketrightvertical}{FE40}
\pdfglyphtounicode{angleleft}{2329}
\pdfglyphtounicode{angleright}{232A}
\pdfglyphtounicode{angstrom}{212B}
\pdfglyphtounicode{anoteleia}{0387}
\pdfglyphtounicode{anticlockwise}{27F2}
\pdfglyphtounicode{anudattadeva}{0952}
\pdfglyphtounicode{anusvarabengali}{0982}
\pdfglyphtounicode{anusvaradeva}{0902}
\pdfglyphtounicode{anusvaragujarati}{0A82}
\pdfglyphtounicode{aogonek}{0105}
\pdfglyphtounicode{apaatosquare}{3300}
\pdfglyphtounicode{aparen}{249C}
\pdfglyphtounicode{apostrophearmenian}{055A}
\pdfglyphtounicode{apostrophemod}{02BC}
\pdfglyphtounicode{apple}{F8FF}
\pdfglyphtounicode{approaches}{2250}
\pdfglyphtounicode{approxequal}{2248}
\pdfglyphtounicode{approxequalorimage}{2252}
\pdfglyphtounicode{approximatelyequal}{2245}
\pdfglyphtounicode{approxorequal}{224A}
\pdfglyphtounicode{araeaekorean}{318E}
\pdfglyphtounicode{araeakorean}{318D}
\pdfglyphtounicode{arc}{2312}
\pdfglyphtounicode{archleftdown}{21B6}
\pdfglyphtounicode{archrightdown}{21B7}
\pdfglyphtounicode{arighthalfring}{1E9A}
\pdfglyphtounicode{aring}{00E5}
\pdfglyphtounicode{aringacute}{01FB}
\pdfglyphtounicode{aringbelow}{1E01}
\pdfglyphtounicode{arrowboth}{2194}
\pdfglyphtounicode{arrowbothv}{2195}
\pdfglyphtounicode{arrowdashdown}{21E3}
\pdfglyphtounicode{arrowdashleft}{21E0}
\pdfglyphtounicode{arrowdashright}{21E2}
\pdfglyphtounicode{arrowdashup}{21E1}
\pdfglyphtounicode{arrowdblboth}{21D4}
\pdfglyphtounicode{arrowdblbothv}{21D5}
\pdfglyphtounicode{arrowdbldown}{21D3}
\pdfglyphtounicode{arrowdblleft}{21D0}
\pdfglyphtounicode{arrowdblright}{21D2}
\pdfglyphtounicode{arrowdblup}{21D1}
\pdfglyphtounicode{arrowdown}{2193}
\pdfglyphtounicode{arrowdownleft}{2199}
\pdfglyphtounicode{arrowdownright}{2198}
\pdfglyphtounicode{arrowdownwhite}{21E9}
\pdfglyphtounicode{arrowheaddownmod}{02C5}
\pdfglyphtounicode{arrowheadleftmod}{02C2}
\pdfglyphtounicode{arrowheadrightmod}{02C3}
\pdfglyphtounicode{arrowheadupmod}{02C4}
\pdfglyphtounicode{arrowhorizex}{F8E7}
\pdfglyphtounicode{arrowleft}{2190}
\pdfglyphtounicode{arrowleftbothalf}{21BD}
\pdfglyphtounicode{arrowleftdbl}{21D0}
\pdfglyphtounicode{arrowleftdblstroke}{21CD}
\pdfglyphtounicode{arrowleftoverright}{21C6}
\pdfglyphtounicode{arrowlefttophalf}{21BC}
\pdfglyphtounicode{arrowleftwhite}{21E6}
\pdfglyphtounicode{arrownortheast}{2197}
\pdfglyphtounicode{arrownorthwest}{2196}
\pdfglyphtounicode{arrowparrleftright}{21C6}
\pdfglyphtounicode{arrowparrrightleft}{21C4}
\pdfglyphtounicode{arrowright}{2192}
\pdfglyphtounicode{arrowrightbothalf}{21C1}
\pdfglyphtounicode{arrowrightdblstroke}{21CF}
\pdfglyphtounicode{arrowrightheavy}{279E}
\pdfglyphtounicode{arrowrightoverleft}{21C4}
\pdfglyphtounicode{arrowrighttophalf}{21C0}
\pdfglyphtounicode{arrowrightwhite}{21E8}
\pdfglyphtounicode{arrowsoutheast}{2198}
\pdfglyphtounicode{arrowsouthwest}{2199}
\pdfglyphtounicode{arrowtableft}{21E4}
\pdfglyphtounicode{arrowtabright}{21E5}
\pdfglyphtounicode{arrowtailleft}{21A2}
\pdfglyphtounicode{arrowtailright}{21A3}
\pdfglyphtounicode{arrowtripleleft}{21DA}
\pdfglyphtounicode{arrowtripleright}{21DB}
\pdfglyphtounicode{arrowup}{2191}
\pdfglyphtounicode{arrowupdn}{2195}
\pdfglyphtounicode{arrowupdnbse}{21A8}
\pdfglyphtounicode{arrowupdownbase}{21A8}
\pdfglyphtounicode{arrowupleft}{2196}
\pdfglyphtounicode{arrowupleftofdown}{21C5}
\pdfglyphtounicode{arrowupright}{2197}
\pdfglyphtounicode{arrowupwhite}{21E7}
\pdfglyphtounicode{arrowvertex}{F8E6}
\pdfglyphtounicode{asciicircum}{005E}
\pdfglyphtounicode{asciicircummonospace}{FF3E}
\pdfglyphtounicode{asciitilde}{007E}
\pdfglyphtounicode{asciitildemonospace}{FF5E}
\pdfglyphtounicode{ascript}{0251}
\pdfglyphtounicode{ascriptturned}{0252}
\pdfglyphtounicode{asmallhiragana}{3041}
\pdfglyphtounicode{asmallkatakana}{30A1}
\pdfglyphtounicode{asmallkatakanahalfwidth}{FF67}
\pdfglyphtounicode{asterisk}{002A}
\pdfglyphtounicode{asteriskaltonearabic}{066D}
\pdfglyphtounicode{asteriskarabic}{066D}
\pdfglyphtounicode{asteriskcentered}{2217}
\pdfglyphtounicode{asteriskmath}{2217}
\pdfglyphtounicode{asteriskmonospace}{FF0A}
\pdfglyphtounicode{asterisksmall}{FE61}
\pdfglyphtounicode{asterism}{2042}
\pdfglyphtounicode{asuperior}{0061}
\pdfglyphtounicode{asymptoticallyequal}{2243}
\pdfglyphtounicode{at}{0040}
\pdfglyphtounicode{atilde}{00E3}
\pdfglyphtounicode{atmonospace}{FF20}
\pdfglyphtounicode{atsmall}{FE6B}
\pdfglyphtounicode{aturned}{0250}
\pdfglyphtounicode{aubengali}{0994}
\pdfglyphtounicode{aubopomofo}{3120}
\pdfglyphtounicode{audeva}{0914}
\pdfglyphtounicode{augujarati}{0A94}
\pdfglyphtounicode{augurmukhi}{0A14}
\pdfglyphtounicode{aulengthmarkbengali}{09D7}
\pdfglyphtounicode{aumatragurmukhi}{0A4C}
\pdfglyphtounicode{auvowelsignbengali}{09CC}
\pdfglyphtounicode{auvowelsigndeva}{094C}
\pdfglyphtounicode{auvowelsigngujarati}{0ACC}
\pdfglyphtounicode{avagrahadeva}{093D}
\pdfglyphtounicode{aybarmenian}{0561}
\pdfglyphtounicode{ayin}{05E2}
\pdfglyphtounicode{ayinaltonehebrew}{FB20}
\pdfglyphtounicode{ayinhebrew}{05E2}
\pdfglyphtounicode{b}{0062}
\pdfglyphtounicode{babengali}{09AC}
\pdfglyphtounicode{backslash}{005C}
\pdfglyphtounicode{backslashmonospace}{FF3C}
\pdfglyphtounicode{badeva}{092C}
\pdfglyphtounicode{bagujarati}{0AAC}
\pdfglyphtounicode{bagurmukhi}{0A2C}
\pdfglyphtounicode{bahiragana}{3070}
\pdfglyphtounicode{bahtthai}{0E3F}
\pdfglyphtounicode{bakatakana}{30D0}
\pdfglyphtounicode{bar}{007C}
\pdfglyphtounicode{bardbl}{2225}
\pdfglyphtounicode{barmonospace}{FF5C}
\pdfglyphtounicode{bbopomofo}{3105}
\pdfglyphtounicode{bcircle}{24D1}
\pdfglyphtounicode{bdotaccent}{1E03}
\pdfglyphtounicode{bdotbelow}{1E05}
\pdfglyphtounicode{beamedsixteenthnotes}{266C}
\pdfglyphtounicode{because}{2235}
\pdfglyphtounicode{becyrillic}{0431}
\pdfglyphtounicode{beharabic}{0628}
\pdfglyphtounicode{behfinalarabic}{FE90}
\pdfglyphtounicode{behinitialarabic}{FE91}
\pdfglyphtounicode{behiragana}{3079}
\pdfglyphtounicode{behmedialarabic}{FE92}
\pdfglyphtounicode{behmeeminitialarabic}{FC9F}
\pdfglyphtounicode{behmeemisolatedarabic}{FC08}
\pdfglyphtounicode{behnoonfinalarabic}{FC6D}
\pdfglyphtounicode{bekatakana}{30D9}
\pdfglyphtounicode{benarmenian}{0562}
\pdfglyphtounicode{bet}{05D1}
\pdfglyphtounicode{beta}{03B2}
\pdfglyphtounicode{betasymbolgreek}{03D0}
\pdfglyphtounicode{betdagesh}{FB31}
\pdfglyphtounicode{betdageshhebrew}{FB31}
\pdfglyphtounicode{beth}{2136}
\pdfglyphtounicode{bethebrew}{05D1}
\pdfglyphtounicode{betrafehebrew}{FB4C}
\pdfglyphtounicode{between}{226C}
\pdfglyphtounicode{bhabengali}{09AD}
\pdfglyphtounicode{bhadeva}{092D}
\pdfglyphtounicode{bhagujarati}{0AAD}
\pdfglyphtounicode{bhagurmukhi}{0A2D}
\pdfglyphtounicode{bhook}{0253}
\pdfglyphtounicode{bihiragana}{3073}
\pdfglyphtounicode{bikatakana}{30D3}
\pdfglyphtounicode{bilabialclick}{0298}
\pdfglyphtounicode{bindigurmukhi}{0A02}
\pdfglyphtounicode{birusquare}{3331}
\pdfglyphtounicode{blackcircle}{25CF}
\pdfglyphtounicode{blackdiamond}{25C6}
\pdfglyphtounicode{blackdownpointingtriangle}{25BC}
\pdfglyphtounicode{blackleftpointingpointer}{25C4}
\pdfglyphtounicode{blackleftpointingtriangle}{25C0}
\pdfglyphtounicode{blacklenticularbracketleft}{3010}
\pdfglyphtounicode{blacklenticularbracketleftvertical}{FE3B}
\pdfglyphtounicode{blacklenticularbracketright}{3011}
\pdfglyphtounicode{blacklenticularbracketrightvertical}{FE3C}
\pdfglyphtounicode{blacklowerlefttriangle}{25E3}
\pdfglyphtounicode{blacklowerrighttriangle}{25E2}
\pdfglyphtounicode{blackrectangle}{25AC}
\pdfglyphtounicode{blackrightpointingpointer}{25BA}
\pdfglyphtounicode{blackrightpointingtriangle}{25B6}
\pdfglyphtounicode{blacksmallsquare}{25AA}
\pdfglyphtounicode{blacksmilingface}{263B}
\pdfglyphtounicode{blacksquare}{25A0}
\pdfglyphtounicode{blackstar}{2605}
\pdfglyphtounicode{blackupperlefttriangle}{25E4}
\pdfglyphtounicode{blackupperrighttriangle}{25E5}
\pdfglyphtounicode{blackuppointingsmalltriangle}{25B4}
\pdfglyphtounicode{blackuppointingtriangle}{25B2}
\pdfglyphtounicode{blank}{2423}
\pdfglyphtounicode{blinebelow}{1E07}
\pdfglyphtounicode{block}{2588}
\pdfglyphtounicode{bmonospace}{FF42}
\pdfglyphtounicode{bobaimaithai}{0E1A}
\pdfglyphtounicode{bohiragana}{307C}
\pdfglyphtounicode{bokatakana}{30DC}
\pdfglyphtounicode{bparen}{249D}
\pdfglyphtounicode{bqsquare}{33C3}
\pdfglyphtounicode{braceex}{F8F4}
\pdfglyphtounicode{braceleft}{007B}
\pdfglyphtounicode{braceleftbt}{F8F3}
\pdfglyphtounicode{braceleftmid}{F8F2}
\pdfglyphtounicode{braceleftmonospace}{FF5B}
\pdfglyphtounicode{braceleftsmall}{FE5B}
\pdfglyphtounicode{bracelefttp}{F8F1}
\pdfglyphtounicode{braceleftvertical}{FE37}
\pdfglyphtounicode{braceright}{007D}
\pdfglyphtounicode{bracerightbt}{F8FE}
\pdfglyphtounicode{bracerightmid}{F8FD}
\pdfglyphtounicode{bracerightmonospace}{FF5D}
\pdfglyphtounicode{bracerightsmall}{FE5C}
\pdfglyphtounicode{bracerighttp}{F8FC}
\pdfglyphtounicode{bracerightvertical}{FE38}
\pdfglyphtounicode{bracketleft}{005B}
\pdfglyphtounicode{bracketleftbt}{F8F0}
\pdfglyphtounicode{bracketleftex}{F8EF}
\pdfglyphtounicode{bracketleftmonospace}{FF3B}
\pdfglyphtounicode{bracketlefttp}{F8EE}
\pdfglyphtounicode{bracketright}{005D}
\pdfglyphtounicode{bracketrightbt}{F8FB}
\pdfglyphtounicode{bracketrightex}{F8FA}
\pdfglyphtounicode{bracketrightmonospace}{FF3D}
\pdfglyphtounicode{bracketrighttp}{F8F9}
\pdfglyphtounicode{breve}{02D8}
\pdfglyphtounicode{brevebelowcmb}{032E}
\pdfglyphtounicode{brevecmb}{0306}
\pdfglyphtounicode{breveinvertedbelowcmb}{032F}
\pdfglyphtounicode{breveinvertedcmb}{0311}
\pdfglyphtounicode{breveinverteddoublecmb}{0361}
\pdfglyphtounicode{bridgebelowcmb}{032A}
\pdfglyphtounicode{bridgeinvertedbelowcmb}{033A}
\pdfglyphtounicode{brokenbar}{00A6}
\pdfglyphtounicode{bstroke}{0180}
\pdfglyphtounicode{bsuperior}{0062}
\pdfglyphtounicode{btopbar}{0183}
\pdfglyphtounicode{buhiragana}{3076}
\pdfglyphtounicode{bukatakana}{30D6}
\pdfglyphtounicode{bullet}{2022}
\pdfglyphtounicode{bulletinverse}{25D8}
\pdfglyphtounicode{bulletoperator}{2219}
\pdfglyphtounicode{bullseye}{25CE}
\pdfglyphtounicode{c}{0063}
\pdfglyphtounicode{caarmenian}{056E}
\pdfglyphtounicode{cabengali}{099A}
\pdfglyphtounicode{cacute}{0107}
\pdfglyphtounicode{cadeva}{091A}
\pdfglyphtounicode{cagujarati}{0A9A}
\pdfglyphtounicode{cagurmukhi}{0A1A}
\pdfglyphtounicode{calsquare}{3388}
\pdfglyphtounicode{candrabindubengali}{0981}
\pdfglyphtounicode{candrabinducmb}{0310}
\pdfglyphtounicode{candrabindudeva}{0901}
\pdfglyphtounicode{candrabindugujarati}{0A81}
\pdfglyphtounicode{capslock}{21EA}
\pdfglyphtounicode{careof}{2105}
\pdfglyphtounicode{caron}{02C7}
\pdfglyphtounicode{caronbelowcmb}{032C}
\pdfglyphtounicode{caroncmb}{030C}
\pdfglyphtounicode{carriagereturn}{21B5}
\pdfglyphtounicode{cbopomofo}{3118}
\pdfglyphtounicode{ccaron}{010D}
\pdfglyphtounicode{ccedilla}{00E7}
\pdfglyphtounicode{ccedillaacute}{1E09}
\pdfglyphtounicode{ccircle}{24D2}
\pdfglyphtounicode{ccircumflex}{0109}
\pdfglyphtounicode{ccurl}{0255}
\pdfglyphtounicode{cdot}{010B}
\pdfglyphtounicode{cdotaccent}{010B}
\pdfglyphtounicode{cdsquare}{33C5}
\pdfglyphtounicode{cedilla}{00B8}
\pdfglyphtounicode{cedillacmb}{0327}
\pdfglyphtounicode{ceilingleft}{2308}
\pdfglyphtounicode{ceilingright}{2309}
\pdfglyphtounicode{cent}{00A2}
\pdfglyphtounicode{centigrade}{2103}
\pdfglyphtounicode{centinferior}{00A2}
\pdfglyphtounicode{centmonospace}{FFE0}
\pdfglyphtounicode{centoldstyle}{00A2}
\pdfglyphtounicode{centsuperior}{00A2}
\pdfglyphtounicode{chaarmenian}{0579}
\pdfglyphtounicode{chabengali}{099B}
\pdfglyphtounicode{chadeva}{091B}
\pdfglyphtounicode{chagujarati}{0A9B}
\pdfglyphtounicode{chagurmukhi}{0A1B}
\pdfglyphtounicode{chbopomofo}{3114}
\pdfglyphtounicode{cheabkhasiancyrillic}{04BD}
\pdfglyphtounicode{check}{2713}
\pdfglyphtounicode{checkmark}{2713}
\pdfglyphtounicode{checyrillic}{0447}
\pdfglyphtounicode{chedescenderabkhasiancyrillic}{04BF}
\pdfglyphtounicode{chedescendercyrillic}{04B7}
\pdfglyphtounicode{chedieresiscyrillic}{04F5}
\pdfglyphtounicode{cheharmenian}{0573}
\pdfglyphtounicode{chekhakassiancyrillic}{04CC}
\pdfglyphtounicode{cheverticalstrokecyrillic}{04B9}
\pdfglyphtounicode{chi}{03C7}
\pdfglyphtounicode{chieuchacirclekorean}{3277}
\pdfglyphtounicode{chieuchaparenkorean}{3217}
\pdfglyphtounicode{chieuchcirclekorean}{3269}
\pdfglyphtounicode{chieuchkorean}{314A}
\pdfglyphtounicode{chieuchparenkorean}{3209}
\pdfglyphtounicode{chochangthai}{0E0A}
\pdfglyphtounicode{chochanthai}{0E08}
\pdfglyphtounicode{chochingthai}{0E09}
\pdfglyphtounicode{chochoethai}{0E0C}
\pdfglyphtounicode{chook}{0188}
\pdfglyphtounicode{cieucacirclekorean}{3276}
\pdfglyphtounicode{cieucaparenkorean}{3216}
\pdfglyphtounicode{cieuccirclekorean}{3268}
\pdfglyphtounicode{cieuckorean}{3148}
\pdfglyphtounicode{cieucparenkorean}{3208}
\pdfglyphtounicode{cieucuparenkorean}{321C}
\pdfglyphtounicode{circle}{25CB}
\pdfglyphtounicode{circleR}{00AE}
\pdfglyphtounicode{circleS}{24C8}
\pdfglyphtounicode{circleasterisk}{229B}
\pdfglyphtounicode{circlecopyrt}{20DD}
\pdfglyphtounicode{circledivide}{2298}
\pdfglyphtounicode{circledot}{2299}
\pdfglyphtounicode{circleequal}{229C}
\pdfglyphtounicode{circleminus}{2296}
\pdfglyphtounicode{circlemultiply}{2297}
\pdfglyphtounicode{circleot}{2299}
\pdfglyphtounicode{circleplus}{2295}
\pdfglyphtounicode{circlepostalmark}{3036}
\pdfglyphtounicode{circlering}{229A}
\pdfglyphtounicode{circlewithlefthalfblack}{25D0}
\pdfglyphtounicode{circlewithrighthalfblack}{25D1}
\pdfglyphtounicode{circumflex}{02C6}
\pdfglyphtounicode{circumflexbelowcmb}{032D}
\pdfglyphtounicode{circumflexcmb}{0302}
\pdfglyphtounicode{clear}{2327}
\pdfglyphtounicode{clickalveolar}{01C2}
\pdfglyphtounicode{clickdental}{01C0}
\pdfglyphtounicode{clicklateral}{01C1}
\pdfglyphtounicode{clickretroflex}{01C3}
\pdfglyphtounicode{clockwise}{27F3}
\pdfglyphtounicode{club}{2663}
\pdfglyphtounicode{clubsuitblack}{2663}
\pdfglyphtounicode{clubsuitwhite}{2667}
\pdfglyphtounicode{cmcubedsquare}{33A4}
\pdfglyphtounicode{cmonospace}{FF43}
\pdfglyphtounicode{cmsquaredsquare}{33A0}
\pdfglyphtounicode{coarmenian}{0581}
\pdfglyphtounicode{colon}{003A}
\pdfglyphtounicode{colonmonetary}{20A1}
\pdfglyphtounicode{colonmonospace}{FF1A}
\pdfglyphtounicode{colonsign}{20A1}
\pdfglyphtounicode{colonsmall}{FE55}
\pdfglyphtounicode{colontriangularhalfmod}{02D1}
\pdfglyphtounicode{colontriangularmod}{02D0}
\pdfglyphtounicode{comma}{002C}
\pdfglyphtounicode{commaabovecmb}{0313}
\pdfglyphtounicode{commaaboverightcmb}{0315}
\pdfglyphtounicode{commaaccent}{F6C3}
\pdfglyphtounicode{commaarabic}{060C}
\pdfglyphtounicode{commaarmenian}{055D}
\pdfglyphtounicode{commainferior}{002C}
\pdfglyphtounicode{commamonospace}{FF0C}
\pdfglyphtounicode{commareversedabovecmb}{0314}
\pdfglyphtounicode{commareversedmod}{02BD}
\pdfglyphtounicode{commasmall}{FE50}
\pdfglyphtounicode{commasuperior}{002C}
\pdfglyphtounicode{commaturnedabovecmb}{0312}
\pdfglyphtounicode{commaturnedmod}{02BB}
\pdfglyphtounicode{compass}{263C}
\pdfglyphtounicode{complement}{2201}
\pdfglyphtounicode{compwordmark}{200C}
\pdfglyphtounicode{congruent}{2245}
\pdfglyphtounicode{contourintegral}{222E}
\pdfglyphtounicode{control}{2303}
\pdfglyphtounicode{controlACK}{0006}
\pdfglyphtounicode{controlBEL}{0007}
\pdfglyphtounicode{controlBS}{0008}
\pdfglyphtounicode{controlCAN}{0018}
\pdfglyphtounicode{controlCR}{000D}
\pdfglyphtounicode{controlDC1}{0011}
\pdfglyphtounicode{controlDC2}{0012}
\pdfglyphtounicode{controlDC3}{0013}
\pdfglyphtounicode{controlDC4}{0014}
\pdfglyphtounicode{controlDEL}{007F}
\pdfglyphtounicode{controlDLE}{0010}
\pdfglyphtounicode{controlEM}{0019}
\pdfglyphtounicode{controlENQ}{0005}
\pdfglyphtounicode{controlEOT}{0004}
\pdfglyphtounicode{controlESC}{001B}
\pdfglyphtounicode{controlETB}{0017}
\pdfglyphtounicode{controlETX}{0003}
\pdfglyphtounicode{controlFF}{000C}
\pdfglyphtounicode{controlFS}{001C}
\pdfglyphtounicode{controlGS}{001D}
\pdfglyphtounicode{controlHT}{0009}
\pdfglyphtounicode{controlLF}{000A}
\pdfglyphtounicode{controlNAK}{0015}
\pdfglyphtounicode{controlRS}{001E}
\pdfglyphtounicode{controlSI}{000F}
\pdfglyphtounicode{controlSO}{000E}
\pdfglyphtounicode{controlSOT}{0002}
\pdfglyphtounicode{controlSTX}{0001}
\pdfglyphtounicode{controlSUB}{001A}
\pdfglyphtounicode{controlSYN}{0016}
\pdfglyphtounicode{controlUS}{001F}
\pdfglyphtounicode{controlVT}{000B}
\pdfglyphtounicode{coproduct}{2A3F}
\pdfglyphtounicode{copyright}{00A9}
\pdfglyphtounicode{copyrightsans}{00A9}
\pdfglyphtounicode{copyrightserif}{00A9}
\pdfglyphtounicode{cornerbracketleft}{300C}
\pdfglyphtounicode{cornerbracketlefthalfwidth}{FF62}
\pdfglyphtounicode{cornerbracketleftvertical}{FE41}
\pdfglyphtounicode{cornerbracketright}{300D}
\pdfglyphtounicode{cornerbracketrighthalfwidth}{FF63}
\pdfglyphtounicode{cornerbracketrightvertical}{FE42}
\pdfglyphtounicode{corporationsquare}{337F}
\pdfglyphtounicode{cosquare}{33C7}
\pdfglyphtounicode{coverkgsquare}{33C6}
\pdfglyphtounicode{cparen}{249E}
\pdfglyphtounicode{cruzeiro}{20A2}
\pdfglyphtounicode{cstretched}{0297}
\pdfglyphtounicode{ct}{0063 0074}
\pdfglyphtounicode{curlyand}{22CF}
\pdfglyphtounicode{curlyleft}{21AB}
\pdfglyphtounicode{curlyor}{22CE}
\pdfglyphtounicode{curlyright}{21AC}
\pdfglyphtounicode{currency}{00A4}
\pdfglyphtounicode{cwm}{200C}
\pdfglyphtounicode{cyrBreve}{02D8}
\pdfglyphtounicode{cyrFlex}{00A0 0311}
\pdfglyphtounicode{cyrbreve}{02D8}
\pdfglyphtounicode{cyrflex}{00A0 0311}
\pdfglyphtounicode{d}{0064}
\pdfglyphtounicode{daarmenian}{0564}
\pdfglyphtounicode{dabengali}{09A6}
\pdfglyphtounicode{dadarabic}{0636}
\pdfglyphtounicode{dadeva}{0926}
\pdfglyphtounicode{dadfinalarabic}{FEBE}
\pdfglyphtounicode{dadinitialarabic}{FEBF}
\pdfglyphtounicode{dadmedialarabic}{FEC0}
\pdfglyphtounicode{dagesh}{05BC}
\pdfglyphtounicode{dageshhebrew}{05BC}
\pdfglyphtounicode{dagger}{2020}
\pdfglyphtounicode{daggerdbl}{2021}
\pdfglyphtounicode{dagujarati}{0AA6}
\pdfglyphtounicode{dagurmukhi}{0A26}
\pdfglyphtounicode{dahiragana}{3060}
\pdfglyphtounicode{dakatakana}{30C0}
\pdfglyphtounicode{dalarabic}{062F}
\pdfglyphtounicode{dalet}{05D3}
\pdfglyphtounicode{daletdagesh}{FB33}
\pdfglyphtounicode{daletdageshhebrew}{FB33}
\pdfglyphtounicode{daleth}{2138}
\pdfglyphtounicode{dalethatafpatah}{05D3 05B2}
\pdfglyphtounicode{dalethatafpatahhebrew}{05D3 05B2}
\pdfglyphtounicode{dalethatafsegol}{05D3 05B1}
\pdfglyphtounicode{dalethatafsegolhebrew}{05D3 05B1}
\pdfglyphtounicode{dalethebrew}{05D3}
\pdfglyphtounicode{dalethiriq}{05D3 05B4}
\pdfglyphtounicode{dalethiriqhebrew}{05D3 05B4}
\pdfglyphtounicode{daletholam}{05D3 05B9}
\pdfglyphtounicode{daletholamhebrew}{05D3 05B9}
\pdfglyphtounicode{daletpatah}{05D3 05B7}
\pdfglyphtounicode{daletpatahhebrew}{05D3 05B7}
\pdfglyphtounicode{daletqamats}{05D3 05B8}
\pdfglyphtounicode{daletqamatshebrew}{05D3 05B8}
\pdfglyphtounicode{daletqubuts}{05D3 05BB}
\pdfglyphtounicode{daletqubutshebrew}{05D3 05BB}
\pdfglyphtounicode{daletsegol}{05D3 05B6}
\pdfglyphtounicode{daletsegolhebrew}{05D3 05B6}
\pdfglyphtounicode{daletsheva}{05D3 05B0}
\pdfglyphtounicode{daletshevahebrew}{05D3 05B0}
\pdfglyphtounicode{dalettsere}{05D3 05B5}
\pdfglyphtounicode{dalettserehebrew}{05D3 05B5}
\pdfglyphtounicode{dalfinalarabic}{FEAA}
\pdfglyphtounicode{dammaarabic}{064F}
\pdfglyphtounicode{dammalowarabic}{064F}
\pdfglyphtounicode{dammatanaltonearabic}{064C}
\pdfglyphtounicode{dammatanarabic}{064C}
\pdfglyphtounicode{danda}{0964}
\pdfglyphtounicode{dargahebrew}{05A7}
\pdfglyphtounicode{dargalefthebrew}{05A7}
\pdfglyphtounicode{dasiapneumatacyrilliccmb}{0485}
\pdfglyphtounicode{dbar}{0111}
\pdfglyphtounicode{dblGrave}{00A0 030F}
\pdfglyphtounicode{dblanglebracketleft}{300A}
\pdfglyphtounicode{dblanglebracketleftvertical}{FE3D}
\pdfglyphtounicode{dblanglebracketright}{300B}
\pdfglyphtounicode{dblanglebracketrightvertical}{FE3E}
\pdfglyphtounicode{dblarchinvertedbelowcmb}{032B}
\pdfglyphtounicode{dblarrowdwn}{21CA}
\pdfglyphtounicode{dblarrowheadleft}{219E}
\pdfglyphtounicode{dblarrowheadright}{21A0}
\pdfglyphtounicode{dblarrowleft}{21D4}
\pdfglyphtounicode{dblarrowright}{21D2}
\pdfglyphtounicode{dblarrowup}{21C8}
\pdfglyphtounicode{dblbracketleft}{27E6}
\pdfglyphtounicode{dblbracketright}{27E7}
\pdfglyphtounicode{dbldanda}{0965}
\pdfglyphtounicode{dblgrave}{00A0 030F}
\pdfglyphtounicode{dblgravecmb}{030F}
\pdfglyphtounicode{dblintegral}{222C}
\pdfglyphtounicode{dbllowline}{2017}
\pdfglyphtounicode{dbllowlinecmb}{0333}
\pdfglyphtounicode{dbloverlinecmb}{033F}
\pdfglyphtounicode{dblprimemod}{02BA}
\pdfglyphtounicode{dblverticalbar}{2016}
\pdfglyphtounicode{dblverticallineabovecmb}{030E}
\pdfglyphtounicode{dbopomofo}{3109}
\pdfglyphtounicode{dbsquare}{33C8}
\pdfglyphtounicode{dcaron}{010F}
\pdfglyphtounicode{dcedilla}{1E11}
\pdfglyphtounicode{dcircle}{24D3}
\pdfglyphtounicode{dcircumflexbelow}{1E13}
\pdfglyphtounicode{dcroat}{0111}
\pdfglyphtounicode{ddabengali}{09A1}
\pdfglyphtounicode{ddadeva}{0921}
\pdfglyphtounicode{ddagujarati}{0AA1}
\pdfglyphtounicode{ddagurmukhi}{0A21}
\pdfglyphtounicode{ddalarabic}{0688}
\pdfglyphtounicode{ddalfinalarabic}{FB89}
\pdfglyphtounicode{dddhadeva}{095C}
\pdfglyphtounicode{ddhabengali}{09A2}
\pdfglyphtounicode{ddhadeva}{0922}
\pdfglyphtounicode{ddhagujarati}{0AA2}
\pdfglyphtounicode{ddhagurmukhi}{0A22}
\pdfglyphtounicode{ddotaccent}{1E0B}
\pdfglyphtounicode{ddotbelow}{1E0D}
\pdfglyphtounicode{decimalseparatorarabic}{066B}
\pdfglyphtounicode{decimalseparatorpersian}{066B}
\pdfglyphtounicode{decyrillic}{0434}
\pdfglyphtounicode{defines}{225C}
\pdfglyphtounicode{degree}{00B0}
\pdfglyphtounicode{dehihebrew}{05AD}
\pdfglyphtounicode{dehiragana}{3067}
\pdfglyphtounicode{deicoptic}{03EF}
\pdfglyphtounicode{dekatakana}{30C7}
\pdfglyphtounicode{deleteleft}{232B}
\pdfglyphtounicode{deleteright}{2326}
\pdfglyphtounicode{delta}{03B4}
\pdfglyphtounicode{deltaturned}{018D}
\pdfglyphtounicode{denominatorminusonenumeratorbengali}{09F8}
\pdfglyphtounicode{dezh}{02A4}
\pdfglyphtounicode{dhabengali}{09A7}
\pdfglyphtounicode{dhadeva}{0927}
\pdfglyphtounicode{dhagujarati}{0AA7}
\pdfglyphtounicode{dhagurmukhi}{0A27}
\pdfglyphtounicode{dhook}{0257}
\pdfglyphtounicode{dialytikatonos}{0385}
\pdfglyphtounicode{dialytikatonoscmb}{0344}
\pdfglyphtounicode{diamond}{2666}
\pdfglyphtounicode{diamondmath}{22C4}
\pdfglyphtounicode{diamondsolid}{2666}
\pdfglyphtounicode{diamondsuitwhite}{2662}
\pdfglyphtounicode{dieresis}{00A8}
\pdfglyphtounicode{dieresisacute}{00A0 0308 0301}
\pdfglyphtounicode{dieresisbelowcmb}{0324}
\pdfglyphtounicode{dieresiscmb}{0308}
\pdfglyphtounicode{dieresisgrave}{00A0 0308 0300}
\pdfglyphtounicode{dieresistonos}{0385}
\pdfglyphtounicode{difference}{224F}
\pdfglyphtounicode{dihiragana}{3062}
\pdfglyphtounicode{dikatakana}{30C2}
\pdfglyphtounicode{dittomark}{3003}
\pdfglyphtounicode{divide}{00F7}
\pdfglyphtounicode{dividemultiply}{22C7}
\pdfglyphtounicode{divides}{2223}
\pdfglyphtounicode{divisionslash}{2215}
\pdfglyphtounicode{djecyrillic}{0452}
\pdfglyphtounicode{dkshade}{2593}
\pdfglyphtounicode{dlinebelow}{1E0F}
\pdfglyphtounicode{dlsquare}{3397}
\pdfglyphtounicode{dmacron}{0111}
\pdfglyphtounicode{dmonospace}{FF44}
\pdfglyphtounicode{dnblock}{2584}
\pdfglyphtounicode{dochadathai}{0E0E}
\pdfglyphtounicode{dodekthai}{0E14}
\pdfglyphtounicode{dohiragana}{3069}
\pdfglyphtounicode{dokatakana}{30C9}
\pdfglyphtounicode{dollar}{0024}
\pdfglyphtounicode{dollarinferior}{0024}
\pdfglyphtounicode{dollarmonospace}{FF04}
\pdfglyphtounicode{dollaroldstyle}{0024}
\pdfglyphtounicode{dollarsmall}{FE69}
\pdfglyphtounicode{dollarsuperior}{0024}
\pdfglyphtounicode{dong}{20AB}
\pdfglyphtounicode{dorusquare}{3326}
\pdfglyphtounicode{dotaccent}{02D9}
\pdfglyphtounicode{dotaccentcmb}{0307}
\pdfglyphtounicode{dotbelowcmb}{0323}
\pdfglyphtounicode{dotbelowcomb}{0323}
\pdfglyphtounicode{dotkatakana}{30FB}
\pdfglyphtounicode{dotlessi}{0131}
\pdfglyphtounicode{dotlessj}{0237}
\pdfglyphtounicode{dotlessjstrokehook}{0284}
\pdfglyphtounicode{dotmath}{22C5}
\pdfglyphtounicode{dotplus}{2214}
\pdfglyphtounicode{dottedcircle}{25CC}
\pdfglyphtounicode{doubleyodpatah}{FB1F}
\pdfglyphtounicode{doubleyodpatahhebrew}{FB1F}
\pdfglyphtounicode{downfall}{22CE}
\pdfglyphtounicode{downslope}{29F9}
\pdfglyphtounicode{downtackbelowcmb}{031E}
\pdfglyphtounicode{downtackmod}{02D5}
\pdfglyphtounicode{dparen}{249F}
\pdfglyphtounicode{dsuperior}{0064}
\pdfglyphtounicode{dtail}{0256}
\pdfglyphtounicode{dtopbar}{018C}
\pdfglyphtounicode{duhiragana}{3065}
\pdfglyphtounicode{dukatakana}{30C5}
\pdfglyphtounicode{dz}{01F3}
\pdfglyphtounicode{dzaltone}{02A3}
\pdfglyphtounicode{dzcaron}{01C6}
\pdfglyphtounicode{dzcurl}{02A5}
\pdfglyphtounicode{dzeabkhasiancyrillic}{04E1}
\pdfglyphtounicode{dzecyrillic}{0455}
\pdfglyphtounicode{dzhecyrillic}{045F}
\pdfglyphtounicode{e}{0065}
\pdfglyphtounicode{eacute}{00E9}
\pdfglyphtounicode{earth}{2641}
\pdfglyphtounicode{ebengali}{098F}
\pdfglyphtounicode{ebopomofo}{311C}
\pdfglyphtounicode{ebreve}{0115}
\pdfglyphtounicode{ecandradeva}{090D}
\pdfglyphtounicode{ecandragujarati}{0A8D}
\pdfglyphtounicode{ecandravowelsigndeva}{0945}
\pdfglyphtounicode{ecandravowelsigngujarati}{0AC5}
\pdfglyphtounicode{ecaron}{011B}
\pdfglyphtounicode{ecedillabreve}{1E1D}
\pdfglyphtounicode{echarmenian}{0565}
\pdfglyphtounicode{echyiwnarmenian}{0587}
\pdfglyphtounicode{ecircle}{24D4}
\pdfglyphtounicode{ecircumflex}{00EA}
\pdfglyphtounicode{ecircumflexacute}{1EBF}
\pdfglyphtounicode{ecircumflexbelow}{1E19}
\pdfglyphtounicode{ecircumflexdotbelow}{1EC7}
\pdfglyphtounicode{ecircumflexgrave}{1EC1}
\pdfglyphtounicode{ecircumflexhookabove}{1EC3}
\pdfglyphtounicode{ecircumflextilde}{1EC5}
\pdfglyphtounicode{ecyrillic}{0454}
\pdfglyphtounicode{edblgrave}{0205}
\pdfglyphtounicode{edeva}{090F}
\pdfglyphtounicode{edieresis}{00EB}
\pdfglyphtounicode{edot}{0117}
\pdfglyphtounicode{edotaccent}{0117}
\pdfglyphtounicode{edotbelow}{1EB9}
\pdfglyphtounicode{eegurmukhi}{0A0F}
\pdfglyphtounicode{eematragurmukhi}{0A47}
\pdfglyphtounicode{efcyrillic}{0444}
\pdfglyphtounicode{egrave}{00E8}
\pdfglyphtounicode{egujarati}{0A8F}
\pdfglyphtounicode{eharmenian}{0567}
\pdfglyphtounicode{ehbopomofo}{311D}
\pdfglyphtounicode{ehiragana}{3048}
\pdfglyphtounicode{ehookabove}{1EBB}
\pdfglyphtounicode{eibopomofo}{311F}
\pdfglyphtounicode{eight}{0038}
\pdfglyphtounicode{eightarabic}{0668}
\pdfglyphtounicode{eightbengali}{09EE}
\pdfglyphtounicode{eightcircle}{2467}
\pdfglyphtounicode{eightcircleinversesansserif}{2791}
\pdfglyphtounicode{eightdeva}{096E}
\pdfglyphtounicode{eighteencircle}{2471}
\pdfglyphtounicode{eighteenparen}{2485}
\pdfglyphtounicode{eighteenperiod}{2499}
\pdfglyphtounicode{eightgujarati}{0AEE}
\pdfglyphtounicode{eightgurmukhi}{0A6E}
\pdfglyphtounicode{eighthackarabic}{0668}
\pdfglyphtounicode{eighthangzhou}{3028}
\pdfglyphtounicode{eighthnotebeamed}{266B}
\pdfglyphtounicode{eightideographicparen}{3227}
\pdfglyphtounicode{eightinferior}{2088}
\pdfglyphtounicode{eightmonospace}{FF18}
\pdfglyphtounicode{eightoldstyle}{0038}
\pdfglyphtounicode{eightparen}{247B}
\pdfglyphtounicode{eightperiod}{248F}
\pdfglyphtounicode{eightpersian}{06F8}
\pdfglyphtounicode{eightroman}{2177}
\pdfglyphtounicode{eightsuperior}{2078}
\pdfglyphtounicode{eightthai}{0E58}
\pdfglyphtounicode{einvertedbreve}{0207}
\pdfglyphtounicode{eiotifiedcyrillic}{0465}
\pdfglyphtounicode{ekatakana}{30A8}
\pdfglyphtounicode{ekatakanahalfwidth}{FF74}
\pdfglyphtounicode{ekonkargurmukhi}{0A74}
\pdfglyphtounicode{ekorean}{3154}
\pdfglyphtounicode{elcyrillic}{043B}
\pdfglyphtounicode{element}{2208}
\pdfglyphtounicode{elevencircle}{246A}
\pdfglyphtounicode{elevenparen}{247E}
\pdfglyphtounicode{elevenperiod}{2492}
\pdfglyphtounicode{elevenroman}{217A}
\pdfglyphtounicode{ellipsis}{2026}
\pdfglyphtounicode{ellipsisvertical}{22EE}
\pdfglyphtounicode{emacron}{0113}
\pdfglyphtounicode{emacronacute}{1E17}
\pdfglyphtounicode{emacrongrave}{1E15}
\pdfglyphtounicode{emcyrillic}{043C}
\pdfglyphtounicode{emdash}{2014}
\pdfglyphtounicode{emdashvertical}{FE31}
\pdfglyphtounicode{emonospace}{FF45}
\pdfglyphtounicode{emphasismarkarmenian}{055B}
\pdfglyphtounicode{emptyset}{2205}
\pdfglyphtounicode{enbopomofo}{3123}
\pdfglyphtounicode{encyrillic}{043D}
\pdfglyphtounicode{endash}{2013}
\pdfglyphtounicode{endashvertical}{FE32}
\pdfglyphtounicode{endescendercyrillic}{04A3}
\pdfglyphtounicode{eng}{014B}
\pdfglyphtounicode{engbopomofo}{3125}
\pdfglyphtounicode{enghecyrillic}{04A5}
\pdfglyphtounicode{enhookcyrillic}{04C8}
\pdfglyphtounicode{enspace}{2002}
\pdfglyphtounicode{eogonek}{0119}
\pdfglyphtounicode{eokorean}{3153}
\pdfglyphtounicode{eopen}{025B}
\pdfglyphtounicode{eopenclosed}{029A}
\pdfglyphtounicode{eopenreversed}{025C}
\pdfglyphtounicode{eopenreversedclosed}{025E}
\pdfglyphtounicode{eopenreversedhook}{025D}
\pdfglyphtounicode{eparen}{24A0}
\pdfglyphtounicode{epsilon}{03B5}
\pdfglyphtounicode{epsilon1}{03F5}
\pdfglyphtounicode{epsiloninv}{03F6}
\pdfglyphtounicode{epsilontonos}{03AD}
\pdfglyphtounicode{equal}{003D}
\pdfglyphtounicode{equaldotleftright}{2252}
\pdfglyphtounicode{equaldotrightleft}{2253}
\pdfglyphtounicode{equalmonospace}{FF1D}
\pdfglyphtounicode{equalorfollows}{22DF}
\pdfglyphtounicode{equalorgreater}{2A96}
\pdfglyphtounicode{equalorless}{2A95}
\pdfglyphtounicode{equalorprecedes}{22DE}
\pdfglyphtounicode{equalorsimilar}{2242}
\pdfglyphtounicode{equalsdots}{2251}
\pdfglyphtounicode{equalsmall}{FE66}
\pdfglyphtounicode{equalsuperior}{207C}
\pdfglyphtounicode{equivalence}{2261}
\pdfglyphtounicode{equivasymptotic}{224D}
\pdfglyphtounicode{erbopomofo}{3126}
\pdfglyphtounicode{ercyrillic}{0440}
\pdfglyphtounicode{ereversed}{0258}
\pdfglyphtounicode{ereversedcyrillic}{044D}
\pdfglyphtounicode{escyrillic}{0441}
\pdfglyphtounicode{esdescendercyrillic}{04AB}
\pdfglyphtounicode{esh}{0283}
\pdfglyphtounicode{eshcurl}{0286}
\pdfglyphtounicode{eshortdeva}{090E}
\pdfglyphtounicode{eshortvowelsigndeva}{0946}
\pdfglyphtounicode{eshreversedloop}{01AA}
\pdfglyphtounicode{eshsquatreversed}{0285}
\pdfglyphtounicode{esmallhiragana}{3047}
\pdfglyphtounicode{esmallkatakana}{30A7}
\pdfglyphtounicode{esmallkatakanahalfwidth}{FF6A}
\pdfglyphtounicode{estimated}{212E}
\pdfglyphtounicode{esuperior}{0065}
\pdfglyphtounicode{eta}{03B7}
\pdfglyphtounicode{etarmenian}{0568}
\pdfglyphtounicode{etatonos}{03AE}
\pdfglyphtounicode{eth}{00F0}
\pdfglyphtounicode{etilde}{1EBD}
\pdfglyphtounicode{etildebelow}{1E1B}
\pdfglyphtounicode{etnahtafoukhhebrew}{0591}
\pdfglyphtounicode{etnahtafoukhlefthebrew}{0591}
\pdfglyphtounicode{etnahtahebrew}{0591}
\pdfglyphtounicode{etnahtalefthebrew}{0591}
\pdfglyphtounicode{eturned}{01DD}
\pdfglyphtounicode{eukorean}{3161}
\pdfglyphtounicode{euro}{20AC}
\pdfglyphtounicode{evowelsignbengali}{09C7}
\pdfglyphtounicode{evowelsigndeva}{0947}
\pdfglyphtounicode{evowelsigngujarati}{0AC7}
\pdfglyphtounicode{exclam}{0021}
\pdfglyphtounicode{exclamarmenian}{055C}
\pdfglyphtounicode{exclamdbl}{203C}
\pdfglyphtounicode{exclamdown}{00A1}
\pdfglyphtounicode{exclamdownsmall}{00A1}
\pdfglyphtounicode{exclammonospace}{FF01}
\pdfglyphtounicode{exclamsmall}{0021}
\pdfglyphtounicode{existential}{2203}
\pdfglyphtounicode{ezh}{0292}
\pdfglyphtounicode{ezhcaron}{01EF}
\pdfglyphtounicode{ezhcurl}{0293}
\pdfglyphtounicode{ezhreversed}{01B9}
\pdfglyphtounicode{ezhtail}{01BA}
\pdfglyphtounicode{f}{0066}
\pdfglyphtounicode{fadeva}{095E}
\pdfglyphtounicode{fagurmukhi}{0A5E}
\pdfglyphtounicode{fahrenheit}{2109}
\pdfglyphtounicode{fathaarabic}{064E}
\pdfglyphtounicode{fathalowarabic}{064E}
\pdfglyphtounicode{fathatanarabic}{064B}
\pdfglyphtounicode{fbopomofo}{3108}
\pdfglyphtounicode{fcircle}{24D5}
\pdfglyphtounicode{fdotaccent}{1E1F}
\pdfglyphtounicode{feharabic}{0641}
\pdfglyphtounicode{feharmenian}{0586}
\pdfglyphtounicode{fehfinalarabic}{FED2}
\pdfglyphtounicode{fehinitialarabic}{FED3}
\pdfglyphtounicode{fehmedialarabic}{FED4}
\pdfglyphtounicode{feicoptic}{03E5}
\pdfglyphtounicode{female}{2640}
\pdfglyphtounicode{ff}{0066 0066}
\pdfglyphtounicode{ffi}{0066 0066 0069}
\pdfglyphtounicode{ffl}{0066 0066 006C}
\pdfglyphtounicode{fi}{0066 0069}
\pdfglyphtounicode{fifteencircle}{246E}
\pdfglyphtounicode{fifteenparen}{2482}
\pdfglyphtounicode{fifteenperiod}{2496}
\pdfglyphtounicode{figuredash}{2012}
\pdfglyphtounicode{filledbox}{25A0}
\pdfglyphtounicode{filledrect}{25AC}
\pdfglyphtounicode{finalkaf}{05DA}
\pdfglyphtounicode{finalkafdagesh}{FB3A}
\pdfglyphtounicode{finalkafdageshhebrew}{FB3A}
\pdfglyphtounicode{finalkafhebrew}{05DA}
\pdfglyphtounicode{finalkafqamats}{05DA 05B8}
\pdfglyphtounicode{finalkafqamatshebrew}{05DA 05B8}
\pdfglyphtounicode{finalkafsheva}{05DA 05B0}
\pdfglyphtounicode{finalkafshevahebrew}{05DA 05B0}
\pdfglyphtounicode{finalmem}{05DD}
\pdfglyphtounicode{finalmemhebrew}{05DD}
\pdfglyphtounicode{finalnun}{05DF}
\pdfglyphtounicode{finalnunhebrew}{05DF}
\pdfglyphtounicode{finalpe}{05E3}
\pdfglyphtounicode{finalpehebrew}{05E3}
\pdfglyphtounicode{finaltsadi}{05E5}
\pdfglyphtounicode{finaltsadihebrew}{05E5}
\pdfglyphtounicode{firsttonechinese}{02C9}
\pdfglyphtounicode{fisheye}{25C9}
\pdfglyphtounicode{fitacyrillic}{0473}
\pdfglyphtounicode{five}{0035}
\pdfglyphtounicode{fivearabic}{0665}
\pdfglyphtounicode{fivebengali}{09EB}
\pdfglyphtounicode{fivecircle}{2464}
\pdfglyphtounicode{fivecircleinversesansserif}{278E}
\pdfglyphtounicode{fivedeva}{096B}
\pdfglyphtounicode{fiveeighths}{215D}
\pdfglyphtounicode{fivegujarati}{0AEB}
\pdfglyphtounicode{fivegurmukhi}{0A6B}
\pdfglyphtounicode{fivehackarabic}{0665}
\pdfglyphtounicode{fivehangzhou}{3025}
\pdfglyphtounicode{fiveideographicparen}{3224}
\pdfglyphtounicode{fiveinferior}{2085}
\pdfglyphtounicode{fivemonospace}{FF15}
\pdfglyphtounicode{fiveoldstyle}{0035}
\pdfglyphtounicode{fiveparen}{2478}
\pdfglyphtounicode{fiveperiod}{248C}
\pdfglyphtounicode{fivepersian}{06F5}
\pdfglyphtounicode{fiveroman}{2174}
\pdfglyphtounicode{fivesuperior}{2075}
\pdfglyphtounicode{fivethai}{0E55}
\pdfglyphtounicode{fl}{0066 006C}
\pdfglyphtounicode{flat}{266D}
\pdfglyphtounicode{floorleft}{230A}
\pdfglyphtounicode{floorright}{230B}
\pdfglyphtounicode{florin}{0192}
\pdfglyphtounicode{fmonospace}{FF46}
\pdfglyphtounicode{fmsquare}{3399}
\pdfglyphtounicode{fofanthai}{0E1F}
\pdfglyphtounicode{fofathai}{0E1D}
\pdfglyphtounicode{follownotdbleqv}{2ABA}
\pdfglyphtounicode{follownotslnteql}{2AB6}
\pdfglyphtounicode{followornoteqvlnt}{22E9}
\pdfglyphtounicode{follows}{227B}
\pdfglyphtounicode{followsequal}{2AB0}
\pdfglyphtounicode{followsorcurly}{227D}
\pdfglyphtounicode{followsorequal}{227F}
\pdfglyphtounicode{fongmanthai}{0E4F}
\pdfglyphtounicode{forall}{2200}
\pdfglyphtounicode{forces}{22A9}
\pdfglyphtounicode{forcesbar}{22AA}
\pdfglyphtounicode{fork}{22D4}
\pdfglyphtounicode{four}{0034}
\pdfglyphtounicode{fourarabic}{0664}
\pdfglyphtounicode{fourbengali}{09EA}
\pdfglyphtounicode{fourcircle}{2463}
\pdfglyphtounicode{fourcircleinversesansserif}{278D}
\pdfglyphtounicode{fourdeva}{096A}
\pdfglyphtounicode{fourgujarati}{0AEA}
\pdfglyphtounicode{fourgurmukhi}{0A6A}
\pdfglyphtounicode{fourhackarabic}{0664}
\pdfglyphtounicode{fourhangzhou}{3024}
\pdfglyphtounicode{fourideographicparen}{3223}
\pdfglyphtounicode{fourinferior}{2084}
\pdfglyphtounicode{fourmonospace}{FF14}
\pdfglyphtounicode{fournumeratorbengali}{09F7}
\pdfglyphtounicode{fouroldstyle}{0034}
\pdfglyphtounicode{fourparen}{2477}
\pdfglyphtounicode{fourperiod}{248B}
\pdfglyphtounicode{fourpersian}{06F4}
\pdfglyphtounicode{fourroman}{2173}
\pdfglyphtounicode{foursuperior}{2074}
\pdfglyphtounicode{fourteencircle}{246D}
\pdfglyphtounicode{fourteenparen}{2481}
\pdfglyphtounicode{fourteenperiod}{2495}
\pdfglyphtounicode{fourthai}{0E54}
\pdfglyphtounicode{fourthtonechinese}{02CB}
\pdfglyphtounicode{fparen}{24A1}
\pdfglyphtounicode{fraction}{2044}
\pdfglyphtounicode{franc}{20A3}
\pdfglyphtounicode{frown}{2322}
\pdfglyphtounicode{g}{0067}
\pdfglyphtounicode{gabengali}{0997}
\pdfglyphtounicode{gacute}{01F5}
\pdfglyphtounicode{gadeva}{0917}
\pdfglyphtounicode{gafarabic}{06AF}
\pdfglyphtounicode{gaffinalarabic}{FB93}
\pdfglyphtounicode{gafinitialarabic}{FB94}
\pdfglyphtounicode{gafmedialarabic}{FB95}
\pdfglyphtounicode{gagujarati}{0A97}
\pdfglyphtounicode{gagurmukhi}{0A17}
\pdfglyphtounicode{gahiragana}{304C}
\pdfglyphtounicode{gakatakana}{30AC}
\pdfglyphtounicode{gamma}{03B3}
\pdfglyphtounicode{gammalatinsmall}{0263}
\pdfglyphtounicode{gammasuperior}{02E0}
\pdfglyphtounicode{gangiacoptic}{03EB}
\pdfglyphtounicode{gbopomofo}{310D}
\pdfglyphtounicode{gbreve}{011F}
\pdfglyphtounicode{gcaron}{01E7}
\pdfglyphtounicode{gcedilla}{0123}
\pdfglyphtounicode{gcircle}{24D6}
\pdfglyphtounicode{gcircumflex}{011D}
\pdfglyphtounicode{gcommaaccent}{0123}
\pdfglyphtounicode{gdot}{0121}
\pdfglyphtounicode{gdotaccent}{0121}
\pdfglyphtounicode{gecyrillic}{0433}
\pdfglyphtounicode{gehiragana}{3052}
\pdfglyphtounicode{gekatakana}{30B2}
\pdfglyphtounicode{geomequivalent}{224E}
\pdfglyphtounicode{geometricallyequal}{2251}
\pdfglyphtounicode{gereshaccenthebrew}{059C}
\pdfglyphtounicode{gereshhebrew}{05F3}
\pdfglyphtounicode{gereshmuqdamhebrew}{059D}
\pdfglyphtounicode{germandbls}{00DF}
\pdfglyphtounicode{gershayimaccenthebrew}{059E}
\pdfglyphtounicode{gershayimhebrew}{05F4}
\pdfglyphtounicode{getamark}{3013}
\pdfglyphtounicode{ghabengali}{0998}
\pdfglyphtounicode{ghadarmenian}{0572}
\pdfglyphtounicode{ghadeva}{0918}
\pdfglyphtounicode{ghagujarati}{0A98}
\pdfglyphtounicode{ghagurmukhi}{0A18}
\pdfglyphtounicode{ghainarabic}{063A}
\pdfglyphtounicode{ghainfinalarabic}{FECE}
\pdfglyphtounicode{ghaininitialarabic}{FECF}
\pdfglyphtounicode{ghainmedialarabic}{FED0}
\pdfglyphtounicode{ghemiddlehookcyrillic}{0495}
\pdfglyphtounicode{ghestrokecyrillic}{0493}
\pdfglyphtounicode{gheupturncyrillic}{0491}
\pdfglyphtounicode{ghhadeva}{095A}
\pdfglyphtounicode{ghhagurmukhi}{0A5A}
\pdfglyphtounicode{ghook}{0260}
\pdfglyphtounicode{ghzsquare}{3393}
\pdfglyphtounicode{gihiragana}{304E}
\pdfglyphtounicode{gikatakana}{30AE}
\pdfglyphtounicode{gimarmenian}{0563}
\pdfglyphtounicode{gimel}{05D2}
\pdfglyphtounicode{gimeldagesh}{FB32}
\pdfglyphtounicode{gimeldageshhebrew}{FB32}
\pdfglyphtounicode{gimelhebrew}{05D2}
\pdfglyphtounicode{gjecyrillic}{0453}
\pdfglyphtounicode{glottalinvertedstroke}{01BE}
\pdfglyphtounicode{glottalstop}{0294}
\pdfglyphtounicode{glottalstopinverted}{0296}
\pdfglyphtounicode{glottalstopmod}{02C0}
\pdfglyphtounicode{glottalstopreversed}{0295}
\pdfglyphtounicode{glottalstopreversedmod}{02C1}
\pdfglyphtounicode{glottalstopreversedsuperior}{02E4}
\pdfglyphtounicode{glottalstopstroke}{02A1}
\pdfglyphtounicode{glottalstopstrokereversed}{02A2}
\pdfglyphtounicode{gmacron}{1E21}
\pdfglyphtounicode{gmonospace}{FF47}
\pdfglyphtounicode{gohiragana}{3054}
\pdfglyphtounicode{gokatakana}{30B4}
\pdfglyphtounicode{gparen}{24A2}
\pdfglyphtounicode{gpasquare}{33AC}
\pdfglyphtounicode{gradient}{2207}
\pdfglyphtounicode{grave}{0060}
\pdfglyphtounicode{gravebelowcmb}{0316}
\pdfglyphtounicode{gravecmb}{0300}
\pdfglyphtounicode{gravecomb}{0300}
\pdfglyphtounicode{gravedeva}{0953}
\pdfglyphtounicode{gravelowmod}{02CE}
\pdfglyphtounicode{gravemonospace}{FF40}
\pdfglyphtounicode{gravetonecmb}{0340}
\pdfglyphtounicode{greater}{003E}
\pdfglyphtounicode{greaterdbleqlless}{2A8C}
\pdfglyphtounicode{greaterdblequal}{2267}
\pdfglyphtounicode{greaterdot}{22D7}
\pdfglyphtounicode{greaterequal}{2265}
\pdfglyphtounicode{greaterequalorless}{22DB}
\pdfglyphtounicode{greaterlessequal}{22DB}
\pdfglyphtounicode{greatermonospace}{FF1E}
\pdfglyphtounicode{greatermuch}{226B}
\pdfglyphtounicode{greaternotdblequal}{2A8A}
\pdfglyphtounicode{greaternotequal}{2A88}
\pdfglyphtounicode{greaterorapproxeql}{2A86}
\pdfglyphtounicode{greaterorequalslant}{2A7E}
\pdfglyphtounicode{greaterorequivalent}{2273}
\pdfglyphtounicode{greaterorless}{2277}
\pdfglyphtounicode{greaterornotdbleql}{2269}
\pdfglyphtounicode{greaterornotequal}{2269}
\pdfglyphtounicode{greaterorsimilar}{2273}
\pdfglyphtounicode{greateroverequal}{2267}
\pdfglyphtounicode{greatersmall}{FE65}
\pdfglyphtounicode{gscript}{0261}
\pdfglyphtounicode{gstroke}{01E5}
\pdfglyphtounicode{guhiragana}{3050}
\pdfglyphtounicode{guillemotleft}{00AB}
\pdfglyphtounicode{guillemotright}{00BB}
\pdfglyphtounicode{guilsinglleft}{2039}
\pdfglyphtounicode{guilsinglright}{203A}
\pdfglyphtounicode{gukatakana}{30B0}
\pdfglyphtounicode{guramusquare}{3318}
\pdfglyphtounicode{gysquare}{33C9}
\pdfglyphtounicode{h}{0068}
\pdfglyphtounicode{haabkhasiancyrillic}{04A9}
\pdfglyphtounicode{haaltonearabic}{06C1}
\pdfglyphtounicode{habengali}{09B9}
\pdfglyphtounicode{hadescendercyrillic}{04B3}
\pdfglyphtounicode{hadeva}{0939}
\pdfglyphtounicode{hagujarati}{0AB9}
\pdfglyphtounicode{hagurmukhi}{0A39}
\pdfglyphtounicode{haharabic}{062D}
\pdfglyphtounicode{hahfinalarabic}{FEA2}
\pdfglyphtounicode{hahinitialarabic}{FEA3}
\pdfglyphtounicode{hahiragana}{306F}
\pdfglyphtounicode{hahmedialarabic}{FEA4}
\pdfglyphtounicode{haitusquare}{332A}
\pdfglyphtounicode{hakatakana}{30CF}
\pdfglyphtounicode{hakatakanahalfwidth}{FF8A}
\pdfglyphtounicode{halantgurmukhi}{0A4D}
\pdfglyphtounicode{hamzaarabic}{0621}
\pdfglyphtounicode{hamzadammaarabic}{0621 064F}
\pdfglyphtounicode{hamzadammatanarabic}{0621 064C}
\pdfglyphtounicode{hamzafathaarabic}{0621 064E}
\pdfglyphtounicode{hamzafathatanarabic}{0621 064B}
\pdfglyphtounicode{hamzalowarabic}{0621}
\pdfglyphtounicode{hamzalowkasraarabic}{0621 0650}
\pdfglyphtounicode{hamzalowkasratanarabic}{0621 064D}
\pdfglyphtounicode{hamzasukunarabic}{0621 0652}
\pdfglyphtounicode{hangulfiller}{3164}
\pdfglyphtounicode{hardsigncyrillic}{044A}
\pdfglyphtounicode{harpoondownleft}{21C3}
\pdfglyphtounicode{harpoondownright}{21C2}
\pdfglyphtounicode{harpoonleftbarbup}{21BC}
\pdfglyphtounicode{harpoonleftright}{21CC}
\pdfglyphtounicode{harpoonrightbarbup}{21C0}
\pdfglyphtounicode{harpoonrightleft}{21CB}
\pdfglyphtounicode{harpoonupleft}{21BF}
\pdfglyphtounicode{harpoonupright}{21BE}
\pdfglyphtounicode{hasquare}{33CA}
\pdfglyphtounicode{hatafpatah}{05B2}
\pdfglyphtounicode{hatafpatah16}{05B2}
\pdfglyphtounicode{hatafpatah23}{05B2}
\pdfglyphtounicode{hatafpatah2f}{05B2}
\pdfglyphtounicode{hatafpatahhebrew}{05B2}
\pdfglyphtounicode{hatafpatahnarrowhebrew}{05B2}
\pdfglyphtounicode{hatafpatahquarterhebrew}{05B2}
\pdfglyphtounicode{hatafpatahwidehebrew}{05B2}
\pdfglyphtounicode{hatafqamats}{05B3}
\pdfglyphtounicode{hatafqamats1b}{05B3}
\pdfglyphtounicode{hatafqamats28}{05B3}
\pdfglyphtounicode{hatafqamats34}{05B3}
\pdfglyphtounicode{hatafqamatshebrew}{05B3}
\pdfglyphtounicode{hatafqamatsnarrowhebrew}{05B3}
\pdfglyphtounicode{hatafqamatsquarterhebrew}{05B3}
\pdfglyphtounicode{hatafqamatswidehebrew}{05B3}
\pdfglyphtounicode{hatafsegol}{05B1}
\pdfglyphtounicode{hatafsegol17}{05B1}
\pdfglyphtounicode{hatafsegol24}{05B1}
\pdfglyphtounicode{hatafsegol30}{05B1}
\pdfglyphtounicode{hatafsegolhebrew}{05B1}
\pdfglyphtounicode{hatafsegolnarrowhebrew}{05B1}
\pdfglyphtounicode{hatafsegolquarterhebrew}{05B1}
\pdfglyphtounicode{hatafsegolwidehebrew}{05B1}
\pdfglyphtounicode{hbar}{0127}
\pdfglyphtounicode{hbopomofo}{310F}
\pdfglyphtounicode{hbrevebelow}{1E2B}
\pdfglyphtounicode{hcedilla}{1E29}
\pdfglyphtounicode{hcircle}{24D7}
\pdfglyphtounicode{hcircumflex}{0125}
\pdfglyphtounicode{hdieresis}{1E27}
\pdfglyphtounicode{hdotaccent}{1E23}
\pdfglyphtounicode{hdotbelow}{1E25}
\pdfglyphtounicode{he}{05D4}
\pdfglyphtounicode{heart}{2665}
\pdfglyphtounicode{heartsuitblack}{2665}
\pdfglyphtounicode{heartsuitwhite}{2661}
\pdfglyphtounicode{hedagesh}{FB34}
\pdfglyphtounicode{hedageshhebrew}{FB34}
\pdfglyphtounicode{hehaltonearabic}{06C1}
\pdfglyphtounicode{heharabic}{0647}
\pdfglyphtounicode{hehebrew}{05D4}
\pdfglyphtounicode{hehfinalaltonearabic}{FBA7}
\pdfglyphtounicode{hehfinalalttwoarabic}{FEEA}
\pdfglyphtounicode{hehfinalarabic}{FEEA}
\pdfglyphtounicode{hehhamzaabovefinalarabic}{FBA5}
\pdfglyphtounicode{hehhamzaaboveisolatedarabic}{FBA4}
\pdfglyphtounicode{hehinitialaltonearabic}{FBA8}
\pdfglyphtounicode{hehinitialarabic}{FEEB}
\pdfglyphtounicode{hehiragana}{3078}
\pdfglyphtounicode{hehmedialaltonearabic}{FBA9}
\pdfglyphtounicode{hehmedialarabic}{FEEC}
\pdfglyphtounicode{heiseierasquare}{337B}
\pdfglyphtounicode{hekatakana}{30D8}
\pdfglyphtounicode{hekatakanahalfwidth}{FF8D}
\pdfglyphtounicode{hekutaarusquare}{3336}
\pdfglyphtounicode{henghook}{0267}
\pdfglyphtounicode{herutusquare}{3339}
\pdfglyphtounicode{het}{05D7}
\pdfglyphtounicode{hethebrew}{05D7}
\pdfglyphtounicode{hhook}{0266}
\pdfglyphtounicode{hhooksuperior}{02B1}
\pdfglyphtounicode{hieuhacirclekorean}{327B}
\pdfglyphtounicode{hieuhaparenkorean}{321B}
\pdfglyphtounicode{hieuhcirclekorean}{326D}
\pdfglyphtounicode{hieuhkorean}{314E}
\pdfglyphtounicode{hieuhparenkorean}{320D}
\pdfglyphtounicode{hihiragana}{3072}
\pdfglyphtounicode{hikatakana}{30D2}
\pdfglyphtounicode{hikatakanahalfwidth}{FF8B}
\pdfglyphtounicode{hiriq}{05B4}
\pdfglyphtounicode{hiriq14}{05B4}
\pdfglyphtounicode{hiriq21}{05B4}
\pdfglyphtounicode{hiriq2d}{05B4}
\pdfglyphtounicode{hiriqhebrew}{05B4}
\pdfglyphtounicode{hiriqnarrowhebrew}{05B4}
\pdfglyphtounicode{hiriqquarterhebrew}{05B4}
\pdfglyphtounicode{hiriqwidehebrew}{05B4}
\pdfglyphtounicode{hlinebelow}{1E96}
\pdfglyphtounicode{hmonospace}{FF48}
\pdfglyphtounicode{hoarmenian}{0570}
\pdfglyphtounicode{hohipthai}{0E2B}
\pdfglyphtounicode{hohiragana}{307B}
\pdfglyphtounicode{hokatakana}{30DB}
\pdfglyphtounicode{hokatakanahalfwidth}{FF8E}
\pdfglyphtounicode{holam}{05B9}
\pdfglyphtounicode{holam19}{05B9}
\pdfglyphtounicode{holam26}{05B9}
\pdfglyphtounicode{holam32}{05B9}
\pdfglyphtounicode{holamhebrew}{05B9}
\pdfglyphtounicode{holamnarrowhebrew}{05B9}
\pdfglyphtounicode{holamquarterhebrew}{05B9}
\pdfglyphtounicode{holamwidehebrew}{05B9}
\pdfglyphtounicode{honokhukthai}{0E2E}
\pdfglyphtounicode{hookabovecomb}{0309}
\pdfglyphtounicode{hookcmb}{0309}
\pdfglyphtounicode{hookpalatalizedbelowcmb}{0321}
\pdfglyphtounicode{hookretroflexbelowcmb}{0322}
\pdfglyphtounicode{hoonsquare}{3342}
\pdfglyphtounicode{horicoptic}{03E9}
\pdfglyphtounicode{horizontalbar}{2015}
\pdfglyphtounicode{horncmb}{031B}
\pdfglyphtounicode{hotsprings}{2668}
\pdfglyphtounicode{house}{2302}
\pdfglyphtounicode{hparen}{24A3}
\pdfglyphtounicode{hsuperior}{02B0}
\pdfglyphtounicode{hturned}{0265}
\pdfglyphtounicode{huhiragana}{3075}
\pdfglyphtounicode{huiitosquare}{3333}
\pdfglyphtounicode{hukatakana}{30D5}
\pdfglyphtounicode{hukatakanahalfwidth}{FF8C}
\pdfglyphtounicode{hungarumlaut}{02DD}
\pdfglyphtounicode{hungarumlautcmb}{030B}
\pdfglyphtounicode{hv}{0195}
\pdfglyphtounicode{hyphen}{002D}
\pdfglyphtounicode{hyphenchar}{002D}
\pdfglyphtounicode{hypheninferior}{002D}
\pdfglyphtounicode{hyphenmonospace}{FF0D}
\pdfglyphtounicode{hyphensmall}{FE63}
\pdfglyphtounicode{hyphensuperior}{002D}
\pdfglyphtounicode{hyphentwo}{2010}
\pdfglyphtounicode{i}{0069}
\pdfglyphtounicode{iacute}{00ED}
\pdfglyphtounicode{iacyrillic}{044F}
\pdfglyphtounicode{ibengali}{0987}
\pdfglyphtounicode{ibopomofo}{3127}
\pdfglyphtounicode{ibreve}{012D}
\pdfglyphtounicode{icaron}{01D0}
\pdfglyphtounicode{icircle}{24D8}
\pdfglyphtounicode{icircumflex}{00EE}
\pdfglyphtounicode{icyrillic}{0456}
\pdfglyphtounicode{idblgrave}{0209}
\pdfglyphtounicode{ideographearthcircle}{328F}
\pdfglyphtounicode{ideographfirecircle}{328B}
\pdfglyphtounicode{ideographicallianceparen}{323F}
\pdfglyphtounicode{ideographiccallparen}{323A}
\pdfglyphtounicode{ideographiccentrecircle}{32A5}
\pdfglyphtounicode{ideographicclose}{3006}
\pdfglyphtounicode{ideographiccomma}{3001}
\pdfglyphtounicode{ideographiccommaleft}{FF64}
\pdfglyphtounicode{ideographiccongratulationparen}{3237}
\pdfglyphtounicode{ideographiccorrectcircle}{32A3}
\pdfglyphtounicode{ideographicearthparen}{322F}
\pdfglyphtounicode{ideographicenterpriseparen}{323D}
\pdfglyphtounicode{ideographicexcellentcircle}{329D}
\pdfglyphtounicode{ideographicfestivalparen}{3240}
\pdfglyphtounicode{ideographicfinancialcircle}{3296}
\pdfglyphtounicode{ideographicfinancialparen}{3236}
\pdfglyphtounicode{ideographicfireparen}{322B}
\pdfglyphtounicode{ideographichaveparen}{3232}
\pdfglyphtounicode{ideographichighcircle}{32A4}
\pdfglyphtounicode{ideographiciterationmark}{3005}
\pdfglyphtounicode{ideographiclaborcircle}{3298}
\pdfglyphtounicode{ideographiclaborparen}{3238}
\pdfglyphtounicode{ideographicleftcircle}{32A7}
\pdfglyphtounicode{ideographiclowcircle}{32A6}
\pdfglyphtounicode{ideographicmedicinecircle}{32A9}
\pdfglyphtounicode{ideographicmetalparen}{322E}
\pdfglyphtounicode{ideographicmoonparen}{322A}
\pdfglyphtounicode{ideographicnameparen}{3234}
\pdfglyphtounicode{ideographicperiod}{3002}
\pdfglyphtounicode{ideographicprintcircle}{329E}
\pdfglyphtounicode{ideographicreachparen}{3243}
\pdfglyphtounicode{ideographicrepresentparen}{3239}
\pdfglyphtounicode{ideographicresourceparen}{323E}
\pdfglyphtounicode{ideographicrightcircle}{32A8}
\pdfglyphtounicode{ideographicsecretcircle}{3299}
\pdfglyphtounicode{ideographicselfparen}{3242}
\pdfglyphtounicode{ideographicsocietyparen}{3233}
\pdfglyphtounicode{ideographicspace}{3000}
\pdfglyphtounicode{ideographicspecialparen}{3235}
\pdfglyphtounicode{ideographicstockparen}{3231}
\pdfglyphtounicode{ideographicstudyparen}{323B}
\pdfglyphtounicode{ideographicsunparen}{3230}
\pdfglyphtounicode{ideographicsuperviseparen}{323C}
\pdfglyphtounicode{ideographicwaterparen}{322C}
\pdfglyphtounicode{ideographicwoodparen}{322D}
\pdfglyphtounicode{ideographiczero}{3007}
\pdfglyphtounicode{ideographmetalcircle}{328E}
\pdfglyphtounicode{ideographmooncircle}{328A}
\pdfglyphtounicode{ideographnamecircle}{3294}
\pdfglyphtounicode{ideographsuncircle}{3290}
\pdfglyphtounicode{ideographwatercircle}{328C}
\pdfglyphtounicode{ideographwoodcircle}{328D}
\pdfglyphtounicode{ideva}{0907}
\pdfglyphtounicode{idieresis}{00EF}
\pdfglyphtounicode{idieresisacute}{1E2F}
\pdfglyphtounicode{idieresiscyrillic}{04E5}
\pdfglyphtounicode{idotbelow}{1ECB}
\pdfglyphtounicode{iebrevecyrillic}{04D7}
\pdfglyphtounicode{iecyrillic}{0435}
\pdfglyphtounicode{ieungacirclekorean}{3275}
\pdfglyphtounicode{ieungaparenkorean}{3215}
\pdfglyphtounicode{ieungcirclekorean}{3267}
\pdfglyphtounicode{ieungkorean}{3147}
\pdfglyphtounicode{ieungparenkorean}{3207}
\pdfglyphtounicode{igrave}{00EC}
\pdfglyphtounicode{igujarati}{0A87}
\pdfglyphtounicode{igurmukhi}{0A07}
\pdfglyphtounicode{ihiragana}{3044}
\pdfglyphtounicode{ihookabove}{1EC9}
\pdfglyphtounicode{iibengali}{0988}
\pdfglyphtounicode{iicyrillic}{0438}
\pdfglyphtounicode{iideva}{0908}
\pdfglyphtounicode{iigujarati}{0A88}
\pdfglyphtounicode{iigurmukhi}{0A08}
\pdfglyphtounicode{iimatragurmukhi}{0A40}
\pdfglyphtounicode{iinvertedbreve}{020B}
\pdfglyphtounicode{iishortcyrillic}{0439}
\pdfglyphtounicode{iivowelsignbengali}{09C0}
\pdfglyphtounicode{iivowelsigndeva}{0940}
\pdfglyphtounicode{iivowelsigngujarati}{0AC0}
\pdfglyphtounicode{ij}{0133}
\pdfglyphtounicode{ikatakana}{30A4}
\pdfglyphtounicode{ikatakanahalfwidth}{FF72}
\pdfglyphtounicode{ikorean}{3163}
\pdfglyphtounicode{ilde}{02DC}
\pdfglyphtounicode{iluyhebrew}{05AC}
\pdfglyphtounicode{imacron}{012B}
\pdfglyphtounicode{imacroncyrillic}{04E3}
\pdfglyphtounicode{imageorapproximatelyequal}{2253}
\pdfglyphtounicode{imatragurmukhi}{0A3F}
\pdfglyphtounicode{imonospace}{FF49}
\pdfglyphtounicode{increment}{2206}
\pdfglyphtounicode{infinity}{221E}
\pdfglyphtounicode{iniarmenian}{056B}
\pdfglyphtounicode{integerdivide}{2216}
\pdfglyphtounicode{integral}{222B}
\pdfglyphtounicode{integralbottom}{2321}
\pdfglyphtounicode{integralbt}{2321}
\pdfglyphtounicode{integralex}{F8F5}
\pdfglyphtounicode{integraltop}{2320}
\pdfglyphtounicode{integraltp}{2320}
\pdfglyphtounicode{intercal}{22BA}
\pdfglyphtounicode{interrobang}{203D}
\pdfglyphtounicode{interrobangdown}{2E18}
\pdfglyphtounicode{intersection}{2229}
\pdfglyphtounicode{intersectiondbl}{22D2}
\pdfglyphtounicode{intersectionsq}{2293}
\pdfglyphtounicode{intisquare}{3305}
\pdfglyphtounicode{invbullet}{25D8}
\pdfglyphtounicode{invcircle}{25D9}
\pdfglyphtounicode{invsmileface}{263B}
\pdfglyphtounicode{iocyrillic}{0451}
\pdfglyphtounicode{iogonek}{012F}
\pdfglyphtounicode{iota}{03B9}
\pdfglyphtounicode{iotadieresis}{03CA}
\pdfglyphtounicode{iotadieresistonos}{0390}
\pdfglyphtounicode{iotalatin}{0269}
\pdfglyphtounicode{iotatonos}{03AF}
\pdfglyphtounicode{iparen}{24A4}
\pdfglyphtounicode{irigurmukhi}{0A72}
\pdfglyphtounicode{ismallhiragana}{3043}
\pdfglyphtounicode{ismallkatakana}{30A3}
\pdfglyphtounicode{ismallkatakanahalfwidth}{FF68}
\pdfglyphtounicode{issharbengali}{09FA}
\pdfglyphtounicode{istroke}{0268}
\pdfglyphtounicode{isuperior}{0069}
\pdfglyphtounicode{iterationhiragana}{309D}
\pdfglyphtounicode{iterationkatakana}{30FD}
\pdfglyphtounicode{itilde}{0129}
\pdfglyphtounicode{itildebelow}{1E2D}
\pdfglyphtounicode{iubopomofo}{3129}
\pdfglyphtounicode{iucyrillic}{044E}
\pdfglyphtounicode{ivowelsignbengali}{09BF}
\pdfglyphtounicode{ivowelsigndeva}{093F}
\pdfglyphtounicode{ivowelsigngujarati}{0ABF}
\pdfglyphtounicode{izhitsacyrillic}{0475}
\pdfglyphtounicode{izhitsadblgravecyrillic}{0477}
\pdfglyphtounicode{j}{006A}
\pdfglyphtounicode{jaarmenian}{0571}
\pdfglyphtounicode{jabengali}{099C}
\pdfglyphtounicode{jadeva}{091C}
\pdfglyphtounicode{jagujarati}{0A9C}
\pdfglyphtounicode{jagurmukhi}{0A1C}
\pdfglyphtounicode{jbopomofo}{3110}
\pdfglyphtounicode{jcaron}{01F0}
\pdfglyphtounicode{jcircle}{24D9}
\pdfglyphtounicode{jcircumflex}{0135}
\pdfglyphtounicode{jcrossedtail}{029D}
\pdfglyphtounicode{jdotlessstroke}{025F}
\pdfglyphtounicode{jecyrillic}{0458}
\pdfglyphtounicode{jeemarabic}{062C}
\pdfglyphtounicode{jeemfinalarabic}{FE9E}
\pdfglyphtounicode{jeeminitialarabic}{FE9F}
\pdfglyphtounicode{jeemmedialarabic}{FEA0}
\pdfglyphtounicode{jeharabic}{0698}
\pdfglyphtounicode{jehfinalarabic}{FB8B}
\pdfglyphtounicode{jhabengali}{099D}
\pdfglyphtounicode{jhadeva}{091D}
\pdfglyphtounicode{jhagujarati}{0A9D}
\pdfglyphtounicode{jhagurmukhi}{0A1D}
\pdfglyphtounicode{jheharmenian}{057B}
\pdfglyphtounicode{jis}{3004}
\pdfglyphtounicode{jmonospace}{FF4A}
\pdfglyphtounicode{jparen}{24A5}
\pdfglyphtounicode{jsuperior}{02B2}
\pdfglyphtounicode{k}{006B}
\pdfglyphtounicode{kabashkircyrillic}{04A1}
\pdfglyphtounicode{kabengali}{0995}
\pdfglyphtounicode{kacute}{1E31}
\pdfglyphtounicode{kacyrillic}{043A}
\pdfglyphtounicode{kadescendercyrillic}{049B}
\pdfglyphtounicode{kadeva}{0915}
\pdfglyphtounicode{kaf}{05DB}
\pdfglyphtounicode{kafarabic}{0643}
\pdfglyphtounicode{kafdagesh}{FB3B}
\pdfglyphtounicode{kafdageshhebrew}{FB3B}
\pdfglyphtounicode{kaffinalarabic}{FEDA}
\pdfglyphtounicode{kafhebrew}{05DB}
\pdfglyphtounicode{kafinitialarabic}{FEDB}
\pdfglyphtounicode{kafmedialarabic}{FEDC}
\pdfglyphtounicode{kafrafehebrew}{FB4D}
\pdfglyphtounicode{kagujarati}{0A95}
\pdfglyphtounicode{kagurmukhi}{0A15}
\pdfglyphtounicode{kahiragana}{304B}
\pdfglyphtounicode{kahookcyrillic}{04C4}
\pdfglyphtounicode{kakatakana}{30AB}
\pdfglyphtounicode{kakatakanahalfwidth}{FF76}
\pdfglyphtounicode{kappa}{03BA}
\pdfglyphtounicode{kappasymbolgreek}{03F0}
\pdfglyphtounicode{kapyeounmieumkorean}{3171}
\pdfglyphtounicode{kapyeounphieuphkorean}{3184}
\pdfglyphtounicode{kapyeounpieupkorean}{3178}
\pdfglyphtounicode{kapyeounssangpieupkorean}{3179}
\pdfglyphtounicode{karoriisquare}{330D}
\pdfglyphtounicode{kashidaautoarabic}{0640}
\pdfglyphtounicode{kashidaautonosidebearingarabic}{0640}
\pdfglyphtounicode{kasmallkatakana}{30F5}
\pdfglyphtounicode{kasquare}{3384}
\pdfglyphtounicode{kasraarabic}{0650}
\pdfglyphtounicode{kasratanarabic}{064D}
\pdfglyphtounicode{kastrokecyrillic}{049F}
\pdfglyphtounicode{katahiraprolongmarkhalfwidth}{FF70}
\pdfglyphtounicode{kaverticalstrokecyrillic}{049D}
\pdfglyphtounicode{kbopomofo}{310E}
\pdfglyphtounicode{kcalsquare}{3389}
\pdfglyphtounicode{kcaron}{01E9}
\pdfglyphtounicode{kcedilla}{0137}
\pdfglyphtounicode{kcircle}{24DA}
\pdfglyphtounicode{kcommaaccent}{0137}
\pdfglyphtounicode{kdotbelow}{1E33}
\pdfglyphtounicode{keharmenian}{0584}
\pdfglyphtounicode{kehiragana}{3051}
\pdfglyphtounicode{kekatakana}{30B1}
\pdfglyphtounicode{kekatakanahalfwidth}{FF79}
\pdfglyphtounicode{kenarmenian}{056F}
\pdfglyphtounicode{kesmallkatakana}{30F6}
\pdfglyphtounicode{kgreenlandic}{0138}
\pdfglyphtounicode{khabengali}{0996}
\pdfglyphtounicode{khacyrillic}{0445}
\pdfglyphtounicode{khadeva}{0916}
\pdfglyphtounicode{khagujarati}{0A96}
\pdfglyphtounicode{khagurmukhi}{0A16}
\pdfglyphtounicode{khaharabic}{062E}
\pdfglyphtounicode{khahfinalarabic}{FEA6}
\pdfglyphtounicode{khahinitialarabic}{FEA7}
\pdfglyphtounicode{khahmedialarabic}{FEA8}
\pdfglyphtounicode{kheicoptic}{03E7}
\pdfglyphtounicode{khhadeva}{0959}
\pdfglyphtounicode{khhagurmukhi}{0A59}
\pdfglyphtounicode{khieukhacirclekorean}{3278}
\pdfglyphtounicode{khieukhaparenkorean}{3218}
\pdfglyphtounicode{khieukhcirclekorean}{326A}
\pdfglyphtounicode{khieukhkorean}{314B}
\pdfglyphtounicode{khieukhparenkorean}{320A}
\pdfglyphtounicode{khokhaithai}{0E02}
\pdfglyphtounicode{khokhonthai}{0E05}
\pdfglyphtounicode{khokhuatthai}{0E03}
\pdfglyphtounicode{khokhwaithai}{0E04}
\pdfglyphtounicode{khomutthai}{0E5B}
\pdfglyphtounicode{khook}{0199}
\pdfglyphtounicode{khorakhangthai}{0E06}
\pdfglyphtounicode{khzsquare}{3391}
\pdfglyphtounicode{kihiragana}{304D}
\pdfglyphtounicode{kikatakana}{30AD}
\pdfglyphtounicode{kikatakanahalfwidth}{FF77}
\pdfglyphtounicode{kiroguramusquare}{3315}
\pdfglyphtounicode{kiromeetorusquare}{3316}
\pdfglyphtounicode{kirosquare}{3314}
\pdfglyphtounicode{kiyeokacirclekorean}{326E}
\pdfglyphtounicode{kiyeokaparenkorean}{320E}
\pdfglyphtounicode{kiyeokcirclekorean}{3260}
\pdfglyphtounicode{kiyeokkorean}{3131}
\pdfglyphtounicode{kiyeokparenkorean}{3200}
\pdfglyphtounicode{kiyeoksioskorean}{3133}
\pdfglyphtounicode{kjecyrillic}{045C}
\pdfglyphtounicode{klinebelow}{1E35}
\pdfglyphtounicode{klsquare}{3398}
\pdfglyphtounicode{kmcubedsquare}{33A6}
\pdfglyphtounicode{kmonospace}{FF4B}
\pdfglyphtounicode{kmsquaredsquare}{33A2}
\pdfglyphtounicode{kohiragana}{3053}
\pdfglyphtounicode{kohmsquare}{33C0}
\pdfglyphtounicode{kokaithai}{0E01}
\pdfglyphtounicode{kokatakana}{30B3}
\pdfglyphtounicode{kokatakanahalfwidth}{FF7A}
\pdfglyphtounicode{kooposquare}{331E}
\pdfglyphtounicode{koppacyrillic}{0481}
\pdfglyphtounicode{koreanstandardsymbol}{327F}
\pdfglyphtounicode{koroniscmb}{0343}
\pdfglyphtounicode{kparen}{24A6}
\pdfglyphtounicode{kpasquare}{33AA}
\pdfglyphtounicode{ksicyrillic}{046F}
\pdfglyphtounicode{ktsquare}{33CF}
\pdfglyphtounicode{kturned}{029E}
\pdfglyphtounicode{kuhiragana}{304F}
\pdfglyphtounicode{kukatakana}{30AF}
\pdfglyphtounicode{kukatakanahalfwidth}{FF78}
\pdfglyphtounicode{kvsquare}{33B8}
\pdfglyphtounicode{kwsquare}{33BE}
\pdfglyphtounicode{l}{006C}
\pdfglyphtounicode{labengali}{09B2}
\pdfglyphtounicode{lacute}{013A}
\pdfglyphtounicode{ladeva}{0932}
\pdfglyphtounicode{lagujarati}{0AB2}
\pdfglyphtounicode{lagurmukhi}{0A32}
\pdfglyphtounicode{lakkhangyaothai}{0E45}
\pdfglyphtounicode{lamaleffinalarabic}{FEFC}
\pdfglyphtounicode{lamalefhamzaabovefinalarabic}{FEF8}
\pdfglyphtounicode{lamalefhamzaaboveisolatedarabic}{FEF7}
\pdfglyphtounicode{lamalefhamzabelowfinalarabic}{FEFA}
\pdfglyphtounicode{lamalefhamzabelowisolatedarabic}{FEF9}
\pdfglyphtounicode{lamalefisolatedarabic}{FEFB}
\pdfglyphtounicode{lamalefmaddaabovefinalarabic}{FEF6}
\pdfglyphtounicode{lamalefmaddaaboveisolatedarabic}{FEF5}
\pdfglyphtounicode{lamarabic}{0644}
\pdfglyphtounicode{lambda}{03BB}
\pdfglyphtounicode{lambdastroke}{019B}
\pdfglyphtounicode{lamed}{05DC}
\pdfglyphtounicode{lameddagesh}{FB3C}
\pdfglyphtounicode{lameddageshhebrew}{FB3C}
\pdfglyphtounicode{lamedhebrew}{05DC}
\pdfglyphtounicode{lamedholam}{05DC 05B9}
\pdfglyphtounicode{lamedholamdagesh}{05DC 05B9 05BC}
\pdfglyphtounicode{lamedholamdageshhebrew}{05DC 05B9 05BC}
\pdfglyphtounicode{lamedholamhebrew}{05DC 05B9}
\pdfglyphtounicode{lamfinalarabic}{FEDE}
\pdfglyphtounicode{lamhahinitialarabic}{FCCA}
\pdfglyphtounicode{laminitialarabic}{FEDF}
\pdfglyphtounicode{lamjeeminitialarabic}{FCC9}
\pdfglyphtounicode{lamkhahinitialarabic}{FCCB}
\pdfglyphtounicode{lamlamhehisolatedarabic}{FDF2}
\pdfglyphtounicode{lammedialarabic}{FEE0}
\pdfglyphtounicode{lammeemhahinitialarabic}{FD88}
\pdfglyphtounicode{lammeeminitialarabic}{FCCC}
\pdfglyphtounicode{lammeemjeeminitialarabic}{FEDF FEE4 FEA0}
\pdfglyphtounicode{lammeemkhahinitialarabic}{FEDF FEE4 FEA8}
\pdfglyphtounicode{largecircle}{25EF}
\pdfglyphtounicode{latticetop}{22A4}
\pdfglyphtounicode{lbar}{019A}
\pdfglyphtounicode{lbelt}{026C}
\pdfglyphtounicode{lbopomofo}{310C}
\pdfglyphtounicode{lcaron}{013E}
\pdfglyphtounicode{lcedilla}{013C}
\pdfglyphtounicode{lcircle}{24DB}
\pdfglyphtounicode{lcircumflexbelow}{1E3D}
\pdfglyphtounicode{lcommaaccent}{013C}
\pdfglyphtounicode{ldot}{0140}
\pdfglyphtounicode{ldotaccent}{0140}
\pdfglyphtounicode{ldotbelow}{1E37}
\pdfglyphtounicode{ldotbelowmacron}{1E39}
\pdfglyphtounicode{leftangleabovecmb}{031A}
\pdfglyphtounicode{lefttackbelowcmb}{0318}
\pdfglyphtounicode{less}{003C}
\pdfglyphtounicode{lessdbleqlgreater}{2A8B}
\pdfglyphtounicode{lessdblequal}{2266}
\pdfglyphtounicode{lessdot}{22D6}
\pdfglyphtounicode{lessequal}{2264}
\pdfglyphtounicode{lessequalgreater}{22DA}
\pdfglyphtounicode{lessequalorgreater}{22DA}
\pdfglyphtounicode{lessmonospace}{FF1C}
\pdfglyphtounicode{lessmuch}{226A}
\pdfglyphtounicode{lessnotdblequal}{2A89}
\pdfglyphtounicode{lessnotequal}{2A87}
\pdfglyphtounicode{lessorapproxeql}{2A85}
\pdfglyphtounicode{lessorequalslant}{2A7D}
\pdfglyphtounicode{lessorequivalent}{2272}
\pdfglyphtounicode{lessorgreater}{2276}
\pdfglyphtounicode{lessornotdbleql}{2268}
\pdfglyphtounicode{lessornotequal}{2268}
\pdfglyphtounicode{lessorsimilar}{2272}
\pdfglyphtounicode{lessoverequal}{2266}
\pdfglyphtounicode{lesssmall}{FE64}
\pdfglyphtounicode{lezh}{026E}
\pdfglyphtounicode{lfblock}{258C}
\pdfglyphtounicode{lhookretroflex}{026D}
\pdfglyphtounicode{lira}{20A4}
\pdfglyphtounicode{liwnarmenian}{056C}
\pdfglyphtounicode{lj}{01C9}
\pdfglyphtounicode{ljecyrillic}{0459}
\pdfglyphtounicode{ll}{006C 006C}
\pdfglyphtounicode{lladeva}{0933}
\pdfglyphtounicode{llagujarati}{0AB3}
\pdfglyphtounicode{llinebelow}{1E3B}
\pdfglyphtounicode{llladeva}{0934}
\pdfglyphtounicode{llvocalicbengali}{09E1}
\pdfglyphtounicode{llvocalicdeva}{0961}
\pdfglyphtounicode{llvocalicvowelsignbengali}{09E3}
\pdfglyphtounicode{llvocalicvowelsigndeva}{0963}
\pdfglyphtounicode{lmiddletilde}{026B}
\pdfglyphtounicode{lmonospace}{FF4C}
\pdfglyphtounicode{lmsquare}{33D0}
\pdfglyphtounicode{lochulathai}{0E2C}
\pdfglyphtounicode{logicaland}{2227}
\pdfglyphtounicode{logicalnot}{00AC}
\pdfglyphtounicode{logicalnotreversed}{2310}
\pdfglyphtounicode{logicalor}{2228}
\pdfglyphtounicode{lolingthai}{0E25}
\pdfglyphtounicode{longdbls}{017F 017F}
\pdfglyphtounicode{longs}{017F}
\pdfglyphtounicode{longsh}{017F 0068}
\pdfglyphtounicode{longsi}{017F 0069}
\pdfglyphtounicode{longsl}{017F 006C}
\pdfglyphtounicode{longst}{017F 0074}
\pdfglyphtounicode{lowlinecenterline}{FE4E}
\pdfglyphtounicode{lowlinecmb}{0332}
\pdfglyphtounicode{lowlinedashed}{FE4D}
\pdfglyphtounicode{lozenge}{25CA}
\pdfglyphtounicode{lparen}{24A7}
\pdfglyphtounicode{lscript}{2113}
\pdfglyphtounicode{lslash}{0142}
\pdfglyphtounicode{lsquare}{2113}
\pdfglyphtounicode{lsuperior}{006C}
\pdfglyphtounicode{ltshade}{2591}
\pdfglyphtounicode{luthai}{0E26}
\pdfglyphtounicode{lvocalicbengali}{098C}
\pdfglyphtounicode{lvocalicdeva}{090C}
\pdfglyphtounicode{lvocalicvowelsignbengali}{09E2}
\pdfglyphtounicode{lvocalicvowelsigndeva}{0962}
\pdfglyphtounicode{lxsquare}{33D3}
\pdfglyphtounicode{m}{006D}
\pdfglyphtounicode{mabengali}{09AE}
\pdfglyphtounicode{macron}{00AF}
\pdfglyphtounicode{macronbelowcmb}{0331}
\pdfglyphtounicode{macroncmb}{0304}
\pdfglyphtounicode{macronlowmod}{02CD}
\pdfglyphtounicode{macronmonospace}{FFE3}
\pdfglyphtounicode{macute}{1E3F}
\pdfglyphtounicode{madeva}{092E}
\pdfglyphtounicode{magujarati}{0AAE}
\pdfglyphtounicode{magurmukhi}{0A2E}
\pdfglyphtounicode{mahapakhhebrew}{05A4}
\pdfglyphtounicode{mahapakhlefthebrew}{05A4}
\pdfglyphtounicode{mahiragana}{307E}
\pdfglyphtounicode{maichattawalowleftthai}{F895}
\pdfglyphtounicode{maichattawalowrightthai}{F894}
\pdfglyphtounicode{maichattawathai}{0E4B}
\pdfglyphtounicode{maichattawaupperleftthai}{F893}
\pdfglyphtounicode{maieklowleftthai}{F88C}
\pdfglyphtounicode{maieklowrightthai}{F88B}
\pdfglyphtounicode{maiekthai}{0E48}
\pdfglyphtounicode{maiekupperleftthai}{F88A}
\pdfglyphtounicode{maihanakatleftthai}{F884}
\pdfglyphtounicode{maihanakatthai}{0E31}
\pdfglyphtounicode{maitaikhuleftthai}{F889}
\pdfglyphtounicode{maitaikhuthai}{0E47}
\pdfglyphtounicode{maitholowleftthai}{F88F}
\pdfglyphtounicode{maitholowrightthai}{F88E}
\pdfglyphtounicode{maithothai}{0E49}
\pdfglyphtounicode{maithoupperleftthai}{F88D}
\pdfglyphtounicode{maitrilowleftthai}{F892}
\pdfglyphtounicode{maitrilowrightthai}{F891}
\pdfglyphtounicode{maitrithai}{0E4A}
\pdfglyphtounicode{maitriupperleftthai}{F890}
\pdfglyphtounicode{maiyamokthai}{0E46}
\pdfglyphtounicode{makatakana}{30DE}
\pdfglyphtounicode{makatakanahalfwidth}{FF8F}
\pdfglyphtounicode{male}{2642}
\pdfglyphtounicode{maltesecross}{2720}
\pdfglyphtounicode{mansyonsquare}{3347}
\pdfglyphtounicode{maqafhebrew}{05BE}
\pdfglyphtounicode{mars}{2642}
\pdfglyphtounicode{masoracirclehebrew}{05AF}
\pdfglyphtounicode{masquare}{3383}
\pdfglyphtounicode{mbopomofo}{3107}
\pdfglyphtounicode{mbsquare}{33D4}
\pdfglyphtounicode{mcircle}{24DC}
\pdfglyphtounicode{mcubedsquare}{33A5}
\pdfglyphtounicode{mdotaccent}{1E41}
\pdfglyphtounicode{mdotbelow}{1E43}
\pdfglyphtounicode{measuredangle}{2221}
\pdfglyphtounicode{meemarabic}{0645}
\pdfglyphtounicode{meemfinalarabic}{FEE2}
\pdfglyphtounicode{meeminitialarabic}{FEE3}
\pdfglyphtounicode{meemmedialarabic}{FEE4}
\pdfglyphtounicode{meemmeeminitialarabic}{FCD1}
\pdfglyphtounicode{meemmeemisolatedarabic}{FC48}
\pdfglyphtounicode{meetorusquare}{334D}
\pdfglyphtounicode{mehiragana}{3081}
\pdfglyphtounicode{meizierasquare}{337E}
\pdfglyphtounicode{mekatakana}{30E1}
\pdfglyphtounicode{mekatakanahalfwidth}{FF92}
\pdfglyphtounicode{mem}{05DE}
\pdfglyphtounicode{memdagesh}{FB3E}
\pdfglyphtounicode{memdageshhebrew}{FB3E}
\pdfglyphtounicode{memhebrew}{05DE}
\pdfglyphtounicode{menarmenian}{0574}
\pdfglyphtounicode{merkhahebrew}{05A5}
\pdfglyphtounicode{merkhakefulahebrew}{05A6}
\pdfglyphtounicode{merkhakefulalefthebrew}{05A6}
\pdfglyphtounicode{merkhalefthebrew}{05A5}
\pdfglyphtounicode{mhook}{0271}
\pdfglyphtounicode{mhzsquare}{3392}
\pdfglyphtounicode{middledotkatakanahalfwidth}{FF65}
\pdfglyphtounicode{middot}{00B7}
\pdfglyphtounicode{mieumacirclekorean}{3272}
\pdfglyphtounicode{mieumaparenkorean}{3212}
\pdfglyphtounicode{mieumcirclekorean}{3264}
\pdfglyphtounicode{mieumkorean}{3141}
\pdfglyphtounicode{mieumpansioskorean}{3170}
\pdfglyphtounicode{mieumparenkorean}{3204}
\pdfglyphtounicode{mieumpieupkorean}{316E}
\pdfglyphtounicode{mieumsioskorean}{316F}
\pdfglyphtounicode{mihiragana}{307F}
\pdfglyphtounicode{mikatakana}{30DF}
\pdfglyphtounicode{mikatakanahalfwidth}{FF90}
\pdfglyphtounicode{minus}{2212}
\pdfglyphtounicode{minusbelowcmb}{0320}
\pdfglyphtounicode{minuscircle}{2296}
\pdfglyphtounicode{minusmod}{02D7}
\pdfglyphtounicode{minusplus}{2213}
\pdfglyphtounicode{minute}{2032}
\pdfglyphtounicode{miribaarusquare}{334A}
\pdfglyphtounicode{mirisquare}{3349}
\pdfglyphtounicode{mlonglegturned}{0270}
\pdfglyphtounicode{mlsquare}{3396}
\pdfglyphtounicode{mmcubedsquare}{33A3}
\pdfglyphtounicode{mmonospace}{FF4D}
\pdfglyphtounicode{mmsquaredsquare}{339F}
\pdfglyphtounicode{mohiragana}{3082}
\pdfglyphtounicode{mohmsquare}{33C1}
\pdfglyphtounicode{mokatakana}{30E2}
\pdfglyphtounicode{mokatakanahalfwidth}{FF93}
\pdfglyphtounicode{molsquare}{33D6}
\pdfglyphtounicode{momathai}{0E21}
\pdfglyphtounicode{moverssquare}{33A7}
\pdfglyphtounicode{moverssquaredsquare}{33A8}
\pdfglyphtounicode{mparen}{24A8}
\pdfglyphtounicode{mpasquare}{33AB}
\pdfglyphtounicode{mssquare}{33B3}
\pdfglyphtounicode{msuperior}{006D}
\pdfglyphtounicode{mturned}{026F}
\pdfglyphtounicode{mu}{00B5}
\pdfglyphtounicode{mu1}{00B5}
\pdfglyphtounicode{muasquare}{3382}
\pdfglyphtounicode{muchgreater}{226B}
\pdfglyphtounicode{muchless}{226A}
\pdfglyphtounicode{mufsquare}{338C}
\pdfglyphtounicode{mugreek}{03BC}
\pdfglyphtounicode{mugsquare}{338D}
\pdfglyphtounicode{muhiragana}{3080}
\pdfglyphtounicode{mukatakana}{30E0}
\pdfglyphtounicode{mukatakanahalfwidth}{FF91}
\pdfglyphtounicode{mulsquare}{3395}
\pdfglyphtounicode{multicloseleft}{22C9}
\pdfglyphtounicode{multicloseright}{22CA}
\pdfglyphtounicode{multimap}{22B8}
\pdfglyphtounicode{multiopenleft}{22CB}
\pdfglyphtounicode{multiopenright}{22CC}
\pdfglyphtounicode{multiply}{00D7}
\pdfglyphtounicode{mumsquare}{339B}
\pdfglyphtounicode{munahhebrew}{05A3}
\pdfglyphtounicode{munahlefthebrew}{05A3}
\pdfglyphtounicode{musicalnote}{266A}
\pdfglyphtounicode{musicalnotedbl}{266B}
\pdfglyphtounicode{musicflatsign}{266D}
\pdfglyphtounicode{musicsharpsign}{266F}
\pdfglyphtounicode{mussquare}{33B2}
\pdfglyphtounicode{muvsquare}{33B6}
\pdfglyphtounicode{muwsquare}{33BC}
\pdfglyphtounicode{mvmegasquare}{33B9}
\pdfglyphtounicode{mvsquare}{33B7}
\pdfglyphtounicode{mwmegasquare}{33BF}
\pdfglyphtounicode{mwsquare}{33BD}
\pdfglyphtounicode{n}{006E}
\pdfglyphtounicode{nabengali}{09A8}
\pdfglyphtounicode{nabla}{2207}
\pdfglyphtounicode{nacute}{0144}
\pdfglyphtounicode{nadeva}{0928}
\pdfglyphtounicode{nagujarati}{0AA8}
\pdfglyphtounicode{nagurmukhi}{0A28}
\pdfglyphtounicode{nahiragana}{306A}
\pdfglyphtounicode{nakatakana}{30CA}
\pdfglyphtounicode{nakatakanahalfwidth}{FF85}
\pdfglyphtounicode{nand}{22BC}
\pdfglyphtounicode{napostrophe}{0149}
\pdfglyphtounicode{nasquare}{3381}
\pdfglyphtounicode{natural}{266E}
\pdfglyphtounicode{nbopomofo}{310B}
\pdfglyphtounicode{nbspace}{00A0}
\pdfglyphtounicode{ncaron}{0148}
\pdfglyphtounicode{ncedilla}{0146}
\pdfglyphtounicode{ncircle}{24DD}
\pdfglyphtounicode{ncircumflexbelow}{1E4B}
\pdfglyphtounicode{ncommaaccent}{0146}
\pdfglyphtounicode{ndotaccent}{1E45}
\pdfglyphtounicode{ndotbelow}{1E47}
\pdfglyphtounicode{negationslash}{0338}
\pdfglyphtounicode{nehiragana}{306D}
\pdfglyphtounicode{nekatakana}{30CD}
\pdfglyphtounicode{nekatakanahalfwidth}{FF88}
\pdfglyphtounicode{newsheqelsign}{20AA}
\pdfglyphtounicode{nfsquare}{338B}
\pdfglyphtounicode{ng}{014B}
\pdfglyphtounicode{ngabengali}{0999}
\pdfglyphtounicode{ngadeva}{0919}
\pdfglyphtounicode{ngagujarati}{0A99}
\pdfglyphtounicode{ngagurmukhi}{0A19}
\pdfglyphtounicode{ngonguthai}{0E07}
\pdfglyphtounicode{nhiragana}{3093}
\pdfglyphtounicode{nhookleft}{0272}
\pdfglyphtounicode{nhookretroflex}{0273}
\pdfglyphtounicode{nieunacirclekorean}{326F}
\pdfglyphtounicode{nieunaparenkorean}{320F}
\pdfglyphtounicode{nieuncieuckorean}{3135}
\pdfglyphtounicode{nieuncirclekorean}{3261}
\pdfglyphtounicode{nieunhieuhkorean}{3136}
\pdfglyphtounicode{nieunkorean}{3134}
\pdfglyphtounicode{nieunpansioskorean}{3168}
\pdfglyphtounicode{nieunparenkorean}{3201}
\pdfglyphtounicode{nieunsioskorean}{3167}
\pdfglyphtounicode{nieuntikeutkorean}{3166}
\pdfglyphtounicode{nihiragana}{306B}
\pdfglyphtounicode{nikatakana}{30CB}
\pdfglyphtounicode{nikatakanahalfwidth}{FF86}
\pdfglyphtounicode{nikhahitleftthai}{F899}
\pdfglyphtounicode{nikhahitthai}{0E4D}
\pdfglyphtounicode{nine}{0039}
\pdfglyphtounicode{ninearabic}{0669}
\pdfglyphtounicode{ninebengali}{09EF}
\pdfglyphtounicode{ninecircle}{2468}
\pdfglyphtounicode{ninecircleinversesansserif}{2792}
\pdfglyphtounicode{ninedeva}{096F}
\pdfglyphtounicode{ninegujarati}{0AEF}
\pdfglyphtounicode{ninegurmukhi}{0A6F}
\pdfglyphtounicode{ninehackarabic}{0669}
\pdfglyphtounicode{ninehangzhou}{3029}
\pdfglyphtounicode{nineideographicparen}{3228}
\pdfglyphtounicode{nineinferior}{2089}
\pdfglyphtounicode{ninemonospace}{FF19}
\pdfglyphtounicode{nineoldstyle}{0039}
\pdfglyphtounicode{nineparen}{247C}
\pdfglyphtounicode{nineperiod}{2490}
\pdfglyphtounicode{ninepersian}{06F9}
\pdfglyphtounicode{nineroman}{2178}
\pdfglyphtounicode{ninesuperior}{2079}
\pdfglyphtounicode{nineteencircle}{2472}
\pdfglyphtounicode{nineteenparen}{2486}
\pdfglyphtounicode{nineteenperiod}{249A}
\pdfglyphtounicode{ninethai}{0E59}
\pdfglyphtounicode{nj}{01CC}
\pdfglyphtounicode{njecyrillic}{045A}
\pdfglyphtounicode{nkatakana}{30F3}
\pdfglyphtounicode{nkatakanahalfwidth}{FF9D}
\pdfglyphtounicode{nlegrightlong}{019E}
\pdfglyphtounicode{nlinebelow}{1E49}
\pdfglyphtounicode{nmonospace}{FF4E}
\pdfglyphtounicode{nmsquare}{339A}
\pdfglyphtounicode{nnabengali}{09A3}
\pdfglyphtounicode{nnadeva}{0923}
\pdfglyphtounicode{nnagujarati}{0AA3}
\pdfglyphtounicode{nnagurmukhi}{0A23}
\pdfglyphtounicode{nnnadeva}{0929}
\pdfglyphtounicode{nohiragana}{306E}
\pdfglyphtounicode{nokatakana}{30CE}
\pdfglyphtounicode{nokatakanahalfwidth}{FF89}
\pdfglyphtounicode{nonbreakingspace}{00A0}
\pdfglyphtounicode{nonenthai}{0E13}
\pdfglyphtounicode{nonuthai}{0E19}
\pdfglyphtounicode{noonarabic}{0646}
\pdfglyphtounicode{noonfinalarabic}{FEE6}
\pdfglyphtounicode{noonghunnaarabic}{06BA}
\pdfglyphtounicode{noonghunnafinalarabic}{FB9F}
\pdfglyphtounicode{noonhehinitialarabic}{FEE7 FEEC}
\pdfglyphtounicode{nooninitialarabic}{FEE7}
\pdfglyphtounicode{noonjeeminitialarabic}{FCD2}
\pdfglyphtounicode{noonjeemisolatedarabic}{FC4B}
\pdfglyphtounicode{noonmedialarabic}{FEE8}
\pdfglyphtounicode{noonmeeminitialarabic}{FCD5}
\pdfglyphtounicode{noonmeemisolatedarabic}{FC4E}
\pdfglyphtounicode{noonnoonfinalarabic}{FC8D}
\pdfglyphtounicode{notapproxequal}{2247}
\pdfglyphtounicode{notarrowboth}{21AE}
\pdfglyphtounicode{notarrowleft}{219A}
\pdfglyphtounicode{notarrowright}{219B}
\pdfglyphtounicode{notbar}{2224}
\pdfglyphtounicode{notcontains}{220C}
\pdfglyphtounicode{notdblarrowboth}{21CE}
\pdfglyphtounicode{notdblarrowleft}{21CD}
\pdfglyphtounicode{notdblarrowright}{21CF}
\pdfglyphtounicode{notelement}{2209}
\pdfglyphtounicode{notelementof}{2209}
\pdfglyphtounicode{notequal}{2260}
\pdfglyphtounicode{notexistential}{2204}
\pdfglyphtounicode{notfollows}{2281}
\pdfglyphtounicode{notfollowsoreql}{2AB0 0338}
\pdfglyphtounicode{notforces}{22AE}
\pdfglyphtounicode{notforcesextra}{22AF}
\pdfglyphtounicode{notgreater}{226F}
\pdfglyphtounicode{notgreaterdblequal}{2267 0338}
\pdfglyphtounicode{notgreaterequal}{2271}
\pdfglyphtounicode{notgreaternorequal}{2271}
\pdfglyphtounicode{notgreaternorless}{2279}
\pdfglyphtounicode{notgreaterorslnteql}{2A7E 0338}
\pdfglyphtounicode{notidentical}{2262}
\pdfglyphtounicode{notless}{226E}
\pdfglyphtounicode{notlessdblequal}{2266 0338}
\pdfglyphtounicode{notlessequal}{2270}
\pdfglyphtounicode{notlessnorequal}{2270}
\pdfglyphtounicode{notlessorslnteql}{2A7D 0338}
\pdfglyphtounicode{notparallel}{2226}
\pdfglyphtounicode{notprecedes}{2280}
\pdfglyphtounicode{notprecedesoreql}{2AAF 0338}
\pdfglyphtounicode{notsatisfies}{22AD}
\pdfglyphtounicode{notsimilar}{2241}
\pdfglyphtounicode{notsubset}{2284}
\pdfglyphtounicode{notsubseteql}{2288}
\pdfglyphtounicode{notsubsetordbleql}{2AC5 0338}
\pdfglyphtounicode{notsubsetoreql}{228A}
\pdfglyphtounicode{notsucceeds}{2281}
\pdfglyphtounicode{notsuperset}{2285}
\pdfglyphtounicode{notsuperseteql}{2289}
\pdfglyphtounicode{notsupersetordbleql}{2AC6 0338}
\pdfglyphtounicode{notsupersetoreql}{228B}
\pdfglyphtounicode{nottriangeqlleft}{22EC}
\pdfglyphtounicode{nottriangeqlright}{22ED}
\pdfglyphtounicode{nottriangleleft}{22EA}
\pdfglyphtounicode{nottriangleright}{22EB}
\pdfglyphtounicode{notturnstile}{22AC}
\pdfglyphtounicode{nowarmenian}{0576}
\pdfglyphtounicode{nparen}{24A9}
\pdfglyphtounicode{nssquare}{33B1}
\pdfglyphtounicode{nsuperior}{207F}
\pdfglyphtounicode{ntilde}{00F1}
\pdfglyphtounicode{nu}{03BD}
\pdfglyphtounicode{nuhiragana}{306C}
\pdfglyphtounicode{nukatakana}{30CC}
\pdfglyphtounicode{nukatakanahalfwidth}{FF87}
\pdfglyphtounicode{nuktabengali}{09BC}
\pdfglyphtounicode{nuktadeva}{093C}
\pdfglyphtounicode{nuktagujarati}{0ABC}
\pdfglyphtounicode{nuktagurmukhi}{0A3C}
\pdfglyphtounicode{numbersign}{0023}
\pdfglyphtounicode{numbersignmonospace}{FF03}
\pdfglyphtounicode{numbersignsmall}{FE5F}
\pdfglyphtounicode{numeralsigngreek}{0374}
\pdfglyphtounicode{numeralsignlowergreek}{0375}
\pdfglyphtounicode{numero}{2116}
\pdfglyphtounicode{nun}{05E0}
\pdfglyphtounicode{nundagesh}{FB40}
\pdfglyphtounicode{nundageshhebrew}{FB40}
\pdfglyphtounicode{nunhebrew}{05E0}
\pdfglyphtounicode{nvsquare}{33B5}
\pdfglyphtounicode{nwsquare}{33BB}
\pdfglyphtounicode{nyabengali}{099E}
\pdfglyphtounicode{nyadeva}{091E}
\pdfglyphtounicode{nyagujarati}{0A9E}
\pdfglyphtounicode{nyagurmukhi}{0A1E}
\pdfglyphtounicode{o}{006F}
\pdfglyphtounicode{oacute}{00F3}
\pdfglyphtounicode{oangthai}{0E2D}
\pdfglyphtounicode{obarred}{0275}
\pdfglyphtounicode{obarredcyrillic}{04E9}
\pdfglyphtounicode{obarreddieresiscyrillic}{04EB}
\pdfglyphtounicode{obengali}{0993}
\pdfglyphtounicode{obopomofo}{311B}
\pdfglyphtounicode{obreve}{014F}
\pdfglyphtounicode{ocandradeva}{0911}
\pdfglyphtounicode{ocandragujarati}{0A91}
\pdfglyphtounicode{ocandravowelsigndeva}{0949}
\pdfglyphtounicode{ocandravowelsigngujarati}{0AC9}
\pdfglyphtounicode{ocaron}{01D2}
\pdfglyphtounicode{ocircle}{24DE}
\pdfglyphtounicode{ocircumflex}{00F4}
\pdfglyphtounicode{ocircumflexacute}{1ED1}
\pdfglyphtounicode{ocircumflexdotbelow}{1ED9}
\pdfglyphtounicode{ocircumflexgrave}{1ED3}
\pdfglyphtounicode{ocircumflexhookabove}{1ED5}
\pdfglyphtounicode{ocircumflextilde}{1ED7}
\pdfglyphtounicode{ocyrillic}{043E}
\pdfglyphtounicode{odblacute}{0151}
\pdfglyphtounicode{odblgrave}{020D}
\pdfglyphtounicode{odeva}{0913}
\pdfglyphtounicode{odieresis}{00F6}
\pdfglyphtounicode{odieresiscyrillic}{04E7}
\pdfglyphtounicode{odotbelow}{1ECD}
\pdfglyphtounicode{oe}{0153}
\pdfglyphtounicode{oekorean}{315A}
\pdfglyphtounicode{ogonek}{02DB}
\pdfglyphtounicode{ogonekcmb}{0328}
\pdfglyphtounicode{ograve}{00F2}
\pdfglyphtounicode{ogujarati}{0A93}
\pdfglyphtounicode{oharmenian}{0585}
\pdfglyphtounicode{ohiragana}{304A}
\pdfglyphtounicode{ohookabove}{1ECF}
\pdfglyphtounicode{ohorn}{01A1}
\pdfglyphtounicode{ohornacute}{1EDB}
\pdfglyphtounicode{ohorndotbelow}{1EE3}
\pdfglyphtounicode{ohorngrave}{1EDD}
\pdfglyphtounicode{ohornhookabove}{1EDF}
\pdfglyphtounicode{ohorntilde}{1EE1}
\pdfglyphtounicode{ohungarumlaut}{0151}
\pdfglyphtounicode{oi}{01A3}
\pdfglyphtounicode{oinvertedbreve}{020F}
\pdfglyphtounicode{okatakana}{30AA}
\pdfglyphtounicode{okatakanahalfwidth}{FF75}
\pdfglyphtounicode{okorean}{3157}
\pdfglyphtounicode{olehebrew}{05AB}
\pdfglyphtounicode{omacron}{014D}
\pdfglyphtounicode{omacronacute}{1E53}
\pdfglyphtounicode{omacrongrave}{1E51}
\pdfglyphtounicode{omdeva}{0950}
\pdfglyphtounicode{omega}{03C9}
\pdfglyphtounicode{omega1}{03D6}
\pdfglyphtounicode{omegacyrillic}{0461}
\pdfglyphtounicode{omegalatinclosed}{0277}
\pdfglyphtounicode{omegaroundcyrillic}{047B}
\pdfglyphtounicode{omegatitlocyrillic}{047D}
\pdfglyphtounicode{omegatonos}{03CE}
\pdfglyphtounicode{omgujarati}{0AD0}
\pdfglyphtounicode{omicron}{03BF}
\pdfglyphtounicode{omicrontonos}{03CC}
\pdfglyphtounicode{omonospace}{FF4F}
\pdfglyphtounicode{one}{0031}
\pdfglyphtounicode{onearabic}{0661}
\pdfglyphtounicode{onebengali}{09E7}
\pdfglyphtounicode{onecircle}{2460}
\pdfglyphtounicode{onecircleinversesansserif}{278A}
\pdfglyphtounicode{onedeva}{0967}
\pdfglyphtounicode{onedotenleader}{2024}
\pdfglyphtounicode{oneeighth}{215B}
\pdfglyphtounicode{onefitted}{0031}
\pdfglyphtounicode{onegujarati}{0AE7}
\pdfglyphtounicode{onegurmukhi}{0A67}
\pdfglyphtounicode{onehackarabic}{0661}
\pdfglyphtounicode{onehalf}{00BD}
\pdfglyphtounicode{onehangzhou}{3021}
\pdfglyphtounicode{oneideographicparen}{3220}
\pdfglyphtounicode{oneinferior}{2081}
\pdfglyphtounicode{onemonospace}{FF11}
\pdfglyphtounicode{onenumeratorbengali}{09F4}
\pdfglyphtounicode{oneoldstyle}{0031}
\pdfglyphtounicode{oneparen}{2474}
\pdfglyphtounicode{oneperiod}{2488}
\pdfglyphtounicode{onepersian}{06F1}
\pdfglyphtounicode{onequarter}{00BC}
\pdfglyphtounicode{oneroman}{2170}
\pdfglyphtounicode{onesuperior}{00B9}
\pdfglyphtounicode{onethai}{0E51}
\pdfglyphtounicode{onethird}{2153}
\pdfglyphtounicode{oogonek}{01EB}
\pdfglyphtounicode{oogonekmacron}{01ED}
\pdfglyphtounicode{oogurmukhi}{0A13}
\pdfglyphtounicode{oomatragurmukhi}{0A4B}
\pdfglyphtounicode{oopen}{0254}
\pdfglyphtounicode{oparen}{24AA}
\pdfglyphtounicode{openbullet}{25E6}
\pdfglyphtounicode{option}{2325}
\pdfglyphtounicode{ordfeminine}{00AA}
\pdfglyphtounicode{ordmasculine}{00BA}
\pdfglyphtounicode{orthogonal}{221F}
\pdfglyphtounicode{orunderscore}{22BB}
\pdfglyphtounicode{oshortdeva}{0912}
\pdfglyphtounicode{oshortvowelsigndeva}{094A}
\pdfglyphtounicode{oslash}{00F8}
\pdfglyphtounicode{oslashacute}{01FF}
\pdfglyphtounicode{osmallhiragana}{3049}
\pdfglyphtounicode{osmallkatakana}{30A9}
\pdfglyphtounicode{osmallkatakanahalfwidth}{FF6B}
\pdfglyphtounicode{ostrokeacute}{01FF}
\pdfglyphtounicode{osuperior}{006F}
\pdfglyphtounicode{otcyrillic}{047F}
\pdfglyphtounicode{otilde}{00F5}
\pdfglyphtounicode{otildeacute}{1E4D}
\pdfglyphtounicode{otildedieresis}{1E4F}
\pdfglyphtounicode{oubopomofo}{3121}
\pdfglyphtounicode{overline}{203E}
\pdfglyphtounicode{overlinecenterline}{FE4A}
\pdfglyphtounicode{overlinecmb}{0305}
\pdfglyphtounicode{overlinedashed}{FE49}
\pdfglyphtounicode{overlinedblwavy}{FE4C}
\pdfglyphtounicode{overlinewavy}{FE4B}
\pdfglyphtounicode{overscore}{00AF}
\pdfglyphtounicode{ovowelsignbengali}{09CB}
\pdfglyphtounicode{ovowelsigndeva}{094B}
\pdfglyphtounicode{ovowelsigngujarati}{0ACB}
\pdfglyphtounicode{owner}{220B}
\pdfglyphtounicode{p}{0070}
\pdfglyphtounicode{paampssquare}{3380}
\pdfglyphtounicode{paasentosquare}{332B}
\pdfglyphtounicode{pabengali}{09AA}
\pdfglyphtounicode{pacute}{1E55}
\pdfglyphtounicode{padeva}{092A}
\pdfglyphtounicode{pagedown}{21DF}
\pdfglyphtounicode{pageup}{21DE}
\pdfglyphtounicode{pagujarati}{0AAA}
\pdfglyphtounicode{pagurmukhi}{0A2A}
\pdfglyphtounicode{pahiragana}{3071}
\pdfglyphtounicode{paiyannoithai}{0E2F}
\pdfglyphtounicode{pakatakana}{30D1}
\pdfglyphtounicode{palatalizationcyrilliccmb}{0484}
\pdfglyphtounicode{palochkacyrillic}{04C0}
\pdfglyphtounicode{pansioskorean}{317F}
\pdfglyphtounicode{paragraph}{00B6}
\pdfglyphtounicode{parallel}{2225}
\pdfglyphtounicode{parenleft}{0028}
\pdfglyphtounicode{parenleftaltonearabic}{FD3E}
\pdfglyphtounicode{parenleftbt}{F8ED}
\pdfglyphtounicode{parenleftex}{F8EC}
\pdfglyphtounicode{parenleftinferior}{208D}
\pdfglyphtounicode{parenleftmonospace}{FF08}
\pdfglyphtounicode{parenleftsmall}{FE59}
\pdfglyphtounicode{parenleftsuperior}{207D}
\pdfglyphtounicode{parenlefttp}{F8EB}
\pdfglyphtounicode{parenleftvertical}{FE35}
\pdfglyphtounicode{parenright}{0029}
\pdfglyphtounicode{parenrightaltonearabic}{FD3F}
\pdfglyphtounicode{parenrightbt}{F8F8}
\pdfglyphtounicode{parenrightex}{F8F7}
\pdfglyphtounicode{parenrightinferior}{208E}
\pdfglyphtounicode{parenrightmonospace}{FF09}
\pdfglyphtounicode{parenrightsmall}{FE5A}
\pdfglyphtounicode{parenrightsuperior}{207E}
\pdfglyphtounicode{parenrighttp}{F8F6}
\pdfglyphtounicode{parenrightvertical}{FE36}
\pdfglyphtounicode{partialdiff}{2202}
\pdfglyphtounicode{paseqhebrew}{05C0}
\pdfglyphtounicode{pashtahebrew}{0599}
\pdfglyphtounicode{pasquare}{33A9}
\pdfglyphtounicode{patah}{05B7}
\pdfglyphtounicode{patah11}{05B7}
\pdfglyphtounicode{patah1d}{05B7}
\pdfglyphtounicode{patah2a}{05B7}
\pdfglyphtounicode{patahhebrew}{05B7}
\pdfglyphtounicode{patahnarrowhebrew}{05B7}
\pdfglyphtounicode{patahquarterhebrew}{05B7}
\pdfglyphtounicode{patahwidehebrew}{05B7}
\pdfglyphtounicode{pazerhebrew}{05A1}
\pdfglyphtounicode{pbopomofo}{3106}
\pdfglyphtounicode{pcircle}{24DF}
\pdfglyphtounicode{pdotaccent}{1E57}
\pdfglyphtounicode{pe}{05E4}
\pdfglyphtounicode{pecyrillic}{043F}
\pdfglyphtounicode{pedagesh}{FB44}
\pdfglyphtounicode{pedageshhebrew}{FB44}
\pdfglyphtounicode{peezisquare}{333B}
\pdfglyphtounicode{pefinaldageshhebrew}{FB43}
\pdfglyphtounicode{peharabic}{067E}
\pdfglyphtounicode{peharmenian}{057A}
\pdfglyphtounicode{pehebrew}{05E4}
\pdfglyphtounicode{pehfinalarabic}{FB57}
\pdfglyphtounicode{pehinitialarabic}{FB58}
\pdfglyphtounicode{pehiragana}{307A}
\pdfglyphtounicode{pehmedialarabic}{FB59}
\pdfglyphtounicode{pekatakana}{30DA}
\pdfglyphtounicode{pemiddlehookcyrillic}{04A7}
\pdfglyphtounicode{perafehebrew}{FB4E}
\pdfglyphtounicode{percent}{0025}
\pdfglyphtounicode{percentarabic}{066A}
\pdfglyphtounicode{percentmonospace}{FF05}
\pdfglyphtounicode{percentsmall}{FE6A}
\pdfglyphtounicode{period}{002E}
\pdfglyphtounicode{periodarmenian}{0589}
\pdfglyphtounicode{periodcentered}{00B7}
\pdfglyphtounicode{periodhalfwidth}{FF61}
\pdfglyphtounicode{periodinferior}{002E}
\pdfglyphtounicode{periodmonospace}{FF0E}
\pdfglyphtounicode{periodsmall}{FE52}
\pdfglyphtounicode{periodsuperior}{002E}
\pdfglyphtounicode{perispomenigreekcmb}{0342}
\pdfglyphtounicode{perpcorrespond}{2A5E}
\pdfglyphtounicode{perpendicular}{22A5}
\pdfglyphtounicode{pertenthousand}{2031}
\pdfglyphtounicode{perthousand}{2030}
\pdfglyphtounicode{peseta}{20A7}
\pdfglyphtounicode{pfsquare}{338A}
\pdfglyphtounicode{phabengali}{09AB}
\pdfglyphtounicode{phadeva}{092B}
\pdfglyphtounicode{phagujarati}{0AAB}
\pdfglyphtounicode{phagurmukhi}{0A2B}
\pdfglyphtounicode{phi}{03C6}
\pdfglyphtounicode{phi1}{03D5}
\pdfglyphtounicode{phieuphacirclekorean}{327A}
\pdfglyphtounicode{phieuphaparenkorean}{321A}
\pdfglyphtounicode{phieuphcirclekorean}{326C}
\pdfglyphtounicode{phieuphkorean}{314D}
\pdfglyphtounicode{phieuphparenkorean}{320C}
\pdfglyphtounicode{philatin}{0278}
\pdfglyphtounicode{phinthuthai}{0E3A}
\pdfglyphtounicode{phisymbolgreek}{03D5}
\pdfglyphtounicode{phook}{01A5}
\pdfglyphtounicode{phophanthai}{0E1E}
\pdfglyphtounicode{phophungthai}{0E1C}
\pdfglyphtounicode{phosamphaothai}{0E20}
\pdfglyphtounicode{pi}{03C0}
\pdfglyphtounicode{pi1}{03D6}
\pdfglyphtounicode{pieupacirclekorean}{3273}
\pdfglyphtounicode{pieupaparenkorean}{3213}
\pdfglyphtounicode{pieupcieuckorean}{3176}
\pdfglyphtounicode{pieupcirclekorean}{3265}
\pdfglyphtounicode{pieupkiyeokkorean}{3172}
\pdfglyphtounicode{pieupkorean}{3142}
\pdfglyphtounicode{pieupparenkorean}{3205}
\pdfglyphtounicode{pieupsioskiyeokkorean}{3174}
\pdfglyphtounicode{pieupsioskorean}{3144}
\pdfglyphtounicode{pieupsiostikeutkorean}{3175}
\pdfglyphtounicode{pieupthieuthkorean}{3177}
\pdfglyphtounicode{pieuptikeutkorean}{3173}
\pdfglyphtounicode{pihiragana}{3074}
\pdfglyphtounicode{pikatakana}{30D4}
\pdfglyphtounicode{pisymbolgreek}{03D6}
\pdfglyphtounicode{piwrarmenian}{0583}
\pdfglyphtounicode{planckover2pi}{210F}
\pdfglyphtounicode{planckover2pi1}{210F}
\pdfglyphtounicode{plus}{002B}
\pdfglyphtounicode{plusbelowcmb}{031F}
\pdfglyphtounicode{pluscircle}{2295}
\pdfglyphtounicode{plusminus}{00B1}
\pdfglyphtounicode{plusmod}{02D6}
\pdfglyphtounicode{plusmonospace}{FF0B}
\pdfglyphtounicode{plussmall}{FE62}
\pdfglyphtounicode{plussuperior}{207A}
\pdfglyphtounicode{pmonospace}{FF50}
\pdfglyphtounicode{pmsquare}{33D8}
\pdfglyphtounicode{pohiragana}{307D}
\pdfglyphtounicode{pointingindexdownwhite}{261F}
\pdfglyphtounicode{pointingindexleftwhite}{261C}
\pdfglyphtounicode{pointingindexrightwhite}{261E}
\pdfglyphtounicode{pointingindexupwhite}{261D}
\pdfglyphtounicode{pokatakana}{30DD}
\pdfglyphtounicode{poplathai}{0E1B}
\pdfglyphtounicode{postalmark}{3012}
\pdfglyphtounicode{postalmarkface}{3020}
\pdfglyphtounicode{pparen}{24AB}
\pdfglyphtounicode{precedenotdbleqv}{2AB9}
\pdfglyphtounicode{precedenotslnteql}{2AB5}
\pdfglyphtounicode{precedeornoteqvlnt}{22E8}
\pdfglyphtounicode{precedes}{227A}
\pdfglyphtounicode{precedesequal}{2AAF}
\pdfglyphtounicode{precedesorcurly}{227C}
\pdfglyphtounicode{precedesorequal}{227E}
\pdfglyphtounicode{prescription}{211E}
\pdfglyphtounicode{prime}{2032}
\pdfglyphtounicode{primemod}{02B9}
\pdfglyphtounicode{primereverse}{2035}
\pdfglyphtounicode{primereversed}{2035}
\pdfglyphtounicode{product}{220F}
\pdfglyphtounicode{projective}{2305}
\pdfglyphtounicode{prolongedkana}{30FC}
\pdfglyphtounicode{propellor}{2318}
\pdfglyphtounicode{propersubset}{2282}
\pdfglyphtounicode{propersuperset}{2283}
\pdfglyphtounicode{proportion}{2237}
\pdfglyphtounicode{proportional}{221D}
\pdfglyphtounicode{psi}{03C8}
\pdfglyphtounicode{psicyrillic}{0471}
\pdfglyphtounicode{psilipneumatacyrilliccmb}{0486}
\pdfglyphtounicode{pssquare}{33B0}
\pdfglyphtounicode{puhiragana}{3077}
\pdfglyphtounicode{pukatakana}{30D7}
\pdfglyphtounicode{punctdash}{2014}
\pdfglyphtounicode{pvsquare}{33B4}
\pdfglyphtounicode{pwsquare}{33BA}
\pdfglyphtounicode{q}{0071}
\pdfglyphtounicode{qadeva}{0958}
\pdfglyphtounicode{qadmahebrew}{05A8}
\pdfglyphtounicode{qafarabic}{0642}
\pdfglyphtounicode{qaffinalarabic}{FED6}
\pdfglyphtounicode{qafinitialarabic}{FED7}
\pdfglyphtounicode{qafmedialarabic}{FED8}
\pdfglyphtounicode{qamats}{05B8}
\pdfglyphtounicode{qamats10}{05B8}
\pdfglyphtounicode{qamats1a}{05B8}
\pdfglyphtounicode{qamats1c}{05B8}
\pdfglyphtounicode{qamats27}{05B8}
\pdfglyphtounicode{qamats29}{05B8}
\pdfglyphtounicode{qamats33}{05B8}
\pdfglyphtounicode{qamatsde}{05B8}
\pdfglyphtounicode{qamatshebrew}{05B8}
\pdfglyphtounicode{qamatsnarrowhebrew}{05B8}
\pdfglyphtounicode{qamatsqatanhebrew}{05B8}
\pdfglyphtounicode{qamatsqatannarrowhebrew}{05B8}
\pdfglyphtounicode{qamatsqatanquarterhebrew}{05B8}
\pdfglyphtounicode{qamatsqatanwidehebrew}{05B8}
\pdfglyphtounicode{qamatsquarterhebrew}{05B8}
\pdfglyphtounicode{qamatswidehebrew}{05B8}
\pdfglyphtounicode{qarneyparahebrew}{059F}
\pdfglyphtounicode{qbopomofo}{3111}
\pdfglyphtounicode{qcircle}{24E0}
\pdfglyphtounicode{qhook}{02A0}
\pdfglyphtounicode{qmonospace}{FF51}
\pdfglyphtounicode{qof}{05E7}
\pdfglyphtounicode{qofdagesh}{FB47}
\pdfglyphtounicode{qofdageshhebrew}{FB47}
\pdfglyphtounicode{qofhatafpatah}{05E7 05B2}
\pdfglyphtounicode{qofhatafpatahhebrew}{05E7 05B2}
\pdfglyphtounicode{qofhatafsegol}{05E7 05B1}
\pdfglyphtounicode{qofhatafsegolhebrew}{05E7 05B1}
\pdfglyphtounicode{qofhebrew}{05E7}
\pdfglyphtounicode{qofhiriq}{05E7 05B4}
\pdfglyphtounicode{qofhiriqhebrew}{05E7 05B4}
\pdfglyphtounicode{qofholam}{05E7 05B9}
\pdfglyphtounicode{qofholamhebrew}{05E7 05B9}
\pdfglyphtounicode{qofpatah}{05E7 05B7}
\pdfglyphtounicode{qofpatahhebrew}{05E7 05B7}
\pdfglyphtounicode{qofqamats}{05E7 05B8}
\pdfglyphtounicode{qofqamatshebrew}{05E7 05B8}
\pdfglyphtounicode{qofqubuts}{05E7 05BB}
\pdfglyphtounicode{qofqubutshebrew}{05E7 05BB}
\pdfglyphtounicode{qofsegol}{05E7 05B6}
\pdfglyphtounicode{qofsegolhebrew}{05E7 05B6}
\pdfglyphtounicode{qofsheva}{05E7 05B0}
\pdfglyphtounicode{qofshevahebrew}{05E7 05B0}
\pdfglyphtounicode{qoftsere}{05E7 05B5}
\pdfglyphtounicode{qoftserehebrew}{05E7 05B5}
\pdfglyphtounicode{qparen}{24AC}
\pdfglyphtounicode{quarternote}{2669}
\pdfglyphtounicode{qubuts}{05BB}
\pdfglyphtounicode{qubuts18}{05BB}
\pdfglyphtounicode{qubuts25}{05BB}
\pdfglyphtounicode{qubuts31}{05BB}
\pdfglyphtounicode{qubutshebrew}{05BB}
\pdfglyphtounicode{qubutsnarrowhebrew}{05BB}
\pdfglyphtounicode{qubutsquarterhebrew}{05BB}
\pdfglyphtounicode{qubutswidehebrew}{05BB}
\pdfglyphtounicode{question}{003F}
\pdfglyphtounicode{questionarabic}{061F}
\pdfglyphtounicode{questionarmenian}{055E}
\pdfglyphtounicode{questiondown}{00BF}
\pdfglyphtounicode{questiondownsmall}{00BF}
\pdfglyphtounicode{questiongreek}{037E}
\pdfglyphtounicode{questionmonospace}{FF1F}
\pdfglyphtounicode{questionsmall}{003F}
\pdfglyphtounicode{quotedbl}{0022}
\pdfglyphtounicode{quotedblbase}{201E}
\pdfglyphtounicode{quotedblleft}{201C}
\pdfglyphtounicode{quotedblmonospace}{FF02}
\pdfglyphtounicode{quotedblprime}{301E}
\pdfglyphtounicode{quotedblprimereversed}{301D}
\pdfglyphtounicode{quotedblright}{201D}
\pdfglyphtounicode{quoteleft}{2018}
\pdfglyphtounicode{quoteleftreversed}{201B}
\pdfglyphtounicode{quotereversed}{201B}
\pdfglyphtounicode{quoteright}{2019}
\pdfglyphtounicode{quoterightn}{0149}
\pdfglyphtounicode{quotesinglbase}{201A}
\pdfglyphtounicode{quotesingle}{0027}
\pdfglyphtounicode{quotesinglemonospace}{FF07}
\pdfglyphtounicode{r}{0072}
\pdfglyphtounicode{raarmenian}{057C}
\pdfglyphtounicode{rabengali}{09B0}
\pdfglyphtounicode{racute}{0155}
\pdfglyphtounicode{radeva}{0930}
\pdfglyphtounicode{radical}{221A}
\pdfglyphtounicode{radicalex}{F8E5}
\pdfglyphtounicode{radoverssquare}{33AE}
\pdfglyphtounicode{radoverssquaredsquare}{33AF}
\pdfglyphtounicode{radsquare}{33AD}
\pdfglyphtounicode{rafe}{05BF}
\pdfglyphtounicode{rafehebrew}{05BF}
\pdfglyphtounicode{ragujarati}{0AB0}
\pdfglyphtounicode{ragurmukhi}{0A30}
\pdfglyphtounicode{rahiragana}{3089}
\pdfglyphtounicode{rakatakana}{30E9}
\pdfglyphtounicode{rakatakanahalfwidth}{FF97}
\pdfglyphtounicode{ralowerdiagonalbengali}{09F1}
\pdfglyphtounicode{ramiddlediagonalbengali}{09F0}
\pdfglyphtounicode{ramshorn}{0264}
\pdfglyphtounicode{rangedash}{2013}
\pdfglyphtounicode{ratio}{2236}
\pdfglyphtounicode{rbopomofo}{3116}
\pdfglyphtounicode{rcaron}{0159}
\pdfglyphtounicode{rcedilla}{0157}
\pdfglyphtounicode{rcircle}{24E1}
\pdfglyphtounicode{rcommaaccent}{0157}
\pdfglyphtounicode{rdblgrave}{0211}
\pdfglyphtounicode{rdotaccent}{1E59}
\pdfglyphtounicode{rdotbelow}{1E5B}
\pdfglyphtounicode{rdotbelowmacron}{1E5D}
\pdfglyphtounicode{referencemark}{203B}
\pdfglyphtounicode{reflexsubset}{2286}
\pdfglyphtounicode{reflexsuperset}{2287}
\pdfglyphtounicode{registered}{00AE}
\pdfglyphtounicode{registersans}{00AE}
\pdfglyphtounicode{registerserif}{00AE}
\pdfglyphtounicode{reharabic}{0631}
\pdfglyphtounicode{reharmenian}{0580}
\pdfglyphtounicode{rehfinalarabic}{FEAE}
\pdfglyphtounicode{rehiragana}{308C}
\pdfglyphtounicode{rehyehaleflamarabic}{0631 FEF3 FE8E 0644}
\pdfglyphtounicode{rekatakana}{30EC}
\pdfglyphtounicode{rekatakanahalfwidth}{FF9A}
\pdfglyphtounicode{resh}{05E8}
\pdfglyphtounicode{reshdageshhebrew}{FB48}
\pdfglyphtounicode{reshhatafpatah}{05E8 05B2}
\pdfglyphtounicode{reshhatafpatahhebrew}{05E8 05B2}
\pdfglyphtounicode{reshhatafsegol}{05E8 05B1}
\pdfglyphtounicode{reshhatafsegolhebrew}{05E8 05B1}
\pdfglyphtounicode{reshhebrew}{05E8}
\pdfglyphtounicode{reshhiriq}{05E8 05B4}
\pdfglyphtounicode{reshhiriqhebrew}{05E8 05B4}
\pdfglyphtounicode{reshholam}{05E8 05B9}
\pdfglyphtounicode{reshholamhebrew}{05E8 05B9}
\pdfglyphtounicode{reshpatah}{05E8 05B7}
\pdfglyphtounicode{reshpatahhebrew}{05E8 05B7}
\pdfglyphtounicode{reshqamats}{05E8 05B8}
\pdfglyphtounicode{reshqamatshebrew}{05E8 05B8}
\pdfglyphtounicode{reshqubuts}{05E8 05BB}
\pdfglyphtounicode{reshqubutshebrew}{05E8 05BB}
\pdfglyphtounicode{reshsegol}{05E8 05B6}
\pdfglyphtounicode{reshsegolhebrew}{05E8 05B6}
\pdfglyphtounicode{reshsheva}{05E8 05B0}
\pdfglyphtounicode{reshshevahebrew}{05E8 05B0}
\pdfglyphtounicode{reshtsere}{05E8 05B5}
\pdfglyphtounicode{reshtserehebrew}{05E8 05B5}
\pdfglyphtounicode{revasymptequal}{22CD}
\pdfglyphtounicode{reversedtilde}{223D}
\pdfglyphtounicode{reviahebrew}{0597}
\pdfglyphtounicode{reviamugrashhebrew}{0597}
\pdfglyphtounicode{revlogicalnot}{2310}
\pdfglyphtounicode{revsimilar}{223D}
\pdfglyphtounicode{rfishhook}{027E}
\pdfglyphtounicode{rfishhookreversed}{027F}
\pdfglyphtounicode{rhabengali}{09DD}
\pdfglyphtounicode{rhadeva}{095D}
\pdfglyphtounicode{rho}{03C1}
\pdfglyphtounicode{rho1}{03F1}
\pdfglyphtounicode{rhook}{027D}
\pdfglyphtounicode{rhookturned}{027B}
\pdfglyphtounicode{rhookturnedsuperior}{02B5}
\pdfglyphtounicode{rhosymbolgreek}{03F1}
\pdfglyphtounicode{rhotichookmod}{02DE}
\pdfglyphtounicode{rieulacirclekorean}{3271}
\pdfglyphtounicode{rieulaparenkorean}{3211}
\pdfglyphtounicode{rieulcirclekorean}{3263}
\pdfglyphtounicode{rieulhieuhkorean}{3140}
\pdfglyphtounicode{rieulkiyeokkorean}{313A}
\pdfglyphtounicode{rieulkiyeoksioskorean}{3169}
\pdfglyphtounicode{rieulkorean}{3139}
\pdfglyphtounicode{rieulmieumkorean}{313B}
\pdfglyphtounicode{rieulpansioskorean}{316C}
\pdfglyphtounicode{rieulparenkorean}{3203}
\pdfglyphtounicode{rieulphieuphkorean}{313F}
\pdfglyphtounicode{rieulpieupkorean}{313C}
\pdfglyphtounicode{rieulpieupsioskorean}{316B}
\pdfglyphtounicode{rieulsioskorean}{313D}
\pdfglyphtounicode{rieulthieuthkorean}{313E}
\pdfglyphtounicode{rieultikeutkorean}{316A}
\pdfglyphtounicode{rieulyeorinhieuhkorean}{316D}
\pdfglyphtounicode{rightangle}{221F}
\pdfglyphtounicode{rightanglene}{231D}
\pdfglyphtounicode{rightanglenw}{231C}
\pdfglyphtounicode{rightanglese}{231F}
\pdfglyphtounicode{rightanglesw}{231E}
\pdfglyphtounicode{righttackbelowcmb}{0319}
\pdfglyphtounicode{righttriangle}{22BF}
\pdfglyphtounicode{rihiragana}{308A}
\pdfglyphtounicode{rikatakana}{30EA}
\pdfglyphtounicode{rikatakanahalfwidth}{FF98}
\pdfglyphtounicode{ring}{02DA}
\pdfglyphtounicode{ringbelowcmb}{0325}
\pdfglyphtounicode{ringcmb}{030A}
\pdfglyphtounicode{ringhalfleft}{02BF}
\pdfglyphtounicode{ringhalfleftarmenian}{0559}
\pdfglyphtounicode{ringhalfleftbelowcmb}{031C}
\pdfglyphtounicode{ringhalfleftcentered}{02D3}
\pdfglyphtounicode{ringhalfright}{02BE}
\pdfglyphtounicode{ringhalfrightbelowcmb}{0339}
\pdfglyphtounicode{ringhalfrightcentered}{02D2}
\pdfglyphtounicode{ringinequal}{2256}
\pdfglyphtounicode{rinvertedbreve}{0213}
\pdfglyphtounicode{rittorusquare}{3351}
\pdfglyphtounicode{rlinebelow}{1E5F}
\pdfglyphtounicode{rlongleg}{027C}
\pdfglyphtounicode{rlonglegturned}{027A}
\pdfglyphtounicode{rmonospace}{FF52}
\pdfglyphtounicode{rohiragana}{308D}
\pdfglyphtounicode{rokatakana}{30ED}
\pdfglyphtounicode{rokatakanahalfwidth}{FF9B}
\pdfglyphtounicode{roruathai}{0E23}
\pdfglyphtounicode{rparen}{24AD}
\pdfglyphtounicode{rrabengali}{09DC}
\pdfglyphtounicode{rradeva}{0931}
\pdfglyphtounicode{rragurmukhi}{0A5C}
\pdfglyphtounicode{rreharabic}{0691}
\pdfglyphtounicode{rrehfinalarabic}{FB8D}
\pdfglyphtounicode{rrvocalicbengali}{09E0}
\pdfglyphtounicode{rrvocalicdeva}{0960}
\pdfglyphtounicode{rrvocalicgujarati}{0AE0}
\pdfglyphtounicode{rrvocalicvowelsignbengali}{09C4}
\pdfglyphtounicode{rrvocalicvowelsigndeva}{0944}
\pdfglyphtounicode{rrvocalicvowelsigngujarati}{0AC4}
\pdfglyphtounicode{rsuperior}{0072}
\pdfglyphtounicode{rtblock}{2590}
\pdfglyphtounicode{rturned}{0279}
\pdfglyphtounicode{rturnedsuperior}{02B4}
\pdfglyphtounicode{ruhiragana}{308B}
\pdfglyphtounicode{rukatakana}{30EB}
\pdfglyphtounicode{rukatakanahalfwidth}{FF99}
\pdfglyphtounicode{rupeemarkbengali}{09F2}
\pdfglyphtounicode{rupeesignbengali}{09F3}
\pdfglyphtounicode{rupiah}{20A8}
\pdfglyphtounicode{ruthai}{0E24}
\pdfglyphtounicode{rvocalicbengali}{098B}
\pdfglyphtounicode{rvocalicdeva}{090B}
\pdfglyphtounicode{rvocalicgujarati}{0A8B}
\pdfglyphtounicode{rvocalicvowelsignbengali}{09C3}
\pdfglyphtounicode{rvocalicvowelsigndeva}{0943}
\pdfglyphtounicode{rvocalicvowelsigngujarati}{0AC3}
\pdfglyphtounicode{s}{0073}
\pdfglyphtounicode{sabengali}{09B8}
\pdfglyphtounicode{sacute}{015B}
\pdfglyphtounicode{sacutedotaccent}{1E65}
\pdfglyphtounicode{sadarabic}{0635}
\pdfglyphtounicode{sadeva}{0938}
\pdfglyphtounicode{sadfinalarabic}{FEBA}
\pdfglyphtounicode{sadinitialarabic}{FEBB}
\pdfglyphtounicode{sadmedialarabic}{FEBC}
\pdfglyphtounicode{sagujarati}{0AB8}
\pdfglyphtounicode{sagurmukhi}{0A38}
\pdfglyphtounicode{sahiragana}{3055}
\pdfglyphtounicode{sakatakana}{30B5}
\pdfglyphtounicode{sakatakanahalfwidth}{FF7B}
\pdfglyphtounicode{sallallahoualayhewasallamarabic}{FDFA}
\pdfglyphtounicode{samekh}{05E1}
\pdfglyphtounicode{samekhdagesh}{FB41}
\pdfglyphtounicode{samekhdageshhebrew}{FB41}
\pdfglyphtounicode{samekhhebrew}{05E1}
\pdfglyphtounicode{saraaathai}{0E32}
\pdfglyphtounicode{saraaethai}{0E41}
\pdfglyphtounicode{saraaimaimalaithai}{0E44}
\pdfglyphtounicode{saraaimaimuanthai}{0E43}
\pdfglyphtounicode{saraamthai}{0E33}
\pdfglyphtounicode{saraathai}{0E30}
\pdfglyphtounicode{saraethai}{0E40}
\pdfglyphtounicode{saraiileftthai}{F886}
\pdfglyphtounicode{saraiithai}{0E35}
\pdfglyphtounicode{saraileftthai}{F885}
\pdfglyphtounicode{saraithai}{0E34}
\pdfglyphtounicode{saraothai}{0E42}
\pdfglyphtounicode{saraueeleftthai}{F888}
\pdfglyphtounicode{saraueethai}{0E37}
\pdfglyphtounicode{saraueleftthai}{F887}
\pdfglyphtounicode{sarauethai}{0E36}
\pdfglyphtounicode{sarauthai}{0E38}
\pdfglyphtounicode{sarauuthai}{0E39}
\pdfglyphtounicode{satisfies}{22A8}
\pdfglyphtounicode{sbopomofo}{3119}
\pdfglyphtounicode{scaron}{0161}
\pdfglyphtounicode{scarondotaccent}{1E67}
\pdfglyphtounicode{scedilla}{015F}
\pdfglyphtounicode{schwa}{0259}
\pdfglyphtounicode{schwacyrillic}{04D9}
\pdfglyphtounicode{schwadieresiscyrillic}{04DB}
\pdfglyphtounicode{schwahook}{025A}
\pdfglyphtounicode{scircle}{24E2}
\pdfglyphtounicode{scircumflex}{015D}
\pdfglyphtounicode{scommaaccent}{0219}
\pdfglyphtounicode{sdotaccent}{1E61}
\pdfglyphtounicode{sdotbelow}{1E63}
\pdfglyphtounicode{sdotbelowdotaccent}{1E69}
\pdfglyphtounicode{seagullbelowcmb}{033C}
\pdfglyphtounicode{second}{2033}
\pdfglyphtounicode{secondtonechinese}{02CA}
\pdfglyphtounicode{section}{00A7}
\pdfglyphtounicode{seenarabic}{0633}
\pdfglyphtounicode{seenfinalarabic}{FEB2}
\pdfglyphtounicode{seeninitialarabic}{FEB3}
\pdfglyphtounicode{seenmedialarabic}{FEB4}
\pdfglyphtounicode{segol}{05B6}
\pdfglyphtounicode{segol13}{05B6}
\pdfglyphtounicode{segol1f}{05B6}
\pdfglyphtounicode{segol2c}{05B6}
\pdfglyphtounicode{segolhebrew}{05B6}
\pdfglyphtounicode{segolnarrowhebrew}{05B6}
\pdfglyphtounicode{segolquarterhebrew}{05B6}
\pdfglyphtounicode{segoltahebrew}{0592}
\pdfglyphtounicode{segolwidehebrew}{05B6}
\pdfglyphtounicode{seharmenian}{057D}
\pdfglyphtounicode{sehiragana}{305B}
\pdfglyphtounicode{sekatakana}{30BB}
\pdfglyphtounicode{sekatakanahalfwidth}{FF7E}
\pdfglyphtounicode{semicolon}{003B}
\pdfglyphtounicode{semicolonarabic}{061B}
\pdfglyphtounicode{semicolonmonospace}{FF1B}
\pdfglyphtounicode{semicolonsmall}{FE54}
\pdfglyphtounicode{semivoicedmarkkana}{309C}
\pdfglyphtounicode{semivoicedmarkkanahalfwidth}{FF9F}
\pdfglyphtounicode{sentisquare}{3322}
\pdfglyphtounicode{sentosquare}{3323}
\pdfglyphtounicode{seven}{0037}
\pdfglyphtounicode{sevenarabic}{0667}
\pdfglyphtounicode{sevenbengali}{09ED}
\pdfglyphtounicode{sevencircle}{2466}
\pdfglyphtounicode{sevencircleinversesansserif}{2790}
\pdfglyphtounicode{sevendeva}{096D}
\pdfglyphtounicode{seveneighths}{215E}
\pdfglyphtounicode{sevengujarati}{0AED}
\pdfglyphtounicode{sevengurmukhi}{0A6D}
\pdfglyphtounicode{sevenhackarabic}{0667}
\pdfglyphtounicode{sevenhangzhou}{3027}
\pdfglyphtounicode{sevenideographicparen}{3226}
\pdfglyphtounicode{seveninferior}{2087}
\pdfglyphtounicode{sevenmonospace}{FF17}
\pdfglyphtounicode{sevenoldstyle}{0037}
\pdfglyphtounicode{sevenparen}{247A}
\pdfglyphtounicode{sevenperiod}{248E}
\pdfglyphtounicode{sevenpersian}{06F7}
\pdfglyphtounicode{sevenroman}{2176}
\pdfglyphtounicode{sevensuperior}{2077}
\pdfglyphtounicode{seventeencircle}{2470}
\pdfglyphtounicode{seventeenparen}{2484}
\pdfglyphtounicode{seventeenperiod}{2498}
\pdfglyphtounicode{seventhai}{0E57}
\pdfglyphtounicode{sfthyphen}{00AD}
\pdfglyphtounicode{shaarmenian}{0577}
\pdfglyphtounicode{shabengali}{09B6}
\pdfglyphtounicode{shacyrillic}{0448}
\pdfglyphtounicode{shaddaarabic}{0651}
\pdfglyphtounicode{shaddadammaarabic}{FC61}
\pdfglyphtounicode{shaddadammatanarabic}{FC5E}
\pdfglyphtounicode{shaddafathaarabic}{FC60}
\pdfglyphtounicode{shaddafathatanarabic}{0651 064B}
\pdfglyphtounicode{shaddakasraarabic}{FC62}
\pdfglyphtounicode{shaddakasratanarabic}{FC5F}
\pdfglyphtounicode{shade}{2592}
\pdfglyphtounicode{shadedark}{2593}
\pdfglyphtounicode{shadelight}{2591}
\pdfglyphtounicode{shademedium}{2592}
\pdfglyphtounicode{shadeva}{0936}
\pdfglyphtounicode{shagujarati}{0AB6}
\pdfglyphtounicode{shagurmukhi}{0A36}
\pdfglyphtounicode{shalshelethebrew}{0593}
\pdfglyphtounicode{sharp}{266F}
\pdfglyphtounicode{shbopomofo}{3115}
\pdfglyphtounicode{shchacyrillic}{0449}
\pdfglyphtounicode{sheenarabic}{0634}
\pdfglyphtounicode{sheenfinalarabic}{FEB6}
\pdfglyphtounicode{sheeninitialarabic}{FEB7}
\pdfglyphtounicode{sheenmedialarabic}{FEB8}
\pdfglyphtounicode{sheicoptic}{03E3}
\pdfglyphtounicode{sheqel}{20AA}
\pdfglyphtounicode{sheqelhebrew}{20AA}
\pdfglyphtounicode{sheva}{05B0}
\pdfglyphtounicode{sheva115}{05B0}
\pdfglyphtounicode{sheva15}{05B0}
\pdfglyphtounicode{sheva22}{05B0}
\pdfglyphtounicode{sheva2e}{05B0}
\pdfglyphtounicode{shevahebrew}{05B0}
\pdfglyphtounicode{shevanarrowhebrew}{05B0}
\pdfglyphtounicode{shevaquarterhebrew}{05B0}
\pdfglyphtounicode{shevawidehebrew}{05B0}
\pdfglyphtounicode{shhacyrillic}{04BB}
\pdfglyphtounicode{shiftleft}{21B0}
\pdfglyphtounicode{shiftright}{21B1}
\pdfglyphtounicode{shimacoptic}{03ED}
\pdfglyphtounicode{shin}{05E9}
\pdfglyphtounicode{shindagesh}{FB49}
\pdfglyphtounicode{shindageshhebrew}{FB49}
\pdfglyphtounicode{shindageshshindot}{FB2C}
\pdfglyphtounicode{shindageshshindothebrew}{FB2C}
\pdfglyphtounicode{shindageshsindot}{FB2D}
\pdfglyphtounicode{shindageshsindothebrew}{FB2D}
\pdfglyphtounicode{shindothebrew}{05C1}
\pdfglyphtounicode{shinhebrew}{05E9}
\pdfglyphtounicode{shinshindot}{FB2A}
\pdfglyphtounicode{shinshindothebrew}{FB2A}
\pdfglyphtounicode{shinsindot}{FB2B}
\pdfglyphtounicode{shinsindothebrew}{FB2B}
\pdfglyphtounicode{shook}{0282}
\pdfglyphtounicode{sigma}{03C3}
\pdfglyphtounicode{sigma1}{03C2}
\pdfglyphtounicode{sigmafinal}{03C2}
\pdfglyphtounicode{sigmalunatesymbolgreek}{03F2}
\pdfglyphtounicode{sihiragana}{3057}
\pdfglyphtounicode{sikatakana}{30B7}
\pdfglyphtounicode{sikatakanahalfwidth}{FF7C}
\pdfglyphtounicode{siluqhebrew}{05BD}
\pdfglyphtounicode{siluqlefthebrew}{05BD}
\pdfglyphtounicode{similar}{223C}
\pdfglyphtounicode{similarequal}{2243}
\pdfglyphtounicode{sindothebrew}{05C2}
\pdfglyphtounicode{siosacirclekorean}{3274}
\pdfglyphtounicode{siosaparenkorean}{3214}
\pdfglyphtounicode{sioscieuckorean}{317E}
\pdfglyphtounicode{sioscirclekorean}{3266}
\pdfglyphtounicode{sioskiyeokkorean}{317A}
\pdfglyphtounicode{sioskorean}{3145}
\pdfglyphtounicode{siosnieunkorean}{317B}
\pdfglyphtounicode{siosparenkorean}{3206}
\pdfglyphtounicode{siospieupkorean}{317D}
\pdfglyphtounicode{siostikeutkorean}{317C}
\pdfglyphtounicode{six}{0036}
\pdfglyphtounicode{sixarabic}{0666}
\pdfglyphtounicode{sixbengali}{09EC}
\pdfglyphtounicode{sixcircle}{2465}
\pdfglyphtounicode{sixcircleinversesansserif}{278F}
\pdfglyphtounicode{sixdeva}{096C}
\pdfglyphtounicode{sixgujarati}{0AEC}
\pdfglyphtounicode{sixgurmukhi}{0A6C}
\pdfglyphtounicode{sixhackarabic}{0666}
\pdfglyphtounicode{sixhangzhou}{3026}
\pdfglyphtounicode{sixideographicparen}{3225}
\pdfglyphtounicode{sixinferior}{2086}
\pdfglyphtounicode{sixmonospace}{FF16}
\pdfglyphtounicode{sixoldstyle}{0036}
\pdfglyphtounicode{sixparen}{2479}
\pdfglyphtounicode{sixperiod}{248D}
\pdfglyphtounicode{sixpersian}{06F6}
\pdfglyphtounicode{sixroman}{2175}
\pdfglyphtounicode{sixsuperior}{2076}
\pdfglyphtounicode{sixteencircle}{246F}
\pdfglyphtounicode{sixteencurrencydenominatorbengali}{09F9}
\pdfglyphtounicode{sixteenparen}{2483}
\pdfglyphtounicode{sixteenperiod}{2497}
\pdfglyphtounicode{sixthai}{0E56}
\pdfglyphtounicode{slash}{002F}
\pdfglyphtounicode{slashmonospace}{FF0F}
\pdfglyphtounicode{slong}{017F}
\pdfglyphtounicode{slongdotaccent}{1E9B}
\pdfglyphtounicode{slurabove}{2322}
\pdfglyphtounicode{slurbelow}{2323}
\pdfglyphtounicode{smile}{2323}
\pdfglyphtounicode{smileface}{263A}
\pdfglyphtounicode{smonospace}{FF53}
\pdfglyphtounicode{sofpasuqhebrew}{05C3}
\pdfglyphtounicode{softhyphen}{00AD}
\pdfglyphtounicode{softsigncyrillic}{044C}
\pdfglyphtounicode{sohiragana}{305D}
\pdfglyphtounicode{sokatakana}{30BD}
\pdfglyphtounicode{sokatakanahalfwidth}{FF7F}
\pdfglyphtounicode{soliduslongoverlaycmb}{0338}
\pdfglyphtounicode{solidusshortoverlaycmb}{0337}
\pdfglyphtounicode{sorusithai}{0E29}
\pdfglyphtounicode{sosalathai}{0E28}
\pdfglyphtounicode{sosothai}{0E0B}
\pdfglyphtounicode{sosuathai}{0E2A}
\pdfglyphtounicode{space}{0020}
\pdfglyphtounicode{spacehackarabic}{0020}
\pdfglyphtounicode{spade}{2660}
\pdfglyphtounicode{spadesuitblack}{2660}
\pdfglyphtounicode{spadesuitwhite}{2664}
\pdfglyphtounicode{sparen}{24AE}
\pdfglyphtounicode{sphericalangle}{2222}
\pdfglyphtounicode{square}{25A1}
\pdfglyphtounicode{squarebelowcmb}{033B}
\pdfglyphtounicode{squarecc}{33C4}
\pdfglyphtounicode{squarecm}{339D}
\pdfglyphtounicode{squarediagonalcrosshatchfill}{25A9}
\pdfglyphtounicode{squaredot}{22A1}
\pdfglyphtounicode{squarehorizontalfill}{25A4}
\pdfglyphtounicode{squareimage}{228F}
\pdfglyphtounicode{squarekg}{338F}
\pdfglyphtounicode{squarekm}{339E}
\pdfglyphtounicode{squarekmcapital}{33CE}
\pdfglyphtounicode{squareln}{33D1}
\pdfglyphtounicode{squarelog}{33D2}
\pdfglyphtounicode{squaremg}{338E}
\pdfglyphtounicode{squaremil}{33D5}
\pdfglyphtounicode{squareminus}{229F}
\pdfglyphtounicode{squaremm}{339C}
\pdfglyphtounicode{squaremsquared}{33A1}
\pdfglyphtounicode{squaremultiply}{22A0}
\pdfglyphtounicode{squareoriginal}{2290}
\pdfglyphtounicode{squareorthogonalcrosshatchfill}{25A6}
\pdfglyphtounicode{squareplus}{229E}
\pdfglyphtounicode{squaresolid}{25A0}
\pdfglyphtounicode{squareupperlefttolowerrightfill}{25A7}
\pdfglyphtounicode{squareupperrighttolowerleftfill}{25A8}
\pdfglyphtounicode{squareverticalfill}{25A5}
\pdfglyphtounicode{squarewhitewithsmallblack}{25A3}
\pdfglyphtounicode{squiggleleftright}{21AD}
\pdfglyphtounicode{squiggleright}{21DD}
\pdfglyphtounicode{srsquare}{33DB}
\pdfglyphtounicode{ssabengali}{09B7}
\pdfglyphtounicode{ssadeva}{0937}
\pdfglyphtounicode{ssagujarati}{0AB7}
\pdfglyphtounicode{ssangcieuckorean}{3149}
\pdfglyphtounicode{ssanghieuhkorean}{3185}
\pdfglyphtounicode{ssangieungkorean}{3180}
\pdfglyphtounicode{ssangkiyeokkorean}{3132}
\pdfglyphtounicode{ssangnieunkorean}{3165}
\pdfglyphtounicode{ssangpieupkorean}{3143}
\pdfglyphtounicode{ssangsioskorean}{3146}
\pdfglyphtounicode{ssangtikeutkorean}{3138}
\pdfglyphtounicode{ssuperior}{0073}
\pdfglyphtounicode{st}{0073 0074}
\pdfglyphtounicode{star}{22C6}
\pdfglyphtounicode{sterling}{00A3}
\pdfglyphtounicode{sterlingmonospace}{FFE1}
\pdfglyphtounicode{strokelongoverlaycmb}{0336}
\pdfglyphtounicode{strokeshortoverlaycmb}{0335}
\pdfglyphtounicode{subset}{2282}
\pdfglyphtounicode{subsetdbl}{22D0}
\pdfglyphtounicode{subsetdblequal}{2AC5}
\pdfglyphtounicode{subsetnoteql}{228A}
\pdfglyphtounicode{subsetnotequal}{228A}
\pdfglyphtounicode{subsetorequal}{2286}
\pdfglyphtounicode{subsetornotdbleql}{2ACB}
\pdfglyphtounicode{subsetsqequal}{2291}
\pdfglyphtounicode{succeeds}{227B}
\pdfglyphtounicode{suchthat}{220B}
\pdfglyphtounicode{suhiragana}{3059}
\pdfglyphtounicode{sukatakana}{30B9}
\pdfglyphtounicode{sukatakanahalfwidth}{FF7D}
\pdfglyphtounicode{sukunarabic}{0652}
\pdfglyphtounicode{summation}{2211}
\pdfglyphtounicode{sun}{263C}
\pdfglyphtounicode{superset}{2283}
\pdfglyphtounicode{supersetdbl}{22D1}
\pdfglyphtounicode{supersetdblequal}{2AC6}
\pdfglyphtounicode{supersetnoteql}{228B}
\pdfglyphtounicode{supersetnotequal}{228B}
\pdfglyphtounicode{supersetorequal}{2287}
\pdfglyphtounicode{supersetornotdbleql}{2ACC}
\pdfglyphtounicode{supersetsqequal}{2292}
\pdfglyphtounicode{svsquare}{33DC}
\pdfglyphtounicode{syouwaerasquare}{337C}
\pdfglyphtounicode{t}{0074}
\pdfglyphtounicode{tabengali}{09A4}
\pdfglyphtounicode{tackdown}{22A4}
\pdfglyphtounicode{tackleft}{22A3}
\pdfglyphtounicode{tadeva}{0924}
\pdfglyphtounicode{tagujarati}{0AA4}
\pdfglyphtounicode{tagurmukhi}{0A24}
\pdfglyphtounicode{taharabic}{0637}
\pdfglyphtounicode{tahfinalarabic}{FEC2}
\pdfglyphtounicode{tahinitialarabic}{FEC3}
\pdfglyphtounicode{tahiragana}{305F}
\pdfglyphtounicode{tahmedialarabic}{FEC4}
\pdfglyphtounicode{taisyouerasquare}{337D}
\pdfglyphtounicode{takatakana}{30BF}
\pdfglyphtounicode{takatakanahalfwidth}{FF80}
\pdfglyphtounicode{tatweelarabic}{0640}
\pdfglyphtounicode{tau}{03C4}
\pdfglyphtounicode{tav}{05EA}
\pdfglyphtounicode{tavdages}{FB4A}
\pdfglyphtounicode{tavdagesh}{FB4A}
\pdfglyphtounicode{tavdageshhebrew}{FB4A}
\pdfglyphtounicode{tavhebrew}{05EA}
\pdfglyphtounicode{tbar}{0167}
\pdfglyphtounicode{tbopomofo}{310A}
\pdfglyphtounicode{tcaron}{0165}
\pdfglyphtounicode{tccurl}{02A8}
\pdfglyphtounicode{tcedilla}{0163}
\pdfglyphtounicode{tcheharabic}{0686}
\pdfglyphtounicode{tchehfinalarabic}{FB7B}
\pdfglyphtounicode{tchehinitialarabic}{FB7C}
\pdfglyphtounicode{tchehmedialarabic}{FB7D}
\pdfglyphtounicode{tchehmeeminitialarabic}{FB7C FEE4}
\pdfglyphtounicode{tcircle}{24E3}
\pdfglyphtounicode{tcircumflexbelow}{1E71}
\pdfglyphtounicode{tcommaaccent}{0163}
\pdfglyphtounicode{tdieresis}{1E97}
\pdfglyphtounicode{tdotaccent}{1E6B}
\pdfglyphtounicode{tdotbelow}{1E6D}
\pdfglyphtounicode{tecyrillic}{0442}
\pdfglyphtounicode{tedescendercyrillic}{04AD}
\pdfglyphtounicode{teharabic}{062A}
\pdfglyphtounicode{tehfinalarabic}{FE96}
\pdfglyphtounicode{tehhahinitialarabic}{FCA2}
\pdfglyphtounicode{tehhahisolatedarabic}{FC0C}
\pdfglyphtounicode{tehinitialarabic}{FE97}
\pdfglyphtounicode{tehiragana}{3066}
\pdfglyphtounicode{tehjeeminitialarabic}{FCA1}
\pdfglyphtounicode{tehjeemisolatedarabic}{FC0B}
\pdfglyphtounicode{tehmarbutaarabic}{0629}
\pdfglyphtounicode{tehmarbutafinalarabic}{FE94}
\pdfglyphtounicode{tehmedialarabic}{FE98}
\pdfglyphtounicode{tehmeeminitialarabic}{FCA4}
\pdfglyphtounicode{tehmeemisolatedarabic}{FC0E}
\pdfglyphtounicode{tehnoonfinalarabic}{FC73}
\pdfglyphtounicode{tekatakana}{30C6}
\pdfglyphtounicode{tekatakanahalfwidth}{FF83}
\pdfglyphtounicode{telephone}{2121}
\pdfglyphtounicode{telephoneblack}{260E}
\pdfglyphtounicode{telishagedolahebrew}{05A0}
\pdfglyphtounicode{telishaqetanahebrew}{05A9}
\pdfglyphtounicode{tencircle}{2469}
\pdfglyphtounicode{tenideographicparen}{3229}
\pdfglyphtounicode{tenparen}{247D}
\pdfglyphtounicode{tenperiod}{2491}
\pdfglyphtounicode{tenroman}{2179}
\pdfglyphtounicode{tesh}{02A7}
\pdfglyphtounicode{tet}{05D8}
\pdfglyphtounicode{tetdagesh}{FB38}
\pdfglyphtounicode{tetdageshhebrew}{FB38}
\pdfglyphtounicode{tethebrew}{05D8}
\pdfglyphtounicode{tetsecyrillic}{04B5}
\pdfglyphtounicode{tevirhebrew}{059B}
\pdfglyphtounicode{tevirlefthebrew}{059B}
\pdfglyphtounicode{tfm:cmbsy10/diamond}{2662}
\pdfglyphtounicode{tfm:cmbsy10/heart}{2661}
\pdfglyphtounicode{tfm:cmbsy5/diamond}{2662}
\pdfglyphtounicode{tfm:cmbsy5/heart}{2661}
\pdfglyphtounicode{tfm:cmbsy6/diamond}{2662}
\pdfglyphtounicode{tfm:cmbsy6/heart}{2661}
\pdfglyphtounicode{tfm:cmbsy7/diamond}{2662}
\pdfglyphtounicode{tfm:cmbsy7/heart}{2661}
\pdfglyphtounicode{tfm:cmbsy8/diamond}{2662}
\pdfglyphtounicode{tfm:cmbsy8/heart}{2661}
\pdfglyphtounicode{tfm:cmbsy9/diamond}{2662}
\pdfglyphtounicode{tfm:cmbsy9/heart}{2661}
\pdfglyphtounicode{tfm:cmmi10/phi}{03D5}
\pdfglyphtounicode{tfm:cmmi10/phi1}{03C6}
\pdfglyphtounicode{tfm:cmmi12/phi}{03D5}
\pdfglyphtounicode{tfm:cmmi12/phi1}{03C6}
\pdfglyphtounicode{tfm:cmmi5/phi}{03D5}
\pdfglyphtounicode{tfm:cmmi5/phi1}{03C6}
\pdfglyphtounicode{tfm:cmmi6/phi}{03D5}
\pdfglyphtounicode{tfm:cmmi6/phi1}{03C6}
\pdfglyphtounicode{tfm:cmmi7/phi}{03D5}
\pdfglyphtounicode{tfm:cmmi7/phi1}{03C6}
\pdfglyphtounicode{tfm:cmmi8/phi}{03D5}
\pdfglyphtounicode{tfm:cmmi8/phi1}{03C6}
\pdfglyphtounicode{tfm:cmmi9/phi}{03D5}
\pdfglyphtounicode{tfm:cmmi9/phi1}{03C6}
\pdfglyphtounicode{tfm:cmmib10/phi}{03D5}
\pdfglyphtounicode{tfm:cmmib10/phi1}{03C6}
\pdfglyphtounicode{tfm:cmmib5/phi}{03D5}
\pdfglyphtounicode{tfm:cmmib5/phi1}{03C6}
\pdfglyphtounicode{tfm:cmmib6/phi}{03D5}
\pdfglyphtounicode{tfm:cmmib6/phi1}{03C6}
\pdfglyphtounicode{tfm:cmmib7/phi}{03D5}
\pdfglyphtounicode{tfm:cmmib7/phi1}{03C6}
\pdfglyphtounicode{tfm:cmmib8/phi}{03D5}
\pdfglyphtounicode{tfm:cmmib8/phi1}{03C6}
\pdfglyphtounicode{tfm:cmmib9/phi}{03D5}
\pdfglyphtounicode{tfm:cmmib9/phi1}{03C6}
\pdfglyphtounicode{tfm:cmsy10/diamond}{2662}
\pdfglyphtounicode{tfm:cmsy10/heart}{2661}
\pdfglyphtounicode{tfm:cmsy5/heart}{2661}
\pdfglyphtounicode{tfm:cmsy6/diamond}{2662}
\pdfglyphtounicode{tfm:cmsy6/heart}{2661}
\pdfglyphtounicode{tfm:cmsy7/diamond}{2662}
\pdfglyphtounicode{tfm:cmsy7/heart}{2661}
\pdfglyphtounicode{tfm:cmsy8/diamond}{2662}
\pdfglyphtounicode{tfm:cmsy8/heart}{2661}
\pdfglyphtounicode{tfm:cmsy9/diamond}{2662}
\pdfglyphtounicode{tfm:cmsy9/heart}{2661}
\pdfglyphtounicode{tfm:eurb10/phi}{03D5}
\pdfglyphtounicode{tfm:eurb10/phi1}{03C6}
\pdfglyphtounicode{tfm:eurb5/phi}{03D5}
\pdfglyphtounicode{tfm:eurb5/phi1}{03C6}
\pdfglyphtounicode{tfm:eurb6/phi}{03D5}
\pdfglyphtounicode{tfm:eurb6/phi1}{03C6}
\pdfglyphtounicode{tfm:eurb7/phi}{03D5}
\pdfglyphtounicode{tfm:eurb7/phi1}{03C6}
\pdfglyphtounicode{tfm:eurb8/phi}{03D5}
\pdfglyphtounicode{tfm:eurb8/phi1}{03C6}
\pdfglyphtounicode{tfm:eurb9/phi}{03D5}
\pdfglyphtounicode{tfm:eurb9/phi1}{03C6}
\pdfglyphtounicode{tfm:eurm10/phi}{03D5}
\pdfglyphtounicode{tfm:eurm10/phi1}{03C6}
\pdfglyphtounicode{tfm:eurm5/phi}{03D5}
\pdfglyphtounicode{tfm:eurm5/phi1}{03C6}
\pdfglyphtounicode{tfm:eurm6/phi}{03D5}
\pdfglyphtounicode{tfm:eurm6/phi1}{03C6}
\pdfglyphtounicode{tfm:eurm7/phi}{03D5}
\pdfglyphtounicode{tfm:eurm7/phi1}{03C6}
\pdfglyphtounicode{tfm:eurm8/phi}{03D5}
\pdfglyphtounicode{tfm:eurm8/phi1}{03C6}
\pdfglyphtounicode{tfm:eurm9/phi}{03D5}
\pdfglyphtounicode{tfm:eurm9/phi1}{03C6}
\pdfglyphtounicode{tfm:fplmbi/phi}{03D5}
\pdfglyphtounicode{tfm:fplmbi/phi1}{03C6}
\pdfglyphtounicode{tfm:fplmri/phi}{03D5}
\pdfglyphtounicode{tfm:fplmri/phi1}{03C6}
\pdfglyphtounicode{tfm:lmbsy10/diamond}{2662}
\pdfglyphtounicode{tfm:lmbsy10/heart}{2661}
\pdfglyphtounicode{tfm:lmbsy5/diamond}{2662}
\pdfglyphtounicode{tfm:lmbsy5/heart}{2661}
\pdfglyphtounicode{tfm:lmbsy7/diamond}{2662}
\pdfglyphtounicode{tfm:lmbsy7/heart}{2661}
\pdfglyphtounicode{tfm:lmmi10/phi}{03D5}
\pdfglyphtounicode{tfm:lmmi10/phi1}{03C6}
\pdfglyphtounicode{tfm:lmmi12/phi}{03D5}
\pdfglyphtounicode{tfm:lmmi12/phi1}{03C6}
\pdfglyphtounicode{tfm:lmmi5/phi}{03D5}
\pdfglyphtounicode{tfm:lmmi5/phi1}{03C6}
\pdfglyphtounicode{tfm:lmmi6/phi}{03D5}
\pdfglyphtounicode{tfm:lmmi6/phi1}{03C6}
\pdfglyphtounicode{tfm:lmmi7/phi}{03D5}
\pdfglyphtounicode{tfm:lmmi7/phi1}{03C6}
\pdfglyphtounicode{tfm:lmmi8/phi}{03D5}
\pdfglyphtounicode{tfm:lmmi8/phi1}{03C6}
\pdfglyphtounicode{tfm:lmmi9/phi}{03D5}
\pdfglyphtounicode{tfm:lmmi9/phi1}{03C6}
\pdfglyphtounicode{tfm:lmmib10/phi}{03D5}
\pdfglyphtounicode{tfm:lmmib10/phi1}{03C6}
\pdfglyphtounicode{tfm:lmmib5/phi}{03D5}
\pdfglyphtounicode{tfm:lmmib5/phi1}{03C6}
\pdfglyphtounicode{tfm:lmmib7/phi}{03D5}
\pdfglyphtounicode{tfm:lmmib7/phi1}{03C6}
\pdfglyphtounicode{tfm:lmsy10/diamond}{2662}
\pdfglyphtounicode{tfm:lmsy10/heart}{2661}
\pdfglyphtounicode{tfm:lmsy5/diamond}{2662}
\pdfglyphtounicode{tfm:lmsy5/heart}{2661}
\pdfglyphtounicode{tfm:lmsy6/diamond}{2662}
\pdfglyphtounicode{tfm:lmsy6/heart}{2661}
\pdfglyphtounicode{tfm:lmsy7/diamond}{2662}
\pdfglyphtounicode{tfm:lmsy7/heart}{2661}
\pdfglyphtounicode{tfm:lmsy8/diamond}{2662}
\pdfglyphtounicode{tfm:lmsy8/heart}{2661}
\pdfglyphtounicode{tfm:lmsy9/diamond}{2662}
\pdfglyphtounicode{tfm:lmsy9/heart}{2661}
\pdfglyphtounicode{tfm:msam10/diamond}{2662}
\pdfglyphtounicode{tfm:msam5/diamond}{2662}
\pdfglyphtounicode{tfm:msam6/diamond}{2662}
\pdfglyphtounicode{tfm:msam7/diamond}{2662}
\pdfglyphtounicode{tfm:msam8/diamond}{2662}
\pdfglyphtounicode{tfm:msam9/diamond}{2662}
\pdfglyphtounicode{tfm:pxbmia/phi}{03D5}
\pdfglyphtounicode{tfm:pxbmia/phi1}{03C6}
\pdfglyphtounicode{tfm:pxbsy/diamond}{2662}
\pdfglyphtounicode{tfm:pxbsy/heart}{2661}
\pdfglyphtounicode{tfm:pxbsya/diamond}{2662}
\pdfglyphtounicode{tfm:pxmia/phi}{03D5}
\pdfglyphtounicode{tfm:pxmia/phi1}{03C6}
\pdfglyphtounicode{tfm:pxsy/diamond}{2662}
\pdfglyphtounicode{tfm:pxsy/heart}{2661}
\pdfglyphtounicode{tfm:pxsya/diamond}{2662}
\pdfglyphtounicode{tfm:pzdr/a1}{2701}
\pdfglyphtounicode{tfm:pzdr/a10}{2721}
\pdfglyphtounicode{tfm:pzdr/a100}{275E}
\pdfglyphtounicode{tfm:pzdr/a101}{2761}
\pdfglyphtounicode{tfm:pzdr/a102}{2762}
\pdfglyphtounicode{tfm:pzdr/a103}{2763}
\pdfglyphtounicode{tfm:pzdr/a104}{2764}
\pdfglyphtounicode{tfm:pzdr/a105}{2710}
\pdfglyphtounicode{tfm:pzdr/a106}{2765}
\pdfglyphtounicode{tfm:pzdr/a107}{2766}
\pdfglyphtounicode{tfm:pzdr/a108}{2767}
\pdfglyphtounicode{tfm:pzdr/a109}{2660}
\pdfglyphtounicode{tfm:pzdr/a11}{261B}
\pdfglyphtounicode{tfm:pzdr/a110}{2665}
\pdfglyphtounicode{tfm:pzdr/a111}{2666}
\pdfglyphtounicode{tfm:pzdr/a112}{2663}
\pdfglyphtounicode{tfm:pzdr/a117}{2709}
\pdfglyphtounicode{tfm:pzdr/a118}{2708}
\pdfglyphtounicode{tfm:pzdr/a119}{2707}
\pdfglyphtounicode{tfm:pzdr/a12}{261E}
\pdfglyphtounicode{tfm:pzdr/a120}{2460}
\pdfglyphtounicode{tfm:pzdr/a121}{2461}
\pdfglyphtounicode{tfm:pzdr/a122}{2462}
\pdfglyphtounicode{tfm:pzdr/a123}{2463}
\pdfglyphtounicode{tfm:pzdr/a124}{2464}
\pdfglyphtounicode{tfm:pzdr/a125}{2465}
\pdfglyphtounicode{tfm:pzdr/a126}{2466}
\pdfglyphtounicode{tfm:pzdr/a127}{2467}
\pdfglyphtounicode{tfm:pzdr/a128}{2468}
\pdfglyphtounicode{tfm:pzdr/a129}{2469}
\pdfglyphtounicode{tfm:pzdr/a13}{270C}
\pdfglyphtounicode{tfm:pzdr/a130}{2776}
\pdfglyphtounicode{tfm:pzdr/a131}{2777}
\pdfglyphtounicode{tfm:pzdr/a132}{2778}
\pdfglyphtounicode{tfm:pzdr/a133}{2779}
\pdfglyphtounicode{tfm:pzdr/a134}{277A}
\pdfglyphtounicode{tfm:pzdr/a135}{277B}
\pdfglyphtounicode{tfm:pzdr/a136}{277C}
\pdfglyphtounicode{tfm:pzdr/a137}{277D}
\pdfglyphtounicode{tfm:pzdr/a138}{277E}
\pdfglyphtounicode{tfm:pzdr/a139}{277F}
\pdfglyphtounicode{tfm:pzdr/a14}{270D}
\pdfglyphtounicode{tfm:pzdr/a140}{2780}
\pdfglyphtounicode{tfm:pzdr/a141}{2781}
\pdfglyphtounicode{tfm:pzdr/a142}{2782}
\pdfglyphtounicode{tfm:pzdr/a143}{2783}
\pdfglyphtounicode{tfm:pzdr/a144}{2784}
\pdfglyphtounicode{tfm:pzdr/a145}{2785}
\pdfglyphtounicode{tfm:pzdr/a146}{2786}
\pdfglyphtounicode{tfm:pzdr/a147}{2787}
\pdfglyphtounicode{tfm:pzdr/a148}{2788}
\pdfglyphtounicode{tfm:pzdr/a149}{2789}
\pdfglyphtounicode{tfm:pzdr/a15}{270E}
\pdfglyphtounicode{tfm:pzdr/a150}{278A}
\pdfglyphtounicode{tfm:pzdr/a151}{278B}
\pdfglyphtounicode{tfm:pzdr/a152}{278C}
\pdfglyphtounicode{tfm:pzdr/a153}{278D}
\pdfglyphtounicode{tfm:pzdr/a154}{278E}
\pdfglyphtounicode{tfm:pzdr/a155}{278F}
\pdfglyphtounicode{tfm:pzdr/a156}{2790}
\pdfglyphtounicode{tfm:pzdr/a157}{2791}
\pdfglyphtounicode{tfm:pzdr/a158}{2792}
\pdfglyphtounicode{tfm:pzdr/a159}{2793}
\pdfglyphtounicode{tfm:pzdr/a16}{270F}
\pdfglyphtounicode{tfm:pzdr/a160}{2794}
\pdfglyphtounicode{tfm:pzdr/a161}{2192}
\pdfglyphtounicode{tfm:pzdr/a162}{27A3}
\pdfglyphtounicode{tfm:pzdr/a163}{2194}
\pdfglyphtounicode{tfm:pzdr/a164}{2195}
\pdfglyphtounicode{tfm:pzdr/a165}{2799}
\pdfglyphtounicode{tfm:pzdr/a166}{279B}
\pdfglyphtounicode{tfm:pzdr/a167}{279C}
\pdfglyphtounicode{tfm:pzdr/a168}{279D}
\pdfglyphtounicode{tfm:pzdr/a169}{279E}
\pdfglyphtounicode{tfm:pzdr/a17}{2711}
\pdfglyphtounicode{tfm:pzdr/a170}{279F}
\pdfglyphtounicode{tfm:pzdr/a171}{27A0}
\pdfglyphtounicode{tfm:pzdr/a172}{27A1}
\pdfglyphtounicode{tfm:pzdr/a173}{27A2}
\pdfglyphtounicode{tfm:pzdr/a174}{27A4}
\pdfglyphtounicode{tfm:pzdr/a175}{27A5}
\pdfglyphtounicode{tfm:pzdr/a176}{27A6}
\pdfglyphtounicode{tfm:pzdr/a177}{27A7}
\pdfglyphtounicode{tfm:pzdr/a178}{27A8}
\pdfglyphtounicode{tfm:pzdr/a179}{27A9}
\pdfglyphtounicode{tfm:pzdr/a18}{2712}
\pdfglyphtounicode{tfm:pzdr/a180}{27AB}
\pdfglyphtounicode{tfm:pzdr/a181}{27AD}
\pdfglyphtounicode{tfm:pzdr/a182}{27AF}
\pdfglyphtounicode{tfm:pzdr/a183}{27B2}
\pdfglyphtounicode{tfm:pzdr/a184}{27B3}
\pdfglyphtounicode{tfm:pzdr/a185}{27B5}
\pdfglyphtounicode{tfm:pzdr/a186}{27B8}
\pdfglyphtounicode{tfm:pzdr/a187}{27BA}
\pdfglyphtounicode{tfm:pzdr/a188}{27BB}
\pdfglyphtounicode{tfm:pzdr/a189}{27BC}
\pdfglyphtounicode{tfm:pzdr/a19}{2713}
\pdfglyphtounicode{tfm:pzdr/a190}{27BD}
\pdfglyphtounicode{tfm:pzdr/a191}{27BE}
\pdfglyphtounicode{tfm:pzdr/a192}{279A}
\pdfglyphtounicode{tfm:pzdr/a193}{27AA}
\pdfglyphtounicode{tfm:pzdr/a194}{27B6}
\pdfglyphtounicode{tfm:pzdr/a195}{27B9}
\pdfglyphtounicode{tfm:pzdr/a196}{2798}
\pdfglyphtounicode{tfm:pzdr/a197}{27B4}
\pdfglyphtounicode{tfm:pzdr/a198}{27B7}
\pdfglyphtounicode{tfm:pzdr/a199}{27AC}
\pdfglyphtounicode{tfm:pzdr/a2}{2702}
\pdfglyphtounicode{tfm:pzdr/a20}{2714}
\pdfglyphtounicode{tfm:pzdr/a200}{27AE}
\pdfglyphtounicode{tfm:pzdr/a201}{27B1}
\pdfglyphtounicode{tfm:pzdr/a202}{2703}
\pdfglyphtounicode{tfm:pzdr/a203}{2750}
\pdfglyphtounicode{tfm:pzdr/a204}{2752}
\pdfglyphtounicode{tfm:pzdr/a205}{276E}
\pdfglyphtounicode{tfm:pzdr/a206}{2770}
\pdfglyphtounicode{tfm:pzdr/a21}{2715}
\pdfglyphtounicode{tfm:pzdr/a22}{2716}
\pdfglyphtounicode{tfm:pzdr/a23}{2717}
\pdfglyphtounicode{tfm:pzdr/a24}{2718}
\pdfglyphtounicode{tfm:pzdr/a25}{2719}
\pdfglyphtounicode{tfm:pzdr/a26}{271A}
\pdfglyphtounicode{tfm:pzdr/a27}{271B}
\pdfglyphtounicode{tfm:pzdr/a28}{271C}
\pdfglyphtounicode{tfm:pzdr/a29}{2722}
\pdfglyphtounicode{tfm:pzdr/a3}{2704}
\pdfglyphtounicode{tfm:pzdr/a30}{2723}
\pdfglyphtounicode{tfm:pzdr/a31}{2724}
\pdfglyphtounicode{tfm:pzdr/a32}{2725}
\pdfglyphtounicode{tfm:pzdr/a33}{2726}
\pdfglyphtounicode{tfm:pzdr/a34}{2727}
\pdfglyphtounicode{tfm:pzdr/a35}{2605}
\pdfglyphtounicode{tfm:pzdr/a36}{2729}
\pdfglyphtounicode{tfm:pzdr/a37}{272A}
\pdfglyphtounicode{tfm:pzdr/a38}{272B}
\pdfglyphtounicode{tfm:pzdr/a39}{272C}
\pdfglyphtounicode{tfm:pzdr/a4}{260E}
\pdfglyphtounicode{tfm:pzdr/a40}{272D}
\pdfglyphtounicode{tfm:pzdr/a41}{272E}
\pdfglyphtounicode{tfm:pzdr/a42}{272F}
\pdfglyphtounicode{tfm:pzdr/a43}{2730}
\pdfglyphtounicode{tfm:pzdr/a44}{2731}
\pdfglyphtounicode{tfm:pzdr/a45}{2732}
\pdfglyphtounicode{tfm:pzdr/a46}{2733}
\pdfglyphtounicode{tfm:pzdr/a47}{2734}
\pdfglyphtounicode{tfm:pzdr/a48}{2735}
\pdfglyphtounicode{tfm:pzdr/a49}{2736}
\pdfglyphtounicode{tfm:pzdr/a5}{2706}
\pdfglyphtounicode{tfm:pzdr/a50}{2737}
\pdfglyphtounicode{tfm:pzdr/a51}{2738}
\pdfglyphtounicode{tfm:pzdr/a52}{2739}
\pdfglyphtounicode{tfm:pzdr/a53}{273A}
\pdfglyphtounicode{tfm:pzdr/a54}{273B}
\pdfglyphtounicode{tfm:pzdr/a55}{273C}
\pdfglyphtounicode{tfm:pzdr/a56}{273D}
\pdfglyphtounicode{tfm:pzdr/a57}{273E}
\pdfglyphtounicode{tfm:pzdr/a58}{273F}
\pdfglyphtounicode{tfm:pzdr/a59}{2740}
\pdfglyphtounicode{tfm:pzdr/a6}{271D}
\pdfglyphtounicode{tfm:pzdr/a60}{2741}
\pdfglyphtounicode{tfm:pzdr/a61}{2742}
\pdfglyphtounicode{tfm:pzdr/a62}{2743}
\pdfglyphtounicode{tfm:pzdr/a63}{2744}
\pdfglyphtounicode{tfm:pzdr/a64}{2745}
\pdfglyphtounicode{tfm:pzdr/a65}{2746}
\pdfglyphtounicode{tfm:pzdr/a66}{2747}
\pdfglyphtounicode{tfm:pzdr/a67}{2748}
\pdfglyphtounicode{tfm:pzdr/a68}{2749}
\pdfglyphtounicode{tfm:pzdr/a69}{274A}
\pdfglyphtounicode{tfm:pzdr/a7}{271E}
\pdfglyphtounicode{tfm:pzdr/a70}{274B}
\pdfglyphtounicode{tfm:pzdr/a71}{25CF}
\pdfglyphtounicode{tfm:pzdr/a72}{274D}
\pdfglyphtounicode{tfm:pzdr/a73}{25A0}
\pdfglyphtounicode{tfm:pzdr/a74}{274F}
\pdfglyphtounicode{tfm:pzdr/a75}{2751}
\pdfglyphtounicode{tfm:pzdr/a76}{25B2}
\pdfglyphtounicode{tfm:pzdr/a77}{25BC}
\pdfglyphtounicode{tfm:pzdr/a78}{25C6}
\pdfglyphtounicode{tfm:pzdr/a79}{2756}
\pdfglyphtounicode{tfm:pzdr/a8}{271F}
\pdfglyphtounicode{tfm:pzdr/a81}{25D7}
\pdfglyphtounicode{tfm:pzdr/a82}{2758}
\pdfglyphtounicode{tfm:pzdr/a83}{2759}
\pdfglyphtounicode{tfm:pzdr/a84}{275A}
\pdfglyphtounicode{tfm:pzdr/a85}{276F}
\pdfglyphtounicode{tfm:pzdr/a86}{2771}
\pdfglyphtounicode{tfm:pzdr/a87}{2772}
\pdfglyphtounicode{tfm:pzdr/a88}{2773}
\pdfglyphtounicode{tfm:pzdr/a89}{2768}
\pdfglyphtounicode{tfm:pzdr/a9}{2720}
\pdfglyphtounicode{tfm:pzdr/a90}{2769}
\pdfglyphtounicode{tfm:pzdr/a91}{276C}
\pdfglyphtounicode{tfm:pzdr/a92}{276D}
\pdfglyphtounicode{tfm:pzdr/a93}{276A}
\pdfglyphtounicode{tfm:pzdr/a94}{276B}
\pdfglyphtounicode{tfm:pzdr/a95}{2774}
\pdfglyphtounicode{tfm:pzdr/a96}{2775}
\pdfglyphtounicode{tfm:pzdr/a97}{275B}
\pdfglyphtounicode{tfm:pzdr/a98}{275C}
\pdfglyphtounicode{tfm:pzdr/a99}{275D}
\pdfglyphtounicode{tfm:rpxbmi/phi}{03D5}
\pdfglyphtounicode{tfm:rpxbmi/phi1}{03C6}
\pdfglyphtounicode{tfm:rpxmi/phi}{03D5}
\pdfglyphtounicode{tfm:rpxmi/phi1}{03C6}
\pdfglyphtounicode{tfm:rpzdr/a1}{2701}
\pdfglyphtounicode{tfm:rpzdr/a10}{2721}
\pdfglyphtounicode{tfm:rpzdr/a100}{275E}
\pdfglyphtounicode{tfm:rpzdr/a101}{2761}
\pdfglyphtounicode{tfm:rpzdr/a102}{2762}
\pdfglyphtounicode{tfm:rpzdr/a103}{2763}
\pdfglyphtounicode{tfm:rpzdr/a104}{2764}
\pdfglyphtounicode{tfm:rpzdr/a105}{2710}
\pdfglyphtounicode{tfm:rpzdr/a106}{2765}
\pdfglyphtounicode{tfm:rpzdr/a107}{2766}
\pdfglyphtounicode{tfm:rpzdr/a108}{2767}
\pdfglyphtounicode{tfm:rpzdr/a109}{2660}
\pdfglyphtounicode{tfm:rpzdr/a11}{261B}
\pdfglyphtounicode{tfm:rpzdr/a110}{2665}
\pdfglyphtounicode{tfm:rpzdr/a111}{2666}
\pdfglyphtounicode{tfm:rpzdr/a112}{2663}
\pdfglyphtounicode{tfm:rpzdr/a117}{2709}
\pdfglyphtounicode{tfm:rpzdr/a118}{2708}
\pdfglyphtounicode{tfm:rpzdr/a119}{2707}
\pdfglyphtounicode{tfm:rpzdr/a12}{261E}
\pdfglyphtounicode{tfm:rpzdr/a120}{2460}
\pdfglyphtounicode{tfm:rpzdr/a121}{2461}
\pdfglyphtounicode{tfm:rpzdr/a122}{2462}
\pdfglyphtounicode{tfm:rpzdr/a123}{2463}
\pdfglyphtounicode{tfm:rpzdr/a124}{2464}
\pdfglyphtounicode{tfm:rpzdr/a125}{2465}
\pdfglyphtounicode{tfm:rpzdr/a126}{2466}
\pdfglyphtounicode{tfm:rpzdr/a127}{2467}
\pdfglyphtounicode{tfm:rpzdr/a128}{2468}
\pdfglyphtounicode{tfm:rpzdr/a129}{2469}
\pdfglyphtounicode{tfm:rpzdr/a13}{270C}
\pdfglyphtounicode{tfm:rpzdr/a130}{2776}
\pdfglyphtounicode{tfm:rpzdr/a131}{2777}
\pdfglyphtounicode{tfm:rpzdr/a132}{2778}
\pdfglyphtounicode{tfm:rpzdr/a133}{2779}
\pdfglyphtounicode{tfm:rpzdr/a134}{277A}
\pdfglyphtounicode{tfm:rpzdr/a135}{277B}
\pdfglyphtounicode{tfm:rpzdr/a136}{277C}
\pdfglyphtounicode{tfm:rpzdr/a137}{277D}
\pdfglyphtounicode{tfm:rpzdr/a138}{277E}
\pdfglyphtounicode{tfm:rpzdr/a139}{277F}
\pdfglyphtounicode{tfm:rpzdr/a14}{270D}
\pdfglyphtounicode{tfm:rpzdr/a140}{2780}
\pdfglyphtounicode{tfm:rpzdr/a141}{2781}
\pdfglyphtounicode{tfm:rpzdr/a142}{2782}
\pdfglyphtounicode{tfm:rpzdr/a143}{2783}
\pdfglyphtounicode{tfm:rpzdr/a144}{2784}
\pdfglyphtounicode{tfm:rpzdr/a145}{2785}
\pdfglyphtounicode{tfm:rpzdr/a146}{2786}
\pdfglyphtounicode{tfm:rpzdr/a147}{2787}
\pdfglyphtounicode{tfm:rpzdr/a148}{2788}
\pdfglyphtounicode{tfm:rpzdr/a149}{2789}
\pdfglyphtounicode{tfm:rpzdr/a15}{270E}
\pdfglyphtounicode{tfm:rpzdr/a150}{278A}
\pdfglyphtounicode{tfm:rpzdr/a151}{278B}
\pdfglyphtounicode{tfm:rpzdr/a152}{278C}
\pdfglyphtounicode{tfm:rpzdr/a153}{278D}
\pdfglyphtounicode{tfm:rpzdr/a154}{278E}
\pdfglyphtounicode{tfm:rpzdr/a155}{278F}
\pdfglyphtounicode{tfm:rpzdr/a156}{2790}
\pdfglyphtounicode{tfm:rpzdr/a157}{2791}
\pdfglyphtounicode{tfm:rpzdr/a158}{2792}
\pdfglyphtounicode{tfm:rpzdr/a159}{2793}
\pdfglyphtounicode{tfm:rpzdr/a16}{270F}
\pdfglyphtounicode{tfm:rpzdr/a160}{2794}
\pdfglyphtounicode{tfm:rpzdr/a161}{2192}
\pdfglyphtounicode{tfm:rpzdr/a162}{27A3}
\pdfglyphtounicode{tfm:rpzdr/a163}{2194}
\pdfglyphtounicode{tfm:rpzdr/a164}{2195}
\pdfglyphtounicode{tfm:rpzdr/a165}{2799}
\pdfglyphtounicode{tfm:rpzdr/a166}{279B}
\pdfglyphtounicode{tfm:rpzdr/a167}{279C}
\pdfglyphtounicode{tfm:rpzdr/a168}{279D}
\pdfglyphtounicode{tfm:rpzdr/a169}{279E}
\pdfglyphtounicode{tfm:rpzdr/a17}{2711}
\pdfglyphtounicode{tfm:rpzdr/a170}{279F}
\pdfglyphtounicode{tfm:rpzdr/a171}{27A0}
\pdfglyphtounicode{tfm:rpzdr/a172}{27A1}
\pdfglyphtounicode{tfm:rpzdr/a173}{27A2}
\pdfglyphtounicode{tfm:rpzdr/a174}{27A4}
\pdfglyphtounicode{tfm:rpzdr/a175}{27A5}
\pdfglyphtounicode{tfm:rpzdr/a176}{27A6}
\pdfglyphtounicode{tfm:rpzdr/a177}{27A7}
\pdfglyphtounicode{tfm:rpzdr/a178}{27A8}
\pdfglyphtounicode{tfm:rpzdr/a179}{27A9}
\pdfglyphtounicode{tfm:rpzdr/a18}{2712}
\pdfglyphtounicode{tfm:rpzdr/a180}{27AB}
\pdfglyphtounicode{tfm:rpzdr/a181}{27AD}
\pdfglyphtounicode{tfm:rpzdr/a182}{27AF}
\pdfglyphtounicode{tfm:rpzdr/a183}{27B2}
\pdfglyphtounicode{tfm:rpzdr/a184}{27B3}
\pdfglyphtounicode{tfm:rpzdr/a185}{27B5}
\pdfglyphtounicode{tfm:rpzdr/a186}{27B8}
\pdfglyphtounicode{tfm:rpzdr/a187}{27BA}
\pdfglyphtounicode{tfm:rpzdr/a188}{27BB}
\pdfglyphtounicode{tfm:rpzdr/a189}{27BC}
\pdfglyphtounicode{tfm:rpzdr/a19}{2713}
\pdfglyphtounicode{tfm:rpzdr/a190}{27BD}
\pdfglyphtounicode{tfm:rpzdr/a191}{27BE}
\pdfglyphtounicode{tfm:rpzdr/a192}{279A}
\pdfglyphtounicode{tfm:rpzdr/a193}{27AA}
\pdfglyphtounicode{tfm:rpzdr/a194}{27B6}
\pdfglyphtounicode{tfm:rpzdr/a195}{27B9}
\pdfglyphtounicode{tfm:rpzdr/a196}{2798}
\pdfglyphtounicode{tfm:rpzdr/a197}{27B4}
\pdfglyphtounicode{tfm:rpzdr/a198}{27B7}
\pdfglyphtounicode{tfm:rpzdr/a199}{27AC}
\pdfglyphtounicode{tfm:rpzdr/a2}{2702}
\pdfglyphtounicode{tfm:rpzdr/a20}{2714}
\pdfglyphtounicode{tfm:rpzdr/a200}{27AE}
\pdfglyphtounicode{tfm:rpzdr/a201}{27B1}
\pdfglyphtounicode{tfm:rpzdr/a202}{2703}
\pdfglyphtounicode{tfm:rpzdr/a203}{2750}
\pdfglyphtounicode{tfm:rpzdr/a204}{2752}
\pdfglyphtounicode{tfm:rpzdr/a205}{276E}
\pdfglyphtounicode{tfm:rpzdr/a206}{2770}
\pdfglyphtounicode{tfm:rpzdr/a21}{2715}
\pdfglyphtounicode{tfm:rpzdr/a22}{2716}
\pdfglyphtounicode{tfm:rpzdr/a23}{2717}
\pdfglyphtounicode{tfm:rpzdr/a24}{2718}
\pdfglyphtounicode{tfm:rpzdr/a25}{2719}
\pdfglyphtounicode{tfm:rpzdr/a26}{271A}
\pdfglyphtounicode{tfm:rpzdr/a27}{271B}
\pdfglyphtounicode{tfm:rpzdr/a28}{271C}
\pdfglyphtounicode{tfm:rpzdr/a29}{2722}
\pdfglyphtounicode{tfm:rpzdr/a3}{2704}
\pdfglyphtounicode{tfm:rpzdr/a30}{2723}
\pdfglyphtounicode{tfm:rpzdr/a31}{2724}
\pdfglyphtounicode{tfm:rpzdr/a32}{2725}
\pdfglyphtounicode{tfm:rpzdr/a33}{2726}
\pdfglyphtounicode{tfm:rpzdr/a34}{2727}
\pdfglyphtounicode{tfm:rpzdr/a35}{2605}
\pdfglyphtounicode{tfm:rpzdr/a36}{2729}
\pdfglyphtounicode{tfm:rpzdr/a37}{272A}
\pdfglyphtounicode{tfm:rpzdr/a38}{272B}
\pdfglyphtounicode{tfm:rpzdr/a39}{272C}
\pdfglyphtounicode{tfm:rpzdr/a4}{260E}
\pdfglyphtounicode{tfm:rpzdr/a40}{272D}
\pdfglyphtounicode{tfm:rpzdr/a41}{272E}
\pdfglyphtounicode{tfm:rpzdr/a42}{272F}
\pdfglyphtounicode{tfm:rpzdr/a43}{2730}
\pdfglyphtounicode{tfm:rpzdr/a44}{2731}
\pdfglyphtounicode{tfm:rpzdr/a45}{2732}
\pdfglyphtounicode{tfm:rpzdr/a46}{2733}
\pdfglyphtounicode{tfm:rpzdr/a47}{2734}
\pdfglyphtounicode{tfm:rpzdr/a48}{2735}
\pdfglyphtounicode{tfm:rpzdr/a49}{2736}
\pdfglyphtounicode{tfm:rpzdr/a5}{2706}
\pdfglyphtounicode{tfm:rpzdr/a50}{2737}
\pdfglyphtounicode{tfm:rpzdr/a51}{2738}
\pdfglyphtounicode{tfm:rpzdr/a52}{2739}
\pdfglyphtounicode{tfm:rpzdr/a53}{273A}
\pdfglyphtounicode{tfm:rpzdr/a54}{273B}
\pdfglyphtounicode{tfm:rpzdr/a55}{273C}
\pdfglyphtounicode{tfm:rpzdr/a56}{273D}
\pdfglyphtounicode{tfm:rpzdr/a57}{273E}
\pdfglyphtounicode{tfm:rpzdr/a58}{273F}
\pdfglyphtounicode{tfm:rpzdr/a59}{2740}
\pdfglyphtounicode{tfm:rpzdr/a6}{271D}
\pdfglyphtounicode{tfm:rpzdr/a60}{2741}
\pdfglyphtounicode{tfm:rpzdr/a61}{2742}
\pdfglyphtounicode{tfm:rpzdr/a62}{2743}
\pdfglyphtounicode{tfm:rpzdr/a63}{2744}
\pdfglyphtounicode{tfm:rpzdr/a64}{2745}
\pdfglyphtounicode{tfm:rpzdr/a65}{2746}
\pdfglyphtounicode{tfm:rpzdr/a66}{2747}
\pdfglyphtounicode{tfm:rpzdr/a67}{2748}
\pdfglyphtounicode{tfm:rpzdr/a68}{2749}
\pdfglyphtounicode{tfm:rpzdr/a69}{274A}
\pdfglyphtounicode{tfm:rpzdr/a7}{271E}
\pdfglyphtounicode{tfm:rpzdr/a70}{274B}
\pdfglyphtounicode{tfm:rpzdr/a71}{25CF}
\pdfglyphtounicode{tfm:rpzdr/a72}{274D}
\pdfglyphtounicode{tfm:rpzdr/a73}{25A0}
\pdfglyphtounicode{tfm:rpzdr/a74}{274F}
\pdfglyphtounicode{tfm:rpzdr/a75}{2751}
\pdfglyphtounicode{tfm:rpzdr/a76}{25B2}
\pdfglyphtounicode{tfm:rpzdr/a77}{25BC}
\pdfglyphtounicode{tfm:rpzdr/a78}{25C6}
\pdfglyphtounicode{tfm:rpzdr/a79}{2756}
\pdfglyphtounicode{tfm:rpzdr/a8}{271F}
\pdfglyphtounicode{tfm:rpzdr/a81}{25D7}
\pdfglyphtounicode{tfm:rpzdr/a82}{2758}
\pdfglyphtounicode{tfm:rpzdr/a83}{2759}
\pdfglyphtounicode{tfm:rpzdr/a84}{275A}
\pdfglyphtounicode{tfm:rpzdr/a85}{276F}
\pdfglyphtounicode{tfm:rpzdr/a86}{2771}
\pdfglyphtounicode{tfm:rpzdr/a87}{2772}
\pdfglyphtounicode{tfm:rpzdr/a88}{2773}
\pdfglyphtounicode{tfm:rpzdr/a89}{2768}
\pdfglyphtounicode{tfm:rpzdr/a9}{2720}
\pdfglyphtounicode{tfm:rpzdr/a90}{2769}
\pdfglyphtounicode{tfm:rpzdr/a91}{276C}
\pdfglyphtounicode{tfm:rpzdr/a92}{276D}
\pdfglyphtounicode{tfm:rpzdr/a93}{276A}
\pdfglyphtounicode{tfm:rpzdr/a94}{276B}
\pdfglyphtounicode{tfm:rpzdr/a95}{2774}
\pdfglyphtounicode{tfm:rpzdr/a96}{2775}
\pdfglyphtounicode{tfm:rpzdr/a97}{275B}
\pdfglyphtounicode{tfm:rpzdr/a98}{275C}
\pdfglyphtounicode{tfm:rpzdr/a99}{275D}
\pdfglyphtounicode{tfm:rtxbmi/phi}{03D5}
\pdfglyphtounicode{tfm:rtxbmi/phi1}{03C6}
\pdfglyphtounicode{tfm:rtxmi/phi}{03D5}
\pdfglyphtounicode{tfm:rtxmi/phi1}{03C6}
\pdfglyphtounicode{tfm:txbmia/phi}{03D5}
\pdfglyphtounicode{tfm:txbmia/phi1}{03C6}
\pdfglyphtounicode{tfm:txbsy/diamond}{2662}
\pdfglyphtounicode{tfm:txbsy/heart}{2661}
\pdfglyphtounicode{tfm:txbsya/diamond}{2662}
\pdfglyphtounicode{tfm:txmia/phi}{03D5}
\pdfglyphtounicode{tfm:txmia/phi1}{03C6}
\pdfglyphtounicode{tfm:txsy/diamond}{2662}
\pdfglyphtounicode{tfm:txsy/heart}{2661}
\pdfglyphtounicode{tfm:txsya/diamond}{2662}
\pdfglyphtounicode{tfm:zd/a1}{2701}
\pdfglyphtounicode{tfm:zd/a10}{2721}
\pdfglyphtounicode{tfm:zd/a100}{275E}
\pdfglyphtounicode{tfm:zd/a101}{2761}
\pdfglyphtounicode{tfm:zd/a102}{2762}
\pdfglyphtounicode{tfm:zd/a103}{2763}
\pdfglyphtounicode{tfm:zd/a104}{2764}
\pdfglyphtounicode{tfm:zd/a105}{2710}
\pdfglyphtounicode{tfm:zd/a106}{2765}
\pdfglyphtounicode{tfm:zd/a107}{2766}
\pdfglyphtounicode{tfm:zd/a108}{2767}
\pdfglyphtounicode{tfm:zd/a109}{2660}
\pdfglyphtounicode{tfm:zd/a11}{261B}
\pdfglyphtounicode{tfm:zd/a110}{2665}
\pdfglyphtounicode{tfm:zd/a111}{2666}
\pdfglyphtounicode{tfm:zd/a112}{2663}
\pdfglyphtounicode{tfm:zd/a117}{2709}
\pdfglyphtounicode{tfm:zd/a118}{2708}
\pdfglyphtounicode{tfm:zd/a119}{2707}
\pdfglyphtounicode{tfm:zd/a12}{261E}
\pdfglyphtounicode{tfm:zd/a120}{2460}
\pdfglyphtounicode{tfm:zd/a121}{2461}
\pdfglyphtounicode{tfm:zd/a122}{2462}
\pdfglyphtounicode{tfm:zd/a123}{2463}
\pdfglyphtounicode{tfm:zd/a124}{2464}
\pdfglyphtounicode{tfm:zd/a125}{2465}
\pdfglyphtounicode{tfm:zd/a126}{2466}
\pdfglyphtounicode{tfm:zd/a127}{2467}
\pdfglyphtounicode{tfm:zd/a128}{2468}
\pdfglyphtounicode{tfm:zd/a129}{2469}
\pdfglyphtounicode{tfm:zd/a13}{270C}
\pdfglyphtounicode{tfm:zd/a130}{2776}
\pdfglyphtounicode{tfm:zd/a131}{2777}
\pdfglyphtounicode{tfm:zd/a132}{2778}
\pdfglyphtounicode{tfm:zd/a133}{2779}
\pdfglyphtounicode{tfm:zd/a134}{277A}
\pdfglyphtounicode{tfm:zd/a135}{277B}
\pdfglyphtounicode{tfm:zd/a136}{277C}
\pdfglyphtounicode{tfm:zd/a137}{277D}
\pdfglyphtounicode{tfm:zd/a138}{277E}
\pdfglyphtounicode{tfm:zd/a139}{277F}
\pdfglyphtounicode{tfm:zd/a14}{270D}
\pdfglyphtounicode{tfm:zd/a140}{2780}
\pdfglyphtounicode{tfm:zd/a141}{2781}
\pdfglyphtounicode{tfm:zd/a142}{2782}
\pdfglyphtounicode{tfm:zd/a143}{2783}
\pdfglyphtounicode{tfm:zd/a144}{2784}
\pdfglyphtounicode{tfm:zd/a145}{2785}
\pdfglyphtounicode{tfm:zd/a146}{2786}
\pdfglyphtounicode{tfm:zd/a147}{2787}
\pdfglyphtounicode{tfm:zd/a148}{2788}
\pdfglyphtounicode{tfm:zd/a149}{2789}
\pdfglyphtounicode{tfm:zd/a15}{270E}
\pdfglyphtounicode{tfm:zd/a150}{278A}
\pdfglyphtounicode{tfm:zd/a151}{278B}
\pdfglyphtounicode{tfm:zd/a152}{278C}
\pdfglyphtounicode{tfm:zd/a153}{278D}
\pdfglyphtounicode{tfm:zd/a154}{278E}
\pdfglyphtounicode{tfm:zd/a155}{278F}
\pdfglyphtounicode{tfm:zd/a156}{2790}
\pdfglyphtounicode{tfm:zd/a157}{2791}
\pdfglyphtounicode{tfm:zd/a158}{2792}
\pdfglyphtounicode{tfm:zd/a159}{2793}
\pdfglyphtounicode{tfm:zd/a16}{270F}
\pdfglyphtounicode{tfm:zd/a160}{2794}
\pdfglyphtounicode{tfm:zd/a161}{2192}
\pdfglyphtounicode{tfm:zd/a162}{27A3}
\pdfglyphtounicode{tfm:zd/a163}{2194}
\pdfglyphtounicode{tfm:zd/a164}{2195}
\pdfglyphtounicode{tfm:zd/a165}{2799}
\pdfglyphtounicode{tfm:zd/a166}{279B}
\pdfglyphtounicode{tfm:zd/a167}{279C}
\pdfglyphtounicode{tfm:zd/a168}{279D}
\pdfglyphtounicode{tfm:zd/a169}{279E}
\pdfglyphtounicode{tfm:zd/a17}{2711}
\pdfglyphtounicode{tfm:zd/a170}{279F}
\pdfglyphtounicode{tfm:zd/a171}{27A0}
\pdfglyphtounicode{tfm:zd/a172}{27A1}
\pdfglyphtounicode{tfm:zd/a173}{27A2}
\pdfglyphtounicode{tfm:zd/a174}{27A4}
\pdfglyphtounicode{tfm:zd/a175}{27A5}
\pdfglyphtounicode{tfm:zd/a176}{27A6}
\pdfglyphtounicode{tfm:zd/a177}{27A7}
\pdfglyphtounicode{tfm:zd/a178}{27A8}
\pdfglyphtounicode{tfm:zd/a179}{27A9}
\pdfglyphtounicode{tfm:zd/a18}{2712}
\pdfglyphtounicode{tfm:zd/a180}{27AB}
\pdfglyphtounicode{tfm:zd/a181}{27AD}
\pdfglyphtounicode{tfm:zd/a182}{27AF}
\pdfglyphtounicode{tfm:zd/a183}{27B2}
\pdfglyphtounicode{tfm:zd/a184}{27B3}
\pdfglyphtounicode{tfm:zd/a185}{27B5}
\pdfglyphtounicode{tfm:zd/a186}{27B8}
\pdfglyphtounicode{tfm:zd/a187}{27BA}
\pdfglyphtounicode{tfm:zd/a188}{27BB}
\pdfglyphtounicode{tfm:zd/a189}{27BC}
\pdfglyphtounicode{tfm:zd/a19}{2713}
\pdfglyphtounicode{tfm:zd/a190}{27BD}
\pdfglyphtounicode{tfm:zd/a191}{27BE}
\pdfglyphtounicode{tfm:zd/a192}{279A}
\pdfglyphtounicode{tfm:zd/a193}{27AA}
\pdfglyphtounicode{tfm:zd/a194}{27B6}
\pdfglyphtounicode{tfm:zd/a195}{27B9}
\pdfglyphtounicode{tfm:zd/a196}{2798}
\pdfglyphtounicode{tfm:zd/a197}{27B4}
\pdfglyphtounicode{tfm:zd/a198}{27B7}
\pdfglyphtounicode{tfm:zd/a199}{27AC}
\pdfglyphtounicode{tfm:zd/a2}{2702}
\pdfglyphtounicode{tfm:zd/a20}{2714}
\pdfglyphtounicode{tfm:zd/a200}{27AE}
\pdfglyphtounicode{tfm:zd/a201}{27B1}
\pdfglyphtounicode{tfm:zd/a202}{2703}
\pdfglyphtounicode{tfm:zd/a203}{2750}
\pdfglyphtounicode{tfm:zd/a204}{2752}
\pdfglyphtounicode{tfm:zd/a205}{276E}
\pdfglyphtounicode{tfm:zd/a206}{2770}
\pdfglyphtounicode{tfm:zd/a21}{2715}
\pdfglyphtounicode{tfm:zd/a22}{2716}
\pdfglyphtounicode{tfm:zd/a23}{2717}
\pdfglyphtounicode{tfm:zd/a24}{2718}
\pdfglyphtounicode{tfm:zd/a25}{2719}
\pdfglyphtounicode{tfm:zd/a26}{271A}
\pdfglyphtounicode{tfm:zd/a27}{271B}
\pdfglyphtounicode{tfm:zd/a28}{271C}
\pdfglyphtounicode{tfm:zd/a29}{2722}
\pdfglyphtounicode{tfm:zd/a3}{2704}
\pdfglyphtounicode{tfm:zd/a30}{2723}
\pdfglyphtounicode{tfm:zd/a31}{2724}
\pdfglyphtounicode{tfm:zd/a32}{2725}
\pdfglyphtounicode{tfm:zd/a33}{2726}
\pdfglyphtounicode{tfm:zd/a34}{2727}
\pdfglyphtounicode{tfm:zd/a35}{2605}
\pdfglyphtounicode{tfm:zd/a36}{2729}
\pdfglyphtounicode{tfm:zd/a37}{272A}
\pdfglyphtounicode{tfm:zd/a38}{272B}
\pdfglyphtounicode{tfm:zd/a39}{272C}
\pdfglyphtounicode{tfm:zd/a4}{260E}
\pdfglyphtounicode{tfm:zd/a40}{272D}
\pdfglyphtounicode{tfm:zd/a41}{272E}
\pdfglyphtounicode{tfm:zd/a42}{272F}
\pdfglyphtounicode{tfm:zd/a43}{2730}
\pdfglyphtounicode{tfm:zd/a44}{2731}
\pdfglyphtounicode{tfm:zd/a45}{2732}
\pdfglyphtounicode{tfm:zd/a46}{2733}
\pdfglyphtounicode{tfm:zd/a47}{2734}
\pdfglyphtounicode{tfm:zd/a48}{2735}
\pdfglyphtounicode{tfm:zd/a49}{2736}
\pdfglyphtounicode{tfm:zd/a5}{2706}
\pdfglyphtounicode{tfm:zd/a50}{2737}
\pdfglyphtounicode{tfm:zd/a51}{2738}
\pdfglyphtounicode{tfm:zd/a52}{2739}
\pdfglyphtounicode{tfm:zd/a53}{273A}
\pdfglyphtounicode{tfm:zd/a54}{273B}
\pdfglyphtounicode{tfm:zd/a55}{273C}
\pdfglyphtounicode{tfm:zd/a56}{273D}
\pdfglyphtounicode{tfm:zd/a57}{273E}
\pdfglyphtounicode{tfm:zd/a58}{273F}
\pdfglyphtounicode{tfm:zd/a59}{2740}
\pdfglyphtounicode{tfm:zd/a6}{271D}
\pdfglyphtounicode{tfm:zd/a60}{2741}
\pdfglyphtounicode{tfm:zd/a61}{2742}
\pdfglyphtounicode{tfm:zd/a62}{2743}
\pdfglyphtounicode{tfm:zd/a63}{2744}
\pdfglyphtounicode{tfm:zd/a64}{2745}
\pdfglyphtounicode{tfm:zd/a65}{2746}
\pdfglyphtounicode{tfm:zd/a66}{2747}
\pdfglyphtounicode{tfm:zd/a67}{2748}
\pdfglyphtounicode{tfm:zd/a68}{2749}
\pdfglyphtounicode{tfm:zd/a69}{274A}
\pdfglyphtounicode{tfm:zd/a7}{271E}
\pdfglyphtounicode{tfm:zd/a70}{274B}
\pdfglyphtounicode{tfm:zd/a71}{25CF}
\pdfglyphtounicode{tfm:zd/a72}{274D}
\pdfglyphtounicode{tfm:zd/a73}{25A0}
\pdfglyphtounicode{tfm:zd/a74}{274F}
\pdfglyphtounicode{tfm:zd/a75}{2751}
\pdfglyphtounicode{tfm:zd/a76}{25B2}
\pdfglyphtounicode{tfm:zd/a77}{25BC}
\pdfglyphtounicode{tfm:zd/a78}{25C6}
\pdfglyphtounicode{tfm:zd/a79}{2756}
\pdfglyphtounicode{tfm:zd/a8}{271F}
\pdfglyphtounicode{tfm:zd/a81}{25D7}
\pdfglyphtounicode{tfm:zd/a82}{2758}
\pdfglyphtounicode{tfm:zd/a83}{2759}
\pdfglyphtounicode{tfm:zd/a84}{275A}
\pdfglyphtounicode{tfm:zd/a85}{276F}
\pdfglyphtounicode{tfm:zd/a86}{2771}
\pdfglyphtounicode{tfm:zd/a87}{2772}
\pdfglyphtounicode{tfm:zd/a88}{2773}
\pdfglyphtounicode{tfm:zd/a89}{2768}
\pdfglyphtounicode{tfm:zd/a9}{2720}
\pdfglyphtounicode{tfm:zd/a90}{2769}
\pdfglyphtounicode{tfm:zd/a91}{276C}
\pdfglyphtounicode{tfm:zd/a92}{276D}
\pdfglyphtounicode{tfm:zd/a93}{276A}
\pdfglyphtounicode{tfm:zd/a94}{276B}
\pdfglyphtounicode{tfm:zd/a95}{2774}
\pdfglyphtounicode{tfm:zd/a96}{2775}
\pdfglyphtounicode{tfm:zd/a97}{275B}
\pdfglyphtounicode{tfm:zd/a98}{275C}
\pdfglyphtounicode{tfm:zd/a99}{275D}
\pdfglyphtounicode{tfm:zpzdr-reversed/a1}{2701}
\pdfglyphtounicode{tfm:zpzdr-reversed/a10}{2721}
\pdfglyphtounicode{tfm:zpzdr-reversed/a100}{275E}
\pdfglyphtounicode{tfm:zpzdr-reversed/a101}{2761}
\pdfglyphtounicode{tfm:zpzdr-reversed/a102}{2762}
\pdfglyphtounicode{tfm:zpzdr-reversed/a103}{2763}
\pdfglyphtounicode{tfm:zpzdr-reversed/a104}{2764}
\pdfglyphtounicode{tfm:zpzdr-reversed/a105}{2710}
\pdfglyphtounicode{tfm:zpzdr-reversed/a106}{2765}
\pdfglyphtounicode{tfm:zpzdr-reversed/a107}{2766}
\pdfglyphtounicode{tfm:zpzdr-reversed/a108}{2767}
\pdfglyphtounicode{tfm:zpzdr-reversed/a109}{2660}
\pdfglyphtounicode{tfm:zpzdr-reversed/a11}{261B}
\pdfglyphtounicode{tfm:zpzdr-reversed/a110}{2665}
\pdfglyphtounicode{tfm:zpzdr-reversed/a111}{2666}
\pdfglyphtounicode{tfm:zpzdr-reversed/a112}{2663}
\pdfglyphtounicode{tfm:zpzdr-reversed/a117}{2709}
\pdfglyphtounicode{tfm:zpzdr-reversed/a118}{2708}
\pdfglyphtounicode{tfm:zpzdr-reversed/a119}{2707}
\pdfglyphtounicode{tfm:zpzdr-reversed/a12}{261E}
\pdfglyphtounicode{tfm:zpzdr-reversed/a120}{2460}
\pdfglyphtounicode{tfm:zpzdr-reversed/a121}{2461}
\pdfglyphtounicode{tfm:zpzdr-reversed/a122}{2462}
\pdfglyphtounicode{tfm:zpzdr-reversed/a123}{2463}
\pdfglyphtounicode{tfm:zpzdr-reversed/a124}{2464}
\pdfglyphtounicode{tfm:zpzdr-reversed/a125}{2465}
\pdfglyphtounicode{tfm:zpzdr-reversed/a126}{2466}
\pdfglyphtounicode{tfm:zpzdr-reversed/a127}{2467}
\pdfglyphtounicode{tfm:zpzdr-reversed/a128}{2468}
\pdfglyphtounicode{tfm:zpzdr-reversed/a129}{2469}
\pdfglyphtounicode{tfm:zpzdr-reversed/a13}{270C}
\pdfglyphtounicode{tfm:zpzdr-reversed/a130}{2776}
\pdfglyphtounicode{tfm:zpzdr-reversed/a131}{2777}
\pdfglyphtounicode{tfm:zpzdr-reversed/a132}{2778}
\pdfglyphtounicode{tfm:zpzdr-reversed/a133}{2779}
\pdfglyphtounicode{tfm:zpzdr-reversed/a134}{277A}
\pdfglyphtounicode{tfm:zpzdr-reversed/a135}{277B}
\pdfglyphtounicode{tfm:zpzdr-reversed/a136}{277C}
\pdfglyphtounicode{tfm:zpzdr-reversed/a137}{277D}
\pdfglyphtounicode{tfm:zpzdr-reversed/a138}{277E}
\pdfglyphtounicode{tfm:zpzdr-reversed/a139}{277F}
\pdfglyphtounicode{tfm:zpzdr-reversed/a14}{270D}
\pdfglyphtounicode{tfm:zpzdr-reversed/a140}{2780}
\pdfglyphtounicode{tfm:zpzdr-reversed/a141}{2781}
\pdfglyphtounicode{tfm:zpzdr-reversed/a142}{2782}
\pdfglyphtounicode{tfm:zpzdr-reversed/a143}{2783}
\pdfglyphtounicode{tfm:zpzdr-reversed/a144}{2784}
\pdfglyphtounicode{tfm:zpzdr-reversed/a145}{2785}
\pdfglyphtounicode{tfm:zpzdr-reversed/a146}{2786}
\pdfglyphtounicode{tfm:zpzdr-reversed/a147}{2787}
\pdfglyphtounicode{tfm:zpzdr-reversed/a148}{2788}
\pdfglyphtounicode{tfm:zpzdr-reversed/a149}{2789}
\pdfglyphtounicode{tfm:zpzdr-reversed/a15}{270E}
\pdfglyphtounicode{tfm:zpzdr-reversed/a150}{278A}
\pdfglyphtounicode{tfm:zpzdr-reversed/a151}{278B}
\pdfglyphtounicode{tfm:zpzdr-reversed/a152}{278C}
\pdfglyphtounicode{tfm:zpzdr-reversed/a153}{278D}
\pdfglyphtounicode{tfm:zpzdr-reversed/a154}{278E}
\pdfglyphtounicode{tfm:zpzdr-reversed/a155}{278F}
\pdfglyphtounicode{tfm:zpzdr-reversed/a156}{2790}
\pdfglyphtounicode{tfm:zpzdr-reversed/a157}{2791}
\pdfglyphtounicode{tfm:zpzdr-reversed/a158}{2792}
\pdfglyphtounicode{tfm:zpzdr-reversed/a159}{2793}
\pdfglyphtounicode{tfm:zpzdr-reversed/a16}{270F}
\pdfglyphtounicode{tfm:zpzdr-reversed/a160}{2794}
\pdfglyphtounicode{tfm:zpzdr-reversed/a161}{2192}
\pdfglyphtounicode{tfm:zpzdr-reversed/a162}{27A3}
\pdfglyphtounicode{tfm:zpzdr-reversed/a163}{2194}
\pdfglyphtounicode{tfm:zpzdr-reversed/a164}{2195}
\pdfglyphtounicode{tfm:zpzdr-reversed/a165}{2799}
\pdfglyphtounicode{tfm:zpzdr-reversed/a166}{279B}
\pdfglyphtounicode{tfm:zpzdr-reversed/a167}{279C}
\pdfglyphtounicode{tfm:zpzdr-reversed/a168}{279D}
\pdfglyphtounicode{tfm:zpzdr-reversed/a169}{279E}
\pdfglyphtounicode{tfm:zpzdr-reversed/a17}{2711}
\pdfglyphtounicode{tfm:zpzdr-reversed/a170}{279F}
\pdfglyphtounicode{tfm:zpzdr-reversed/a171}{27A0}
\pdfglyphtounicode{tfm:zpzdr-reversed/a172}{27A1}
\pdfglyphtounicode{tfm:zpzdr-reversed/a173}{27A2}
\pdfglyphtounicode{tfm:zpzdr-reversed/a174}{27A4}
\pdfglyphtounicode{tfm:zpzdr-reversed/a175}{27A5}
\pdfglyphtounicode{tfm:zpzdr-reversed/a176}{27A6}
\pdfglyphtounicode{tfm:zpzdr-reversed/a177}{27A7}
\pdfglyphtounicode{tfm:zpzdr-reversed/a178}{27A8}
\pdfglyphtounicode{tfm:zpzdr-reversed/a179}{27A9}
\pdfglyphtounicode{tfm:zpzdr-reversed/a18}{2712}
\pdfglyphtounicode{tfm:zpzdr-reversed/a180}{27AB}
\pdfglyphtounicode{tfm:zpzdr-reversed/a181}{27AD}
\pdfglyphtounicode{tfm:zpzdr-reversed/a182}{27AF}
\pdfglyphtounicode{tfm:zpzdr-reversed/a183}{27B2}
\pdfglyphtounicode{tfm:zpzdr-reversed/a184}{27B3}
\pdfglyphtounicode{tfm:zpzdr-reversed/a185}{27B5}
\pdfglyphtounicode{tfm:zpzdr-reversed/a186}{27B8}
\pdfglyphtounicode{tfm:zpzdr-reversed/a187}{27BA}
\pdfglyphtounicode{tfm:zpzdr-reversed/a188}{27BB}
\pdfglyphtounicode{tfm:zpzdr-reversed/a189}{27BC}
\pdfglyphtounicode{tfm:zpzdr-reversed/a19}{2713}
\pdfglyphtounicode{tfm:zpzdr-reversed/a190}{27BD}
\pdfglyphtounicode{tfm:zpzdr-reversed/a191}{27BE}
\pdfglyphtounicode{tfm:zpzdr-reversed/a192}{279A}
\pdfglyphtounicode{tfm:zpzdr-reversed/a193}{27AA}
\pdfglyphtounicode{tfm:zpzdr-reversed/a194}{27B6}
\pdfglyphtounicode{tfm:zpzdr-reversed/a195}{27B9}
\pdfglyphtounicode{tfm:zpzdr-reversed/a196}{2798}
\pdfglyphtounicode{tfm:zpzdr-reversed/a197}{27B4}
\pdfglyphtounicode{tfm:zpzdr-reversed/a198}{27B7}
\pdfglyphtounicode{tfm:zpzdr-reversed/a199}{27AC}
\pdfglyphtounicode{tfm:zpzdr-reversed/a2}{2702}
\pdfglyphtounicode{tfm:zpzdr-reversed/a20}{2714}
\pdfglyphtounicode{tfm:zpzdr-reversed/a200}{27AE}
\pdfglyphtounicode{tfm:zpzdr-reversed/a201}{27B1}
\pdfglyphtounicode{tfm:zpzdr-reversed/a202}{2703}
\pdfglyphtounicode{tfm:zpzdr-reversed/a203}{2750}
\pdfglyphtounicode{tfm:zpzdr-reversed/a204}{2752}
\pdfglyphtounicode{tfm:zpzdr-reversed/a205}{276E}
\pdfglyphtounicode{tfm:zpzdr-reversed/a206}{2770}
\pdfglyphtounicode{tfm:zpzdr-reversed/a21}{2715}
\pdfglyphtounicode{tfm:zpzdr-reversed/a22}{2716}
\pdfglyphtounicode{tfm:zpzdr-reversed/a23}{2717}
\pdfglyphtounicode{tfm:zpzdr-reversed/a24}{2718}
\pdfglyphtounicode{tfm:zpzdr-reversed/a25}{2719}
\pdfglyphtounicode{tfm:zpzdr-reversed/a26}{271A}
\pdfglyphtounicode{tfm:zpzdr-reversed/a27}{271B}
\pdfglyphtounicode{tfm:zpzdr-reversed/a28}{271C}
\pdfglyphtounicode{tfm:zpzdr-reversed/a29}{2722}
\pdfglyphtounicode{tfm:zpzdr-reversed/a3}{2704}
\pdfglyphtounicode{tfm:zpzdr-reversed/a30}{2723}
\pdfglyphtounicode{tfm:zpzdr-reversed/a31}{2724}
\pdfglyphtounicode{tfm:zpzdr-reversed/a32}{2725}
\pdfglyphtounicode{tfm:zpzdr-reversed/a33}{2726}
\pdfglyphtounicode{tfm:zpzdr-reversed/a34}{2727}
\pdfglyphtounicode{tfm:zpzdr-reversed/a35}{2605}
\pdfglyphtounicode{tfm:zpzdr-reversed/a36}{2729}
\pdfglyphtounicode{tfm:zpzdr-reversed/a37}{272A}
\pdfglyphtounicode{tfm:zpzdr-reversed/a38}{272B}
\pdfglyphtounicode{tfm:zpzdr-reversed/a39}{272C}
\pdfglyphtounicode{tfm:zpzdr-reversed/a4}{260E}
\pdfglyphtounicode{tfm:zpzdr-reversed/a40}{272D}
\pdfglyphtounicode{tfm:zpzdr-reversed/a41}{272E}
\pdfglyphtounicode{tfm:zpzdr-reversed/a42}{272F}
\pdfglyphtounicode{tfm:zpzdr-reversed/a43}{2730}
\pdfglyphtounicode{tfm:zpzdr-reversed/a44}{2731}
\pdfglyphtounicode{tfm:zpzdr-reversed/a45}{2732}
\pdfglyphtounicode{tfm:zpzdr-reversed/a46}{2733}
\pdfglyphtounicode{tfm:zpzdr-reversed/a47}{2734}
\pdfglyphtounicode{tfm:zpzdr-reversed/a48}{2735}
\pdfglyphtounicode{tfm:zpzdr-reversed/a49}{2736}
\pdfglyphtounicode{tfm:zpzdr-reversed/a5}{2706}
\pdfglyphtounicode{tfm:zpzdr-reversed/a50}{2737}
\pdfglyphtounicode{tfm:zpzdr-reversed/a51}{2738}
\pdfglyphtounicode{tfm:zpzdr-reversed/a52}{2739}
\pdfglyphtounicode{tfm:zpzdr-reversed/a53}{273A}
\pdfglyphtounicode{tfm:zpzdr-reversed/a54}{273B}
\pdfglyphtounicode{tfm:zpzdr-reversed/a55}{273C}
\pdfglyphtounicode{tfm:zpzdr-reversed/a56}{273D}
\pdfglyphtounicode{tfm:zpzdr-reversed/a57}{273E}
\pdfglyphtounicode{tfm:zpzdr-reversed/a58}{273F}
\pdfglyphtounicode{tfm:zpzdr-reversed/a59}{2740}
\pdfglyphtounicode{tfm:zpzdr-reversed/a6}{271D}
\pdfglyphtounicode{tfm:zpzdr-reversed/a60}{2741}
\pdfglyphtounicode{tfm:zpzdr-reversed/a61}{2742}
\pdfglyphtounicode{tfm:zpzdr-reversed/a62}{2743}
\pdfglyphtounicode{tfm:zpzdr-reversed/a63}{2744}
\pdfglyphtounicode{tfm:zpzdr-reversed/a64}{2745}
\pdfglyphtounicode{tfm:zpzdr-reversed/a65}{2746}
\pdfglyphtounicode{tfm:zpzdr-reversed/a66}{2747}
\pdfglyphtounicode{tfm:zpzdr-reversed/a67}{2748}
\pdfglyphtounicode{tfm:zpzdr-reversed/a68}{2749}
\pdfglyphtounicode{tfm:zpzdr-reversed/a69}{274A}
\pdfglyphtounicode{tfm:zpzdr-reversed/a7}{271E}
\pdfglyphtounicode{tfm:zpzdr-reversed/a70}{274B}
\pdfglyphtounicode{tfm:zpzdr-reversed/a71}{25CF}
\pdfglyphtounicode{tfm:zpzdr-reversed/a72}{274D}
\pdfglyphtounicode{tfm:zpzdr-reversed/a73}{25A0}
\pdfglyphtounicode{tfm:zpzdr-reversed/a74}{274F}
\pdfglyphtounicode{tfm:zpzdr-reversed/a75}{2751}
\pdfglyphtounicode{tfm:zpzdr-reversed/a76}{25B2}
\pdfglyphtounicode{tfm:zpzdr-reversed/a77}{25BC}
\pdfglyphtounicode{tfm:zpzdr-reversed/a78}{25C6}
\pdfglyphtounicode{tfm:zpzdr-reversed/a79}{2756}
\pdfglyphtounicode{tfm:zpzdr-reversed/a8}{271F}
\pdfglyphtounicode{tfm:zpzdr-reversed/a81}{25D7}
\pdfglyphtounicode{tfm:zpzdr-reversed/a82}{2758}
\pdfglyphtounicode{tfm:zpzdr-reversed/a83}{2759}
\pdfglyphtounicode{tfm:zpzdr-reversed/a84}{275A}
\pdfglyphtounicode{tfm:zpzdr-reversed/a85}{276F}
\pdfglyphtounicode{tfm:zpzdr-reversed/a86}{2771}
\pdfglyphtounicode{tfm:zpzdr-reversed/a87}{2772}
\pdfglyphtounicode{tfm:zpzdr-reversed/a88}{2773}
\pdfglyphtounicode{tfm:zpzdr-reversed/a89}{2768}
\pdfglyphtounicode{tfm:zpzdr-reversed/a9}{2720}
\pdfglyphtounicode{tfm:zpzdr-reversed/a90}{2769}
\pdfglyphtounicode{tfm:zpzdr-reversed/a91}{276C}
\pdfglyphtounicode{tfm:zpzdr-reversed/a92}{276D}
\pdfglyphtounicode{tfm:zpzdr-reversed/a93}{276A}
\pdfglyphtounicode{tfm:zpzdr-reversed/a94}{276B}
\pdfglyphtounicode{tfm:zpzdr-reversed/a95}{2774}
\pdfglyphtounicode{tfm:zpzdr-reversed/a96}{2775}
\pdfglyphtounicode{tfm:zpzdr-reversed/a97}{275B}
\pdfglyphtounicode{tfm:zpzdr-reversed/a98}{275C}
\pdfglyphtounicode{tfm:zpzdr-reversed/a99}{275D}
\pdfglyphtounicode{thabengali}{09A5}
\pdfglyphtounicode{thadeva}{0925}
\pdfglyphtounicode{thagujarati}{0AA5}
\pdfglyphtounicode{thagurmukhi}{0A25}
\pdfglyphtounicode{thalarabic}{0630}
\pdfglyphtounicode{thalfinalarabic}{FEAC}
\pdfglyphtounicode{thanthakhatlowleftthai}{F898}
\pdfglyphtounicode{thanthakhatlowrightthai}{F897}
\pdfglyphtounicode{thanthakhatthai}{0E4C}
\pdfglyphtounicode{thanthakhatupperleftthai}{F896}
\pdfglyphtounicode{theharabic}{062B}
\pdfglyphtounicode{thehfinalarabic}{FE9A}
\pdfglyphtounicode{thehinitialarabic}{FE9B}
\pdfglyphtounicode{thehmedialarabic}{FE9C}
\pdfglyphtounicode{thereexists}{2203}
\pdfglyphtounicode{therefore}{2234}
\pdfglyphtounicode{theta}{03B8}
\pdfglyphtounicode{theta1}{03D1}
\pdfglyphtounicode{thetasymbolgreek}{03D1}
\pdfglyphtounicode{thieuthacirclekorean}{3279}
\pdfglyphtounicode{thieuthaparenkorean}{3219}
\pdfglyphtounicode{thieuthcirclekorean}{326B}
\pdfglyphtounicode{thieuthkorean}{314C}
\pdfglyphtounicode{thieuthparenkorean}{320B}
\pdfglyphtounicode{thirteencircle}{246C}
\pdfglyphtounicode{thirteenparen}{2480}
\pdfglyphtounicode{thirteenperiod}{2494}
\pdfglyphtounicode{thonangmonthothai}{0E11}
\pdfglyphtounicode{thook}{01AD}
\pdfglyphtounicode{thophuthaothai}{0E12}
\pdfglyphtounicode{thorn}{00FE}
\pdfglyphtounicode{thothahanthai}{0E17}
\pdfglyphtounicode{thothanthai}{0E10}
\pdfglyphtounicode{thothongthai}{0E18}
\pdfglyphtounicode{thothungthai}{0E16}
\pdfglyphtounicode{thousandcyrillic}{0482}
\pdfglyphtounicode{thousandsseparatorarabic}{066C}
\pdfglyphtounicode{thousandsseparatorpersian}{066C}
\pdfglyphtounicode{three}{0033}
\pdfglyphtounicode{threearabic}{0663}
\pdfglyphtounicode{threebengali}{09E9}
\pdfglyphtounicode{threecircle}{2462}
\pdfglyphtounicode{threecircleinversesansserif}{278C}
\pdfglyphtounicode{threedeva}{0969}
\pdfglyphtounicode{threeeighths}{215C}
\pdfglyphtounicode{threegujarati}{0AE9}
\pdfglyphtounicode{threegurmukhi}{0A69}
\pdfglyphtounicode{threehackarabic}{0663}
\pdfglyphtounicode{threehangzhou}{3023}
\pdfglyphtounicode{threeideographicparen}{3222}
\pdfglyphtounicode{threeinferior}{2083}
\pdfglyphtounicode{threemonospace}{FF13}
\pdfglyphtounicode{threenumeratorbengali}{09F6}
\pdfglyphtounicode{threeoldstyle}{0033}
\pdfglyphtounicode{threeparen}{2476}
\pdfglyphtounicode{threeperiod}{248A}
\pdfglyphtounicode{threepersian}{06F3}
\pdfglyphtounicode{threequarters}{00BE}
\pdfglyphtounicode{threequartersemdash}{F6DE}
\pdfglyphtounicode{threeroman}{2172}
\pdfglyphtounicode{threesuperior}{00B3}
\pdfglyphtounicode{threethai}{0E53}
\pdfglyphtounicode{thzsquare}{3394}
\pdfglyphtounicode{tihiragana}{3061}
\pdfglyphtounicode{tikatakana}{30C1}
\pdfglyphtounicode{tikatakanahalfwidth}{FF81}
\pdfglyphtounicode{tikeutacirclekorean}{3270}
\pdfglyphtounicode{tikeutaparenkorean}{3210}
\pdfglyphtounicode{tikeutcirclekorean}{3262}
\pdfglyphtounicode{tikeutkorean}{3137}
\pdfglyphtounicode{tikeutparenkorean}{3202}
\pdfglyphtounicode{tilde}{02DC}
\pdfglyphtounicode{tildebelowcmb}{0330}
\pdfglyphtounicode{tildecmb}{0303}
\pdfglyphtounicode{tildecomb}{0303}
\pdfglyphtounicode{tildedoublecmb}{0360}
\pdfglyphtounicode{tildeoperator}{223C}
\pdfglyphtounicode{tildeoverlaycmb}{0334}
\pdfglyphtounicode{tildeverticalcmb}{033E}
\pdfglyphtounicode{timescircle}{2297}
\pdfglyphtounicode{tipehahebrew}{0596}
\pdfglyphtounicode{tipehalefthebrew}{0596}
\pdfglyphtounicode{tippigurmukhi}{0A70}
\pdfglyphtounicode{titlocyrilliccmb}{0483}
\pdfglyphtounicode{tiwnarmenian}{057F}
\pdfglyphtounicode{tlinebelow}{1E6F}
\pdfglyphtounicode{tmonospace}{FF54}
\pdfglyphtounicode{toarmenian}{0569}
\pdfglyphtounicode{tohiragana}{3068}
\pdfglyphtounicode{tokatakana}{30C8}
\pdfglyphtounicode{tokatakanahalfwidth}{FF84}
\pdfglyphtounicode{tonebarextrahighmod}{02E5}
\pdfglyphtounicode{tonebarextralowmod}{02E9}
\pdfglyphtounicode{tonebarhighmod}{02E6}
\pdfglyphtounicode{tonebarlowmod}{02E8}
\pdfglyphtounicode{tonebarmidmod}{02E7}
\pdfglyphtounicode{tonefive}{01BD}
\pdfglyphtounicode{tonesix}{0185}
\pdfglyphtounicode{tonetwo}{01A8}
\pdfglyphtounicode{tonos}{0384}
\pdfglyphtounicode{tonsquare}{3327}
\pdfglyphtounicode{topatakthai}{0E0F}
\pdfglyphtounicode{tortoiseshellbracketleft}{3014}
\pdfglyphtounicode{tortoiseshellbracketleftsmall}{FE5D}
\pdfglyphtounicode{tortoiseshellbracketleftvertical}{FE39}
\pdfglyphtounicode{tortoiseshellbracketright}{3015}
\pdfglyphtounicode{tortoiseshellbracketrightsmall}{FE5E}
\pdfglyphtounicode{tortoiseshellbracketrightvertical}{FE3A}
\pdfglyphtounicode{totaothai}{0E15}
\pdfglyphtounicode{tpalatalhook}{01AB}
\pdfglyphtounicode{tparen}{24AF}
\pdfglyphtounicode{trademark}{2122}
\pdfglyphtounicode{trademarksans}{2122}
\pdfglyphtounicode{trademarkserif}{2122}
\pdfglyphtounicode{tretroflexhook}{0288}
\pdfglyphtounicode{triagdn}{25BC}
\pdfglyphtounicode{triaglf}{25C4}
\pdfglyphtounicode{triagrt}{25BA}
\pdfglyphtounicode{triagup}{25B2}
\pdfglyphtounicode{triangle}{25B3}
\pdfglyphtounicode{triangledownsld}{25BC}
\pdfglyphtounicode{triangleinv}{25BD}
\pdfglyphtounicode{triangleleft}{25C1}
\pdfglyphtounicode{triangleleftequal}{22B4}
\pdfglyphtounicode{triangleleftsld}{25C0}
\pdfglyphtounicode{triangleright}{25B7}
\pdfglyphtounicode{trianglerightequal}{22B5}
\pdfglyphtounicode{trianglerightsld}{25B6}
\pdfglyphtounicode{trianglesolid}{25B2}
\pdfglyphtounicode{ts}{02A6}
\pdfglyphtounicode{tsadi}{05E6}
\pdfglyphtounicode{tsadidagesh}{FB46}
\pdfglyphtounicode{tsadidageshhebrew}{FB46}
\pdfglyphtounicode{tsadihebrew}{05E6}
\pdfglyphtounicode{tsecyrillic}{0446}
\pdfglyphtounicode{tsere}{05B5}
\pdfglyphtounicode{tsere12}{05B5}
\pdfglyphtounicode{tsere1e}{05B5}
\pdfglyphtounicode{tsere2b}{05B5}
\pdfglyphtounicode{tserehebrew}{05B5}
\pdfglyphtounicode{tserenarrowhebrew}{05B5}
\pdfglyphtounicode{tserequarterhebrew}{05B5}
\pdfglyphtounicode{tserewidehebrew}{05B5}
\pdfglyphtounicode{tshecyrillic}{045B}
\pdfglyphtounicode{tsuperior}{0074}
\pdfglyphtounicode{ttabengali}{099F}
\pdfglyphtounicode{ttadeva}{091F}
\pdfglyphtounicode{ttagujarati}{0A9F}
\pdfglyphtounicode{ttagurmukhi}{0A1F}
\pdfglyphtounicode{tteharabic}{0679}
\pdfglyphtounicode{ttehfinalarabic}{FB67}
\pdfglyphtounicode{ttehinitialarabic}{FB68}
\pdfglyphtounicode{ttehmedialarabic}{FB69}
\pdfglyphtounicode{tthabengali}{09A0}
\pdfglyphtounicode{tthadeva}{0920}
\pdfglyphtounicode{tthagujarati}{0AA0}
\pdfglyphtounicode{tthagurmukhi}{0A20}
\pdfglyphtounicode{tturned}{0287}
\pdfglyphtounicode{tuhiragana}{3064}
\pdfglyphtounicode{tukatakana}{30C4}
\pdfglyphtounicode{tukatakanahalfwidth}{FF82}
\pdfglyphtounicode{turnstileleft}{22A2}
\pdfglyphtounicode{turnstileright}{22A3}
\pdfglyphtounicode{tusmallhiragana}{3063}
\pdfglyphtounicode{tusmallkatakana}{30C3}
\pdfglyphtounicode{tusmallkatakanahalfwidth}{FF6F}
\pdfglyphtounicode{twelvecircle}{246B}
\pdfglyphtounicode{twelveparen}{247F}
\pdfglyphtounicode{twelveperiod}{2493}
\pdfglyphtounicode{twelveroman}{217B}
\pdfglyphtounicode{twentycircle}{2473}
\pdfglyphtounicode{twentyhangzhou}{5344}
\pdfglyphtounicode{twentyparen}{2487}
\pdfglyphtounicode{twentyperiod}{249B}
\pdfglyphtounicode{two}{0032}
\pdfglyphtounicode{twoarabic}{0662}
\pdfglyphtounicode{twobengali}{09E8}
\pdfglyphtounicode{twocircle}{2461}
\pdfglyphtounicode{twocircleinversesansserif}{278B}
\pdfglyphtounicode{twodeva}{0968}
\pdfglyphtounicode{twodotenleader}{2025}
\pdfglyphtounicode{twodotleader}{2025}
\pdfglyphtounicode{twodotleadervertical}{FE30}
\pdfglyphtounicode{twogujarati}{0AE8}
\pdfglyphtounicode{twogurmukhi}{0A68}
\pdfglyphtounicode{twohackarabic}{0662}
\pdfglyphtounicode{twohangzhou}{3022}
\pdfglyphtounicode{twoideographicparen}{3221}
\pdfglyphtounicode{twoinferior}{2082}
\pdfglyphtounicode{twomonospace}{FF12}
\pdfglyphtounicode{twonumeratorbengali}{09F5}
\pdfglyphtounicode{twooldstyle}{0032}
\pdfglyphtounicode{twoparen}{2475}
\pdfglyphtounicode{twoperiod}{2489}
\pdfglyphtounicode{twopersian}{06F2}
\pdfglyphtounicode{tworoman}{2171}
\pdfglyphtounicode{twostroke}{01BB}
\pdfglyphtounicode{twosuperior}{00B2}
\pdfglyphtounicode{twothai}{0E52}
\pdfglyphtounicode{twothirds}{2154}
\pdfglyphtounicode{u}{0075}
\pdfglyphtounicode{uacute}{00FA}
\pdfglyphtounicode{ubar}{0289}
\pdfglyphtounicode{ubengali}{0989}
\pdfglyphtounicode{ubopomofo}{3128}
\pdfglyphtounicode{ubreve}{016D}
\pdfglyphtounicode{ucaron}{01D4}
\pdfglyphtounicode{ucircle}{24E4}
\pdfglyphtounicode{ucircumflex}{00FB}
\pdfglyphtounicode{ucircumflexbelow}{1E77}
\pdfglyphtounicode{ucyrillic}{0443}
\pdfglyphtounicode{udattadeva}{0951}
\pdfglyphtounicode{udblacute}{0171}
\pdfglyphtounicode{udblgrave}{0215}
\pdfglyphtounicode{udeva}{0909}
\pdfglyphtounicode{udieresis}{00FC}
\pdfglyphtounicode{udieresisacute}{01D8}
\pdfglyphtounicode{udieresisbelow}{1E73}
\pdfglyphtounicode{udieresiscaron}{01DA}
\pdfglyphtounicode{udieresiscyrillic}{04F1}
\pdfglyphtounicode{udieresisgrave}{01DC}
\pdfglyphtounicode{udieresismacron}{01D6}
\pdfglyphtounicode{udotbelow}{1EE5}
\pdfglyphtounicode{ugrave}{00F9}
\pdfglyphtounicode{ugujarati}{0A89}
\pdfglyphtounicode{ugurmukhi}{0A09}
\pdfglyphtounicode{uhiragana}{3046}
\pdfglyphtounicode{uhookabove}{1EE7}
\pdfglyphtounicode{uhorn}{01B0}
\pdfglyphtounicode{uhornacute}{1EE9}
\pdfglyphtounicode{uhorndotbelow}{1EF1}
\pdfglyphtounicode{uhorngrave}{1EEB}
\pdfglyphtounicode{uhornhookabove}{1EED}
\pdfglyphtounicode{uhorntilde}{1EEF}
\pdfglyphtounicode{uhungarumlaut}{0171}
\pdfglyphtounicode{uhungarumlautcyrillic}{04F3}
\pdfglyphtounicode{uinvertedbreve}{0217}
\pdfglyphtounicode{ukatakana}{30A6}
\pdfglyphtounicode{ukatakanahalfwidth}{FF73}
\pdfglyphtounicode{ukcyrillic}{0479}
\pdfglyphtounicode{ukorean}{315C}
\pdfglyphtounicode{umacron}{016B}
\pdfglyphtounicode{umacroncyrillic}{04EF}
\pdfglyphtounicode{umacrondieresis}{1E7B}
\pdfglyphtounicode{umatragurmukhi}{0A41}
\pdfglyphtounicode{umonospace}{FF55}
\pdfglyphtounicode{underscore}{005F}
\pdfglyphtounicode{underscoredbl}{2017}
\pdfglyphtounicode{underscoremonospace}{FF3F}
\pdfglyphtounicode{underscorevertical}{FE33}
\pdfglyphtounicode{underscorewavy}{FE4F}
\pdfglyphtounicode{union}{222A}
\pdfglyphtounicode{uniondbl}{22D3}
\pdfglyphtounicode{unionmulti}{228E}
\pdfglyphtounicode{unionsq}{2294}
\pdfglyphtounicode{universal}{2200}
\pdfglyphtounicode{uogonek}{0173}
\pdfglyphtounicode{uparen}{24B0}
\pdfglyphtounicode{upblock}{2580}
\pdfglyphtounicode{upperdothebrew}{05C4}
\pdfglyphtounicode{uprise}{22CF}
\pdfglyphtounicode{upsilon}{03C5}
\pdfglyphtounicode{upsilondieresis}{03CB}
\pdfglyphtounicode{upsilondieresistonos}{03B0}
\pdfglyphtounicode{upsilonlatin}{028A}
\pdfglyphtounicode{upsilontonos}{03CD}
\pdfglyphtounicode{upslope}{29F8}
\pdfglyphtounicode{uptackbelowcmb}{031D}
\pdfglyphtounicode{uptackmod}{02D4}
\pdfglyphtounicode{uragurmukhi}{0A73}
\pdfglyphtounicode{uring}{016F}
\pdfglyphtounicode{ushortcyrillic}{045E}
\pdfglyphtounicode{usmallhiragana}{3045}
\pdfglyphtounicode{usmallkatakana}{30A5}
\pdfglyphtounicode{usmallkatakanahalfwidth}{FF69}
\pdfglyphtounicode{ustraightcyrillic}{04AF}
\pdfglyphtounicode{ustraightstrokecyrillic}{04B1}
\pdfglyphtounicode{utilde}{0169}
\pdfglyphtounicode{utildeacute}{1E79}
\pdfglyphtounicode{utildebelow}{1E75}
\pdfglyphtounicode{uubengali}{098A}
\pdfglyphtounicode{uudeva}{090A}
\pdfglyphtounicode{uugujarati}{0A8A}
\pdfglyphtounicode{uugurmukhi}{0A0A}
\pdfglyphtounicode{uumatragurmukhi}{0A42}
\pdfglyphtounicode{uuvowelsignbengali}{09C2}
\pdfglyphtounicode{uuvowelsigndeva}{0942}
\pdfglyphtounicode{uuvowelsigngujarati}{0AC2}
\pdfglyphtounicode{uvowelsignbengali}{09C1}
\pdfglyphtounicode{uvowelsigndeva}{0941}
\pdfglyphtounicode{uvowelsigngujarati}{0AC1}
\pdfglyphtounicode{v}{0076}
\pdfglyphtounicode{vadeva}{0935}
\pdfglyphtounicode{vagujarati}{0AB5}
\pdfglyphtounicode{vagurmukhi}{0A35}
\pdfglyphtounicode{vakatakana}{30F7}
\pdfglyphtounicode{vav}{05D5}
\pdfglyphtounicode{vavdagesh}{FB35}
\pdfglyphtounicode{vavdagesh65}{FB35}
\pdfglyphtounicode{vavdageshhebrew}{FB35}
\pdfglyphtounicode{vavhebrew}{05D5}
\pdfglyphtounicode{vavholam}{FB4B}
\pdfglyphtounicode{vavholamhebrew}{FB4B}
\pdfglyphtounicode{vavvavhebrew}{05F0}
\pdfglyphtounicode{vavyodhebrew}{05F1}
\pdfglyphtounicode{vcircle}{24E5}
\pdfglyphtounicode{vdotbelow}{1E7F}
\pdfglyphtounicode{vector}{20D7}
\pdfglyphtounicode{vecyrillic}{0432}
\pdfglyphtounicode{veharabic}{06A4}
\pdfglyphtounicode{vehfinalarabic}{FB6B}
\pdfglyphtounicode{vehinitialarabic}{FB6C}
\pdfglyphtounicode{vehmedialarabic}{FB6D}
\pdfglyphtounicode{vekatakana}{30F9}
\pdfglyphtounicode{venus}{2640}
\pdfglyphtounicode{verticalbar}{007C}
\pdfglyphtounicode{verticallineabovecmb}{030D}
\pdfglyphtounicode{verticallinebelowcmb}{0329}
\pdfglyphtounicode{verticallinelowmod}{02CC}
\pdfglyphtounicode{verticallinemod}{02C8}
\pdfglyphtounicode{vewarmenian}{057E}
\pdfglyphtounicode{vhook}{028B}
\pdfglyphtounicode{vikatakana}{30F8}
\pdfglyphtounicode{viramabengali}{09CD}
\pdfglyphtounicode{viramadeva}{094D}
\pdfglyphtounicode{viramagujarati}{0ACD}
\pdfglyphtounicode{visargabengali}{0983}
\pdfglyphtounicode{visargadeva}{0903}
\pdfglyphtounicode{visargagujarati}{0A83}
\pdfglyphtounicode{visiblespace}{2423}
\pdfglyphtounicode{visualspace}{2423}
\pdfglyphtounicode{vmonospace}{FF56}
\pdfglyphtounicode{voarmenian}{0578}
\pdfglyphtounicode{voicediterationhiragana}{309E}
\pdfglyphtounicode{voicediterationkatakana}{30FE}
\pdfglyphtounicode{voicedmarkkana}{309B}
\pdfglyphtounicode{voicedmarkkanahalfwidth}{FF9E}
\pdfglyphtounicode{vokatakana}{30FA}
\pdfglyphtounicode{vparen}{24B1}
\pdfglyphtounicode{vtilde}{1E7D}
\pdfglyphtounicode{vturned}{028C}
\pdfglyphtounicode{vuhiragana}{3094}
\pdfglyphtounicode{vukatakana}{30F4}
\pdfglyphtounicode{w}{0077}
\pdfglyphtounicode{wacute}{1E83}
\pdfglyphtounicode{waekorean}{3159}
\pdfglyphtounicode{wahiragana}{308F}
\pdfglyphtounicode{wakatakana}{30EF}
\pdfglyphtounicode{wakatakanahalfwidth}{FF9C}
\pdfglyphtounicode{wakorean}{3158}
\pdfglyphtounicode{wasmallhiragana}{308E}
\pdfglyphtounicode{wasmallkatakana}{30EE}
\pdfglyphtounicode{wattosquare}{3357}
\pdfglyphtounicode{wavedash}{301C}
\pdfglyphtounicode{wavyunderscorevertical}{FE34}
\pdfglyphtounicode{wawarabic}{0648}
\pdfglyphtounicode{wawfinalarabic}{FEEE}
\pdfglyphtounicode{wawhamzaabovearabic}{0624}
\pdfglyphtounicode{wawhamzaabovefinalarabic}{FE86}
\pdfglyphtounicode{wbsquare}{33DD}
\pdfglyphtounicode{wcircle}{24E6}
\pdfglyphtounicode{wcircumflex}{0175}
\pdfglyphtounicode{wdieresis}{1E85}
\pdfglyphtounicode{wdotaccent}{1E87}
\pdfglyphtounicode{wdotbelow}{1E89}
\pdfglyphtounicode{wehiragana}{3091}
\pdfglyphtounicode{weierstrass}{2118}
\pdfglyphtounicode{wekatakana}{30F1}
\pdfglyphtounicode{wekorean}{315E}
\pdfglyphtounicode{weokorean}{315D}
\pdfglyphtounicode{wgrave}{1E81}
\pdfglyphtounicode{whitebullet}{25E6}
\pdfglyphtounicode{whitecircle}{25CB}
\pdfglyphtounicode{whitecircleinverse}{25D9}
\pdfglyphtounicode{whitecornerbracketleft}{300E}
\pdfglyphtounicode{whitecornerbracketleftvertical}{FE43}
\pdfglyphtounicode{whitecornerbracketright}{300F}
\pdfglyphtounicode{whitecornerbracketrightvertical}{FE44}
\pdfglyphtounicode{whitediamond}{25C7}
\pdfglyphtounicode{whitediamondcontainingblacksmalldiamond}{25C8}
\pdfglyphtounicode{whitedownpointingsmalltriangle}{25BF}
\pdfglyphtounicode{whitedownpointingtriangle}{25BD}
\pdfglyphtounicode{whiteleftpointingsmalltriangle}{25C3}
\pdfglyphtounicode{whiteleftpointingtriangle}{25C1}
\pdfglyphtounicode{whitelenticularbracketleft}{3016}
\pdfglyphtounicode{whitelenticularbracketright}{3017}
\pdfglyphtounicode{whiterightpointingsmalltriangle}{25B9}
\pdfglyphtounicode{whiterightpointingtriangle}{25B7}
\pdfglyphtounicode{whitesmallsquare}{25AB}
\pdfglyphtounicode{whitesmilingface}{263A}
\pdfglyphtounicode{whitesquare}{25A1}
\pdfglyphtounicode{whitestar}{2606}
\pdfglyphtounicode{whitetelephone}{260F}
\pdfglyphtounicode{whitetortoiseshellbracketleft}{3018}
\pdfglyphtounicode{whitetortoiseshellbracketright}{3019}
\pdfglyphtounicode{whiteuppointingsmalltriangle}{25B5}
\pdfglyphtounicode{whiteuppointingtriangle}{25B3}
\pdfglyphtounicode{wihiragana}{3090}
\pdfglyphtounicode{wikatakana}{30F0}
\pdfglyphtounicode{wikorean}{315F}
\pdfglyphtounicode{wmonospace}{FF57}
\pdfglyphtounicode{wohiragana}{3092}
\pdfglyphtounicode{wokatakana}{30F2}
\pdfglyphtounicode{wokatakanahalfwidth}{FF66}
\pdfglyphtounicode{won}{20A9}
\pdfglyphtounicode{wonmonospace}{FFE6}
\pdfglyphtounicode{wowaenthai}{0E27}
\pdfglyphtounicode{wparen}{24B2}
\pdfglyphtounicode{wreathproduct}{2240}
\pdfglyphtounicode{wring}{1E98}
\pdfglyphtounicode{wsuperior}{02B7}
\pdfglyphtounicode{wturned}{028D}
\pdfglyphtounicode{wynn}{01BF}
\pdfglyphtounicode{x}{0078}
\pdfglyphtounicode{xabovecmb}{033D}
\pdfglyphtounicode{xbopomofo}{3112}
\pdfglyphtounicode{xcircle}{24E7}
\pdfglyphtounicode{xdieresis}{1E8D}
\pdfglyphtounicode{xdotaccent}{1E8B}
\pdfglyphtounicode{xeharmenian}{056D}
\pdfglyphtounicode{xi}{03BE}
\pdfglyphtounicode{xmonospace}{FF58}
\pdfglyphtounicode{xparen}{24B3}
\pdfglyphtounicode{xsuperior}{02E3}
\pdfglyphtounicode{y}{0079}
\pdfglyphtounicode{yaadosquare}{334E}
\pdfglyphtounicode{yabengali}{09AF}
\pdfglyphtounicode{yacute}{00FD}
\pdfglyphtounicode{yadeva}{092F}
\pdfglyphtounicode{yaekorean}{3152}
\pdfglyphtounicode{yagujarati}{0AAF}
\pdfglyphtounicode{yagurmukhi}{0A2F}
\pdfglyphtounicode{yahiragana}{3084}
\pdfglyphtounicode{yakatakana}{30E4}
\pdfglyphtounicode{yakatakanahalfwidth}{FF94}
\pdfglyphtounicode{yakorean}{3151}
\pdfglyphtounicode{yamakkanthai}{0E4E}
\pdfglyphtounicode{yasmallhiragana}{3083}
\pdfglyphtounicode{yasmallkatakana}{30E3}
\pdfglyphtounicode{yasmallkatakanahalfwidth}{FF6C}
\pdfglyphtounicode{yatcyrillic}{0463}
\pdfglyphtounicode{ycircle}{24E8}
\pdfglyphtounicode{ycircumflex}{0177}
\pdfglyphtounicode{ydieresis}{00FF}
\pdfglyphtounicode{ydotaccent}{1E8F}
\pdfglyphtounicode{ydotbelow}{1EF5}
\pdfglyphtounicode{yeharabic}{064A}
\pdfglyphtounicode{yehbarreearabic}{06D2}
\pdfglyphtounicode{yehbarreefinalarabic}{FBAF}
\pdfglyphtounicode{yehfinalarabic}{FEF2}
\pdfglyphtounicode{yehhamzaabovearabic}{0626}
\pdfglyphtounicode{yehhamzaabovefinalarabic}{FE8A}
\pdfglyphtounicode{yehhamzaaboveinitialarabic}{FE8B}
\pdfglyphtounicode{yehhamzaabovemedialarabic}{FE8C}
\pdfglyphtounicode{yehinitialarabic}{FEF3}
\pdfglyphtounicode{yehmedialarabic}{FEF4}
\pdfglyphtounicode{yehmeeminitialarabic}{FCDD}
\pdfglyphtounicode{yehmeemisolatedarabic}{FC58}
\pdfglyphtounicode{yehnoonfinalarabic}{FC94}
\pdfglyphtounicode{yehthreedotsbelowarabic}{06D1}
\pdfglyphtounicode{yekorean}{3156}
\pdfglyphtounicode{yen}{00A5}
\pdfglyphtounicode{yenmonospace}{FFE5}
\pdfglyphtounicode{yeokorean}{3155}
\pdfglyphtounicode{yeorinhieuhkorean}{3186}
\pdfglyphtounicode{yerahbenyomohebrew}{05AA}
\pdfglyphtounicode{yerahbenyomolefthebrew}{05AA}
\pdfglyphtounicode{yericyrillic}{044B}
\pdfglyphtounicode{yerudieresiscyrillic}{04F9}
\pdfglyphtounicode{yesieungkorean}{3181}
\pdfglyphtounicode{yesieungpansioskorean}{3183}
\pdfglyphtounicode{yesieungsioskorean}{3182}
\pdfglyphtounicode{yetivhebrew}{059A}
\pdfglyphtounicode{ygrave}{1EF3}
\pdfglyphtounicode{yhook}{01B4}
\pdfglyphtounicode{yhookabove}{1EF7}
\pdfglyphtounicode{yiarmenian}{0575}
\pdfglyphtounicode{yicyrillic}{0457}
\pdfglyphtounicode{yikorean}{3162}
\pdfglyphtounicode{yinyang}{262F}
\pdfglyphtounicode{yiwnarmenian}{0582}
\pdfglyphtounicode{ymonospace}{FF59}
\pdfglyphtounicode{yod}{05D9}
\pdfglyphtounicode{yoddagesh}{FB39}
\pdfglyphtounicode{yoddageshhebrew}{FB39}
\pdfglyphtounicode{yodhebrew}{05D9}
\pdfglyphtounicode{yodyodhebrew}{05F2}
\pdfglyphtounicode{yodyodpatahhebrew}{FB1F}
\pdfglyphtounicode{yohiragana}{3088}
\pdfglyphtounicode{yoikorean}{3189}
\pdfglyphtounicode{yokatakana}{30E8}
\pdfglyphtounicode{yokatakanahalfwidth}{FF96}
\pdfglyphtounicode{yokorean}{315B}
\pdfglyphtounicode{yosmallhiragana}{3087}
\pdfglyphtounicode{yosmallkatakana}{30E7}
\pdfglyphtounicode{yosmallkatakanahalfwidth}{FF6E}
\pdfglyphtounicode{yotgreek}{03F3}
\pdfglyphtounicode{yoyaekorean}{3188}
\pdfglyphtounicode{yoyakorean}{3187}
\pdfglyphtounicode{yoyakthai}{0E22}
\pdfglyphtounicode{yoyingthai}{0E0D}
\pdfglyphtounicode{yparen}{24B4}
\pdfglyphtounicode{ypogegrammeni}{037A}
\pdfglyphtounicode{ypogegrammenigreekcmb}{0345}
\pdfglyphtounicode{yr}{01A6}
\pdfglyphtounicode{yring}{1E99}
\pdfglyphtounicode{ysuperior}{02B8}
\pdfglyphtounicode{ytilde}{1EF9}
\pdfglyphtounicode{yturned}{028E}
\pdfglyphtounicode{yuhiragana}{3086}
\pdfglyphtounicode{yuikorean}{318C}
\pdfglyphtounicode{yukatakana}{30E6}
\pdfglyphtounicode{yukatakanahalfwidth}{FF95}
\pdfglyphtounicode{yukorean}{3160}
\pdfglyphtounicode{yusbigcyrillic}{046B}
\pdfglyphtounicode{yusbigiotifiedcyrillic}{046D}
\pdfglyphtounicode{yuslittlecyrillic}{0467}
\pdfglyphtounicode{yuslittleiotifiedcyrillic}{0469}
\pdfglyphtounicode{yusmallhiragana}{3085}
\pdfglyphtounicode{yusmallkatakana}{30E5}
\pdfglyphtounicode{yusmallkatakanahalfwidth}{FF6D}
\pdfglyphtounicode{yuyekorean}{318B}
\pdfglyphtounicode{yuyeokorean}{318A}
\pdfglyphtounicode{yyabengali}{09DF}
\pdfglyphtounicode{yyadeva}{095F}
\pdfglyphtounicode{z}{007A}
\pdfglyphtounicode{zaarmenian}{0566}
\pdfglyphtounicode{zacute}{017A}
\pdfglyphtounicode{zadeva}{095B}
\pdfglyphtounicode{zagurmukhi}{0A5B}
\pdfglyphtounicode{zaharabic}{0638}
\pdfglyphtounicode{zahfinalarabic}{FEC6}
\pdfglyphtounicode{zahinitialarabic}{FEC7}
\pdfglyphtounicode{zahiragana}{3056}
\pdfglyphtounicode{zahmedialarabic}{FEC8}
\pdfglyphtounicode{zainarabic}{0632}
\pdfglyphtounicode{zainfinalarabic}{FEB0}
\pdfglyphtounicode{zakatakana}{30B6}
\pdfglyphtounicode{zaqefgadolhebrew}{0595}
\pdfglyphtounicode{zaqefqatanhebrew}{0594}
\pdfglyphtounicode{zarqahebrew}{0598}
\pdfglyphtounicode{zayin}{05D6}
\pdfglyphtounicode{zayindagesh}{FB36}
\pdfglyphtounicode{zayindageshhebrew}{FB36}
\pdfglyphtounicode{zayinhebrew}{05D6}
\pdfglyphtounicode{zbopomofo}{3117}
\pdfglyphtounicode{zcaron}{017E}
\pdfglyphtounicode{zcircle}{24E9}
\pdfglyphtounicode{zcircumflex}{1E91}
\pdfglyphtounicode{zcurl}{0291}
\pdfglyphtounicode{zdot}{017C}
\pdfglyphtounicode{zdotaccent}{017C}
\pdfglyphtounicode{zdotbelow}{1E93}
\pdfglyphtounicode{zecyrillic}{0437}
\pdfglyphtounicode{zedescendercyrillic}{0499}
\pdfglyphtounicode{zedieresiscyrillic}{04DF}
\pdfglyphtounicode{zehiragana}{305C}
\pdfglyphtounicode{zekatakana}{30BC}
\pdfglyphtounicode{zero}{0030}
\pdfglyphtounicode{zeroarabic}{0660}
\pdfglyphtounicode{zerobengali}{09E6}
\pdfglyphtounicode{zerodeva}{0966}
\pdfglyphtounicode{zerogujarati}{0AE6}
\pdfglyphtounicode{zerogurmukhi}{0A66}
\pdfglyphtounicode{zerohackarabic}{0660}
\pdfglyphtounicode{zeroinferior}{2080}
\pdfglyphtounicode{zeromonospace}{FF10}
\pdfglyphtounicode{zerooldstyle}{0030}
\pdfglyphtounicode{zeropersian}{06F0}
\pdfglyphtounicode{zerosuperior}{2070}
\pdfglyphtounicode{zerothai}{0E50}
\pdfglyphtounicode{zerowidthjoiner}{FEFF}
\pdfglyphtounicode{zerowidthnonjoiner}{200C}
\pdfglyphtounicode{zerowidthspace}{200B}
\pdfglyphtounicode{zeta}{03B6}
\pdfglyphtounicode{zhbopomofo}{3113}
\pdfglyphtounicode{zhearmenian}{056A}
\pdfglyphtounicode{zhebrevecyrillic}{04C2}
\pdfglyphtounicode{zhecyrillic}{0436}
\pdfglyphtounicode{zhedescendercyrillic}{0497}
\pdfglyphtounicode{zhedieresiscyrillic}{04DD}
\pdfglyphtounicode{zihiragana}{3058}
\pdfglyphtounicode{zikatakana}{30B8}
\pdfglyphtounicode{zinorhebrew}{05AE}
\pdfglyphtounicode{zlinebelow}{1E95}
\pdfglyphtounicode{zmonospace}{FF5A}
\pdfglyphtounicode{zohiragana}{305E}
\pdfglyphtounicode{zokatakana}{30BE}
\pdfglyphtounicode{zparen}{24B5}
\pdfglyphtounicode{zretroflexhook}{0290}
\pdfglyphtounicode{zstroke}{01B6}
\pdfglyphtounicode{zuhiragana}{305A}
\pdfglyphtounicode{zukatakana}{30BA}

\usepackage{amsmath,graphics,amssymb,epsfig}
\usepackage{url}
\usepackage{color}
\usepackage{verbatim}
\usepackage{amsthm}
\usepackage{tabularx}
\usepackage{float}
\usepackage[above]{placeins}
\usepackage[nocompress]{cite}
\usepackage[caption=false,font=footnotesize]{subfig}

\makeatletter
\newcommand\fs@paperruled{%
  \def\@fs@cfont{\color{black}\bfseries}\let\@fs@capt\floatc@ruled
  \def\@fs@pre{\color{black}\hrule height.8pt depth0pt \kern2pt}%
  \def\@fs@post{\color{black}\kern2pt\hrule\relax}%
  \def\@fs@mid{\color{black}\kern2pt\hrule\kern2pt}%
  \let\@fs@iftopcapt\iftrue}
\makeatother
\floatstyle{paperruled}
\newfloat{algorithm}{t}{loa}
\floatname{algorithm}{Algorithm}
\newcounter{algline}
\newcounter{algdepth}
\newdimen\algindentstep
\newcommand{\algline}{%
  \stepcounter{algline}%
  \makebox[1.55em][r]{\scriptsize\arabic{algline}:}\hspace{0.38em}}
\newcommand{\algprefix}{%
  \algline\hspace*{\dimexpr\value{algdepth}\algindentstep\relax}}
\newenvironment{algorithmic}[1][]{%
  \setcounter{algline}{0}%
  \setcounter{algdepth}{0}%
  \footnotesize\raggedright
  \begin{list}{}{%
    \leftmargin=0pt
    \labelwidth=0pt
    \labelsep=0pt
    \itemindent=0pt
    \itemsep=1.2pt
    \parsep=0pt
    \topsep=2pt
    \rightmargin=0pt}%
}{\end{list}}
\newcommand{\Require}{\item[]\textbf{Input:}\ }
\newcommand{\State}{\item[]\algprefix}

\newcommand{\Repeat}{%
  \item[]\algprefix\textbf{repeat}\stepcounter{algdepth}}
\newcommand{\Until}[1]{%
  \addtocounter{algdepth}{-1}\item[]\algprefix\textbf{until} #1}
\newcommand{\For}[1]{%
  \item[]\algprefix\textbf{for} #1 \textbf{do}\stepcounter{algdepth}}
\newcommand{\EndFor}{%
  \addtocounter{algdepth}{-1}\item[]\algprefix\textbf{end for}}
\newcommand{\Return}{\textbf{return }}

\newcommand{\FlowMean}{0.537}
\newcommand{\FlowP}{1.367}
\newcommand{\GridMatchedMean}{0.917}
\newcommand{\GridMatchedP}{1.607}
\newcommand{\GridMatchedParams}{68.3M}
\newcommand{\FlowParams}{68.2M}
\newcommand{\MLPMean}{1.045}
\newcommand{\MLPP}{2.204}
\newcommand{\MLPParams}{131.7M}
\newcommand{\CNNMean}{0.975}
\newcommand{\CNNP}{2.007}
\newcommand{\CNNParams}{1.3M}
\newcommand{\DDIMMean}{1.139}
\newcommand{\DDIMP}{1.922}
\newcommand{\DDIMParams}{68.2M}
\newcommand{\CTMean}{1.250}
\newcommand{\CTP}{2.244}
\newcommand{\CTParams}{68.2M}
\newcommand{\RSSKNNMean}{2.380}
\newcommand{\RSSKNNP}{9.635}
\newcommand{\SrsAoAMean}{78.238}
\newcommand{\SrsAoAP}{152.411}
\newcommand{\SrsTDoAMean}{64.724}
\newcommand{\SrsTDoAP}{123.933}
\newcommand{\SrsJointMean}{43.287}
\newcommand{\SrsJointP}{81.636}
\newcommand{\PrimaryAoAMean}{77.114}
\newcommand{\PrimaryAoAP}{145.389}
\newcommand{\PrimaryTDoAMean}{58.255}
\newcommand{\PrimaryTDoAP}{122.496}
\newcommand{\PrimaryJointMean}{40.612}
\newcommand{\PrimaryJointP}{77.393}

\DeclareMathOperator{\col}{col}

\begin{document}
\bstctlcite{TSPBibControl}
\setlength{\abovedisplayskip}{5pt}
\setlength{\belowdisplayskip}{5pt}
\setlength{\textfloatsep}{7pt plus 2pt minus 2pt}
\setlength{\floatsep}{6pt plus 2pt minus 2pt}
\setlength{\intextsep}{7pt plus 2pt minus 2pt}
\setlength{\dbltextfloatsep}{7pt plus 2pt minus 2pt}
\setlength{\dblfloatsep}{6pt plus 2pt minus 2pt}
\raggedbottom

\title{\textcolor{black}{Grid-Supervised Flow Refinement for Wireless Localization}}
 \author{\IEEEauthorblockN{
 Taekyun Lee, Tommaso Balercia, Heasung Kim, Hyeji Kim, and Jeffrey G. Andrews
} \\ 
\thanks{
This work was conducted while the first author was at NVIDIA Corporation.
This work was partly supported by NSF Award CNS-2148141 and Keysight Technologies through the 6G@UT center within the Wireless Networking and Communications Group (WNCG) at the University of Texas at Austin, as well as ARO Award W911NF-23-1-0062, ONR Award N00014-24-1-2542, and NSF Award CCF-2443857.
Taekyun Lee, Heasung Kim, Hyeji Kim, and Jeffrey G. Andrews are with
the 6G@UT center in the WNCG at the University of Texas at Austin, Austin, TX 78712, USA (email: taekyun@utexas.edu, heasung.kim@utexas.edu,
hyeji.kim@austin.utexas.edu, jandrews@ece.utexas.edu).

Tommaso Balercia is with NVIDIA Corporation, Santa Clara, CA 95051, USA (email: tbalercia@nvidia.com).
\par\textcolor{black}{An earlier version of this work appeared as the DiffLoc preprint~\cite{Lee2025DiffLoc}.
FlowLoc replaces the earlier method with grid-supervised features,
a learned anchor, and residual flow, with a rebuilt evaluation.}
\par Last revised: October 15, 2026.
}
}
\maketitle

\begin{abstract}
\textcolor{black}{This paper introduces FlowLoc, a framework for high-accuracy
user equipment positioning from channel state information observed
at \textcolor{black}{multiple base stations}. The input is a
magnitude channel impulse response fingerprint \textcolor{black}{derived from channel estimates}.
FlowLoc combines a learned initial position estimate with a
flow model that corrects the remaining position error.
First, FlowLoc trains \textcolor{black}{base station} encoders using grid supervision and then removes
the grid classifiers.
Second, it freezes the encoders and trains a network to estimate an initial
position, called the anchor, from their outputs. Third, it freezes the
anchor network and trains a flow model to correct the remaining position
error. We evaluate
FlowLoc using ray-traced channels from an outdoor urban region, with separate
trajectories for training and testing.
FlowLoc reduces mean localization error from 0.917m for a grid-based
localization baseline adapted from prior work to 0.537m, a 41.4\% reduction. It also
outperforms supervised regression and direct-coordinate generative
baselines. FlowLoc retains sub-meter mean error on independently sampled
channels across mobility settings from pedestrian speeds to 25m/s. Further
experiments quantify sensitivity to channel noise and missing \textcolor{black}{base station} observations.}
\end{abstract}

\noindent\begin{minipage}{\linewidth}
\end{minipage}
\par

\section{Introduction}

Accurate positioning of mobile user equipment (UE) is foundational for numerous \textcolor{black}{wireless applications}, including autonomous vehicles, industrial automation, augmented and virtual reality, smart cities, precision agriculture, and digital twins \cite{trevlakis2023localization, Liu2025DigitalTwin, Morais2024Localization}. Sub-meter localization accuracy is essential for the reliable operation of these applications, enabling capabilities such as proactive resource allocation, predictive maintenance, collision avoidance, and intelligent urban operations \cite{Mohammed2025SupportingGlobalCommunications}.

\textcolor{black}{While UE positioning can be performed by the device or the network,
this work focuses on network-side localization. %
Access to device-derived location depends on privacy
controls, platform policy, and application-layer consent. 
These constraints motivate a
network-side approach, where we infer the UE position solely from signals already observed by the radio access network, without requesting or
receiving a device-derived position. Specifically, we use 
magnitude channel impulse response (CIR) fingerprints obtained from
\textcolor{black}{channel estimates. Uplink Sounding Reference Signal (SRS)
transmissions provide one way to acquire these observations}.
}

\subsection{Related Work}
\label{Sec2}

\textbf{Standard UE positioning methods.}
\textcolor{black}{\textit{Standard UE positioning methods face accuracy challenges in
outdoor environments with multipath and non-line-of-sight propagation.}
5G New Radio (NR) positioning standardizes downlink Positioning Reference
Signal (PRS) and uplink SRS measurements~\cite{3gpp_positioning}.
Received signal strength (RSS) methods infer location from power
fingerprints or path-loss relationships, which are sensitive to these
propagation effects~\cite{bahl:radar,Youssef2005Horus}.
Time-based methods rely on propagation delay and require sufficient
bandwidth and timing synchronization\textcolor{black}{~\cite{Chan1994TDoA}}. Angle of arrival (AoA) \textcolor{black}{and other array-based} methods
require calibrated arrays at each base station (BS) and consistent
coordinate frames for multi-BS \textcolor{black}{localization~\cite{3gpp_aoa,Garcia2017DirectLocalization}.}}

{{\color{black}\textbf{Channel fingerprinting for localization.}}}
\textcolor{black}{\textit{Despite extensive research on channel fingerprinting,
high-accuracy localization remains challenging in the outdoor multi-BS
setting considered here.}
Prior work uses frequency-domain responses, spatial covariance, and
delay-domain features derived from channel state information (CSI) as fingerprints~\cite{wang2021wifi,tian2023deep,gonultas2022csi}.
The CIR magnitude used here is a delay-domain fingerprint.
Early approaches used SVM~\cite{zhou2017device} and kNN~\cite{Song17},
while neural methods learn channel features for position estimation\textcolor{black}{~\cite{wang2017csi,Arnold2018Localization}}.
Subsequent models learn from CSI amplitude and estimated AoA using a residual-sharing network~\cite{wang2021wifi}
or combine predictions from networks trained on different fingerprints~\cite{tian2023deep}.}

\textcolor{black}{The most closely related approach is multi-point probability
fusion~\cite{gonultas2022csi}. It uses soft targets over grid points,
predicts a spatial probability map from each receiver's CSI, and fuses
these maps to estimate position. FlowLoc retains grid supervision for
encoder training, then uses the joint BS features to estimate a continuous
anchor and correct its residual error.}

{{\color{black}\textbf{Diffusion and flow models for wireless systems.}}}
\textcolor{black}{{{\color{black}\textit{Diffusion and flow models are increasingly used
in wireless systems.} FlowLoc introduces a framework for direct UE
localization that combines grid-supervised channel features with flow
refinement of a learned anchor.}}
Denoising diffusion models learn to reverse a process that gradually adds
noise to data, generating samples by iterative denoising~\cite{DDPM}.
Flow matching trains a vector field that transports samples from a source
distribution to a target distribution~\cite{Lipman2023FlowMatching}.
At inference, samples follow this field through numerical integration of
an ordinary differential equation (ODE).}

\textcolor{black}{In wireless systems, diffusion models have been used to synthesize
channel realizations~\cite{Sengupta23}, including user-specific channels
conditioned on UE position~\cite{Lee2024Generating}. They have also served
as learned channel priors for MIMO channel estimation~\cite{Fesl2024DiffusionBased}
and as conditional decoders for CSI compression and prediction~\cite{Kim2025GenerativeCompression}.
More recently, flow-matching models have been applied to CSI feedback,
using the feedback to guide a prior sample or an initial CSI estimate
toward a reconstructed channel~\cite{Park2026ChannelGeometryFlow}.}

\textcolor{black}{Diffusion models have also been explored for localization.
Guo et al.~\cite{Guo2023CDPM} use conditional diffusion to augment CSI
fingerprints for indoor localization. For LiDAR localization,
Li et al.~\cite{Li2024DiffLocLidar} estimate poses through iterative
denoising conditioned on point-cloud features.}

\subsection{Contributions}

\textcolor{black}{Our main contributions are as follows:}

\begin{itemize}
\item \textcolor{black}{We introduce \textbf{FlowLoc}, a localization framework that
couples grid-supervised channel feature learning with anchor-based flow
refinement. BS encoders learn spatially informative channel features
under grid supervision, an anchor network turns their joint latent into
an initial continuous position, and a bounded conditional flow corrects
the remaining position error.
}

\item \textcolor{black}{We show that FlowLoc substantially outperforms conventional
geometric and RSS-based localization, supervised MLP and CNN regression,
grid-based probability fusion, and direct-coordinate generative models
on the same outdoor test set. FlowLoc reduces mean error by 41.4\%
relative to the parameter-matched probability-fusion grid baseline
adapted from prior work~\cite{gonultas2022csi} and by 44.9\% relative to
CNN regression. Its mean error is less than half that of DDIM and
consistency training, which generate coordinates directly from an
observation-independent prior. The same trained model also supports split
inference: each BS can send only its encoder latent to the localization
server instead of its full channel fingerprint, without changing the
position estimate.
}

\item \textcolor{black}{We validate FlowLoc's design through controlled ablations and
demonstrate its accuracy advantage across diverse operating conditions.
Both starting from the learned anchor and correcting its residual improve
localization accuracy. In the SRS receiver experiment, it outperforms the
grid and CNN baselines despite noisy channel estimates. It also achieves
the lowest mean error across the tested mobility and channel-noise
conditions. By randomly masking BS inputs during training and telling the
model which BSs are available, FlowLoc can localize from the remaining BSs
with much lower error than a model trained only on complete observations.
}
\end{itemize}

\subsection{Notation \& Organization}
\textcolor{black}{Bold uppercase letters (e.g., $\mathbf A$) denote matrices or tensors,
bold lowercase letters (e.g., $\mathbf a$) denote vectors, and standard italic
letters denote scalars.  A vector obtained by vectorizing a matrix or tensor
$\mathbf A$ is written as $\underline{\mathbf A}$. Continuous-time arguments
use parentheses, as in $\mathbf x(\tau)$. Calligraphic letters denote sets or
indexed collections. The superscripts $(\cdot)^T$ and $(\cdot)^H$ denote
transpose and Hermitian transpose, respectively.}  The fields of real and complex numbers are denoted by
$\mathbb{R}$ and $\mathbb{C}$, respectively.  Table~\ref{tab:notation}
summarizes the principal notation.

\begin{table}[ht]
\centering
\caption{Key notation and indexing conventions.}
\label{tab:notation}
\renewcommand{\arraystretch}{1.3}
\begin{tabular}{c|c}
\hline
\textbf{Symbol} & \textbf{Description} \\
\hline
\multicolumn{2}{c}{\textbf{Learning and Flow Variables}} \\
\hline
$\mathbf x_{\mathrm{tr}}(\tau),\mathbf x(\tau)$ & Training state and inference state \\
$\textcolor{black}{\mathbf{x}_{\mathrm{pert}}}$ & \textcolor{black}{Locally perturbed training source} \\
$\mathbf{x}_{\mathrm{gt}}$ & Ground-truth UE position (flow target) \\
$\mathbf x_{\mathrm{src}},\widehat{\mathbf x}_{\mathrm{FL}}$ & Anchor and final position estimates \\
\textcolor{black}{$\mathbf v_{\theta}$} & \textcolor{black}{Learned conditional vector field} \\
\textcolor{black}{$\mathbf v^\star$} & \textcolor{black}{Supervised residual-velocity target} \\
$r_{\max}$ & Maximum inference correction radius \\
$E_{\phi_b},\mathbf a_\eta$ & BS encoder and anchor network \\
\textcolor{black}{$\mathbf z^{(b)},\mathbf z$} & \textcolor{black}{Per-BS and standardized joint latents} \\
\textcolor{black}{$P_b, S$} & \textcolor{black}{Per-BS grid posterior and fused log score} \\
\textcolor{black}{$q$} & \textcolor{black}{Soft spatial target on the training grid} \\
\hline
\multicolumn{2}{c}{\textcolor{black}{\textbf{Channel and Base Stations}}} \\
\hline
\textcolor{black}{$\mathbf{H}^{(b)}$} & \textcolor{black}{Complex CIR tensor at BS $b$} \\
\textcolor{black}{$\underline{\mathbf H}_i^{(b)}$} & \textcolor{black}{Standardized fingerprint at sample $i$, BS $b$} \\
\textcolor{black}{$\mathcal{H}_i$} & \textcolor{black}{Ordered BS fingerprints for sample $i$} \\
$B$ & \textcolor{black}{Total number of base stations} \\
\hline
\multicolumn{2}{c}{\textbf{System Parameters}} \\
\hline
$N_r, N_t$ & Receive and transmit antenna-port counts \\
$N_c$ & Number of subcarriers \\
$N_d$ & Number of common delay-grid bins \\
$\Delta f$ & Subcarrier spacing \\
$L$ & CIR truncation length (delay taps) \\
$d,D$ & Fingerprint and per-BS latent dimensions \\
$\mathcal G_0,K_0$ & Training grid and number of grid points \\
\hline
\end{tabular}
\end{table}

\textcolor{black}{Sections~\ref{Sec3}--\ref{Sec5} present the observation model, FlowLoc
formulation, and training and inference procedures. Sections~\ref{Sec6}--\ref{Sec8}
cover the experimental setup, results, and conclusions.}
\suppressfloats[t]
\begin{figure}[t]
 \centering
 \includegraphics[width=\columnwidth]{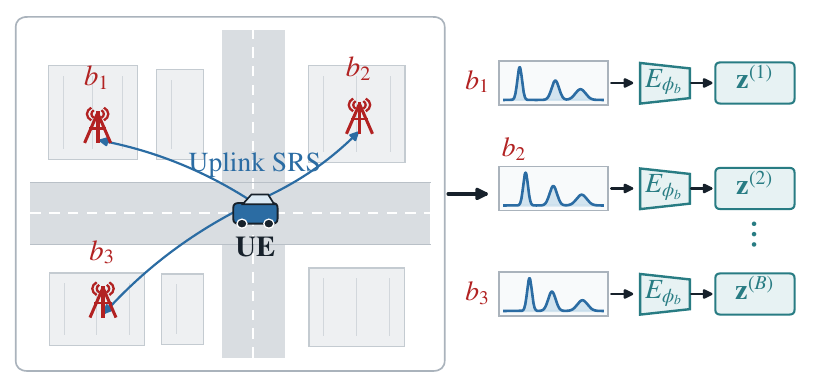}
 \caption{\textcolor{black}{Network-side channel observation used by FlowLoc. Each BS
 encodes its uplink SRS observation locally, producing the shared latent
 representation used for both model training and UE localization.}}
\label{fig:flowloc_overview}
\end{figure}

\begin{figure*}[t]
\centering
\includegraphics[width=\textwidth]{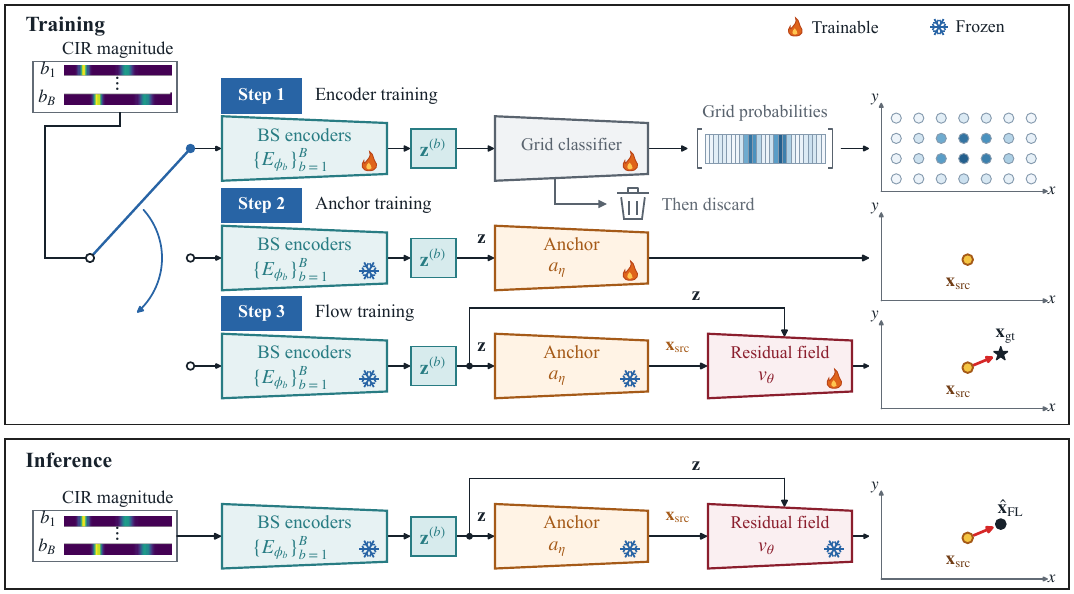}
\caption{\textcolor{black}{FlowLoc training and inference from magnitude-CIR fingerprints.
The switch illustrates the training order. Rows show successive states
of the same modules. Step 1 trains the encoders through temporary grid
classifiers that output probabilities over spatial grid points.
The classifiers are then discarded. Steps 2 and 3 train the anchor
and residual field. Flames and snowflakes denote trainable and frozen
modules. Per-BS latents $\mathbf z^{(b)}$ are concatenated and standardized
into $\mathbf z$. Yellow dots denote anchors, stars denote ground-truth
positions, and red arrows indicate corrections. A black dot marks the final
estimate. Training perturbations and
the field's current-state and time inputs are omitted for clarity.}}
\label{fig:flowloc_train_infer}
\end{figure*}

\section{System Model \& Problem Statement}
\label{Sec3}

In Section~\ref{sec:system_model}, we present our system model, including
the wireless communication setup and the preprocessing procedure for
generating channel fingerprints. In Section~\ref{sec:problem_statement}, we
formally state the positioning problem for both single-BS and multi-BS
scenarios and define our learning objective.

\subsection{System Model}
\label{sec:system_model}

{
\color{black}
{{\color{black}\textbf{Per-subcarrier signal model.}}} Consider an NR uplink in which a
UE transmits SRS pilots through $N_t$ transmit antenna ports to a BS
with $N_r$ receive antenna ports.
Across $Q\geq N_t$ orthogonal SRS symbols on occupied subcarrier $k$, the
received matrix is
\begin{equation}
\mathbf{Y}_k = \mathbf{G}_k \mathbf{S}_k + \mathbf{N}_k,
\label{eq:srs_observation}
\end{equation}
where
\(\mathbf{Y}_k,\mathbf{N}_k\in\mathbb{C}^{N_r\times Q}\),
\(\mathbf{G}_k\in\mathbb{C}^{N_r\times N_t}\), and the known pilot matrix
\(\mathbf{S}_k\in\mathbb{C}^{N_t\times Q}\) satisfies
\(\mathbf{S}_k\mathbf{S}_k^{H}=Q\mathbf I_{N_t}\) with
\(\mathbf I_{N_t}\) denoting the \(N_t\)-dimensional identity matrix.  The
corresponding LS channel
estimate is
\begin{equation}
\widehat{\mathbf G}_k
=\frac{1}{Q}\mathbf Y_k\mathbf S_k^H
=\mathbf G_k+\frac{1}{Q}\mathbf N_k\mathbf S_k^H.
\label{eq:srs_ls_estimate}
\end{equation}
This matrix observation retains individually identifiable responses for
all \(N_t\) transmit ports.

{{\color{black}\textbf{Multi-dimensional channel tensor.}}}
By collecting the CFR across all transmit--receive antenna-port pairs and
subcarriers, we obtain a three-dimensional frequency-domain channel tensor,
\[
\mathbf G_{\mathrm{freq}}\in\mathbb C^{N_r\times N_t\times N_c},
\]
where $N_c$ is the number of frequency samples and
$[\mathbf G_{\mathrm{freq}}]_{r,t,k}=[\mathbf G_k]_{r,t}$ represents the
channel coefficient from transmit port $t$ to receive port $r$ on
subcarrier $k$. Replacing $\mathbf G_k$ with its estimate
$\widehat{\mathbf G}_k$ gives the corresponding estimated CFR tensor.

{{\color{black}\textbf{Frequency-to-delay domain transformation.}}}
The frequency-domain channel can be transformed into a delay-domain CIR
by applying an inverse Fourier transform along the subcarrier dimension.
Let $\mathbf W\in\mathbb C^{N_d\times N_c}$ denote this mapping onto an
$N_d$-point delay grid, including the placement of the frequency samples
on the transform grid. For each antenna-port pair $(r,t)$, the
delay-domain channel is
\begin{equation}
[\mathbf G_{\mathrm{delay}}]_{r,t,\ell}
=\sum_{k=1}^{N_c}[\mathbf W]_{\ell,k}
[\mathbf G_{\mathrm{freq}}]_{r,t,k},
\label{eq:frequency_to_delay}
\end{equation}
where $\ell=0,\ldots,N_d-1$ and
$\mathbf G_{\mathrm{delay}}\in\mathbb C^{N_r\times N_t\times N_d}$.
This operation changes the channel representation from frequency samples
to delay taps while retaining the transmit--receive port indices.

We retain the first $L$ delay taps to form the channel tensor used for
localization,
\begin{equation}
[\mathbf H]_{r,t,\ell}=[\mathbf G_{\mathrm{delay}}]_{r,t,\ell},
\qquad \ell=0,\ldots,L-1,
\label{eq:cir_truncation}
\end{equation}
where $\mathbf H\in\mathbb C^{N_r\times N_t\times L}$.
In the primary experiments, $\mathbf G_{\mathrm{delay}}$ is constructed
directly from ray-traced path delays and complex coefficients, as described
in Section~\ref{sec:experimental_environment}. The SRS receiver experiment
in Section~\ref{sec:srs_receiver_control} instead constructs the CIR from
an estimated CFR after pilot-based channel estimation and interpolation.
Both cases use the same delay-domain fingerprint definition below.

{{\color{black}\textbf{Channel fingerprint generation.}}}
{{\color{black}Let $\mathbf H$ denote the delay-domain channel tensor in
$\mathbb C^{N_r\times N_t\times L}$, with entries $H_{r,t,\ell}$.}} We construct a
real fingerprint vector by vectorizing its magnitude,
\begin{equation}
\label{eq:fingerprint_def}
\underline{\mathbf H}_{\mathrm{mag}}
=\mathrm{vec}\!\left(\lvert\mathbf H\rvert\right)
\in\mathbb R^d,\qquad d=N_rN_tL.
\end{equation}
The vector $\underline{\mathbf H}_{\mathrm{mag}}$ is the channel
fingerprint used by all learned methods. For multiple BSs, the per-BS
fingerprints remain separate inputs and are jointly processed or fused as
specified by each method.

Each magnitude fingerprint is standardized elementwise using the mean and
standard deviation estimated from the training set. These statistics are
fixed for validation and testing. We use
$\underline{\mathbf H}_{\mathrm{mag}}^{(b)}$ for the unstandardized
fingerprint of BS $b$ and $\underline{\mathbf H}^{(b)}$ for the
corresponding network input. Accordingly,
$\underline{\mathbf H}_i^{(b)}$ denotes the input associated with sample
$i$ and BS $b$, while $\mathbf H^{(b)}$ continues to denote the complex
CIR tensor.
}

\subsection{Problem Statement}
\label{sec:problem_statement}

\textbf{Multi–base-station setting.}
Let $\Omega\subset\mathbb{R}^2$ denote the admissible 2D position
domain for the UE.  A single uplink SRS transmission is received by $B$ BSs.  \textcolor{black}{For each BS $b\in\{1,\dots,B\}$,
the magnitude-CIR construction in~\eqref{eq:fingerprint_def}, followed by
training-set standardization, yields a real fingerprint
$\underline{\mathbf H}^{(b)}\in\mathbb{R}^d$.  We collect these as
$\mathcal{H}=(\underline{\mathbf H}^{(1)},\dots,
\underline{\mathbf H}^{(B)})$.}

\textbf{Learning objective.}
\textcolor{black}{The task is to learn
\(f:\mathcal H\mapsto\widehat{\mathbf x}\in\mathbb R^2\) that minimizes held-out
Euclidean localization error.  FlowLoc implements \(f\) with spatially
supervised BS encoders, a continuous anchor, and a bounded residual field.}

\textcolor{black}{The formulation also includes \(B=1\): \(\mathcal H=(\underline{\mathbf H}^{(1)})\), and \(f\) then estimates position from one BS fingerprint.}

\textbf{Grid-localization reference.}
A common surrogate is to discretize $\Omega$ on a finite grid
$\mathcal{G}=\{\mathbf{g}_k\}_{k=1}^{K}\subset\Omega$ and compute the
posterior-weighted coordinate
\begin{equation}
\label{eq:mmse_grid}
\widehat{\mathbf{x}}_{\mathrm{grid}}
= \sum_{k=1}^{K}\mathbf{g}_k\,P_{\mathrm{grid}}(k\mid \mathcal{H}),
\end{equation}
where \(P_{\mathrm{grid}}(k\mid \mathcal{H})\) is the fused discrete
posterior mass at \(\mathbf g_k\) (e.g., \cite{gonultas2022csi}).  The
expectation-decoded estimate is continuous, but the underlying
classification layer grows linearly with \(K\).

\mbox{\textbf{Our formulation.} \textcolor{black}{FlowLoc uses}} \textcolor{black}{the grid only to supervise the BS encoders, removes the spatial-classification layers, and obtains a continuous source \(\mathbf{x}_{\mathrm{src}}\) from their latents. The anchor supplies the global coordinate, while the field learns its residual. Section~\ref{Sec4} formalizes these design principles.}

\begingroup
\color{black}

\begingroup
\color{black}
\section{FlowLoc Formulation and Design Rationale}
\label{Sec4}

\textit{This section presents the high-level formulation and design rationale
of FlowLoc.} Section~\ref{Sec5} then specifies the network architecture and
training and inference procedures that implement this formulation.

FlowLoc estimates an initial position from the channel observations and
then corrects the remaining error with a conditional flow model.
Temporary grid supervision is
used to learn spatial BS representations, but the deployed estimate remains
continuous. The anchor combines these representations to determine the
initial position, and the flow field learns the remaining correction.

\subsection{Spatially Supervised Continuous Anchor}
\label{sec:spatial_supervision_theory}

\textit{Grid supervision trains the BS encoders to extract spatial features
that support continuous position estimation.} Let
$\mathcal H=(\underline{\mathbf H}^{(1)},\ldots,
\underline{\mathbf H}^{(B)})$ denote the ordered collection of BS
fingerprints and let $\mathbf x_{\mathrm{gt}}\in\Omega$ be the
corresponding UE position. FlowLoc first maps the complete observation to a
continuous global estimate,
\begin{equation}
\mathbf x_{\mathrm{src}}=\mathbf a(\mathcal H)\in\mathbb R^2.
\label{eq:theory_continuous_anchor}
\end{equation}
Here, $\mathbf a$ denotes the complete learned mapping from the multi-BS
observation to the source coordinate. During training, a temporary grid task
encourages the underlying BS representations to retain location information.
Each training position supplies a soft target over nearby grid points, so
the supervision represents its spatial neighborhood rather than only the
identity of one grid point. Each BS encoder learns from the same position
target using its own channel observation.

After this training stage, the grid classifiers are removed and the BS
representations are combined into a joint latent. A continuous coordinate
network then estimates the anchor from this latent. Thus, grid supervision
sets the encoder's learning task without constraining the anchor to output
a grid point or to decode a fused grid posterior. The anchor and residual
field both use the learned channel features. Sections~\ref{sec:score_based_fusion}
and~\ref{sec:continuous_anchor} factorize $\mathbf a$ into the per-BS
encoders, joint latent, standardization, and coordinate network.

\subsection{Learning the Remaining Position Error}
\label{sec:conditional_transport_theory}

\textit{The anchor estimates the global position, and the conditional flow
learns to correct the remaining position error.} During training, a local
Gaussian perturbation around the anchor defines a source neighborhood.
The straight path from a perturbed source to the ground-truth position
provides a supervised velocity equal to the displacement between them.
Thus, the field target is the residual vector from the perturbed anchor
to the UE position rather than the full global coordinate.
Section~\ref{sec:flowloc_refiner} gives the perturbed source, rectified
path, and velocity target explicitly.

\textbf{Residual-target energy.}
\textit{An anchor with lower mean squared position error reduces the expected
squared magnitude of the residual target.} Let $\sigma_0$ be the fixed
training perturbation scale
and let $\boldsymbol\epsilon\sim\mathcal N(\mathbf0,\mathbf I_2)$ be
independent of the observations. Taking the expectation over the training
observations and this perturbation, the squared target displacement is
\begin{align}
\mathcal E_{\mathrm{src}}
&=\mathbb E\|\mathbf x_{\mathrm{gt}}-\mathbf x_{\mathrm{src}}
-\sigma_0\boldsymbol\epsilon\|_2^2\nonumber\\
&=\mathbb E\|\mathbf x_{\mathrm{gt}}-\mathbf x_{\mathrm{src}}\|_2^2
+2\sigma_0^2.
\label{eq:conditional_source_energy}
\end{align}
The cross term vanishes because the perturbation is zero mean and
$\mathbb E\|\boldsymbol\epsilon\|_2^2=2$. For an observation-independent
center, the smallest corresponding energy is
$\operatorname{tr}\operatorname{Cov}(\mathbf x_{\mathrm{gt}})+2\sigma_0^2$.
A conditional anchor with lower coordinate MSE therefore reduces the second
moment of the supervised residual before the field is learned.
The design
comparisons in Section~\ref{sec:design_analysis} evaluate the effects
of the starting position and the learned residual correction separately.

\textbf{Conditional flow refinement.}
\textit{The conditional flow refines the anchor using the channel observation
and the current position estimate.} Let $\mathbf z=\mathbf z(\mathcal H)$
be the standardized joint encoder latent defined in
Section~\ref{sec:score_based_fusion}. The field
$\mathbf v_\theta(\mathbf z,\mathbf x,\mathbf x_{\mathrm{src}},\tau)$ has
learned parameters $\theta$ and flow time $\tau\in[0,1]$. Its inference
trajectory starts from the unperturbed anchor,
\begin{equation}
\frac{d\mathbf x(\tau)}{d\tau}
=\mathbf v_\theta(\mathbf z,\mathbf x(\tau),\mathbf x_{\mathrm{src}},\tau),
\quad \mathbf x(0)=\mathbf x_{\mathrm{src}}.
\label{eq:conditional_flow_ode}
\end{equation}
Its terminal coordinate is
\begin{equation}
\mathbf x(1)=\mathbf x_{\mathrm{src}}+
\int_0^1\mathbf v_\theta(\mathbf z,\mathbf x(\tau),
\mathbf x_{\mathrm{src}},\tau)\,d\tau.
\label{eq:theory_flow_integral}
\end{equation}
The initial coordinate supplies the global estimate, while the integrated
field supplies its correction. Section~\ref{sec:flowloc_refiner} specifies
the training loss, and Section~\ref{sec:bounded_flow_inference} approximates
\eqref{eq:theory_flow_integral} with the midpoint rule.
\endgroup

\subsection{\textcolor{black}{Bounding the Local Correction}}
\label{sec:bounded_continuous_refinement}

\textit{Bounding the correction keeps refinement near the anchor and limits
movement along an erroneous predicted direction.} FlowLoc restricts the
final correction to a validation-selected radius $r_{\max}$ around the anchor,
\begin{equation}
\|\widehat{\mathbf x}-\mathbf x_{\mathrm{src}}\|_2\leq r_{\max}.
\label{eq:theory_bounded_constraint}
\end{equation}
For a source $\mathbf a$, target $\mathbf x$, and
radius $r$, the smallest geometrically attainable error is
\begin{equation}
\min_{\|\boldsymbol\delta\|_2\leq r}
\|\mathbf x-(\mathbf a+\boldsymbol\delta)\|_2
=\max(\|\mathbf x-\mathbf a\|_2-r,0).
\label{eq:bounded_attainable_error}
\end{equation}
The reverse triangle inequality gives the lower bound, and a correction
toward the target attains it. Every target within the selected radius is
therefore reachable. Section~\ref{sec:bounded_flow_inference} gives the
numerical update and projection used at inference.

\endgroup

\begingroup
\color{black}

\section{\textcolor{black}{FlowLoc Architecture and Training}}
\label{Sec5}

{\color{black}
\textit{This section specifies the networks, training objectives, and inference
procedure that implement the formulation in Section~\ref{Sec4}.} FlowLoc has three training stages: encoder pretraining
with temporary grid supervision, continuous-anchor estimation with fixed
encoders, and conditional residual-field training with fixed encoders and
anchor. After all networks are fixed, inference evaluates the encoders and
anchor and applies a flow correction. Fig.~\ref{fig:flowloc_train_infer}
shows this training order and the inference procedure. The final subsection
specifies the deployed network dimensions and parameter scaling.
}

\begin{figure*}[t]
\centering
\subfloat[Grid-supervised anchor and fixed-size deployment.]{\includegraphics[width=0.49\textwidth]{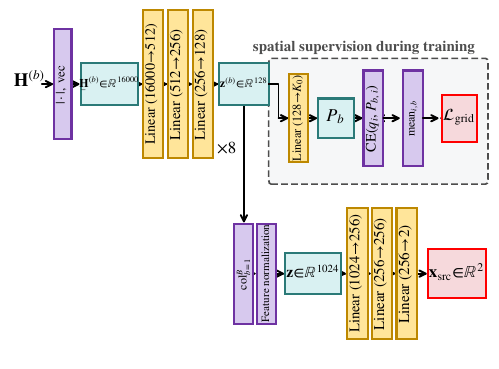}}\hfill
\subfloat[Conditional residual transport and bounded local refinement.]{\includegraphics[width=0.49\textwidth]{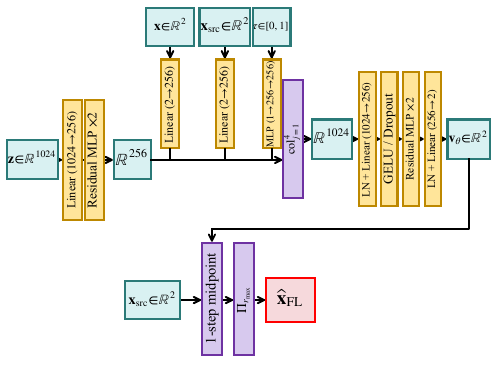}}
\caption{\textcolor{black}{FlowLoc architecture illustrating the design principles in Section~\ref{Sec4}.
In (a), encoder training averages the per-BS grid losses in~\eqref{eq:scale3_grid_loss}. Once the map-sized
heads are removed, the anchor produces a continuous coordinate. In (b), a
conditional residual field is trained and its update is bounded at inference.}}
\label{fig:scale3_flowloc_architecture}
\end{figure*}

\subsection{\textcolor{black}{Step 1: Encoder Pretraining with Grid Supervision}}
\label{sec:score_based_fusion}

\textit{Each BS encoder learns spatially informative channel features by
predicting a soft target over the position grid.} For BS \(b\), let
\(\underline{\mathbf H}^{(b)}\in\mathbb R^{d}\) denote the standardized
magnitude-CIR
fingerprint and let
\[
\mathbf z^{(b)}
=E_{\phi_b}\!\left(\underline{\mathbf H}^{(b)}\right)\in\mathbb R^{D}.
\]
The encoder is followed during pretraining by a linear classification layer
\(C_b:\mathbb R^D\rightarrow\mathbb R^{K_0}\), where \(K_0\) is the number
of points in a coarse training grid
\(\mathcal G_0=\{\mathbf g_k\}_{k=1}^{K_0}\).
Let \(\ell_b(k)=[C_b(\mathbf z^{(b)})]_k\). The resulting BS posterior is
\begin{equation}
P_b(k\mid\underline{\mathbf H}^{(b)})
 =\frac{\exp(\ell_b(k))}
{\sum_{j=1}^{K_0}\exp(\ell_b(j))}.
\label{eq:scale3_bs_posterior}
\end{equation}
For training sample $i$, let $\mathbf x_{i,\mathrm{gt}}$ be its position and
let $\sigma_g>0$ control the spatial width of the soft grid target. Its
probability assigned to grid point $k\in\{1,\ldots,K_0\}$ is
\begin{equation}
q_i(k)=
\frac{\exp\!\left(
-\|\mathbf x_{i,\mathrm{gt}}-\mathbf g_k\|_2^2/(2\sigma_g^2)
\right)}
{\sum_{j=1}^{K_0}\exp\!\left(
-\|\mathbf x_{i,\mathrm{gt}}-\mathbf g_j\|_2^2/(2\sigma_g^2)
\right)}.
\label{eq:scale3_grid_target}
\end{equation}
For compactness, write
$P_{b,i}(k)=P_b(k\mid\underline{\mathbf H}_i^{(b)})$.
For multi-BS grid decoding, we sum the per-BS log probabilities to form
$S_i(k)=\sum_{b=1}^{B}\log P_{b,i}(k)$. Its normalized posterior is
\begin{equation}
P_{\mathrm{grid},i}(k)
=\frac{\exp(S_i(k))}{\sum_{j=1}^{K_0}\exp(S_i(j))}.
\label{eq:scale3_grid_fusion}
\end{equation}
\textcolor{black}{Let $\operatorname{CE}(\cdot,\cdot)$ denote cross-entropy. During
encoder pretraining, every BS classifier is supervised by the same spatial
target. The encoders and classification layers are optimized jointly with}
{\color{black}
\begin{equation}
\mathcal L_{\mathrm{grid}}
=
\mathbb E_i\!\left[
\frac{1}{B}\sum_{b=1}^{B}
\operatorname{CE}\!\left(q_i,P_{b,i}\right)
\right].
\label{eq:scale3_grid_loss}
\end{equation}
}
\textcolor{black}{This objective makes each BS latent spatially informative. For grid
decoding, the learned per-BS log posteriors are subsequently combined using
\eqref{eq:scale3_grid_fusion}.}

\textcolor{black}{The classification matrices are required only for encoder pretraining.
FlowLoc then discards all \(B\) matrices and their biases and forms the
standardized joint latent \(\mathbf z\in\mathbb R^{BD}\):}
\begin{equation}
\mathbf z=\operatorname{Std}_{\mathrm{train}}\!\left[
\col\!\left(\mathbf z^{(1)},\ldots,\mathbf z^{(B)}\right)\right].
\label{eq:scale3_shared_state}
\end{equation}
\textcolor{black}{Here, $\operatorname{Std}_{\mathrm{train}}$ standardizes each component
using its training-set mean and standard deviation, held fixed at inference.
Both coordinate networks receive this same standardized latent. Fixed affine
coordinate normalizations are likewise absorbed into the definitions of the
networks and losses below.}
\subsection{\textcolor{black}{Step 2: Continuous Anchor Network}}
\label{sec:continuous_anchor}

\textcolor{black}{\textit{The anchor network combines the BS latents into an initial
continuous position estimate.} Let
\(\mathbf a_\eta:\mathbb R^{BD}\rightarrow\mathbb R^2\) denote the complete
coordinate mapping, including the fixed inverse coordinate normalization. The
anchor is}
\begin{equation}
\mathbf x_{\mathrm{src}}
=\mathbf a_\eta(\mathbf z).
\label{eq:scale3_anchor}
\end{equation}
\textcolor{black}{Let $\ell_{\mathrm H}$ denote the component-averaged Huber loss under
the fixed coordinate normalization used for optimization.}
The anchor loss is
\begin{equation}
\mathcal L_{\mathrm{anchor}}
=\mathbb E_i\!\left[
\ell_{\mathrm H}\!\left(
\mathbf x_{i,\mathrm{src}},
\mathbf x_{i,\mathrm{gt}}
\right)
\right],
\label{eq:scale3_anchor_loss}
\end{equation}
which trains the fixed-size coordinate head and defines the data-dependent
source for the residual field.

\subsection{\textcolor{black}{Step 3: Conditional Residual-Field Training}}
\label{sec:flowloc_refiner}

\textit{The residual field learns to correct position errors around
the anchor using the joint channel latent.} We form training sources by
perturbing the anchor with local Gaussian noise.
Perturbing the anchor exposes the field to nearby starting positions
during training. Let
$\sigma_0>0$ denote the fixed perturbation scale. For each training example,
we form
\begin{equation}
\mathbf x_{\mathrm{pert}}
=
\mathbf x_{\mathrm{src}}+\sigma_0\boldsymbol\epsilon,
\qquad
\boldsymbol\epsilon\sim\mathcal N(\mathbf0,\mathbf I_2),
\label{eq:scale3_source_perturbation}
\end{equation}
and construct the rectified path
\begin{equation}
\mathbf x_{\mathrm{tr}}(\tau)
=(1-\tau)\mathbf x_{\mathrm{pert}}+\tau\mathbf x_{\mathrm{gt}},
\quad
\mathbf v^\star
=\mathbf x_{\mathrm{gt}}-\mathbf x_{\mathrm{pert}}.
\label{eq:scale3_rectified_path}
\end{equation}
For each training example, \(\tau=0\) is selected independently with probability
\(0.5\), emphasizing predictions at the start of the training path. Otherwise,
\(\tau\) is drawn uniformly from the open unit interval. At inference,
the path starts from the unperturbed anchor \(\mathbf x_{\mathrm{src}}\).

\textcolor{black}{The vector field is conditioned on the joint BS latent, the current
position $\mathbf x$, the anchor, and flow time.} Fixed coordinate
normalization is included in the network mapping. Its two-dimensional output is
\begin{equation}
\mathbf v_\theta\!\left(
\mathbf z,
\mathbf x,
\mathbf x_{\mathrm{src}},
\tau
\right)\in\mathbb R^2.
\label{eq:scale3_vector_field}
\end{equation}
Let $\lambda_s\geq0$ denote the source-error weighting strength and let $c_s>0$
denote its distance cap. For sample $i$, define
\[
w_i
=1+\lambda_s
\min\!\left(
\frac{\|\mathbf x_{i,\mathrm{src}}-\mathbf x_{i,\mathrm{gt}}\|_2}{c_s},
1
\right).
\]
For the indexed quantities in the loss, $\mathbf v_{\theta,i}$ denotes the
field output in \eqref{eq:scale3_vector_field} evaluated at
$(\mathbf z_i,\mathbf x_{i,\mathrm{tr}}(\tau_i),
\mathbf x_{i,\mathrm{src}},\tau_i)$, and $\mathbf v_i^\star$ is the target in
\eqref{eq:scale3_rectified_path}. The training objective is
\begin{align}
\mathcal L_{\mathrm{flow}}
=\mathop{\mathbb E}\limits_{i,\boldsymbol\epsilon,\tau}\Big[
&w_i\,\ell_{\mathrm H}
\left(\mathbf v_{\theta,i},\mathbf v_i^\star\right)
\nonumber\\
&+
w_i\,\ell_{\mathrm H}
\left(
\mathbf x_{i,\mathrm{tr}}(\tau_i)
+(1-\tau_i)\mathbf v_{\theta,i},
\mathbf x_{i,\mathrm{gt}}
\right)
\Big].
\label{eq:scale3_flow_loss}
\end{align}
\textcolor{black}{The endpoint term penalizes position error, while the capped weight
gives greater emphasis to samples with larger anchor errors.}

\subsection{\textcolor{black}{Step 4: Bounded Flow Inference}}
\label{sec:bounded_flow_inference}

\textit{At inference, FlowLoc refines the anchor with a single midpoint
step and bounds the resulting correction.} The number of steps is selected
on the validation set. One explicit midpoint step
comprises two evaluations of the vector-field network. Starting from the
anchor, FlowLoc evaluates
\begin{align}
\mathbf v_0
&=\mathbf v_\theta(\mathbf z,\mathbf x_{\mathrm{src}},
\mathbf x_{\mathrm{src}},0),\\
\mathbf x_{\mathrm{mid}}
&=\mathbf x_{\mathrm{src}}+\tfrac12\mathbf v_0,\\
\boldsymbol\delta
&=\mathbf v_\theta(\mathbf z,\mathbf x_{\mathrm{mid}},
\mathbf x_{\mathrm{src}},\tfrac12).
\end{align}
Let $\Pi_r$ denote Euclidean projection onto the closed ball of radius $r$
centered at the origin. The final estimate is
\begin{equation}
\widehat{\mathbf x}_{\mathrm{FL}}
=
\mathbf x_{\mathrm{src}}
+\Pi_{r_{\max}}(\boldsymbol\delta).
\label{eq:scale3_bounded_inference}
\end{equation}
This projection enforces the movement constraint in
\eqref{eq:theory_bounded_constraint}.

\textcolor{black}{Algorithms~\ref{alg:flowloc_encoder_anchor},
\ref{alg:flowloc_residual_training}, and~\ref{alg:flowloc_inference} present
encoder and anchor training, residual-field training, and inference as three
separate procedures.}

\begingroup
\color{black}
\begin{algorithm}[t]
\color{black}
\caption{Grid-Supervised Encoder and Anchor Training}
\label{alg:flowloc_encoder_anchor}
\begin{algorithmic}[1]
\Require \(\mathcal D=\{(\mathcal H_i,\mathbf x_{i,\mathrm{gt}})\}_{i=1}^{N}\),
training grid \(\mathcal G_0\)
\State Initialize \(\{E_{\phi_b},C_b\}_{b=1}^{B}\)
\Repeat
    \State Sample a mini-batch from \(\mathcal D\)
    \State Evaluate \(q_i\) using \eqref{eq:scale3_grid_target}
    \For{\(b=1,\ldots,B\)}
        \State \(\mathbf z_i^{(b)}\gets
        E_{\phi_b}(\underline{\mathbf H}_i^{(b)})\)
        \State Evaluate \(P_{b,i}\) using \eqref{eq:scale3_bs_posterior}
    \EndFor
    \State Update \(\{\phi_b,C_b\}_{b=1}^{B}\) using
    \(\nabla\mathcal L_{\mathrm{grid}}\) in~\eqref{eq:scale3_grid_loss}
\Until{converged}
\State Remove \(\{C_b\}_{b=1}^{B}\) and fix \(\{\phi_b\}_{b=1}^{B}\)
\State Form the standardized joint latents \(\{\mathbf z_i\}_{i=1}^{N}\)
using \eqref{eq:scale3_shared_state}
\Repeat
    \State Sample a mini-batch from \(\mathcal D\)
    \State \(\mathbf x_{i,\mathrm{src}}\gets
    \mathbf a_\eta(\mathbf z_i)\)
    \State Update \(\eta\) using
    \(\nabla_\eta\mathcal L_{\mathrm{anchor}}\) in~\eqref{eq:scale3_anchor_loss}
\Until{converged}
\State \Return \(\{E_{\phi_b}\}_{b=1}^{B}\) and \(\mathbf a_\eta\)
\end{algorithmic}
\end{algorithm}
\endgroup

\begingroup
\color{black}
\begin{algorithm}[t]
\color{black}
\caption{Conditional Residual-Field Training}
\label{alg:flowloc_residual_training}
\begin{algorithmic}[1]
\Require \(\mathcal D\), \(\{E_{\phi_b}\}_{b=1}^{B}\), \(\mathbf a_\eta\),
and \(\sigma_0\)
\State Initialize \(\theta\)
\Repeat
    \State Sample a mini-batch from \(\mathcal D\)
    \State Form \(\mathbf z_i\) using \eqref{eq:scale3_shared_state}
    \State \(\mathbf x_{i,\mathrm{src}}\gets
    \mathbf a_\eta(\mathbf z_i)\)
    \State \(\boldsymbol\epsilon_i\sim\mathcal N(\mathbf0,\mathbf I_2)\),
    \quad \(\tau_i\sim\tfrac12\delta_0+\tfrac12\mathcal U(0,1)\)
    \State \(\mathbf x_{i,\mathrm{pert}}\gets
    \mathbf x_{i,\mathrm{src}}+\sigma_0\boldsymbol\epsilon_i\)
    \State \(\mathbf x_{i,\mathrm{tr}}(\tau_i)\gets
    (1-\tau_i)\mathbf x_{i,\mathrm{pert}}+
    \tau_i\mathbf x_{i,\mathrm{gt}}\)
    \State \(\mathbf v_i^\star\gets
    \mathbf x_{i,\mathrm{gt}}-\mathbf x_{i,\mathrm{pert}}\)
    \State Update \(\theta\) using
    \(\nabla_\theta\mathcal L_{\mathrm{flow}}\) in~\eqref{eq:scale3_flow_loss}
\Until{converged}
\State \Return \(\mathbf v_\theta\)
\end{algorithmic}
\end{algorithm}
\endgroup

\begingroup
\color{black}
\begin{algorithm}[t]
\color{black}
\caption{FlowLoc Inference}
\label{alg:flowloc_inference}
\begin{algorithmic}[1]
\Require \(\mathcal H\), \(\{E_{\phi_b}\}_{b=1}^{B}\), \(\mathbf a_\eta\),
\(\mathbf v_\theta\), and \(r_{\max}\)
\For{\(b=1,\ldots,B\)}
    \State \(\mathbf z^{(b)}\gets
    E_{\phi_b}(\underline{\mathbf H}^{(b)})\)
\EndFor
\State Form \(\mathbf z\) using \eqref{eq:scale3_shared_state}
\State \(\mathbf x_{\mathrm{src}}\gets
\mathbf a_\eta(\mathbf z)\)
\State \(\mathbf v_0\gets
\mathbf v_\theta(\mathbf z,
\mathbf x_{\mathrm{src}},\mathbf x_{\mathrm{src}},0)\)
\State \(\mathbf x_{\mathrm{mid}}\gets
\mathbf x_{\mathrm{src}}+\tfrac12\mathbf v_0\)
\State \(\boldsymbol\delta\gets
\mathbf v_\theta(\mathbf z,
\mathbf x_{\mathrm{mid}},\mathbf x_{\mathrm{src}},\tfrac12)\)
\State \Return \(\widehat{\mathbf x}_{\mathrm{FL}}=
\mathbf x_{\mathrm{src}}+\Pi_{r_{\max}}(\boldsymbol\delta)\)
\end{algorithmic}
\end{algorithm}
\endgroup

\subsection{\textcolor{black}{Deployed Network Architecture and Scaling}}
\label{sec:architecture_deployment_scaling}

\textcolor{black}{\textit{FlowLoc estimates continuous UE coordinates from compact BS
features, so deployment does not require a separate output for every grid
point.}}

\textcolor{black}{In the evaluated configuration, each BS encoder maps its CIR
fingerprint to a 128-dimensional latent, reducing the number of scalar
values by a factor of 125. The BS latents are concatenated into the joint
representation shared by the anchor and flow networks.}

\textcolor{black}{The anchor uses a fully connected network to predict the continuous
position. The flow network uses a latent projection, residual MLP blocks,
and coordinate and time embeddings to predict the correction field.
Fig.~\ref{fig:scale3_flowloc_architecture} gives the layer dimensions.}

\label{sec:discrete_to_continuous}
Let $N_{\mathrm{enc}}$, $N_{\mathrm{anchor}}$, and $N_{\mathrm{flow}}$
denote the respective parameter counts. With $B$ encoders of output
dimension $D$, retaining affine classifiers for $K$ grid locations gives
\begin{align}
N_{\mathrm{grid}}&=N_{\mathrm{enc}}+B(D+1)K,\nonumber\\
N_{\mathrm{FlowLoc}}&=N_{\mathrm{enc}}+N_{\mathrm{anchor}}+N_{\mathrm{flow}}.
\label{eq:deployment_parameter_counts}
\end{align}
A grid with spacing $h$ over area $A$ has approximately $A/h^2$ locations.
For $s$ bytes per parameter, the grid-head storage is therefore
\begin{equation}
M_{\mathrm{head}}\approx sB(D+1)\frac{A}{h^2}.
\label{eq:grid_area_memory_scaling}
\end{equation}
FlowLoc removes these heads after pretraining. For a fixed BS count and
network architecture, its deployed parameter count is independent of the
training-grid size.

\endgroup
\begingroup
\color{black}

\section{Experimental Setup}
\label{Sec6}

Section~\ref{sec:experimental_environment} details the service region
and data split.  \textcolor{black}{Section~\ref{sec:training_model_selection} specifies
the training and model-selection procedure, and
Section~\ref{sec:baselines} introduces the baseline methods used for
comparison.}

\subsection{Service Region and Data Split}
\label{sec:experimental_environment}

We use channel data generated by the NVIDIA Aerial Omniverse Digital
Twin (AODT) site-specific ray-tracing engine in the Tokyo scenario at 3.6GHz.
UE trajectories traverse detailed 3D city geometry, producing channel
fingerprints with rich multipath propagation.
\textcolor{black}{The primary experiments use CIRs constructed directly from ray-traced
propagation paths, without receiver noise or channel-estimation error.}

\textcolor{black}{For each path $p$, the simulator provides a delay $\tau_p$ and complex
coefficients $\mathbf A_p\in\mathbb C^{N_r\times N_t}$ for the antenna-port
pairs. We assign each path to the nearest delay-grid location
$\ell/(N_d\Delta f)$, $\ell=0,\ldots,N_d-1$, and sum the complex
coefficients assigned to each bin. The grid uses $N_d=4{,}096$ and
$\Delta f=30$kHz. Retaining the first $L=500$ bins gives the CIR tensor
$\mathbf H$ used in~\eqref{eq:fingerprint_def}.}

\textcolor{black}{The service region spans approximately 250m by 300m and contains
eight BSs.} Each BS--UE link has 32 BS antenna ports and four UE antenna
ports. A fixed selection retains 16 BS-side and two UE-side ports for every
sample and method. %
{{\color{black}Each BS contributes a $16\times2\times500$ tensor of CIR
magnitudes.}}

\textcolor{black}{The training, validation, and test sets contain 108,000, 2,000, and
2,000 snapshots, respectively. Each UE trajectory is assigned to one set.}
Hyperparameters and model weights are selected using only the
validation set.

For each snapshot, localization error is the Euclidean distance between
the estimated and true UE positions. Across snapshots, we report the mean
and the empirical 95th percentile (p95). The p95 is the error threshold that
at least 95\% of snapshots do not exceed.

\subsection{Training and Model Selection}
\label{sec:training_model_selection}

{{\color{black}Encoder pretraining uses a grid with 12m spacing, restricted
to the BS convex hull and augmented by the BS coordinates.}} Each soft
target uses $\sigma_g=4\mathrm{m}$ in \eqref{eq:scale3_grid_target}. Each
BS encoder has hidden widths 512, 256, and 128 with ReLU activations,
LayerNorm, and dropout 0.1.
{{\color{black}The spatial classification heads are removed after encoder
training.}}

{{\color{black}The continuous anchor is trained from the fixed encoder
latents, followed by the residual field with both the encoders and anchor
fixed.}} Each stage selects the parameter set minimizing the validation
objective below, where $e_{\mathrm{mean}}$ and $e_{\mathrm{p95}}$ are the
validation-set mean and 95th-percentile localization errors:
\[
\mathcal J_{\mathrm{val}}
=e_{\mathrm{mean}}+0.25e_{\mathrm{p95}}.
\]
During residual-field training, the anchor is perturbed with Gaussian
noise of standard deviation 0.15m per coordinate.  The endpoint term has unit
weight, while the capped source-error term has weight two and a cap of 2m. Half
of the training examples use \(\tau=0\) to emphasize source-end
predictions. Inference starts from the unperturbed anchor. Among solver settings
within 0.5\% of the best validation objective, we select the smallest solver
depth.  This gives one midpoint step with \(r_{\max}=2.25\)m.

{{\color{black}\textbf{Implementation settings.} All three training stages
use AdamW with an initial learning rate of $5\times10^{-4}$ and cosine
decay. Encoder pretraining optimizes \eqref{eq:scale3_grid_loss} with
weight decay $10^{-4}$ and batch size 512. The anchor and residual field
are trained for 300 and 220 epochs with batch sizes 512 and 256,
respectively. Both use weight decay $10^{-5}$ and decay the learning rate
to 5\% of its initial value. We evaluate validation performance every
10 epochs during anchor training and every 20 epochs during residual-field
training.}}

Where confidence intervals are shown and per-sample predictions are available,
they use 5,000 paired bootstrap resamples of the common test snapshots.

\subsection{Baselines}
\label{sec:baselines}

\textcolor{black}{The learned and received-power baselines use the same magnitude-CIR
observations as FlowLoc. The geometric baselines use the corresponding
complex delay-domain channel in the primary comparison. In Section~\ref{sec:srs_receiver_control}, they operate on the estimated complex CFR.}
\begin{itemize}
\setlength{\itemsep}{2pt}
\setlength{\parsep}{0pt}
\setlength{\topsep}{2pt}
\item \textbf{Supervised MLP.} The vectorized eight-BS magnitude CIR passes
through hidden layers of widths 1,024, 512, and 256 with LayerNorm, SiLU, and
dropout 0.1 before the two-coordinate output.  Training uses the Huber loss
and AdamW with an initial learning rate of \(5\times10^{-4}\).  Input
standardization and output calibration use training data only.  This baseline
represents direct neural regression from CSI fingerprints.
\item \textbf{CNN.} The CIR is arranged by BS, transmit--receive port pair, and delay, then
processed by three \(3\times3\) convolutional blocks with 32, 64, and 128
channels.  Adaptive \(2\times8\) pooling feeds a dense coordinate head.  The
remaining training and calibration settings match the supervised MLP.  It
follows the use of deep convolutional feature extraction for massive-MIMO
fingerprint localization~\cite{sun2019fingerprint}.
\item \textbf{DDIM.} A fixed 1,024-dimensional channel latent conditions a
256-dimensional coordinate network that generates the complete position from
an observation-independent coordinate prior.  Training uses 1,000 epochs,
batch size 512, and AdamW with an initial learning rate of
\(5\times10^{-4}\).  Validation is checked every 50 epochs, and inference uses
200 deterministic steps~\cite{DDIM}.
\item \textbf{Consistency training.} This baseline uses the same latent,
coordinate-network family, training schedule, and coordinate prior as DDIM.
It uses four inference steps~\cite{consistency}.
\item \textbf{RSS kNN.} The fingerprint contains one received-power
feature per BS.  The neighbor count and distance weighting are selected on
the validation set~\cite{bahl:radar,Youssef2005Horus}.
\item \textbf{Grid.} The per-BS encoders retain their spatial output layers.
The per-BS posteriors are fused using~\eqref{eq:scale3_grid_fusion}, and the
position estimate is the expectation of the fused posterior. This
probability-fusion baseline follows multi-point CSI localization with BS-specific spatial
posteriors~\cite{gonultas2022csi}.
\item \textbf{AoA.} \textcolor{black}{This baseline estimates a bearing at each BS from
the spatial spectrum of the complex channel and fits the resulting bearing
lines within the service region}~\cite{Niculescu2003AoA}.
\item \textbf{TDoA.} \textcolor{black}{This baseline estimates relative arrival delays
from the channel delay profiles and fits the corresponding range
differences}~\cite{Chan1994TDoA}.
\item \textbf{Joint AoA--TDoA.} \textcolor{black}{This baseline combines the bearing and
relative-delay constraints in a single position estimate.}
\end{itemize}

\textcolor{black}{All three geometric estimators use validation-selected calibration and
are constrained to the known service region.  Their accuracy degrades in the
dense urban channel because reflected paths can dominate the angle and delay
profiles, producing inconsistent geometric constraints across BSs.}

To ensure a fair experimental comparison, all learned methods use the
same training and test partitions, and all model choices are made on the
validation set.

\section{Results and Discussion}
\label{Sec7}

\textcolor{black}{Section~\ref{sec:primary_results_group} compares FlowLoc with the
localization baselines. Section~\ref{sec:design_analysis} examines the
learned starting position, residual refinement, and inference depth.
Section~\ref{sec:robustness_deployment_controls} evaluates performance across
operating conditions and compares centralized and split inference.}

\subsection{\textcolor{black}{Primary Results}}
\label{sec:primary_results_group}

\textcolor{black}{\textit{FlowLoc achieves the lowest mean and p95 localization errors
among all evaluated methods on the common outdoor test set.}
Table~\ref{tab:scale3_primary_results} gives the numerical comparison, and
the spatial error-change map and empirical distributions show where the errors occur.}

\label{sec:primary_scale3_results}
{{\color{black}Relative to the approximately parameter-matched grid
baseline, FlowLoc reduces mean error by 41.4\% and p95 error by 14.9\%.}}

\begin{table}[!tbp]
\centering
\caption{\textcolor{black}{Primary test-set results. Learned-method errors are averaged across three runs. Errors are in meters.}}
\label{tab:scale3_primary_results}
\small
\setlength{\tabcolsep}{3.2pt}
\renewcommand{\arraystretch}{1.12}
\begin{tabular}{lrrr}
\hline
\textbf{Method} & \textbf{Parameters} & \textbf{Mean} & \textbf{p95} \\ \hline
Supervised MLP & \MLPParams & \MLPMean & \MLPP \\
CNN & \CNNParams & \CNNMean & \CNNP \\
DDIM & \DDIMParams & \DDIMMean & \DDIMP \\
Consistency training & \CTParams & \CTMean & \CTP \\
RSS kNN & -- & \RSSKNNMean & \RSSKNNP \\
\textcolor{black}{AoA} & -- & \textcolor{black}{\PrimaryAoAMean} & \textcolor{black}{\PrimaryAoAP} \\
\textcolor{black}{TDoA} & -- & \textcolor{black}{\PrimaryTDoAMean} & \textcolor{black}{\PrimaryTDoAP} \\
\textcolor{black}{Joint AoA--TDoA} & -- & \textcolor{black}{\PrimaryJointMean} & \textcolor{black}{\PrimaryJointP} \\
Grid & \GridMatchedParams & \GridMatchedMean & \GridMatchedP \\
\textbf{FlowLoc (Ours)} & \FlowParams & \textbf{\FlowMean} & \textbf{\FlowP} \\ \hline
\end{tabular}
\end{table}

{{\color{black}FlowLoc reduces mean error by more than a factor of two
relative to both direct-coordinate generative baselines and also achieves
lower p95 error. The received-power baseline discards the antenna and
delay structure retained by the CIR and has substantially larger errors.}}

\IfFileExists{pictures/scale3_grid_flow_spatial_gain.pdf}{
\begin{figure}[!tbp]
\centering\includegraphics[width=\columnwidth]{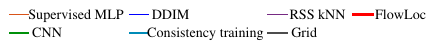}\\[-0.8ex]
\subfloat[Spatial error change.]{\includegraphics[width=0.485\columnwidth]{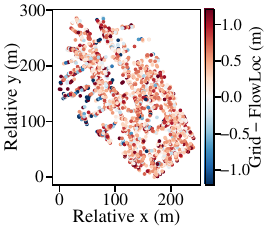}}\hfill
\subfloat[Error distributions.]{\includegraphics[width=0.485\columnwidth]{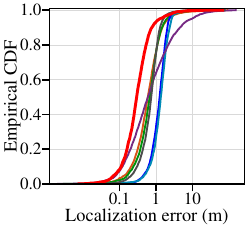}}
\caption{\textcolor{black}{Spatial and distributional effect of FlowLoc on the common
test set. FlowLoc yields lower error than the grid baseline across most of
the service region, and its error distribution is shifted toward lower
values than those of the baselines.}}
\label{fig:scale3_grid_flow_diagnostics}
\end{figure}
}{}

Fig.~\ref{fig:scale3_grid_flow_diagnostics} shows the paired grid-to-FlowLoc
spatial change and the error distributions of the learned and RSS baselines
for fixed trained models. Table~\ref{tab:scale3_primary_results} reports
three-run averages.
\par{{\color{black}FlowLoc improves on the grid baseline at 82.6\% of the
paired test positions, although the spatial gains are not uniform.}}

\subsection{\textcolor{black}{Design Analysis}}
\label{sec:design_analysis}
\textcolor{black}{\textit{Controlled ablations show that both the learned starting
position and residual correction improve localization accuracy.} We examine
each component and then the number of flow-inference steps. The experiments
in Table~\ref{tab:anchor_residual_ablation} use the same 108,000 training,
2,000 validation, and 2,000 test snapshots as the primary comparison.
We retrain a common set of BS encoders using a 12m supervision grid and
soft-target width $\sigma_g=12\mathrm{m}$, then freeze them across all
configurations. Anchor and residual-field training are repeated three
times. Model selection uses $\mathcal J_{\mathrm{val}}$, and each reported
metric is averaged over the three runs.}

\subsubsection{Residual Correction beyond the Anchor}
\label{sec:anchor_residual_ablation}

\textcolor{black}{\textit{Flow refinement reduces both mean and p95 errors compared with
using the learned anchor alone.} We compare each retrained model at its
anchor output and after flow refinement with $r_{\max}=2.25$m. The encoders and anchor are identical
within each pair. {{\color{black}Adding the residual field reduces mean
error by 33.3\% and p95 error by 38.4\%.}} The channel latents therefore support a learned
correction beyond the initial coordinate estimate.}

\begin{table}[!tbp]
\centering
\caption{\textcolor{black}{Controlled ablations with a common retrained encoder bank.
Values are three-run averages on the primary test set. Errors are in meters.
The movement bound is shared within each comparison.}}
\label{tab:anchor_residual_ablation}
\small
\setlength{\tabcolsep}{5pt}
\renewcommand{\arraystretch}{1.10}
\begin{tabular}{lrr}
\hline
\textbf{Configuration} & \textbf{Mean} & \textbf{p95} \\ \hline
\multicolumn{3}{l}{{{\color{black}\textbf{(a) Residual correction, $r_{\max}=2.25$m}}}} \\
Anchor only & 0.817 & 2.185 \\
Anchor with flow refinement & \textbf{0.545} & \textbf{1.347} \\
\hline
\multicolumn{3}{l}{{{\color{black}\textbf{(b) Starting position, $r_{\max}=200$m}}}} \\
Fixed center with flow & 0.733 & 1.292 \\
Learned anchor with flow & \textbf{0.537} & \textbf{1.237} \\
\hline
\end{tabular}
\end{table}

\subsubsection{Learned Anchor versus a Fixed Starting Position}
\label{sec:anchor_source_ablation}

\textcolor{black}{\textit{Starting the flow from a learned anchor improves accuracy over
starting from a fixed coordinate.} To isolate this effect, we replace the
learned starting position with the mean of the training coordinates and
retrain the flow. Both fields use the same channel latents,
network architecture, optimizer settings, and initialization within each
pair. Both are evaluated with one midpoint step and a common 200m movement
bound, so the fixed start is not restricted to a small neighborhood around
the region center. {{\color{black}Starting from the learned anchor reduces
mean error by approximately 27\% and p95 error by approximately 4\%.}}
Thus, even when both fields can move across the
service region, an observation-dependent starting position improves
localization. Together with the residual-correction comparison, this
supports assigning the global position to the anchor and the remaining
error to the conditional flow.}

\subsubsection{Inference Steps}
\label{sec:solver_depth}

\textcolor{black}{\textit{One midpoint step matches the accuracy of additional steps
with the lowest measured solver latency.} For a fixed trained model,
increasing the number of midpoint steps from one to eight leaves the mean and p95 errors unchanged at the
reported precision, at 0.526m and 1.287m, while GPU-resident batch-1 latency
increases from 3.21ms to 9.69ms. These results support the
validation-selected one-step solver.}

\subsection{\textcolor{black}{Performance across Operating Conditions}}
\label{sec:robustness_deployment_controls}

\textcolor{black}{\textit{FlowLoc retains its accuracy advantage under noisy channel
estimation and mobility-conditioned shifts, and can accommodate missing BS
observations through training with BS masks.} The final experiment compares
centralized and BS-side feature extraction using the same trained model.}

\IfFileExists{pictures/scale3_mobility_noise_legend.pdf}{
\begin{figure*}[!t]
\centering
\includegraphics[width=\textwidth]{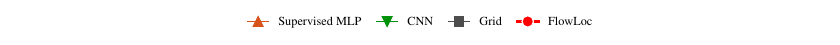}\\[-0.5ex]
\subfloat[Mobility, mean error.]{\includegraphics[width=0.46\textwidth]{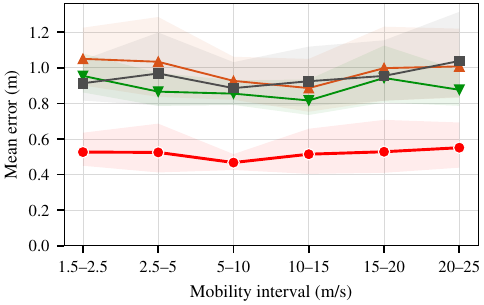}}\hfill
\subfloat[Noise, mean error.]{\includegraphics[width=0.46\textwidth]{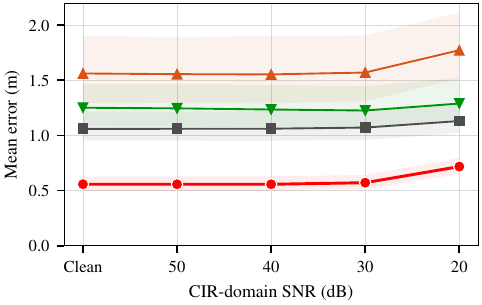}}\\[0.2ex]
\subfloat[Mobility, p95 error.]{\includegraphics[width=0.46\textwidth]{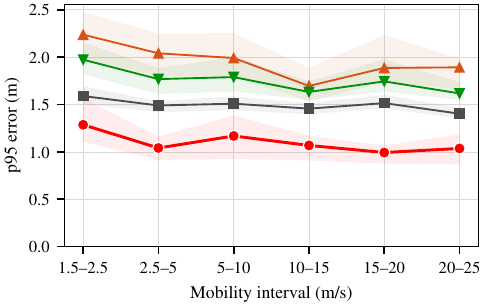}}\hfill
\subfloat[Noise, p95 error.]{\includegraphics[width=0.46\textwidth]{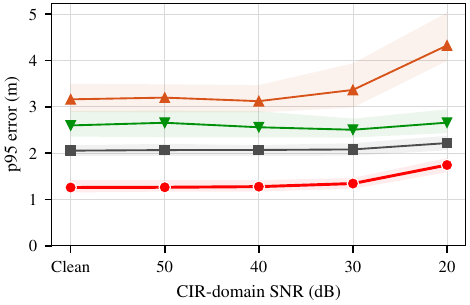}}
\caption{\textcolor{black}{Robustness to mobility-conditioned distribution shift and CIR
noise. FlowLoc retains the lowest mean and p95 errors across all evaluated
mobility intervals and noise levels. Lines show the empirical mean or p95
error, and shaded bands indicate 95\% bootstrap confidence intervals over
test snapshots.}}
\label{fig:scale3_mobility_noise}
\end{figure*}
}{}

\IfFileExists{pictures/scale3_bs_availability_mean.pdf}{
\begin{figure}[!tbp]
\centering
\subfloat[Mean localization error.]{\includegraphics[width=0.86\columnwidth]{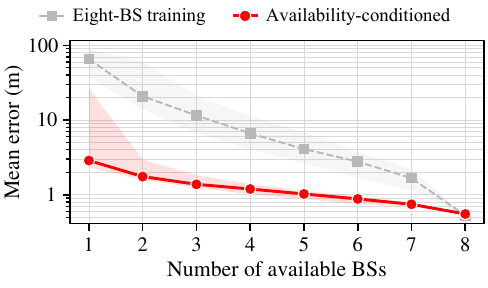}}\\[0.5ex]
\subfloat[p95 localization error.]{\includegraphics[width=0.86\columnwidth]{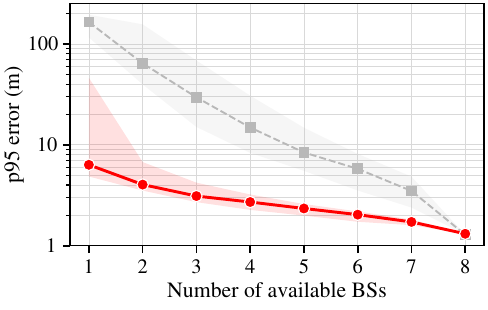}}
\caption{\textcolor{black}{{{\color{black}Robustness to missing BS observations over all
nonempty BS subsets.}} Availability-conditioned training substantially reduces error
when only part of the network is observed while preserving the full eight-BS
accuracy. Lines show median values across BS subsets with the same number
of available BSs, and shaded bands span their 10th--90th percentiles.}}
\label{fig:scale3_bs_availability}
\end{figure}
}{}

\subsubsection{\textcolor{black}{Localization with Estimated Channels}}
\label{sec:srs_receiver_control}
\label{sec:scale3_srs_bandwidth}

\textcolor{black}{\textit{FlowLoc retains the lowest mean and p95 errors among the
evaluated methods when using channel estimates obtained from noisy SRS
observations.} Equation~\eqref{eq:srs_observation} is evaluated over an
occupied bandwidth of 5MHz at a receiver SNR of 30dB. Among 166 occupied
subcarriers, pilots use every other tone and the final occupied tone,
giving 84 pilots. The final pilot supports interpolation at the band edge.
We apply~\eqref{eq:srs_ls_estimate} on the pilot tones and reconstruct the
remaining complex CFR samples by linear interpolation. The estimated CFR
is transformed to the delay domain, and its first 500 magnitude-CIR taps
form the learned-model input. The AoA, TDoA, and joint geometric baselines
use the same estimated complex CFR.}

\textcolor{black}{Each learned method in Table~\ref{tab:scale3_srs_bandwidth} is retrained with the same
108,000 training samples and evaluated on the same test set. FlowLoc's movement bound is
selected on validation data from $\{2.25,3.0,4.5,6.0\}$m, giving
$r_{\max}=6.0$m.}

\begin{table}[!tbp]
\centering
\caption{\textcolor{black}{Localization from estimated channels with an occupied SRS bandwidth of 5MHz. Errors are
in meters.}}
\label{tab:scale3_srs_bandwidth}
\small
\setlength{\tabcolsep}{4.3pt}
\renewcommand{\arraystretch}{1.10}
\begin{tabular}{lrr}
\hline
\textbf{Method} & \textbf{Mean} & \textbf{p95} \\ \hline
Supervised MLP & 2.979 & 7.596 \\
CNN & 1.968 & 4.524 \\
RSS kNN & 4.953 & 21.566 \\
\textcolor{black}{AoA} & \textcolor{black}{\SrsAoAMean} & \textcolor{black}{\SrsAoAP} \\
\textcolor{black}{TDoA} & \textcolor{black}{\SrsTDoAMean} & \textcolor{black}{\SrsTDoAP} \\
\textcolor{black}{Joint AoA--TDoA} & \textcolor{black}{\SrsJointMean} & \textcolor{black}{\SrsJointP} \\
Grid & 3.530 & 7.761 \\
\textbf{FlowLoc (Ours)} & \textbf{1.714} & \textbf{3.985} \\ \hline
\end{tabular}
\end{table}

{{\color{black}Under this receiver condition, FlowLoc approximately halves
both mean and p95 errors relative to the grid baseline, while also
outperforming the supervised, RSS, and geometry-based baselines. The
accuracy advantage therefore persists when pilot subsampling, channel
estimation, interpolation, and receiver noise are introduced together.}}

\subsubsection{\textcolor{black}{Robustness to Mobility and Channel Noise}}
\label{sec:scale3_mobility_noise}

\textcolor{black}{\textit{FlowLoc achieves the lowest mean and p95 errors across all
evaluated mobility intervals and CIR-noise levels.} The fixed primary
models are evaluated on six independently generated mobility-conditioned
channel datasets spanning 1.5--25m/s, each with 2,000
test snapshots.  These curves characterize
mobility-conditioned distribution shift because the instantaneous inputs have
no temporal or Doppler features and the paths are sampled independently.  For the channel-noise experiment, all four methods are trained using both
clean fingerprints and fingerprints formed after adding complex CIR-domain
AWGN. They are evaluated on the same 2,000 test snapshots, on clean inputs
and at SNRs down to 20dB.}

{{\color{black}As shown in Fig.~\ref{fig:scale3_mobility_noise}, FlowLoc's
mean error remains below 0.6m across the mobility-conditioned distributions,
with the lowest mean and p95 errors at every interval. Increasing CIR
noise degrades accuracy, but FlowLoc retains its advantage in both metrics
even at the lowest evaluated SNR.}}

\subsubsection{\textcolor{black}{Robustness to BS Availability}}
\label{sec:scale3_bs_availability}

\textcolor{black}{\textit{Training with randomly masked BS inputs and availability
indicators substantially reduces error when BS observations are missing.}
Let \(\mathbf m=(m_1,\ldots,m_B)\in\{0,1\}^B\) indicate which BSs are
available, with \(m_b=1\) for an available BS. We set each unavailable BS
fingerprint to zero before its encoder:
\begin{equation}
\widetilde{\mathbf H}^{(b)}=m_b\underline{\mathbf H}^{(b)},
\qquad b=1,\ldots,B.
\label{eq:bs_availability_mask}
\end{equation}
The anchor and residual-field networks also receive \(\mathbf m\), allowing
them to identify masked BSs. The pretrained BS encoders remain frozen
while the mask-conditioned anchor and residual field are trained
sequentially, with the anchor fixed during residual-field training.
Half of the training examples retain all eight
BSs. The other half use randomly sampled availability masks.
{{\color{black}At evaluation, we test all nonempty BS subsets.
Fig.~\ref{fig:scale3_bs_availability} shows the errors across subsets for
each number of available BSs. When half the BSs are available, training
with random masks reduces the median mean error across subsets by 82.0\%
relative to training only on complete observations. The median p95 error
also decreases substantially. With all BSs available, the mask-trained
model retains accuracy close to that of the fixed eight-BS reference
model.}}}

\subsubsection{\textcolor{black}{Networked Split-Inference Scenario}}
\label{sec:inference_complexity}

\textit{BS-side feature extraction reduces the uncompressed payload
sent to the localization server by a factor of 125 while preserving the
position estimate.}  \textcolor{black}{We consider a networked deployment in which eight BSs are connected to
a central localization server, and apply the primary FlowLoc model to this
setting.  In the centralized path, the BSs
forward their magnitude-CIR tensors and the complete model is executed at the
server.  In the split path, BS \(b\) executes its trained encoder and
forwards the resulting 128-dimensional latent \(\mathbf z^{(b)}\).
The same anchor and flow networks run at the server. Both
paths use the same test inputs, model parameters,
and uncompressed 32-bit floating-point representation.  A single fixed model
therefore serves both paths, which produce the same position
estimate up to numerical precision.}

\textcolor{black}{For $s_{\mathrm{fp}}$ bytes per transmitted value, the centralized CIR
and BS-side latent payloads are $Q_{\mathrm{CIR}}=s_{\mathrm{fp}}BN_rN_tL$
and $Q_{\mathrm{latent}}=s_{\mathrm{fp}}BD$, respectively.  Hence,}
\begin{equation}
\frac{Q_{\mathrm{CIR}}}{Q_{\mathrm{latent}}}
=\frac{N_rN_tL}{D}.
\label{eq:split_payload_scaling}
\end{equation}
\textcolor{black}{For the retained CIR and $D=128$, this ratio is 125, matching
Table~\ref{tab:split_inference}.}

\textcolor{black}{Measurements begin after the magnitude-CIR fingerprints are available
and use an NVIDIA L40S.  Batch sizes
of one and 256 use the same test inputs in both paths.  Each stage
is warmed up with 20 executions and then measured over 200 executions. The
table reports the 95th percentile.  The centralized path includes transfer
of the complete model input to the accelerator.  The split processing time
combines the slowest of the eight encoder measurements with feature transfer
to the central accelerator and head execution, corresponding to parallel
BS-side encoders.  The reported interval covers model-side processing from
the available magnitude-CIR fingerprints to the position estimate.}

\begin{table}[!tbp]
\centering
\caption{\textcolor{black}{Split-inference communication load and measured p95
model-side processing time at a 10-ms reporting interval.}}
\label{tab:split_inference}
\footnotesize
\setlength{\tabcolsep}{5pt}
\renewcommand{\arraystretch}{1.10}
\begin{tabular}{lrr}
\hline
\textcolor{black}{\textbf{Quantity}} &
\textcolor{black}{\shortstack{\textbf{Centralized}\\\textbf{CIR}}} &
\textcolor{black}{\shortstack{\textbf{BS-side}\\\textbf{features}}} \\ \hline
\textcolor{black}{Payload (MB/update)} & \textcolor{black}{0.512} & \textcolor{black}{0.004096} \\
\textcolor{black}{Rate (Mbit/s/UE)} & \textcolor{black}{409.6} & \textcolor{black}{3.28} \\
\textcolor{black}{p95 processing, 1 update (ms)} & \textcolor{black}{3.834} & \textcolor{black}{1.678} \\
\textcolor{black}{p95 processing, 256 updates (ms)} & \textcolor{black}{7.492} & \textcolor{black}{1.861} \\ \hline
\end{tabular}
\end{table}

\textcolor{black}{The split reduces the uncompressed central payload and corresponding
per-UE rate by a factor of 125.  It also lowers measured p95 processing time
for both a single update and 256 simultaneous updates, reducing central
traffic and model-side processing together.  These timings begin after
fingerprint formation and exclude SRS acquisition and BS-to-server transport.
The rate comparison describes communication demand, not measured end-to-end
localization delay.}

\FloatBarrier
\section{Conclusion}
\label{Sec8}

\textcolor{black}{We introduced FlowLoc, which couples grid-supervised channel feature
learning with anchor-based flow refinement for UE localization. The BS
encoders learn spatially informative representations, the anchor converts
their joint latent into an initial continuous position, and the conditional
flow corrects the remaining position error. Controlled ablations confirm
the benefits of both the learned anchor and residual correction.}

\textcolor{black}{FlowLoc lowers mean and p95 errors relative to the grid baseline on the
primary test set. BS-side feature extraction also reduces the uncompressed
central input by a factor of 125. Across independently sampled
mobility-conditioned channel distributions, the mean error remains below
0.56m, while noise-aware training preserves FlowLoc's advantage at SNRs
down to 20dB.
{{\color{black}A single availability-conditioned model further supports all
nonempty subsets of the deployed BSs.}}}

\textcolor{black}{Future work will assess FlowLoc's feasibility in real deployments
using over-the-air channel measurements. Additional directions include
using temporal channel observations for UE tracking and adapting the
learned representations to new deployment environments.}

\endgroup
\section*{Acknowledgment}
OpenAI's AI tools assisted with text revision, including grammar and clarity,
and figure preparation throughout the manuscript. The authors reviewed the
resulting content and take responsibility for the final manuscript.

\bibliographystyle{IEEEtran}
\begingroup
\let\OriginalTSPBibitem\bibitem
\renewcommand{\bibitem}[1]{\OriginalTSPBibitem{#1}\looseness=-1\relax}
\bibliography{AZREF}

\begin{thebibliography}{10}
\providecommand{\url}[1]{#1}
\csname url@samestyle\endcsname
\providecommand{\newblock}{\relax}
\providecommand{\bibinfo}[2]{#2}
\providecommand{\BIBentrySTDinterwordspacing}{\spaceskip=0pt\relax}
\providecommand{\BIBentryALTinterwordstretchfactor}{4}
\providecommand{\BIBentryALTinterwordspacing}{\spaceskip=\fontdimen2\font plus
\BIBentryALTinterwordstretchfactor\fontdimen3\font minus
  \fontdimen4\font\relax}
\providecommand{\BIBforeignlanguage}[2]{{%
\expandafter\ifx\csname l@#1\endcsname\relax
\typeout{** WARNING: IEEEtran.bst: No hyphenation pattern has been}%
\typeout{** loaded for the language `#1'. Using the pattern for}%
\typeout{** the default language instead.}%
\else
\language=\csname l@#1\endcsname
\fi
#2}}
\providecommand{\BIBdecl}{\relax}
\BIBdecl

\bibitem{Lee2025DiffLoc}
T.~Lee, T.~Balercia, H.~Kim, H.~Kim, and J.~G. Andrews, ``{DiffLoc}: Diffusion
  model-based high-precision positioning for {6G} networks,''
  arXiv:2510.14111v1, 2025.

\bibitem{trevlakis2023localization}
S.~E. Trevlakis, A.-A.~A. Boulogeorgos, D.~Pliatsios, J.~Querol, K.~Ntontin,
  P.~Sarigiannidis, S.~Chatzinotas, and M.~{Di Renzo}, ``Localization as a key
  enabler of 6{G} wireless systems: A comprehensive survey and an outlook,''
  \emph{IEEE Open J. Commun. Soc.}, vol.~4, pp. 2733--2801, Oct. 2023.

\bibitem{Liu2025DigitalTwin}
W.~Liu, Y.~Fu, Z.~Shi, and H.~Wang, ``When digital twin meets 6{G}: Concepts,
  obstacles, and research prospects,'' \emph{IEEE Commun. Mag.}, vol.~63,
  no.~3, pp. 16--22, Mar. 2025.

\bibitem{Morais2024Localization}
J.~Morais and A.~Alkhateeb, ``Localization in digital twin {MIMO} networks: A
  case for massive fingerprinting,'' in \emph{Proc.\ IEEE Int.\ Conf.\ Commun.\
  Workshops (ICC Workshops)}, Jun. 2024, pp. 276--281.

\bibitem{Mohammed2025SupportingGlobalCommunications}
S.~A. Mohammed, S.~S. Murad, H.~J. Albeyboni, M.~D. Soltani, R.~A. Ahmed,
  R.~Badeel, and P.~Chen, ``Supporting global communications of 6{G} networks
  using {AI}, digital twin, hybrid and integrated networks, and cloud:
  Features, challenges, and recommendations,'' \emph{Telecom}, vol.~6, no.~2,
  May 2025, art. no. 35.

\bibitem{3gpp_positioning}
{3GPP}, ``{NG Radio Access Network (NG-RAN): Stage 2 Functional Specification
  of User Equipment (UE) Positioning in NG-RAN},'' 3rd Generation Partnership
  Project (3GPP), Technical Specification TS 38.305, Aug. 2024, v18.2.0,
  Release 18.

\bibitem{bahl:radar}
P.~Bahl and V.~N. Padmanabhan, ``{RADAR}: An in-building {RF}-based user
  location and tracking system,'' in \emph{Proc. IEEE INFOCOM 2000},
  vol.~2.\hskip 1em plus 0.5em minus 0.4em\relax IEEE, 2000, pp. 775--784.

\bibitem{Youssef2005Horus}
M.~Youssef and A.~Agrawala, ``The {Horus} {WLAN} location determination
  system,'' in \emph{Proc.\ 3rd Int.\ Conf.\ Mobile Syst., Appl., and Serv.\
  (MobiSys '05)}, Jun. 2005, pp. 205--218.

\bibitem{Chan1994TDoA}
Y.~T. Chan and K.~C. Ho, ``A simple and efficient estimator for hyperbolic
  location,'' \emph{IEEE Trans. Signal Process.}, vol.~42, no.~8, pp.
  1905--1915, Aug. 1994.

\bibitem{3gpp_aoa}
{3GPP}, ``Study on {NR} positioning support,'' 3GPP, Tech. Rep. TR 38.855
  V16.0.0, Mar. 2019.

\bibitem{Garcia2017DirectLocalization}
N.~Garcia, H.~Wymeersch, E.~G. Larsson, A.~M. Haimovich, and M.~Coulon,
  ``Direct localization for massive {MIMO},'' \emph{IEEE Trans. Signal
  Process.}, vol.~65, no.~10, pp. 2475--2487, May 2017.

\bibitem{wang2021wifi}
X.~Wang, X.~Wang, and S.~Mao, ``Indoor fingerprinting with bimodal {CSI}
  tensors: A deep residual sharing learning approach,'' \emph{IEEE Internet
  Things J.}, vol.~8, no.~6, pp. 4498--4513, 2021.

\bibitem{tian2023deep}
G.~Tian, I.~Yaman, M.~Sandra, X.~Cai, L.~Liu, and F.~Tufvesson,
  ``Deep-learning-based high-precision localization with massive {MIMO},''
  \emph{IEEE Trans. Mach. Learn. Commun. Netw.}, vol.~2, pp. 19--33, 2024.

\bibitem{gonultas2022csi}
E.~{\GonultasSurname}, E.~Lei, J.~Langerman, H.~Huang, and C.~Studer,
  ``{CSI}-based multi-antenna and multi-point indoor positioning using
  probability fusion,'' \emph{IEEE Trans.\ Wireless Commun.}, vol.~21, no.~4,
  pp. 2162--2176, 2022.

\bibitem{zhou2017device}
R.~Zhou, X.~Lu, P.~Zhao, and J.~Chen, ``Device-free presence detection and
  localization with {SVM} and {CSI} fingerprinting,'' \emph{IEEE Sensors J.},
  vol.~17, no.~23, pp. 7990--7999, 2017.

\bibitem{Song17}
Q.~Song, S.~Guo, X.~Liu, and Y.~Yang, ``{CSI} amplitude fingerprinting-based
  {NB-IoT} indoor localization,'' \emph{IEEE Internet Things J.}, vol.~5,
  no.~3, pp. 1494--1504, Jun. 2018.

\bibitem{wang2017csi}
X.~Wang, L.~Gao, S.~Mao, and S.~Pandey, ``{CSI}-based fingerprinting for indoor
  localization: A deep learning approach,'' \emph{IEEE Trans. Veh. Technol.},
  vol.~66, no.~1, pp. 763--776, 2017.

\bibitem{Arnold2018Localization}
M.~Arnold, S.~D{\"o}rner, S.~Cammerer, and S.~{ten Brink}, ``On deep
  learning-based massive {MIMO} indoor user localization,'' in \emph{Proc. IEEE
  Int. Workshop Signal Process. Adv. Wireless Commun. (SPAWC)}, Jun. 2018, pp.
  1--5.

\bibitem{DDPM}
J.~Ho, A.~Jain, and P.~Abbeel, ``Denoising diffusion probabilistic models,'' in
  \emph{{Proc. Adv. Neural Inf. Process. Syst. (NeurIPS)}}, vol.~33, Dec. 2020,
  pp. 6840--6851.

\bibitem{Lipman2023FlowMatching}
Y.~Lipman, R.~T.~Q. Chen, H.~Ben-Hamu, M.~Nickel, and M.~Le, ``Flow matching
  for generative modeling,'' in \emph{Proc. Int. Conf. Learn. Represent.
  (ICLR)}, 2023.

\bibitem{Sengupta23}
U.~Sengupta, C.~Jao, A.~Bernacchia, S.~Vakili, and D.-s. Shiu, ``Generative
  diffusion models for radio wireless channel modelling and sampling,'' in
  \emph{Proc. IEEE Global Commun. Conf. (GLOBECOM)}, Dec. 2023, pp. 4779--4784.

\bibitem{Lee2024Generating}
T.~Lee, J.~Park, H.~Kim, and J.~G. Andrews, ``Generating high dimensional
  user-specific wireless channels using diffusion models,'' \emph{IEEE Trans.
  Wireless Commun.}, vol.~25, pp. 2907--2921, 2026.

\bibitem{Fesl2024DiffusionBased}
B.~Fesl, M.~Baur, F.~Strasser, M.~Joham, and W.~Utschick, ``Diffusion-based
  generative prior for low-complexity {MIMO} channel estimation,'' \emph{IEEE
  Wireless Commun. Lett.}, vol.~13, no.~12, pp. 3493--3497, Dec. 2024.

\bibitem{Kim2025GenerativeCompression}
H.~Kim, T.~Lee, H.~Kim, G.~De~Veciana, M.~A. Arfaoui, A.~Koc, P.~Pietraski,
  G.~Zhang, and J.~Kaewell, ``Generative diffusion model-based compression of
  {MIMO CSI},'' in \emph{Proc. IEEE Int. Conf. Commun. (ICC)}, Jun. 2025, pp.
  6323--6328.

\bibitem{Park2026ChannelGeometryFlow}
J.~Park, T.~Lee, F.~Sohrabi, and J.~G. Andrews, ``Channel geometry preserving
  generative models for {CSI} feedback in {MU-MIMO},'' arXiv preprint
  arXiv:2605.08628, 2026.

\bibitem{Guo2023CDPM}
D.~Guo, D.~Zhang, X.~Jin, and S.~Lu, ``Improve indoor localization accuracy by
  enriching {CSI} fingerprints with {CDPM},'' in \emph{Proc. Chinese Control
  Decis. Conf. (CCDC)}, May 2023, pp. 1451--1457.

\bibitem{Li2024DiffLocLidar}
W.~Li, Y.~Yang, S.~Yu, G.~Hu, C.~Wen, M.~Cheng, and C.~Wang, ``{DiffLoc}:
  Diffusion model for outdoor {LiDAR} localization,'' in \emph{Proc. IEEE/CVF
  Conf. Comput. Vis. Pattern Recognit. (CVPR)}, Jun. 2024, pp.
  15\,045--15\,054.

\bibitem{sun2019fingerprint}
X.~Sun, C.~Wu, X.~Gao, and G.~Y. Li, ``Fingerprint-based localization for
  massive {MIMO-OFDM} system with deep convolutional neural networks,''
  \emph{IEEE Trans. Veh. Technol.}, vol.~68, no.~11, pp. 10\,846--10\,857,
  2019.

\bibitem{DDIM}
J.~Song, C.~Meng, and S.~Ermon, ``Denoising diffusion implicit models,'' in
  \emph{{Proc. Int. Conf. Learn. Represent. (ICLR)}}, May 2021.

\bibitem{consistency}
Y.~Song, P.~Dhariwal, M.~Chen, and I.~Sutskever, ``Consistency models,'' in
  \emph{Proc. Int. Conf. Machine Learn. (ICML)}, ser. Proceedings of Machine
  Learning Research, vol. 202.\hskip 1em plus 0.5em minus 0.4em\relax PMLR,
  Jul. 2023, pp. 32\,211--32\,252.

\bibitem{Niculescu2003AoA}
D.~Niculescu and B.~Nath, ``Ad hoc positioning system ({APS}) using {AoA},'' in
  \emph{Proc. IEEE INFOCOM}, vol.~3, Mar. 2003, pp. 1734--1743.

\end{thebibliography}

\endgroup

\end{document}